\documentclass[11pt]{book} 

\newcommand{\ISBN}{\textrm{\scshape isbn}\xspace}

\usepackage{tocloft}
\usepackage{pageslts}

\usepackage{wrapfig} 

\usepackage{xcolor} 

\usepackage{epigraph} 

\usepackage{ifxetex}

\ifxetex
	\usepackage{mathspec}
	\usepackage[no-math]{fontspec}
	\usepackage{xunicode}
	\usepackage{xltxtra}
	\AtBeginEnvironment{tabular}{\addfontfeatures{Numbers={Monospaced}}}
(Greek,Digits,Latin){Adobe Garamond Pro}
	\usepackage{etoolbox}
	\makeatletter
	\begingroup\lccode`~=`"
	\lowercase{\endgroup
	  \everymath{\let~\eu@active@quote}
	  \everydisplay{\let~\eu@active@quote}
	}
	\makeatother
\else
	 \usepackage[utf8]{inputenc}
	 \DeclareUnicodeCharacter{2212}{-} 
\fi

\usepackage{textgreek}

\usepackage{moresize}

\usepackage{subfig}
\DeclareCaptionFont{ssmall}{\ssmall}
\DeclareCaptionFont{tiny}{\tiny}
\usepackage{floatrow}
\DeclareFloatFont{tiny}{\tiny}
\DeclareFloatFont{ssmall}{\ssmall}
\usepackage{tabularx}

\usepackage{url}

\addtocontents{toc}{\protect\setcounter{tocdepth}{1}}

\usepackage{emptypage} 

\usepackage{xspace}

\usepackage{blindtext}

\usepackage{graphicx}

\usepackage{booktabs}

\usepackage{longtable}
 
\usepackage{amsfonts,amsmath,amssymb}

\usepackage{pdfpages}

\usepackage{auxiliary_texfiles/aas_macros}

\usepackage[swedish,ngerman,british]{babel} 

\usepackage[square,numbers, comma, sort&compress]{natbib}

\usepackage{bbm}

\usepackage[paperwidth=169mm,paperheight=239mm,total={13.3cm,19.6cm}, top=1.8cm, ignorehead, centering, footskip=\footskip+4mm ]{geometry}

\graphicspath{{Figures/}}

\newcommand\sect[1]{%
  \section*{#1}%
  \addcontentsline{toc}{section}{#1}
  \markright{#1}
}

\addto\captionsenglish{ \renewcommand{\listfigurename}{List of plots}}

\usepackage{fancyhdr}
\fancyheadoffset{0cm}
\renewcommand\bibsection{\section{\bibname}\markright{\bibname}}

\usepackage{tcolorbox}
\tcbuselibrary{skins}
\newtcolorbox{thesisbox}[2][]{
oversize=-2.5cm,
  enhanced,
  attach boxed title to top left={yshift=-2ex,xshift=6ex},
  before skip=1em,
  after skip=1em,
  colframe=black,
  colback=white,
  fonttitle=\bfseries, 
  colbacktitle=white,
  coltitle=black,
  boxed title style={
    boxrule=0pt,
    colframe=white,
    },
  title=#2,
  #1}

\newcommand{\myName}{Martin Albertsson}
\newcommand{\myYear}{2021}
\newcommand{\myMainTitle}{Nuclear fission and fusion in a random-walk model}

\newcommand{\myTitle}{\myMainTitle}

\newcommand{\myFaculty}{Faculty of Engineering}
\newcommand{\myDepartment}{Department of Physics}

\newcommand{\myAddress}{Box 118\\SE--221 00 LUND\\Sweden} 

\newcommand{\myDegree}{Thesis for the degree of Doctor of Philosophy in Engineering}
\newcommand{\myAdvisors}{B.~Gillis Carlsson, Sven Åberg, Andrea Idini}
\newcommand{\myOpponent}{Christelle Schmitt}

\newcommand{\myDefenceAnnouncement}{%
	To be presented, with the permission of the
	\myFaculty\xspace of Lund University, for public criticism in the Rydberg 
	lecture hall (Rydbergsalen) at the \myDepartment\xspace
	on Thursday, the 17th of June \myYear\xspace at 09:00. }
\newcommand{\myCoverBack}{%
	{\bf Cover illustrations:} 
	Artistic rendition of the island of stability (front cover) 
	and cartoon describing how nuclear energy is extracted in nuclear reactors (back cover) made by the author.
	}
\newcommand{\myFundingInformation}{%
	{\bf Funding information:} 
	The thesis work was financially supported by Knut and Alice Wallenberg Foundation (Grant No. KAW 2015.0021).}

\newcommand{\myISBNprint}{978-91-7895-869-6}
\newcommand{\myISBNpdf}{978-91-7895-870-2}
\newcommand{\myFormSignDate}{2021-05-10}

\newcommand{\myPages}{\lastpageref{LastPages}}

\newcommand{\myFormDefenceDate}{2021-06-17}
\newcommand{\myFormKeywords}{%
	fission, Brownian shape motion, energy partition, neutron multiplicity, kinetic energy, fusion, quasifission}

\newcommand{\myAbstract}{
This dissertation deals with theoretical descriptions of nuclear fission and
synthesis of superheavy elements via fusion.
The associated shape evolutions are treated using a random-walk approach 
where both the potential energy and the nuclear level density influence the dynamics.
The dissertation consists of seven original research papers, 
and an introductory part providing background information and some additional details of the studies.

Paper I contains results for fission-fragment neutron multiplicities in $^{235}\text{U}(\text{n},\text{f})$ 
using an energy partition based on shape-dependent microscopic level densities.

Paper II gives results regarding the energy dependence of fission-fragment neutron multiplicities in $^{235}\text{U}(\text{n},\text{f})$,
using the same method as in Paper I.

Paper III presents calculations of fission-fragment mass and total-kinetic-energy distributions following fission of the fermium 
isotopes $^{256,258,260}$Fm at low excitation energies. 
A transition from asymmetric fission in $^{256}$Fm to symmetric fission in $^{258}$Fm is obtained
with a correlated large change in total kinetic energy.

Paper IV provides results of fission-fragment mass and total-kinetic-energy distributions following fission of even-even nuclei
in the region $74\leq Z\leq 126$ and $92\leq N\leq 230$.
An island of asymmetric fission is obtained in the superheavy region
where the heavy fragment is found to be close to $^{208}$Pb 
and a corresponding light fragment.

Paper V presents calculations of neutron multiplicities from fission fragments with specified mass numbers for events having
a specified total fragment kinetic energy in $^{235}\text{U}(\text{n},\text{f})$.
With increasing neutron energy a superlong fission mode is found to grow increasingly prominent.

Paper VI studies the persistence of the symmetric super-short fission mode versus both particle number
and excitation energy of even fermium isotopes $^{254-268}$Fm. 

Paper VII investigates the shape dynamics in the fusion process in production of superheavy elements
and how this competes with quasifission.

}%

\newcommand{\PaperIauthor}{\textbf{M. Albertsson}, B.G. Carlsson, T. D{\o}ssing, P. M{\"o}ller, J. Randrup, S. {\AA}berg}
\newcommand{\PaperIIauthor}{\textbf{M. Albertsson}}
\newcommand{\PaperIIIauthor}{\textbf{M. Albertsson}, B.G. Carlsson, T. D{\o}ssing, P. M{\"o}ller, J. Randrup, S. {\AA}berg}
\newcommand{\PaperIVauthor}{\textbf{M. Albertsson}, B.G. Carlsson, T. D{\o}ssing, P. M{\"o}ller, J. Randrup, S. {\AA}berg}
\newcommand{\PaperVauthor}{\textbf{M. Albertsson}, B.G. Carlsson, T. D{\o}ssing, P. M{\"o}ller, J. Randrup, S. {\AA}berg}
\newcommand{\PaperVIauthor}{\textbf{M. Albertsson}, B.G. Carlsson, T. D{\o}ssing, P. M{\"o}ller, J. Randrup, S. {\AA}berg}
\newcommand{\PaperVIIauthor}{\textbf{M. Albertsson}, B.G. Carlsson, T. D{\o}ssing, P. M{\"o}ller, J. Randrup, D. Rudolph, S. {\AA}berg}

\newcommand{\PaperIref}{Phys.\ Lett.\ B \textbf{803}, 135276 (2020)}
\newcommand{\PaperIIref}{Acta Phys.\ Pol.\ B \textbf{12}, 499 (2019)}
\newcommand{\PaperIIIref}{EPJ Web of Conferences \textbf{223}, 01002 (2019)}
\newcommand{\PaperIVref}{Eur.\ Phys.\ J.\ A \textbf{56}, 46 (2020)}
\newcommand{\PaperVref}{Phys.\ Rev.\ C \textbf{103}, 014609 (2021), Editor's suggestion}
\newcommand{\PaperVIref}{Submitted to Phys.\ Rev.\ Lett.\ (2021)}
\newcommand{\PaperVIIref}{Work in progress}

\newcommand{\PaperItitle}{Excitation energy partition in fission}
\newcommand{\PaperIItitle}{Energy dependence of fission-fragment neutron multiplicity in \textsuperscript{235}U(n,f)}
\newcommand{\PaperIIItitle}{Fission modes in fermium isotopes with Brownian shape-motion model}
\newcommand{\PaperIVtitle}{Calculated fission-fragment mass yields and average total kinetic energies of heavy and superheavy nuclei}
\newcommand{\PaperVtitle}{Correlation studies of fission fragment neutron multiplicities}
\newcommand{\PaperVItitle}{The super-short fission mode in fermium isotopes}
\newcommand{\PaperVIItitle}{Fusion-quasifission dynamics in a random-walk model}

\newcommand{\PaperNotIncIauthor}{A. S{\aa}mark-Roth, D.M. Cox, D. Rudolph, L.G. Sarmiento, B.G. Carlsson, J.L. Egido, P. Golubev, 
J. Heery, A. Yakushev, S. {\AA}berg, H.M. Albers, \textbf{M. Albertsson}, M. Block, H. Brand, T. Calverley, R. Cantemir, R.M. Clark, 
Ch.E. D{\"u}llman, J. Eberth, C. Fahlander, U. Forsberg, J.M. Gates, F. Giacoppo, M. G{\"o}tz, S. G{\"o}tz, R.-D. Herzberg, Y. Hrabar, E. J{\"a}ger, D. Judson,
J. Khuyagbaatar, B. Kindler, I. Kojouharov, J.V. Kratz, J. Krier, N. Kurz, L. Lens, J. Ljungberg, B. Lommel, J. Louko, C.-C. Meyer, A. Mistry, 
C. Mokry, P. Papadakis, E. Parr, J.L. Pore, I. Ragnarsson, J. Runke, M. Sch{\"a}del, H. Schaffner, B. Schausten, D.A. Shaughnessy,
P. Th{\"o}rle-Pospiech, N. Trautmann, J. Unusitalo}
\newcommand{\PaperNotIncIref}{Phys.\ Rev.\ Lett.\ \textbf{126}, 032503 (2021)}
\newcommand{\PaperNotIncItitle}{Spectroscopy along flerovium decay chains: Discovery of $^{280}$Ds and an excited state in $^{282}$Cn}

\newcommand{\PaperNotIncIIauthor}{A. S{\aa}mark-Roth, D.M. Cox, D. Rudolph, L.G. Sarmiento, B.G. Carlsson, J.L. Egido, P. Golubev, 
J. Heery, A. Yakushev, S. {\AA}berg, H.M. Albers, \textbf{M. Albertsson}, M. Block, H. Brand, T. Calverley, R. Cantemir, R.M. Clark, 
Ch.E. D{\"u}llman, J. Eberth, C. Fahlander, U. Forsberg, J.M. Gates, F. Giacoppo, M. G{\"o}tz, S. G{\"o}tz, R.-D. Herzberg, Y. Hrabar, E. J{\"a}ger, D. Judson,
J. Khuyagbaatar, B. Kindler, I. Kojouharov, J.V. Kratz, J. Krier, N. Kurz, L. Lens, J. Ljungberg, B. Lommel, J. Louko, C.-C. Meyer, A. Mistry, 
C. Mokry, P. Papadakis, E. Parr, J.L. Pore, I. Ragnarsson, J. Runke, M. Sch{\"a}del, H. Schaffner, B. Schausten, D.A. Shaughnessy,
P. Th{\"o}rle-Pospiech, N. Trautmann, J. Unusitalo}
\newcommand{\PaperNotIncIIref}{To be submitted to Phys.\ Rev.\ C}
\newcommand{\PaperNotIncIItitle}{Spectroscopy along flerovium decay chains}

\newcommand{\PaperNotIncIIIauthor}{D.M. Cox, A. S{\aa}mark-Roth, D. Rudolph, L.G. Sarmiento, B.G. Carlsson, J.L. Egido, P. Golubev, 
J. Heery, A. Yakushev, S. {\AA}berg, H.M. Albers, \textbf{M. Albertsson}, M. Block, H. Brand, T. Calverley, R. Cantemir, R.M. Clark, 
Ch.E. D{\"u}llman, J. Eberth, C. Fahlander, U. Forsberg, J.M. Gates, F. Giacoppo, M. G{\"o}tz, S. G{\"o}tz, R.-D. Herzberg, Y. Hrabar, E. J{\"a}ger, D. Judson,
J. Khuyagbaatar, B. Kindler, I. Kojouharov, J.V. Kratz, J. Krier, N. Kurz, L. Lens, J. Ljungberg, B. Lommel, J. Louko, C.-C. Meyer, A. Mistry, 
C. Mokry, P. Papadakis, E. Parr, J.L. Pore, I. Ragnarsson, J. Runke, M. Sch{\"a}del, H. Schaffner, B. Schausten, D.A. Shaughnessy,
P. Th{\"o}rle-Pospiech, N. Trautmann, J. Unusitalo}
\newcommand{\PaperNotIncIIIref}{To be submitted to Phys.\ Rev.\ Lett.}
\newcommand{\PaperNotIncIIItitle}{Spectroscopy along flerovium decay chains: Fine structure in odd-$A$ $^{289}$Fl and a note on $^{290}$Fl}


\begin{document}



\frontmatter 



| Organization | Document name |
|---|---|
| **LUND UNIVERSITY** | **DOCTORAL DISSERTATION** |
| Department of Physics  Box 118  SE–221 00 LUND  Sweden | Date of disputation  2021-06-17 |
| | Sponsoring organization |

| Author |
|---|
| Martin Albertsson |

| Title |
|---|
| Nuclear fission and fusion in a random-walk model |

**Abstract**

This dissertation deals with theoretical descriptions of nuclear fission and synthesis of superheavy elements via fusion. The associated shape evolutions are treated using a random-walk approach where both the potential energy and the nuclear level density influence the dynamics. The dissertation consists of seven original research papers, and an introductory part providing background information and some additional details of the studies.

Paper I contains results for fission-fragment neutron multiplicities in $^{235}$U(n, f) using an energy partition based on shape-dependent microscopic level densities.

Paper II gives results regarding the energy dependence of fission-fragment neutron multiplicities in $^{235}$U(n, f), using the same method as in Paper I.

Paper III presents calculations of fission-fragment mass and total-kinetic-energy distributions following fission of the fermium isotopes $^{256,258,260}$Fm at low excitation energies. A transition from asymmetric fission in $^{256}$Fm to symmetric fission in $^{258}$Fm is obtained with a correlated large change in total kinetic energy.

Paper IV provides results of fission-fragment mass and total-kinetic-energy distributions following fission of even-even nuclei in the region $74 \leq Z \leq 126$ and $92 \leq N \leq 230$. An island of asymmetric fission is obtained in the superheavy region where the heavy fragment is found to be close to $^{208}$Pb and a corresponding light fragment.

Paper V presents calculations of neutron multiplicities from fission fragments with specified mass numbers for events having a specified total fragment kinetic energy in $^{235}$U(n, f). With increasing neutron energy a superlong fission mode is found to grow increasingly prominent.

Paper VI studies the persistence of the symmetric super-short fission mode versus both particle number and excitation energy of even fermium isotopes $^{254-268}$Fm.

Paper VII investigates the shape dynamics in the fusion process in production of superheavy elements and how this competes with quasifission.

| Key words |
|---|
| fission, Brownian shape motion, energy partition, neutron multiplicity, kinetic energy, fusion, quasifission |

| Classification system and/or index terms (if any) |
|---|
| |

| Supplementary bibliographical information | Language |
|---|---|
| | English |

| ISSN and key title | ISBN |
|---|---|
| | 978-91-7895-869-6 (print)  978-91-7895-870-2 (pdf) |

| Recipient's notes | Number of pages | Price |
|---|---|---|
| | 130 | |
| | Security classification | |

# Nuclear fission and fusion in a random-walk model

by Martin Albertsson

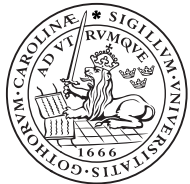


%

\thispagestyle{empty} 
\begin{center}
\vspace*{5cm}
{\Large \myMainTitle}
\end{center}



\cleardoublepage
\thispagestyle{empty} 
~
\vfill
\begin{center}
{\HUGE \myMainTitle}
\\[2mm]

\vfill
{by \myName}

\vfill
\includegraphics[width=0.25\textwidth]{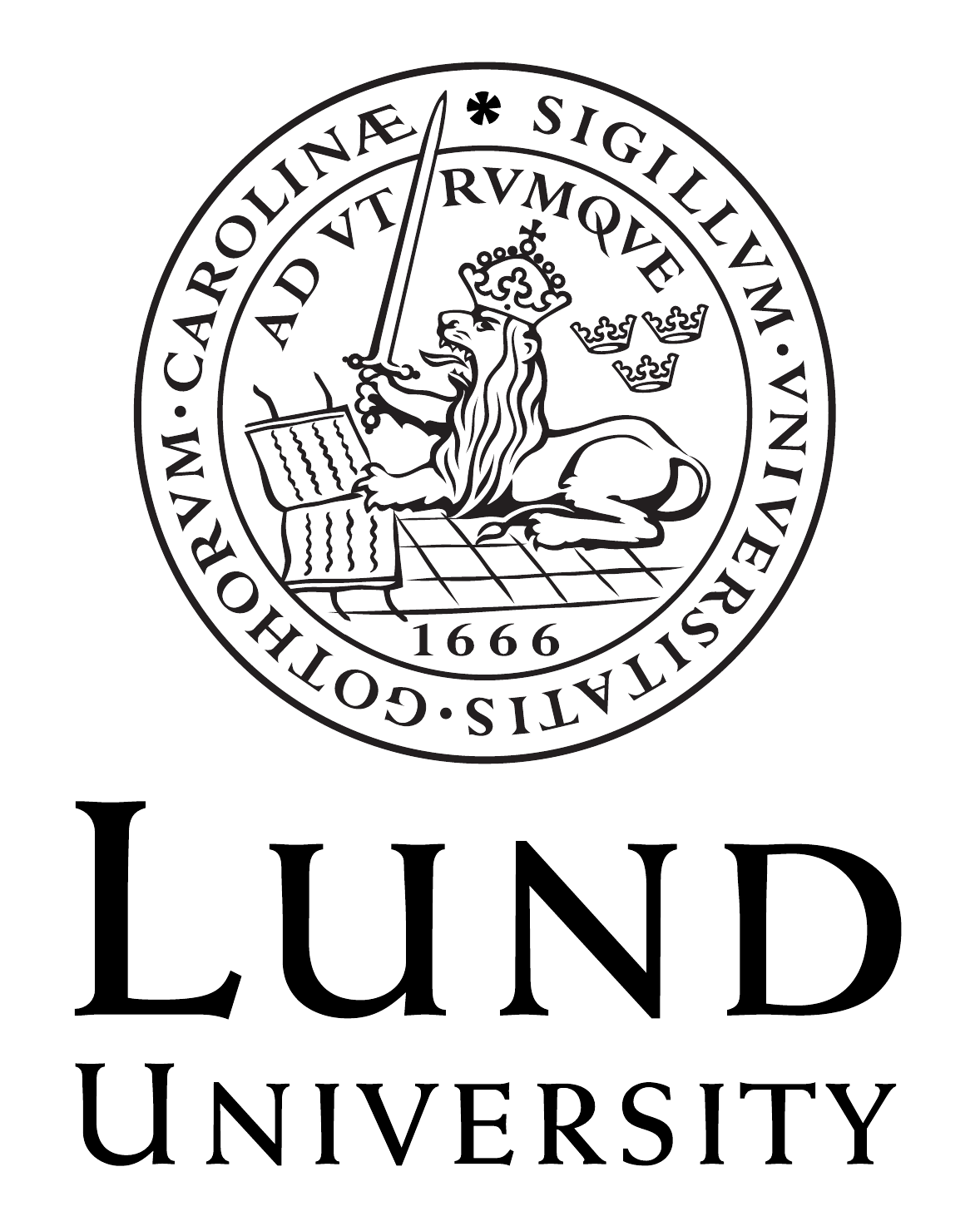}  

\vspace{10mm}
{\large \myDegree}\\
{\large Thesis advisors: \myAdvisors}\\
{\large Faculty opponent: \myOpponent}\\
\vspace{1cm}
{\footnotesize
\myDefenceAnnouncement
}
\\
\end{center}
\vfill

\newpage \thispagestyle{empty} 
\newcounter{aformdx}
\newcounter{aformdy}
\newcounter{aformx}
\newcounter{aformy}
\newcounter{aformtx}
\newcounter{aformtdx}
\newcounter{aformty}
\def\aformw{358}
\def\aformwhalf{\aformw / 2}
\def\aformwquart{\aformw / 4}
\def\aformwthreeq{\aformw / 4 * 3}
\def\aformh{517}
\def\aformb{23}
\def\aformlinew{0.4pt}
\def\aformrowh{20}
\def\aformRowh{24}
\newcommand\aformhline
{\put(\value{aformx},\value{aformy}){\line(1,0){\value{aformdx}}}}
\newcommand\aformvline
{\put(\value{aformx},\value{aformy}){\line(0,-1){\value{aformdy}}}}
\newcommand\aformvlinec
{\put(\value{aformx},\value{aformy}){\line(0,-1){16.25}}}
\newcommand\aformboxbase[5]
{
  \setcounter{aformtx}{\value{aformx}}
  \setcounter{aformtdx}{\value{aformdx}}
  \setcounter{aformty}{\value{aformy}}
  \addtocounter{aformtx}{#1}
  \addtocounter{aformtdx}{-#2}
  \addtocounter{aformty}{-#3}
  \put(\value{aformtx},\value{aformty})
  {
    \parbox[t]{\value{aformtdx}pt}
    {{\sf\tiny #4}{\scriptsize #5\par}}
  }
}
\newcommand\aformbox[2]{\aformboxbase{2}{11}{8}{#1}{#2}}
\newcommand\aformBox[2]{\aformboxbase{2}{11}{8}{#1}{#2}}

\begin{center}
\begin{picture}(\aformw,542)(0,0)
\linethickness{\aformlinew}
\put(0,\aformb){\framebox(\aformw,\aformh){}}
\put(-14,200){\rotatebox{90}{\parbox{200pt}
{\sf\tiny
DOKUMENTDATABLAD
enl SIS 61 41 21\\
}}}
\setcounter{aformx}{0}
\setcounter{aformy}{\aformb}
\addtocounter{aformy}{\aformh}
\setcounter{aformdx}{\aformwhalf}
\aformbox{Organization}
{\\
{\bf LUND UNIVERSITY}\vspace{0.1cm}
\\
{\myDepartment\\
\myAddress}
}
\setcounter{aformdy}{\aformrowh}
\addtocounter{aformdy}{\aformrowh}
\addtocounter{aformdy}{\aformrowh}
\addtocounter{aformdy}{\aformRowh}
\setcounter{aformx}{\aformwhalf}
\aformvline
\setcounter{aformx}{0}
\addtocounter{aformy}{-\aformrowh}
\addtocounter{aformy}{-\aformrowh}
\addtocounter{aformy}{-\aformRowh}
\aformbox{Author}{\\\myName}
\setcounter{aformdy}{\aformrowh}
\aformhline
\addtocounter{aformy}{\aformrowh}
\addtocounter{aformy}{\aformrowh}
\addtocounter{aformy}{\aformRowh}
\setcounter{aformx}{\aformwhalf}
\aformbox{Document name}{\\{\bf DOCTORAL DISSERTATION}}
\addtocounter{aformy}{-\aformrowh}
\aformhline
\aformbox{Date of disputation}{\\\myFormDefenceDate}
\setcounter{aformx}{\aformwhalf}
\aformhline
\addtocounter{aformy}{-\aformrowh}
\addtocounter{aformdy}{\aformRowh}
\aformbox{Sponsoring organization}{}
\aformhline
\addtocounter{aformy}{-\aformRowh}
\addtocounter{aformy}{-\aformrowh}
\setcounter{aformx}{0}
\setcounter{aformdx}{\aformw}
\aformhline
\aformbox{Title }{\\\myTitle}
\addtocounter{aformy}{-\aformRowh}
\setcounter{aformx}{0}
\aformhline
\aformBox{Abstract}
{\\
{\parskip0pt\parindent1em
\myAbstract
}
}
\setcounter{aformy}{\aformb}
\addtocounter{aformy}{\aformRowh}
\addtocounter{aformy}{\aformRowh}
\addtocounter{aformy}{\aformRowh}
\addtocounter{aformy}{\aformRowh}
\addtocounter{aformy}{\aformrowh}
\addtocounter{aformy}{\aformrowh}
\addtocounter{aformy}{10}
\addtocounter{aformy}{10}
\aformhline
\aformbox{Key words}
{
\\\myFormKeywords
}
\addtocounter{aformy}{-\aformRowh}
\addtocounter{aformy}{-10}
\aformhline
\aformbox{Classification system and/or index terms (if any)}{}
\addtocounter{aformy}{-\aformRowh}
\aformhline
\aformbox{Supplementary bibliographical information}{}
\setcounter{aformx}{\aformwthreeq}
\setcounter{aformdx}{\aformwquart}
\setcounter{aformdy}{\aformRowh}
\setcounter{aformx}{255} 
\aformvline
\aformbox{Language}{\\English}
\addtocounter{aformy}{-\aformRowh}
\setcounter{aformx}{0}
\setcounter{aformdx}{\aformw}
\aformhline
\aformbox{ISSN and key title}{}
\setcounter{aformx}{\aformwthreeq}
\setcounter{aformx}{255} 
\addtocounter{aformdy}{10}
\aformvline
\addtocounter{aformdy}{-10}
\aformbox{ISBN}{\\\myISBNprint\xspace(print)\\\myISBNpdf\xspace(pdf)}
\addtocounter{aformy}{-\aformRowh}
\addtocounter{aformy}{-10}
\setcounter{aformx}{0}
\aformhline
\aformbox{Recipient's notes}{}
\setcounter{aformx}{\aformwhalf}
\setcounter{aformdy}{\aformrowh}
\addtocounter{aformdy}{\aformrowh}
\aformvline
\aformbox{Number of pages}{\\\myPages}
\setcounter{aformx}{\aformwthreeq}
\setcounter{aformx}{255} 
\aformbox{Price}{}
\setcounter{aformdy}{\aformrowh}
\aformvline
\addtocounter{aformy}{-\aformrowh}
\setcounter{aformx}{\aformwhalf}
\setcounter{aformdx}{\aformwhalf}
\aformhline
\aformbox{Security classification}{}
\setcounter{aformx}{0}
\setcounter{aformdx}{\aformw}
\addtocounter{aformy}{-\aformrowh}
\addtocounter{aformy}{-4}
\aformbox{}
{\scriptsize I, the undersigned, being the copyright owner of the abstract of the
above-mentioned dissertation, hereby grant to all reference sources
the permission to publish and disseminate the abstract of the
above-mentioned dissertation.
}
\addtocounter{aformy}{-\aformRowh}
\addtocounter{aformy}{-20}
\aformbox{
\hspace{0pt}
}{}
\aformbox{\rm
\scriptsize
Signature \underline{\hskip 140pt}\hfill Date \underline{\hskip 80pt}
}{}
\addtocounter{aformy}{1}
\aformbox{}{\hskip 280pt \scriptsize \myFormSignDate}
\end{picture}
\end{center}



\cleardoublepage
\thispagestyle{empty} 
~
\vfill
\begin{center}
{\HUGE \myMainTitle}
\\[2mm]

\vfill
{by \myName}

\vfill
\includegraphics[width=0.25\textwidth]{LundUniversity_C2line_BLACK-eps-converted-to.pdf}  

\color{white}{
\vspace{10mm}
{\large \myDegree}\\
{\large Thesis advisors: \myAdvisors}\\
{\large Faculty opponent: \myOpponent}\\
\vspace{1cm}
{\footnotesize
\myDefenceAnnouncement
}
}
\\
\end{center}
\vfill

\newpage 
\thispagestyle{empty} 
\vspace{-15mm}


This thesis consists of two parts. An introductory text puts the research work
into context and summarizes the results of the papers. Then, the research 
publications themselves are reproduced, together with a description of my individual
contribution. The research Papers I-V are published, 
Paper VI has been submitted for publication,
and Paper VII is a manuscript in preparation.

\vfill
{\small
\myCoverBack\\
\\
\myFundingInformation

\vspace{5mm}
\copyright\, \myName~\myYear\\
\\
\myFaculty, {\myDepartment}
\\
\\
\ISBN: \myISBNprint~(print)\\ 
\ISBN: \myISBNpdf~(pdf)\\ 
\\
Printed in Sweden by Media-Tryck, Lund~University, Lund~\myYear

\includegraphics[width=0.5\textwidth]{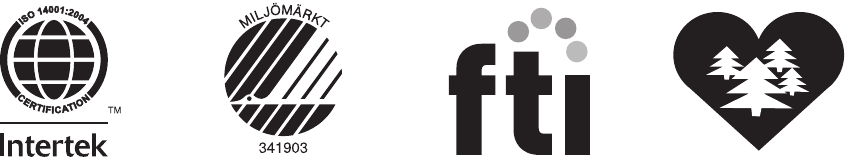}
}

\newpage
\thispagestyle{empty} 
~
\vspace{140pt}
\begin{flushright}
\textit{Harmony makes small things grow,\\lack of it makes great things decay\\ \vspace{3mm} \textnormal{Sallust}}
\end{flushright}

\cleardoublepage

\newpage
\sect{Acknowledgements}

I would like to thank my excellent thesis advisors, Gillis Carlsson, Sven Åberg and Andrea Idini for giving me the opportunity to do research 
with them in the exciting field of nuclear physics. 
I am very grateful to Thomas D{\o}ssing, J{\o}rgen Randrup and Peter Möller for introducing me to low-hanging fruits and sharing their immense knowledge.
Also, I wish to thank all the great people at mathematical physics for providing
an inspiring and uplifting atmosphere.
Lastly, I would like to thank my family for everything.


\newpage
\selectlanguage{swedish}
\sect{Populärvetenskaplig sammanfattning på svenska}

\begin{wrapfigure}{r}{0.34\textwidth}
  \vspace{-10pt}  
    \hspace{-14pt}  
    \includegraphics[width=0.89\textwidth]{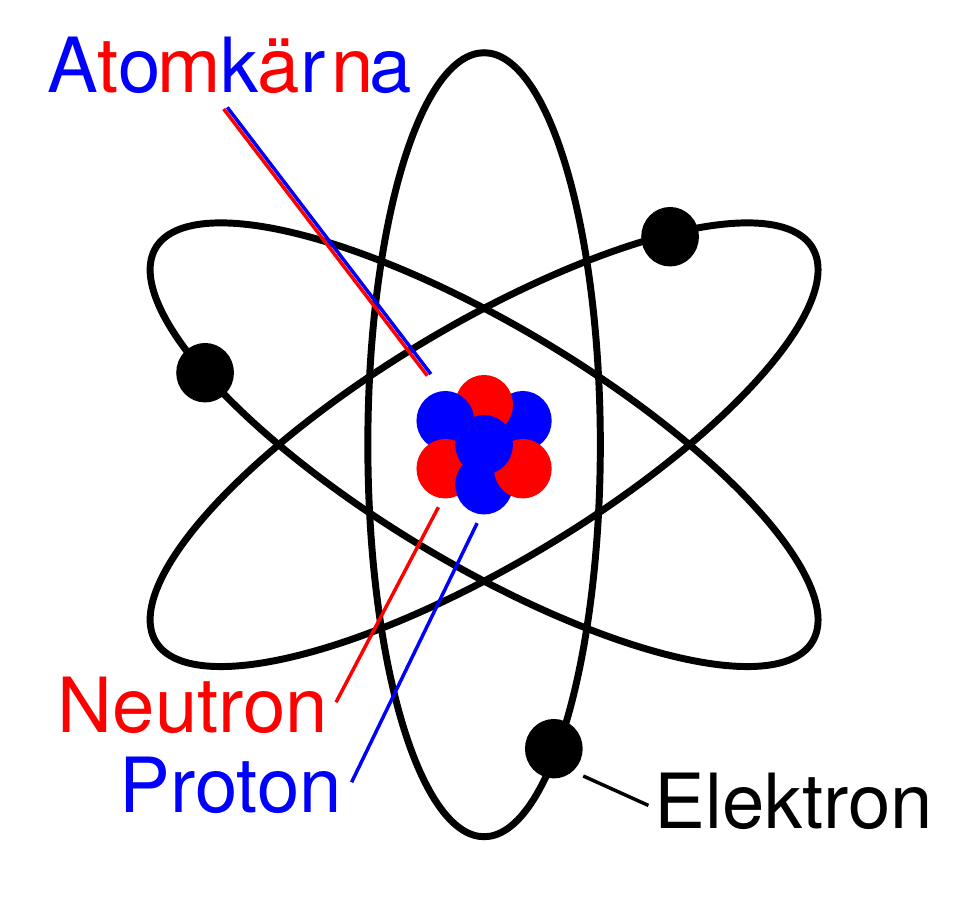}
  \vspace{-15pt}
\end{wrapfigure}

All materia som vi normalt observerar består av atomer. Väte består av väteatomer, syre består av syreatomer och så vidare.
En enskild atom utgörs av cirkulerande elektroner omkring en liten atomkärna.
Om atomen skulle ha en diameter på 50 meter skulle atomkärnan vara lika stor som en ärta.
Atomkärnan är en samling av protoner och neutroner (kärnpartiklar) som sitter ihop. 
Antalet protoner $Z$ i kärnan bestämmer vilket grundämne och antalet neutroner $N$ vilken isotop som atomen utgör.

Många olika kombinationer av antal protoner och neutroner kan sättas samman och bilda en atomkärna, men de flesta
atomkärnor är inte stabila utan sönderfaller efter en tid. Att en atomkärna sönderfaller innebär
att byggstenarna i atomkärnan spontant ändrar om sig till en energimässigt mer gynnsam konfiguration.

\vspace{5pt}
\textbf{Fission}
\vspace{-5pt}

\begin{wrapfigure}{r}{0.48\textwidth}
  \vspace{-18pt}
    \includegraphics[width=0.98\textwidth]{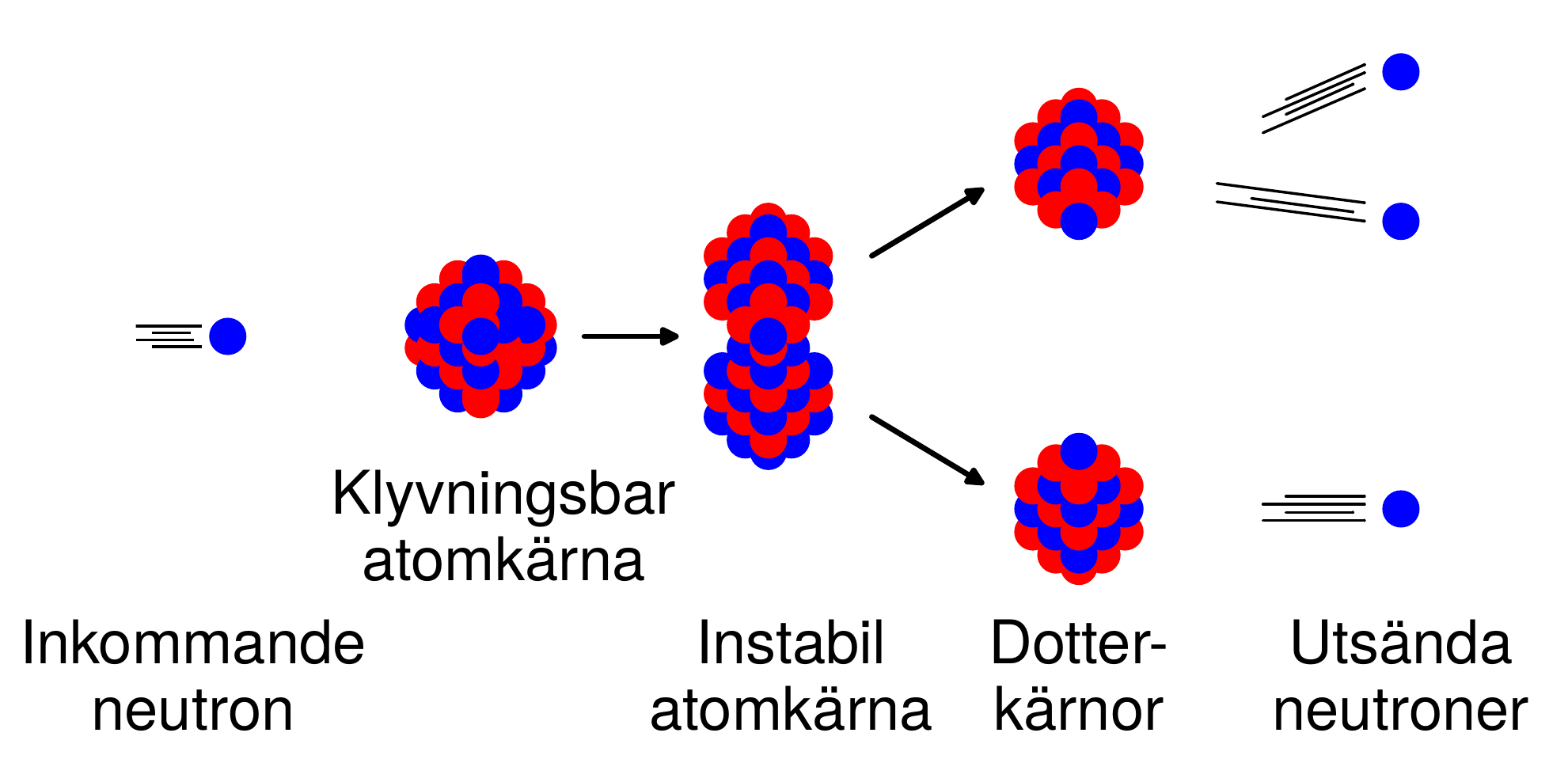}
  \vspace{-5pt}
\end{wrapfigure}

Fission av tunga element är det sönderfall som frigör mest energi
och innebär en delning i två mindre dotterkärnor
som flyger iväg med en fart på ungefär 10000 mil per sekund.
Detta kan användas till att koka vatten, producera ånga, och driva en ångturbin,
vilket i sin tur kan generera elektricitet.
Vid fission av ett gram uran frigörs lika mycket energi som när
3 ton olja bränns eller när 100 000 ton vatten faller 100 meter.
Fission av tunga kärnor kan induceras genom beskjutning med en neutron.
De resulterande dotterkärnorna sänder sedan ut neutroner som kan inducera fission av en ny atomkärna, 
och därmed resultera i en kedjereaktion.
Detta är den grundläggande processen som utnyttjas för att generera energi i dagens kärnkraftsreaktorer.

I den här avhandlingen används en teoretisk modell för beskrivningen av fission
där processen framförallt antas ske väldigt långsamt, ungefär som när en klump sirap delar sig.
Under detta antagande kan formändringen i fissionsprocessen modelleras som en slumpvandring
i ett energilandskap som beror på formen hos kärnan.
Vandringen startar i en punkt motsvarande en sfärisk form och slutar
när man kommit till en punkt som motsvarar en delning i två mindre bitar.

I artikel IV tillämpas modellen på alla atomkärnor som teoretiskt sett kan fissionera
för att beräkna sannolikheten att en kärna delar sig i specifika dotterkärnor och
hur snabbt de flyger iväg.
Det är sedan länge känt att de vanligaste fissionerande kärnorna som uran och plutonium
föredrar att dela sig i en stor och en liten dotterkärna,
där den stora dotterkärnan är ungefär tenn-132.
Denna kärnan är väldigt speciell eftersom kvantmekaniska effekter
gör att den är stabilare än förväntat,
vilket resulterar i en sfärisk form med starkt bundna partiklar.
Denna specifika typ av delning sägs motsvara en ``standard fissionsmod'' vilket
innebär en särskild form hos kärnan precis innan den delar sig och som påverkar 
farten hos dotterkärnorna och antal neutroner de utsänder.
I studien framkom ett liknande fenomen för kärnor med $Z\approx110$ där kärnan
delar sig i en väldigt stor dotterkärna och en väldigt liten dotterkärna,
som därför kallas super-asymmetrisk mod,
där den stora dotterkärnan motsvarar den stabila kärnan bly-208.
För tillfället finns det dock inte tillräckligt med experiment för
att påvisa om detta stämmer.

\begin{wrapfigure}{r}{0.39\textwidth}
  \vspace{-22pt}
    \hspace{-38pt}
    \includegraphics[width=0.98\textwidth]{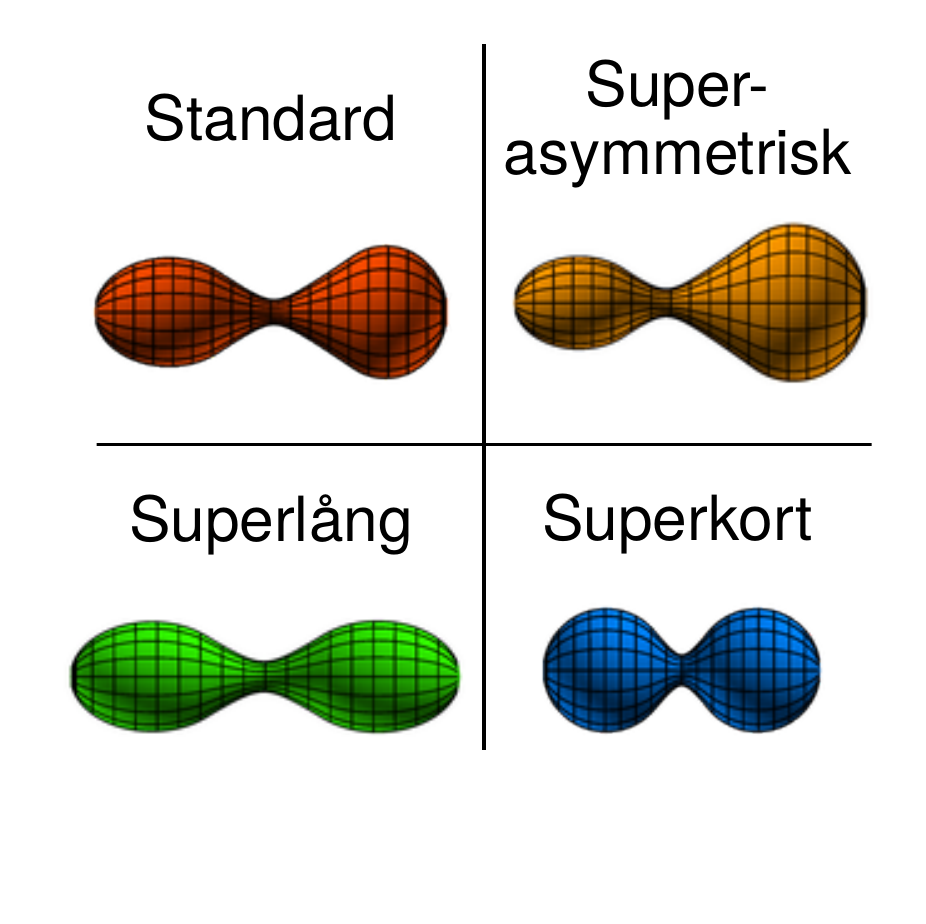}
  \vspace{-38pt}
      \hspace{-22pt}
\end{wrapfigure}

Efter delningen så är de två resulterande dotterkärnorna varma
och de avger denna värmen genom att sända ut neutroner.
Ett mysterium har varit varför den stora dotterkärnan tenn-132 sänder ut färre neutroner
än den motsvarande lilla dotterkärnan,
eftersom detta strider mot enkla statistiska förväntningar.
I artikel I antas att termisk jämvikt hinner uppnås precis innan den delar sig,
vilket innebär att den tillgängliga värmen fördelas baserat på de olika tillstånden i dotterkärnorna.
En kvantmekanisk modell används för att beräkna dessa tillstånd
där stabiliteten hos tenn-132 innebär få tillstånd och därmed mindre andel av värmen i uppdelningen. 
Detta i sin tur
resulterar i färre antal utsända neutroner från tenn-132.
När den inkommande neutronens fart ökar så försvinner effekten av stabiliteten hos tenn-132.
Den får därmed större andel av värmen och sänder därför ut fler neutroner,
vilket även har observerats i experiment.
I artikel II beräknas hur summan av antalet neutroner från de två dotterkärnorna ändras när 
farten hos den inkommande neutronen ändras
och vad sannolikheten är för att sända ut ett specifikt antal neutroner.

Fission är en kvantmekanisk process vilket innebär att identiska experiment kan ge olika utfall.
I experiment utförs därför samma försök många gånger för att på så sätt ta fram sannolikhetsfördelningar
över olika kvantiteter.
Den teoretiska modell som tillämpas i avhandlingen bygger på samma princip och kan därmed även
beskriva korrelationer mellan olika kvantiteter.
Detta studeras för fission av uran-236 i artikel V
där det bland annat beräknas hur många neutroner en specifik dotterkärna utsänder i medeltal då
den åker iväg med en specifik fart.
De beräknade resultaten stämmer väl överens med de experimentella resultaten
då den inkommande neutronen har låg fart.
I det fallet blir tenn-132 nära sfärisk motsvarande standardmoden och sänder ut få neutroner.
När farten på den inkommande neutronen ökar så framkommer istället en ``superlång mod'' kring tenn-132,
där den får en utdragen form och sänder ut många neutroner om den får liten hastighet.
Motsvarande experiment för högre fart har ännu inte utförts.

En unik observation är att vissa isotoper av fermium har möjlighet
att dela sig i två nästan sfäriska tenn-132 kärnor.
Detta innebär att fission sker i en superkort mod där
dotterkärnorna flyger iväg väldigt fort och sänder ut få neutroner.
I artikel III undersöks hur fermium-256 fissionerar i standardmoden, medan fermium-260 fissionerar i den
superkorta moden.
För fermium-258 erhålls att båda moderna kan samexistera, allt i enlighet med experiment.
I artikel VI undersöks vidare hur den superkorta moden övergår till standardmoden
då den initiala energin hos kärnan ökar.

\vspace{5pt}
\textbf{Skapande av nya grundämnen via fusion}
\vspace{-5pt}

I det periodiska systemet är alla kända grundämnen ordnade efter deras ökande antal protoner, 
och även kemiska och fysikaliska egenskaper.
Det tyngsta grundämne som finns i någorlunda kvantiteter i naturen är uran med 92 protoner.
Tyngre grundämnen
har dock skapats i laboratorier vanligtvis genom
fusion där två lättare kärnor slås ihop.
Det tyngsta grundämne som hittills har skapats har 118 protoner och namngavs 2016 till oganesson
efter den ryske kärnfysikern Yuri Oganessian.
På grund av den väldigt låga sannolikheten för att ett nytt grundämne faktiskt ska bildas vid fusion
så krävs pålitliga teoretiska modeller för olika reaktioner och energier
för planeringen av sådana experiment.

\begin{figure}[htbp!] 
\centering    
\includegraphics[width=1.0\textwidth]{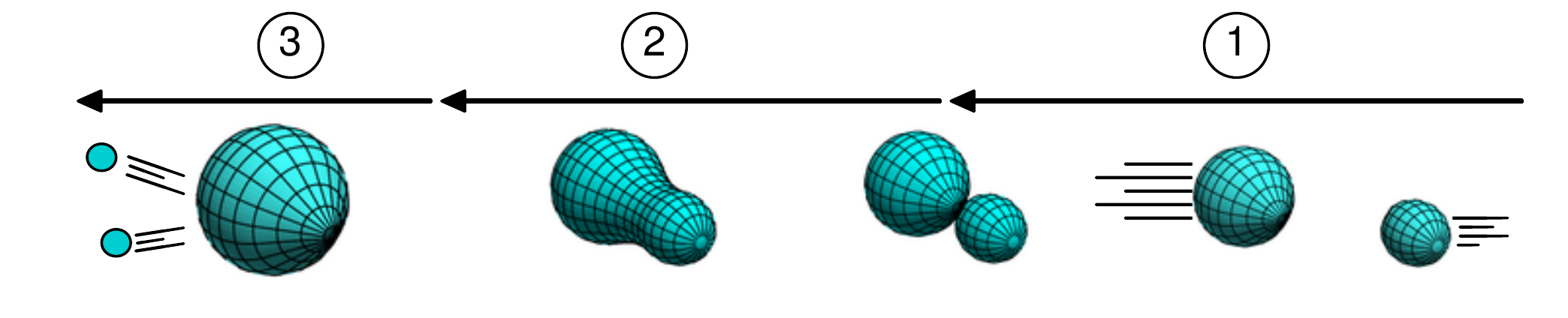}
\end{figure}

Fusionsprocessen kan ungefärligt beskrivas som den motsatta processen till fission och kan schematiskt delas upp i tre steg:
(1) de två kolliderande kärnorna kommer i kontakt med varandra,
(2) de två kärnorna bildar en ny sammansatt kärna, 
(3) den sammansatta atomkärnan kyls av genom utsändning av neutroner.
I artikel VII undersöks hur formändringen i steg (2) sker
genom att tillämpa samma modell som används för beskrivningen av fission.

\newpage
\selectlanguage{british}
\sect{Popular summary in English}

\begin{wrapfigure}{r}{0.34\textwidth}
  \vspace{-10pt}  
    \hspace{-14pt}  
    \includegraphics[width=0.89\textwidth]{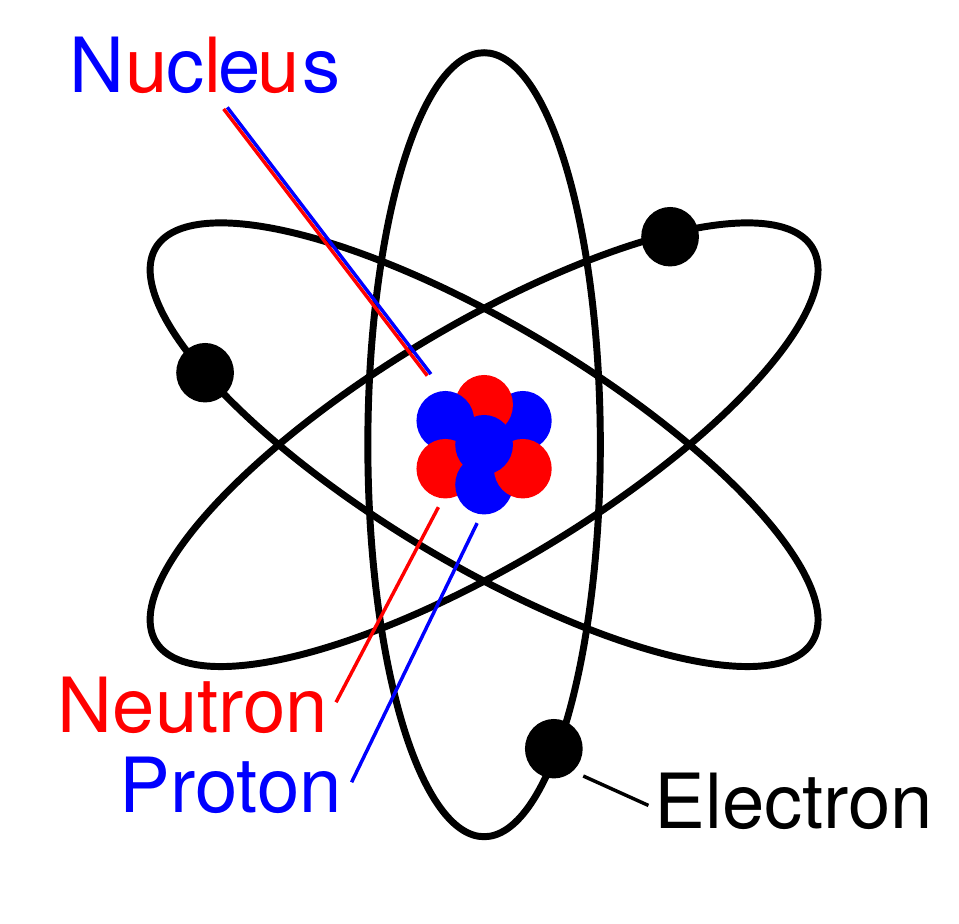}
  \vspace{-15pt}
\end{wrapfigure}

All matter that we normally observe is made of atoms.
An atom consists of orbiting electrons around a tiny nucleus.
If the atom would have a diameter of 50 meter, then the nucleus would be the size of a pea.
The nucleus is a collection of protons and neutrons (nucleons) glued together.
The number of protons $Z$ in the nucleus determines the element and the number of neutrons $N$ determines the isotope.

Many different combinations of the number of protons and neutrons can be combined into an atomic nucleus,
but most atomic nuclei are not stable but decay after some time. 
That a nucleus decays means that the constituents in the nucleus spontaneously rearrange to a more energetically favourable configuration.

\vspace{5pt}
\textbf{Fission}
\vspace{-5pt}

\begin{wrapfigure}{r}{0.48\textwidth}
  \vspace{-18pt}
    \includegraphics[width=0.98\textwidth]{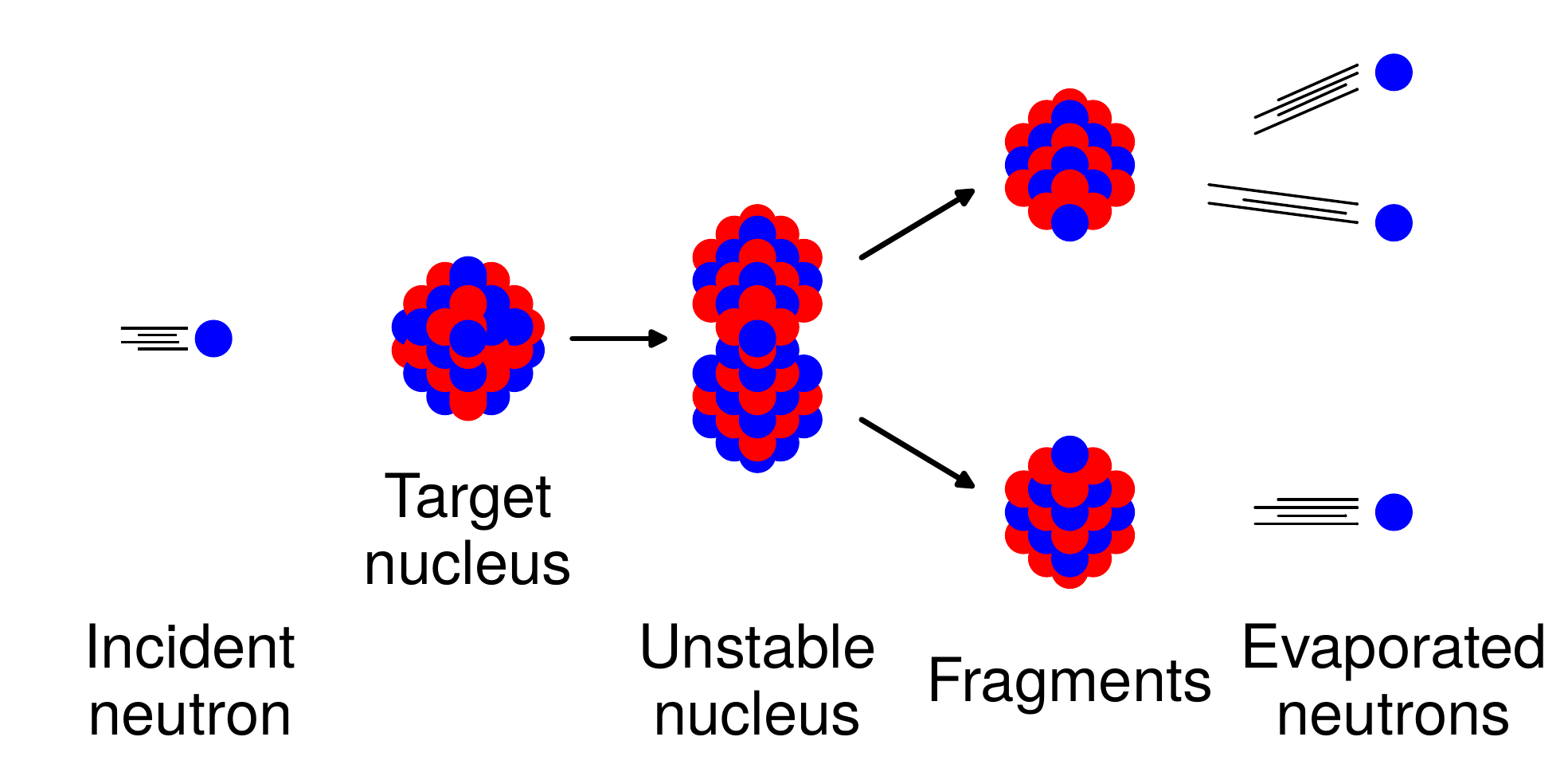}
  \vspace{-5pt}
\end{wrapfigure}

Fission of heavy elements is the decay mode in which the largest amount of energy is released
and corresponds to a split into two smaller daughter nuclei that fly away with a speed of about 100 0000 km per second.
This can be utilized to boil water, produce steam, power a steam turbine,
which in turn can generate electricity. 
Fission of one gram of uranium releases the same amount of energy as when
3 tons of oil is burnt or when 100 000 tons of water falls 100 meters.
Fission of heavy nuclei can be induced by bombardment of a neutron.
The resulting daughter nuclei then emit neutrons that can induce fission of another nucleus,
and thus creating a chain reaction. 
Fission is the process used to generate energy in today's nuclear reactors.

In this thesis a theoretical model is used for the description of fission
where the process is assumed to be very slow,
somewhat like when a chunk of syrup splits.
The shape evolution in the fission process can under this assumption be modeled
as a random walk in an energy landscape which depends on the shape of the nucleus.
The random walk starts at a point corresponding to a spherical shape
and stops when a point is reached 
corresponding to a split into two smaller pieces.

In paper IV the model is applied to all nuclei that theoretically can fission
in order to calculate the probability that a nucleus splits into two specific daughter nuclei
and how fast they fly away.
It has long been known that the most common fissioning nuclei like uranium and plutonium
prefer to split in one heavy and one light daughter nucleus,
where the heavy one is roughly the nucleus tin-132.
This nucleus is very special since quantum mechanical effects make it
more stable than expected, leading to a spherical form with tightly bound particles.
This specific type of split is referred to as a ``standard fission mode''
which corresponds to a certain shape of the nucleus just before it splits
and which affects the speed of the daughter nuclei and the number of neutrons emitted.
In the study a similar phenomenon appeared for nuclei with $Z\approx110$ 
where the nucleus splits into one very heavy and one very light daughter nucleus,
thus called super-asymmetric mode,
where the heavy daughter nucleus corresponds to the stable lead-208.
At the moment there are not enough experimental data to say if this is the case.

\begin{wrapfigure}{r}{0.39\textwidth}
  \vspace{-22pt}
    \hspace{-38pt}
    \includegraphics[width=0.98\textwidth]{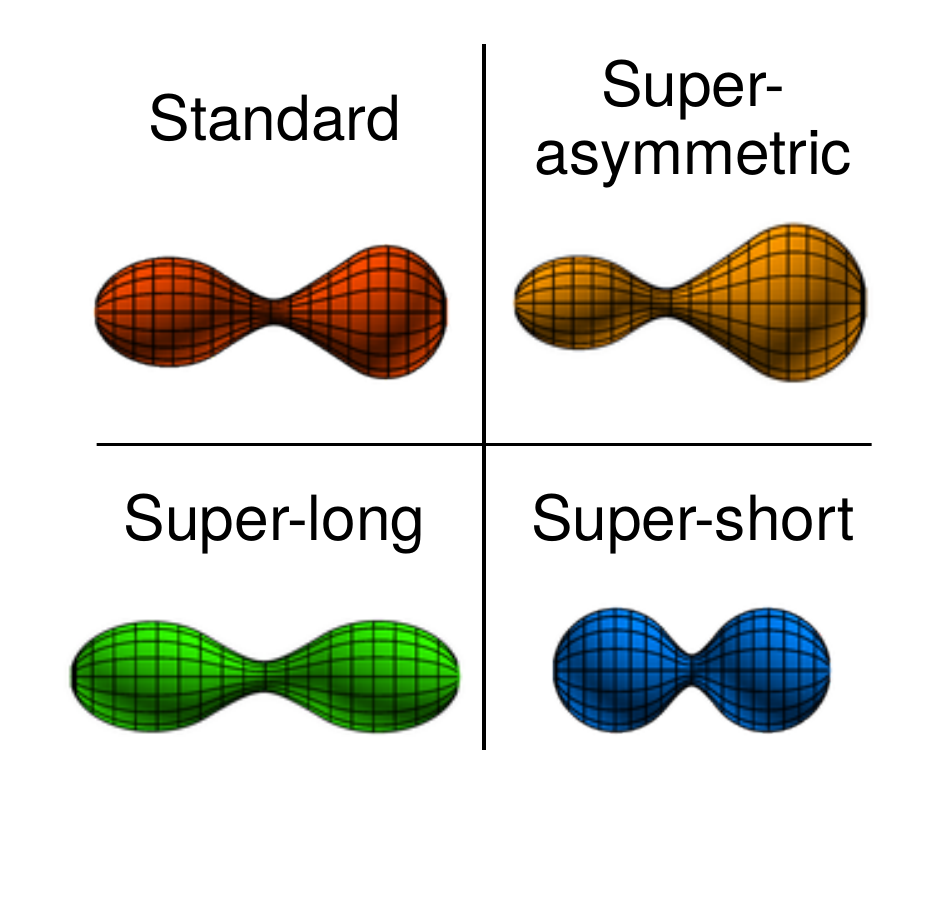}
  \vspace{-38pt}
      \hspace{-22pt}
\end{wrapfigure}

The two resulting daughter nuclei are warm after the split and 
they get rid of this heat by evaporating neutrons.
A mystery has been why tin-132 emits less neutrons than the corresponding light daughter nucleus,
since this appears to differ from simple statistical expectations.
In Paper I it is assumed that thermal equilibrium is achieved
right before it splits,
which implies that the available heat is divided based on 
the different states in the two daughter nuclei.
A quantum-mechanical model is used to calculate these states where the 
stability of tin-132 implies few states and therefore less share of the heat.
This in turn results in less neutrons emitted from tin-132.
When the incident neutron's speed is increased the effect of the stability in tin-132 disappears.
It then obtains a larger share of the heat and thus emits more neutrons,
which have also been observed in experiments.
In Paper II, the sum of the number of neutrons emitted from the two daughter nuclei is calculated
when the speed of the incident neutron changes and what the probability is 
that a specific number of neutrons will be emitted.

Fission is a quantum-mechanical process which means that identical experiments can generate
different outcomes.
An experiment is therefore repeated many times in order to acquire probability distributions
of various quantities.
The theoretical model used in the thesis is based on the same principle and can therefore
describe correlations between quantities.
This is investigated for fission of uranium-236 in Paper V,
where it is studied how many neutrons a specific daughter nucleus emits on average when
it flies away with a specific speed.
The calculated results compare well with the experimental results when the speed of the incident neutron is low.
In that case, tin-132 becomes almost spherical according to the standard mode and emits few neutrons.
When the speed of the incident neutron increases a ``superlong mode'' appears around tin-132,
where it obtains a more elongated shape and emits many neutrons if it flies away slowly.

A unique observation is that certain isotopes of fermium has the possibility
to split into two almost spherical tin-132 nuclei.
This implies that fission occur in a ``super-short'' mode
where the daughter nuclei fly away very fast and emits few neutrons.
In Paper III it is investigated how fermium-256 fission in the standard mode,
while fermium-260 fission in the super-short mode.
For fermium-258 both modes exist, in accordance with experiments.
Paper VI investigates further how the super-short mode transitions to the standard mode
when the initial energy of the nucleus is increased.

\vspace{5pt}
\textbf{Creation of new elements via fusion}
\vspace{-5pt}

The period table is a collection of all the known elements organized according to their proton number,
and also their chemical and physical properties.
The heaviest element occurring in Nature in appreciable quantities is uranium with 92 protons.
Heavier elements have however been produced in laboratories through fusion
where two lighter nuclei collide with each other.
The heaviest element that so for has been produced has 118 protons and
was in 2016 named to oganesson after the Russian nuclear physicist Yuri Oganessian.
Due to the very low probability that a new element will actually be produced in fusion reactions,
reliable theoretical models for different reactions and energies are needed for planning
of these kinds of experiments.

\begin{figure}[htbp!] 
\centering    
\includegraphics[width=1.0\textwidth]{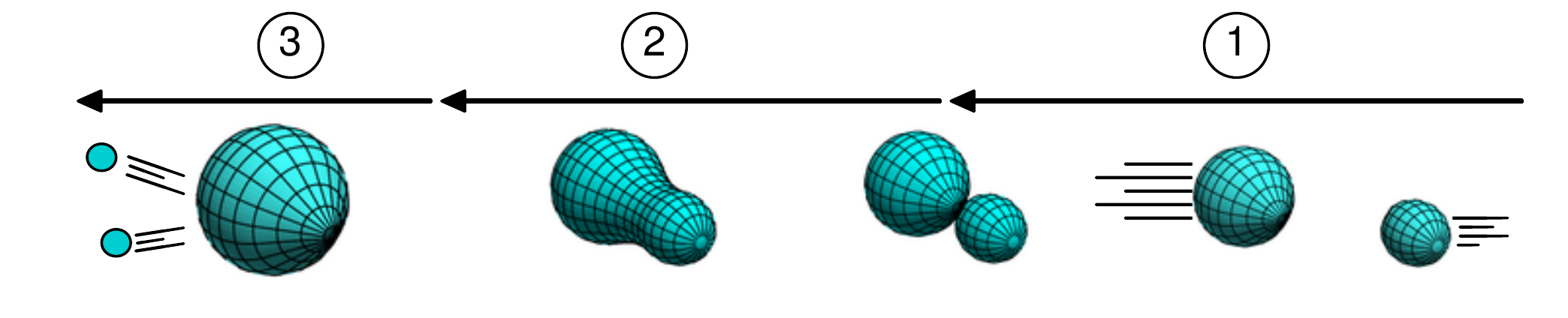}
\end{figure}

The fusion process can roughly be described as the opposite process to fission and can schematically be divided into three steps:
(1) the two colliding nuclei come into contact with each other,
(2) the two nuclei form a compound nucleus,
(3) the compound nucleus cools down by emitting neutrons.
The shape evolution in step (2) is investigated in Paper VII
by applying the same model used in the description of fission.


\newpage
\sect{List of publications\label{sec:paperlist}}
This thesis is based on the following publications, referred to by their Roman numerals.
All papers are reproduced with permission of their respective publishers.
\vspace{2mm}

{
\floatsetup[table]{font={normalsize},position=top}
\begin{tabularx}{\textwidth}{rX}
\normalsize
\hspace{3mm} I	  & {\bf \PaperItitle}\\[2mm]
	  & \PaperIauthor\\[2mm]
          & \PaperIref\\[6mm] 

II	  & {\bf \PaperIItitle}\\[2mm]
	  & \PaperIIauthor\\[2mm]
          & \PaperIIref\\[6mm]

III      & {\bf \PaperIIItitle}\\[2mm]
	  & \PaperIIIauthor\\[2mm]
          & \PaperIIIref\\[6mm]

IV 	  & {\bf \PaperIVtitle}\\[2mm]
	  & \PaperIVauthor\\[2mm]
          & \PaperIVref\\[6mm]
          
V        & {\bf \PaperVtitle}\\[2mm]
	  & \PaperVauthor\\[2mm]
          & \PaperVref\\[6mm]

\end{tabularx}

\begin{tabularx}{\textwidth}{rX}
\normalsize

VI       & {\bf \PaperVItitle}\\[2mm]
	  & \PaperVIauthor\\[2mm]
          & \PaperVIref\\[6mm]
          
VII      & {\bf \PaperVIItitle}\\[2mm]
	  & \PaperVIIauthor\\[2mm]
          & \PaperVIIref\\[6mm]

\end{tabularx}          
          


Publications not included in this thesis:
\vspace{2mm}

 \begin{tabularx}{\textwidth}{rX}
 \normalsize
 
\hspace{6mm}         & {\bf \PaperNotIncItitle}\\[2mm]
 	  & \PaperNotIncIauthor\\[2mm]
          & \PaperNotIncIref\\[6mm]
          
\end{tabularx}
          
 \begin{tabularx}{\textwidth}{rX}
 \normalsize
 
\hspace{6mm}           & {\bf \PaperNotIncIItitle}\\[2mm]
 	  & \PaperNotIncIIauthor\\[2mm]
           & \PaperNotIncIIref\\[6mm]
           
           & {\bf \PaperNotIncIIItitle}\\[2mm]
 	  & \PaperNotIncIIIauthor\\[2mm]
           & \PaperNotIncIIIref\\[6mm]
 
 \end{tabularx}

%

\newpage
\renewcommand{\listfigurename}{}
\sect{List of figures\label{sec:figurelist}}
\listoffigures

\selectlanguage{british}

\mainmatter
\setcounter{table}{0} 
\setcounter{page}{1}

\setcounter{page}{1} 
\setcounter{tocdepth}{1}
\tableofcontents
\addtocontents{toc}{\protect\thispagestyle{empty}}


\part{\color{red}{Introduction and background theory}}\label{part:intro}

\chapter{Introduction}\label{ch:intro}

\section{Brief history of nuclear physics}

\subsubsection*{600 B.C. - 400 B.C.: Classical elements}
The search for the fundamental nature of matter is a long-standing question and was undertaken in many ancient cultures.
One of the earliest known sources of questions regarding this issue date back to the Greek philosophers around 600 B.C.
Leucippus and his student Democritus
suggested that everything in Nature is made of ``atoms'' (from the Greek word atomos meaning uncuttable)
and that different types and combinations of these atoms constituted the various forms of matter \cite{russel04:a}.
They also argued that there must be considerable open space between these atoms, called the void.
This contrasted the theory by Empedocles that the nature of matter could be reduced to
the four classical elements - earth, water, air, and fire -
which were assumed to be continuous.
The theory of the atoms was dismissed by Aristotle and,
due to his large influence at the time,
the theory of the classical elements became the standard dogma for almost two millennia.

\subsubsection*{1661 - 1868: Chemical elements}
The atomic theory of Democritus was not properly reconsidered until the 17th century.
The chemist Robert Boyle proposed that elements are composed of atoms of various types and sizes which can organize themselves into different chemical substances.
He defined an element as a substance that could not be decomposed into other substances.
In 1789 Antoine Lavoisier constructed the first modern list of chemical elements 
which contained 33 elements (of which only 23 are considered chemical elements today).
The atomic theory was placed on more solid ground by John Dalton who argued that
all matter consists of tiny atoms which are indestructible and unchangeable.
He further argued that elements are characterized by the weight of their atoms,
and that the atoms combine to form new compounds when elements react.
In 1828, J{\"o}ns Jakob Berzelius compiled a table of atomic weights relative to oxygen
of all the known elements at the time,
which supported Dalton's atomic theory.

\subsubsection*{1869 - 1895: The periodic table of elements}
Up until the 1860s the chemists had discovered more than 60 different elements,
but they were not sure if there was a system of the elements.
The Russian chemist Dmitri Mendeleev organized the elements according to their atomic mass in increasing order,
which resulted in the period table of elements (see Fig.\ \ref{fig:periodic_table} for the most up-to-date version).
He noticed that elements with similar properties seemed to be repeated with certain intervals.
Although there were gaps in the table, Mendeleev suggested that these gaps corresponded to undiscovered elements.
Sure enough, these elements were later discovered.

\subsubsection*{1896 - 1910: Radioactivity}
In 1896 Henri Becquerel discovered radioactivity \cite{becquerel96:a}, 
i.e. the process in which an unstable atomic nucleus loses energy by emitting radiation.
This indicated that the atom was neither indivisible nor immutable.
In the years that followed, radioactivity was investigated in particular by Marie and Pierre Curie as well as by Ernest Rutherford.
Three types of radiation emanating from atoms were discovered which were named $\alpha$, $\beta$, and $\gamma$ radiation. 
The discovery of the electron by J.J. Thomson in 1897 \cite{thomson1897:a} was a further indication that the atom had internal structure.
This lead to the "plum pudding" model in which the atom was a positively charged ball with smaller negatively charged electrons embedded inside it.

\subsubsection*{1911 - 1932: The atomic nucleus}
Experiments done by Rutherford and his colleagues resulted in the Rutherford model of the atom in 1911 \cite{rutherford11:a}, 
in which the atom consists of a small positively charged massive nucleus surrounded by very distant orbiting negatively charged electrons.
The positive charge of the nucleus is due to $Z$ number of protons,
each carrying one positive unit of electric charge.
It was later discovered in 1932 \cite{chadwick32:a},
that the nucleus also contains $N$ number of particles that are very similar to protons but electrically neutral, hence called neutrons. 

\subsubsection*{1934 - 1939: Fission}
Since the neutron is electrically neutral it can easily enter the nucleus.
This was utilized by Enrico Fermi and his colleagues in 1934 who bombarded ever heavier elements with neutrons.
The bombarded nucleus would typically absorb the neutron
and then undergo $\beta$-decay which would result in an element with higher proton number.
After bombarding uranium ($Z=92$), Fermi concluded that the experiments had created new elements with 93 and 94 protons \cite{fermi34:a}.
Though Fermi received the Nobel price for these discoveries, 
the German chemist Ida Noddack suggested that instead of creating a new heavier element, 
that "it is conceivable that the nucleus breaks up into several large fragments" \cite{noddack34:a}.
However, Noddack's suggestion was not pursued at that time. 

In 1938, Hahn and Strassmann \cite{hahn39:a} also bombarded uranium with neutrons and identified barium ($Z=56$) as one of the products.
This large change in proton number then had to be the result of a new type of nuclear transmutation.
Meitner and Frisch suggested \cite{meitner39:a} 
that the nucleus could be described as a deformable charged liquid drop that had split into two smaller nuclei of roughly equal size.
Based on this model, and making use of the mass-energy equivalence $E=mc^2$ discovered by Albert Einstein,
Meitner calculated that the energy released in each split would be very high, 
which soon after was observed by Frisch \cite{frisch39:a}. 
The process was named ``nuclear fission'' in analogy to binary fission of cells in biology.

The Hungarian physicist Leó Szilárd realized that neutron-induced fission of uranium could be used to create a nuclear chain reaction.
It would thus be a possibility of generating huge amounts of energy; either for civilian purposes as in electric power generation,
but also for military purposes with atomic bombs.
In 1939 Szilárd therefore wrote a letter that was signed by Albert Einstein to President Franklin D. Roosevelt,
advising him to fund research into the possibility of using nuclear fission as a weapon as Nazi Germany may also be conducting such research \cite{rhodes86:a}.
This led to the creation of the Manhattan project that produced the first nuclear weapons.

\subsubsection*{1935 - 1964: The nuclear force and particle zoo}
Since protons repel each other due to the electrical Coulomb force, the nucleus would blow apart if there were not some other force inside the nucleus holding it together.
The idea of a new strong nuclear force was therefore introduced and the first theory was developed by the Japanese physicist Hideki Yukawa in 1935 \cite{yukawa35:a}.
Similar to the theory of the electromagnetic interaction where the interaction is mediated by a massless photon, 
the interaction between the nucleons would be mediated by a massive particle called meson.

The meson was eventually discovered in 1947 in cosmic rays,
which are highly energetic particles
coming from space and entering the earth's atmosphere.
The collisions of cosmic rays with nuclei in the air lead to creations of many new exotic particles in ``cosmic showers''.
These cosmic showers could be reproduced in laboratory settings with the developments of high-energy particle accelerators,
and ever more particles were discovered through the 1950s and 1960s.
Many of them were believed to be elementary particles and the entire collection was nicknamed the ``particle zoo''. 
Most of these particles were eventually explained to be combinations of more fundamental particles called quarks introduced by Gell-Mann and Zwieg in 1964 \cite{gell_mann64:a,zweig64:a}.

\subsubsection*{1911 - present: Modeling the nucleus}
Ever since the nucleus was discovered in 1911
various models have been developed for describing the nucleus.
Nuclei are systems of of strongly interacting particles and 
present many challenging issues.
A few of the most common models for modeling the nucleus are described in the following section.

\section{Structure of the nucleus}
The understanding today is that all ordinary matter in the universe is composed of chemical elements,
where a chemical element is a species of atoms.
The structure of an atom is illustrated in Fig.\ \ref{fig:structure_nucleus}.
The number of protons $Z$ in the nucleus determines the element and the number of neutrons $N$ determines the isotope.
The number of nucleons $A=Z+N$ is called the mass number
and the general notation for an element El is $^A_Z\text{El}_N$.
The properties of the chemical elements are summarized in the periodic table in Fig.\ \ref{fig:periodic_table}.

\begin{figure}[htbp!] 
\centering    
\includegraphics[width=1.0\textwidth]{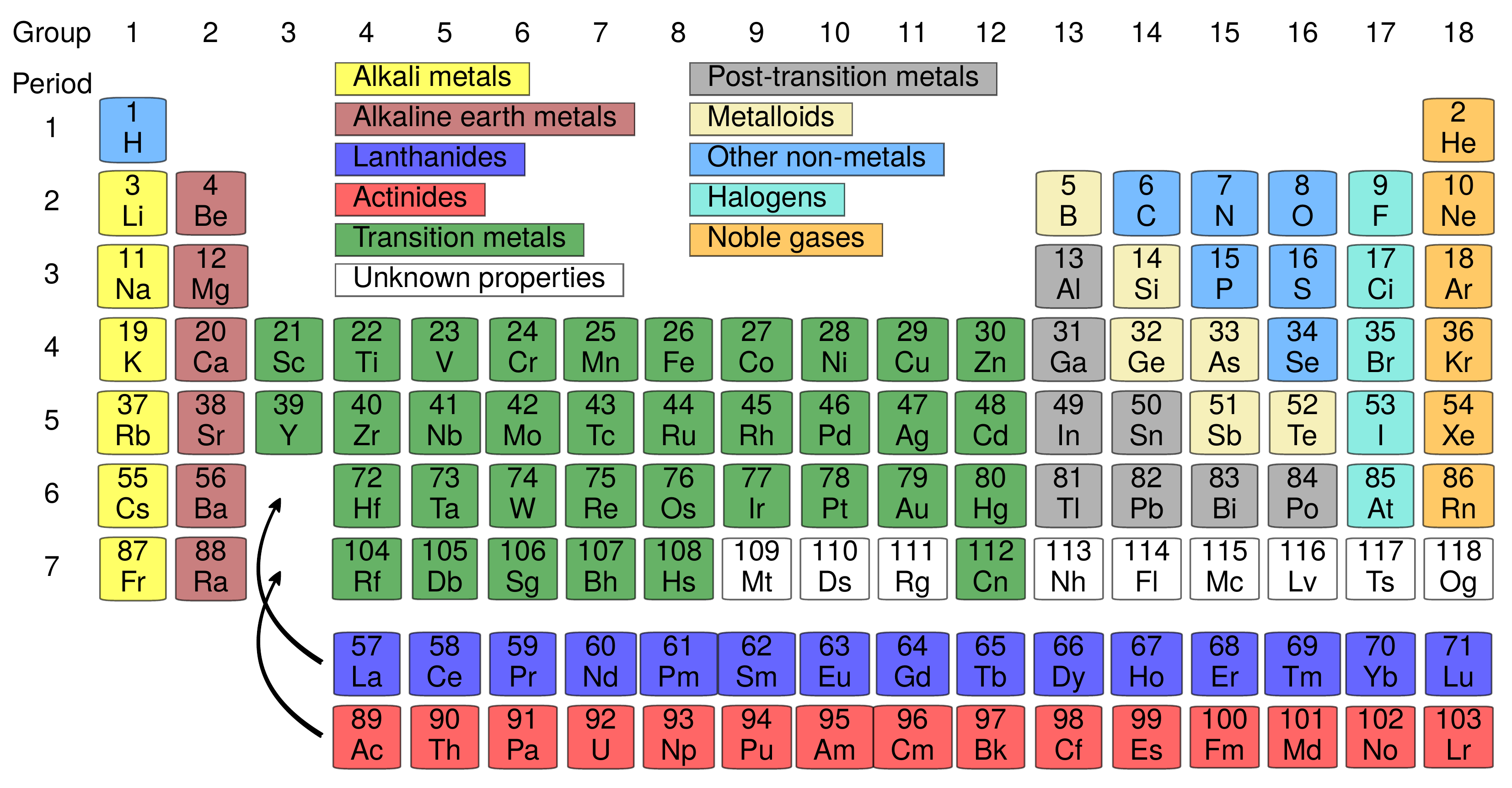}
\caption[Periodic table of elements]{
Periodic table of elements.
}
\label{fig:periodic_table}
\end{figure}

The protons and neutrons are themselves composed of smaller particles called quarks.
The quarks are bound together by the strong interaction described by Quantum Chromodynamics.
Quarks, gluons and their dynamics are mostly confined within nucleons, but residual influences extend slightly beyond 
nucleon boundaries to give rise to the nuclear force.

The large variety of nuclear models can broadly be grouped into four different approaches: ab initio
methods, shell-model theories, self-consistent mean-field models, and macroscopic-microscopic models.

\begin{figure}[htbp!] 
\centering    
\hspace{-1.6cm}
\includegraphics[width=1.0\textwidth]{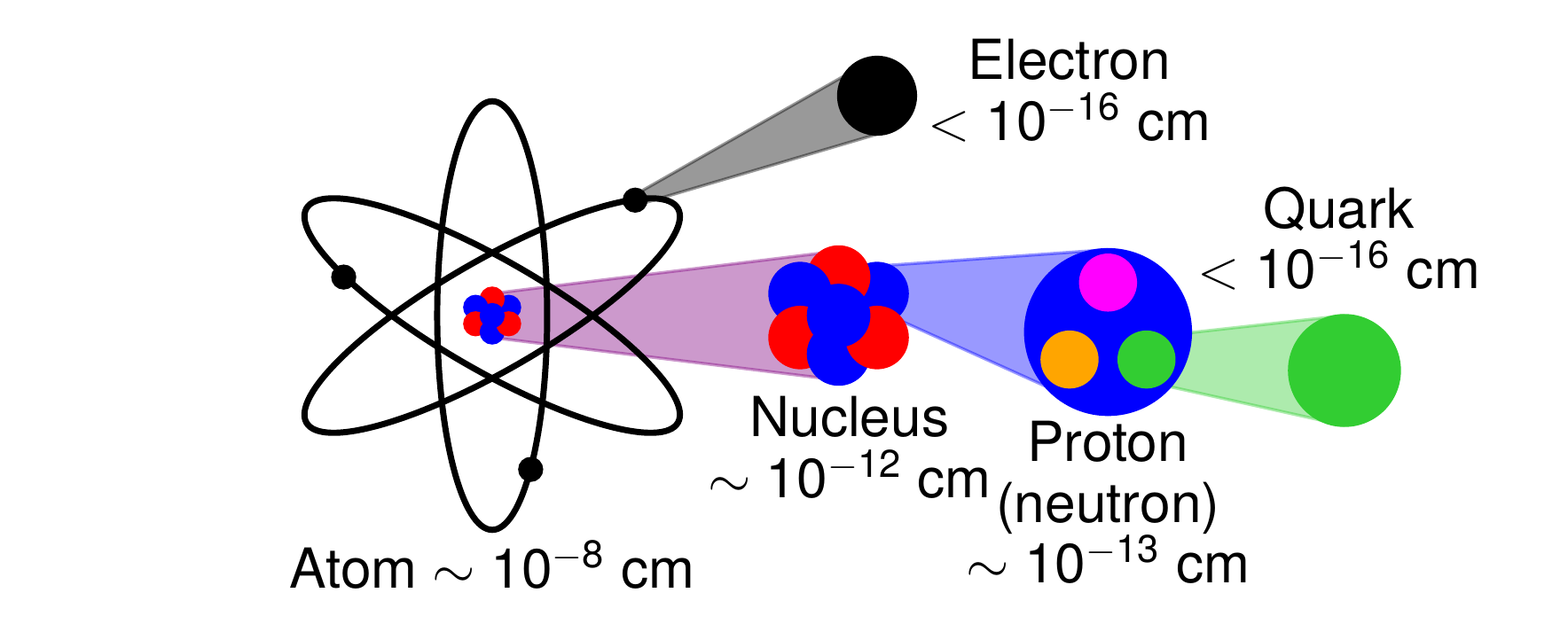}
\caption[Structure of the atom]{
Structure of the atom.
}
\label{fig:structure_nucleus}
\end{figure}

\subsection{Ab initio models}
Ab initio models denote models that try to describe the atomic nucleus by solving the non-relativistic Schr\"{o}dinger equation for 
all constituent nucleons and the forces between them
(see e.g. Ref.\ \cite{hergert20:a} for a review).

The first difficult task in ab initio modeling is to obtain a nuclear interaction from the
theory of Quantum Chromodynamics. 
The most promising methods are based on Chiral effective field theory and Lattice QCD. 

After arriving at a Hamiltonian $\hat{H}$ one must solve the Schr\"{o}dinger equation $\hat{H}|\Psi\rangle=E|\Psi\rangle$, 
where $|\Psi\rangle$ is the many-body wave function of the $A$ nucleons in the nucleus. 
Various ab initio methods have been devised to numerically find solutions to this equation such as
No Core Shell Model or Coupled Cluster.
All of these methods are very computationally intensive and are at the moment limited to relatively light nuclei.

\subsection{Shell models}
Nuclei with certain ``magic numbers'' of protons or neutrons are seen to be especially stable. 
This observation was the origin of the Non-Interacting Shell Model 
developed by Eugene Paul Wigner, Maria Goeppert Mayer and J. Hans D. Jensen \cite{mayer49:a,mayer55:a}, 
who shared the 1963 Nobel Prize. 
Despite the magnitude of the nuclear force, the nucleus is not a very dense system.
The nucleons in a nucleus can therefore be considered as independent particles moving on 
almost unperturbed single-particle orbits
governed by some average potential created by all the other nucleons in the nucleus.

The original shell model employs a spherical potential and can explain many features of spherical nuclei near the magic numbers.
A modification is needed in order to describe nuclei with nucleon numbers between the magic numbers.
This was developed by Sven Gösta Nilsson who considered a deformed potential,
which became very successful and is now called the Nilsson Model \cite{nilsson55:a}.

Another method of describing nuclei between the magic numbers is
to include a residual two-body interaction in
a valence space above a closed shell (the core),
corresponding to the Interacting Shell model.
A configuration-mixing calculation is then performed involving the many-body states
in the valence space.
This is equivalent to the No Core Shell Model if the core is included in the valence space.

\subsection{Self-consistent mean-field models}
The success of the phenomenologically introduced shell model justifies the assumption that the nucleons approximately move independently in an average potential 
produced by all the other nucleons. The task then is to extract such a single-particle potential out of the sum of two-body interactions,
which can be done with the self-consistent Hartree-Fock method.
However, nuclei with an even number of nucleons are systematically more bound than those with
an odd one, which implies that each nucleon binds with another one to form a pair.
Consequently the system cannot be described as independent particles
subjected to a common mean field. 
The nucleons are then subject to both the mean field potential and to the pairing interaction
within the Hartree-Fock-Bogoliubov theory \cite{ring80:a}.

\subsection{Macroscopic-microscopic models}
Another model is the liquid-drop model (LDM), which is based on
observations that a nucleus show similar properties as that of a drop of incompressible fluid.
This model parametrizes the energy of the nucleus in terms of macroscopic
properties such as volume energy and surface energy. The actual parameters are fitted phenomenologically. 
The LDM describes very well the average trends of nuclear quantities,
but is usually augmented by corrections that approximate the quantum-mechanical effects not taken into account in the LDM.
Models which combine the LDM with quantum mechanical corrections are usually called macroscopic-microscopic models.
They have been developed to be both highly descriptive
of many nuclear properties and have high predictive
accuracy for properties not yet measured experimentally.
The macroscopic-microscopic method is described in more detail in Sec.\ \ref{ch:mac_mic_models}.

\section{Nuclear stability and radioactive decay}
Since energy must be added to a nucleus to separate it into its individual protons and neutrons, 
the total rest energy (mass) of the separated nucleons is greater than the rest energy of the nucleons assembled into a nucleus.
The energy that must be added to separate the nucleons is called the binding energy $E_{\rm B}$. 
A measure of how tightly a nucleus is bound is the binding energy per nucleon, $E_{\rm B}/A$. 
Experimental binding energies per nucleon are shown in Fig.\ \ref{fig:binding_energies} as a function of $N$ and $Z$.
The highest value is approximately at the position of iron and nickel ($Z\approx28$, $N\approx34$), which are therefore the most stable nuclei. 
Consequently, energy is released in reactions where the end products are closer to iron and nickel.

\begin{figure}[htbp!] 
\centering    
\includegraphics[width=0.8\textwidth]{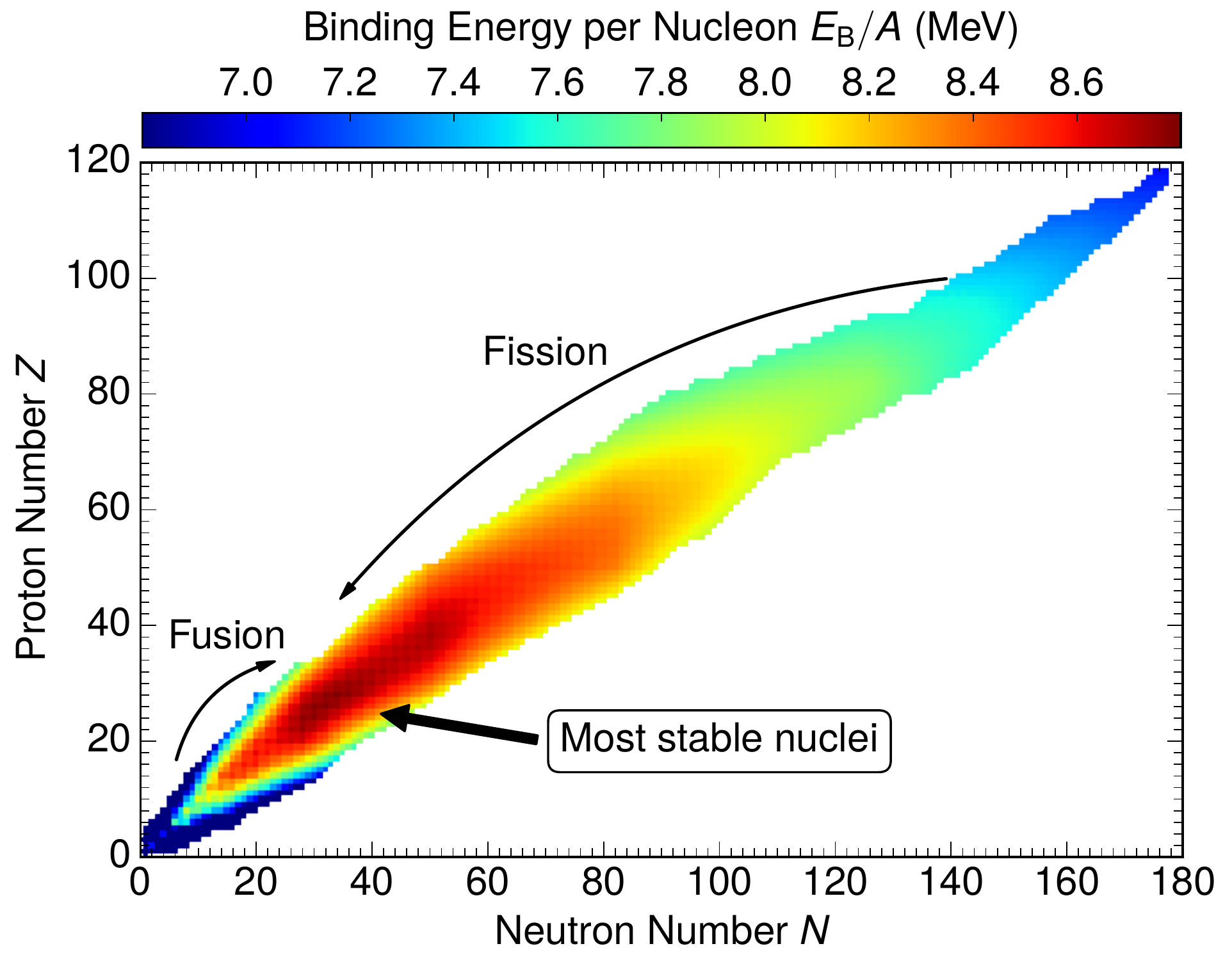}
\caption[Binding energy per nucleon]{Binding energy per nucleon with experimental masses taken from Ref.\ \cite{wang17:a}.}
\label{fig:binding_energies}
\end{figure}

\subsection{Radioactive decay}
Different proton and neutron numbers can be combined to form
an atomic nucleus, but most nuclei are unstable and decay after some time.
Figure \ref{fig:nuclear_chart} shows the typical decay modes for known nuclei and correspond to:
\begin{enumerate}
\item \textit{Proton/neutron emission}: Proton-rich nuclei can obtain a more stable ratio of protons to neutrons via proton emission
(and analogously neutron emission for neutron-rich nuclei).
\item \textit{$\alpha$-decay}: $\alpha$-decay means that an atomic nucleus splits into a Helium nucleus ($\alpha$ particle) and a new daughter nucleus with two less protons and
two less neutrons.
This is typical for heavier nuclei.
\item \textit{$\beta$-decay}: A nucleus with excess neutrons can transform a neutron into a proton by the emission of an electron accompanied by an antineutrino.
Similarly, proton-rich nuclei can transform a proton into a neutron by emitting a positron with a neutrino.
By this process, unstable nuclei obtain a more stable ratio of protons to neutrons by emitting a $\beta$ particle (electron or positron).
\item \textit{Spontaneous fission (SF)}: The higher values of binding energy near $A\approx60$ mean that energy is released when a heavy nucleus 
with $A\approx200$ splits into two lighter nuclei that lie closer to $A\approx60$.
SF occur only for heavy nuclei and competes with $\alpha$-decay.
\end{enumerate}

\begin{figure}[htbp!] 
\centering    
\includegraphics[width=0.8\textwidth]{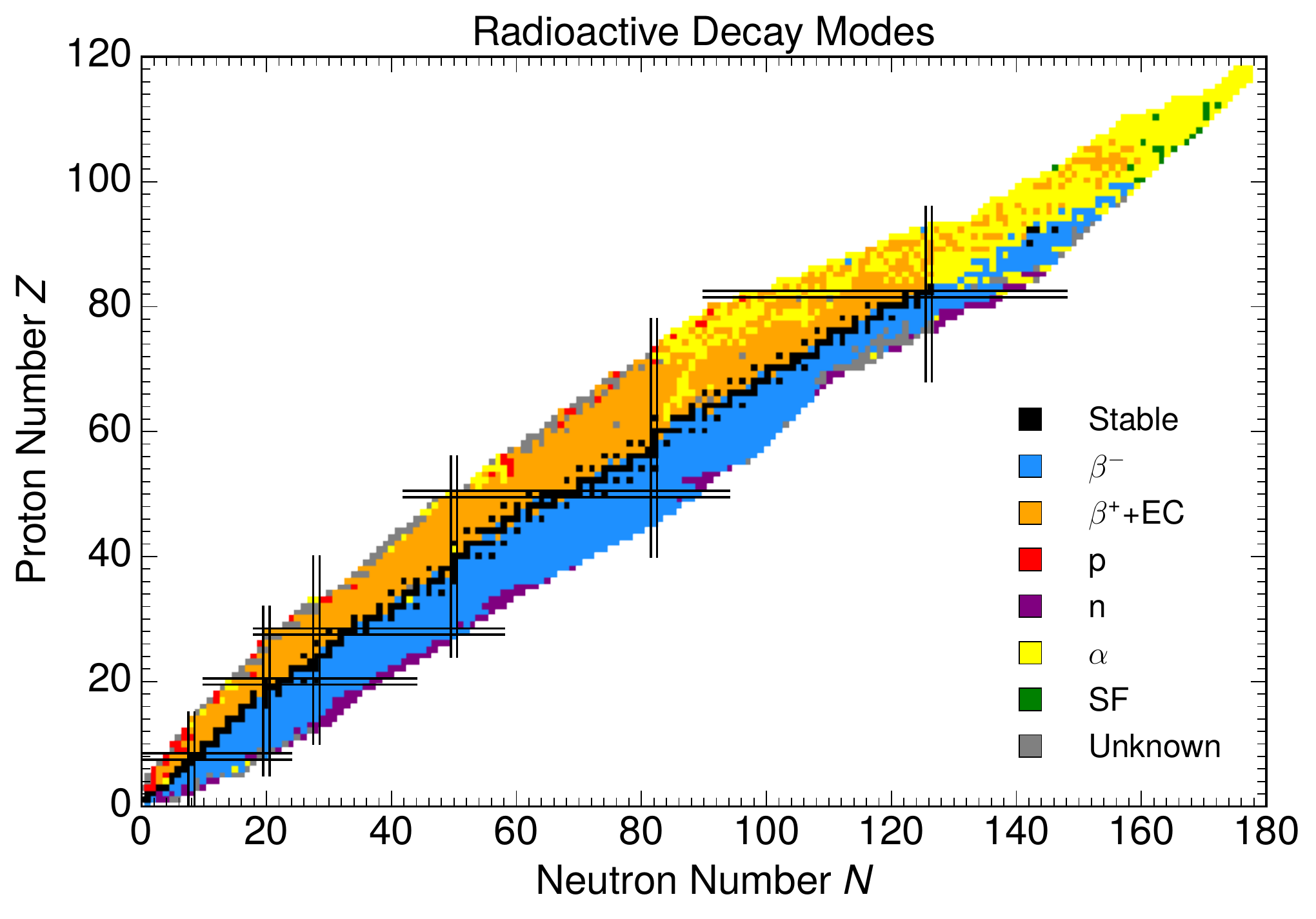}
\caption[Decay modes of nuclei]{
Experimental decay modes of nuclei in the form of a nuclear chart \cite{audi17:a}. 
Pairs of parallel lines indicate magic neutron and proton numbers.
}
\label{fig:nuclear_chart}
\end{figure}

\subsection{Nuclear reactions}
While a radioactive decay is a spontaneous event, a nuclear reaction is an induced reaction between a nucleus and some other particle.
Two common reactions are:

\begin{enumerate}
\item \textit{Induced fission}: In addition to SF, 
fission can also be induced through a nuclear reaction of a target nucleus with projectiles such as neutrons, protons, $\alpha$ particles or $\gamma$ rays.
\item \textit{Fusion}: 
In similar way energy is released when a heavy nucleus fission into two lighter nuclei,
energy is also released when two light nuclei with $A\leq 20$ are combined to form a heavier nucleus.
\end{enumerate}

\subsection{Superheavy elements}
The stability or instability of a particular nucleus is determined by
the competition between the attractive nuclear force among the protons and
neutrons and the repulsive electrical interaction among the protons.
For heavy nuclei, the Coulomb repulsion energy grows rapidly, 
so extra neutrons are required to supply the additional binding energy needed for stability. 
Eventually, the repulsive forces between protons cannot be compensated for by the addition of more neutrons, because of the asymmetry energy.

The heaviest naturally occurring element in appreciable quantities is uranium with proton number $Z=92$.
However, heavier elements have been produced in laboratories via fusion reactions, neutron capture and nucleon transfer.
While energy is released in fusion of light nuclei,
energy must be provided in fusion of heavy elements.
Superheavy nuclei ($Z\geq104$) up to $Z=118$ (oganesson) have been created in experiments.
What are the heaviest atomic nuclei that can exist,
and thus where the periodic table ends, is an active fields of research in nuclear physics
and is discussed more in Ch.\ \ref{ch:fusion}.

\section{Thesis outline}
This thesis presents theoretical descriptions of fission and fusion of heavy elements,
where the system evolves from a near-spherical shape towards two separated fragments in fission,
and vice versa in fusion.
J{\o}rgen Randrup and Peter Möller developed in 2011 a model 
that simulated the dynamics of the fission process as a random walk in the potential-energy landscape of shapes,
based on the assumption that shape evolution is strongly damped \cite{randrup11:a}.
This provided a quantitative description of the shape evolution in fission
with predictive power for the fission-fragment mass yield,
which sparked an increased activity in the field of fission.
The random-walk model was developed further in Ref.\ \cite{ward17:a}
by incorporating how microscopically calculated level densities affect the shape evolution.

The work in this thesis extends the random-walk model by, in addition to the previous description of fragment mass yields,
also simulating how much kinetic energy the fission-fragments obtain and the number of
neutrons they emit, as well as how these two quantities are correlated.
The thesis also presents studies of how different ways of fissioning, called fission modes,
are present in different nuclei and how the presence of these modes depends on the energy of the system.
The model is furthermore applied to the description of the shape evolution in fusion for production of superheavy elements.

The remainder of this part of the thesis, Part \ref{part:intro}, is focused on nuclear
theory relevant for the subsequent description of fission and fusion in Part \ref{part:fission}.
Ch.\ \ref{ch:mac_mic_models} describes the employed parametrization of the nuclear shape
and calculations of nuclear masses within the macroscopic-microscopic model,
while Ch.\ \ref{ch:level_densities} describes calculations of level densities.

Part \ref{part:fission} is mostly devoted to the description of fission
where particular focus will be on three cases; $^{236}$U, fermium isotopes around $^{258}$Fm, and $^{274}$Hs.
Fission barriers and potential-energy surfaces are described in Ch.\ \ref{ch:fission_barriers}.
Fission dynamics within the random-walk model is outlined in Ch.\ \ref{ch:fission_dynamics}.
Scission quantities, such as fragment masses and deformations,
are presented in Ch.\ \ref{ch:scission_quantities}
The energy released in fission in terms of kinetic energy and excitation energy 
is described in Ch.\ \ref{ch:energy_release_fission}.
Ch.\ \ref{ch:particle_emission} covers neutron evaporation from fission-fragments.
Ch.\ \ref{ch:fusion} is devoted to fusion for production of superheavy elements.
An outlook on the research of fission and fusion is given in Ch.\ \ref{ch:outlook_fission_fusion}.

Finally, in Part III 
the scientific publications are included.
See the List of publications in the thesis preamble for an overview of the articles.
A poster is also presented in the appendix.


\chapter{Mass models}\label{ch:mac_mic_models}

The nuclear binding energy $E_{\rm B}$ is defined as the difference in energy between the nucleus and its constituent protons and neutrons:
\begin{equation}
\label{eq:binding_energy}
E_{\rm B}(N,Z,\text{shape})=ZM_{\rm H}+NM_{\rm n}-E(Z,N,\text{shape}),
\end{equation}
where $M_{\rm H}$ and $M_{\rm n}$ are the free neutron and hydrogen atom masses (in units of MeV),
and where $E(Z,N,\text{shape})$ is the potential energy of the nucleus.

In the macroscopic-microscopic method the total energy of the nucleus consists of two parts; macroscopic and microscopic. 
Both parts are calculated at a fixed shape of the nuclear surface. Thus the total nuclear potential energy can be written as
\begin{equation}
E(Z,N,\text{shape})=E_{\text{mac}}(Z,N,\text{shape})+E_{\text{mic}}(Z,N,\text{shape}).
\end{equation}
The macroscopic part accounts for most of the energy and describes the smooth variations in energy when $N,Z$, and the nuclear shape are varied. 
The microscopic part accounts for the fluctuation of the energy around the smooth trends, due to the shell structure. 

A parametrization of the nuclear shape is described in Sec.\ \ref{sec:shape_params}.
Methods for calculating the macroscopic and microscopic terms are discussed in Sec.\ \ref{sec:mac_models} and \ref{sec:mic_models}, respectively.

\section{Shape parametrization}
\label{sec:shape_params}
The fact that the nucleus can be considered to have a surface comes from the properties of the nuclear interaction;
it is repulsive at short distances but attractive when nucleons are just beyond touching.
There is therefore an optimal spacing between neighbouring nucleons to be situated.
As a consequence, nuclei have a fairly uniform interior and a relatively thin surface that reflects the short range of the nuclear force. 
It is therefore reasonable to consider nuclei as incompressible diffuse droplets of nuclear matter.

\subsection{Three-quadratic surface parametrization}
A common shape parametrization for the description of fission
is the three-quadratic-surface (3QS) parametrization \cite{nix1969:a},
in which the shape of the nuclear surface is specified in terms of three smoothly joined portions of quadratic surfaces of revolution.
In terms of a cylindrical coordinate system, the equation for the nuclear surface can be written explicitly as
\begin{equation}
 \rho^2=
 \begin{cases} 
 a_1^2-\frac{a_1^2}{c_1^2}(z-l_1)^2,& l_1-c_1\leq z\leq z_1, \\ 
 a_2^2-\frac{a_2^2}{c_2^2}(z-l_2)^2, & \hspace{7.4mm} z_2\leq z\leq l_2+c_2, \\ 
 a_3^2-\frac{a_3^2}{c_3^2}(z-l_3)^2, & \hspace{7.4mm} z_1\leq z\leq z_2. 
 \end{cases} 
\end{equation}
This expression contains 11 parameters and are illustrated in Fig.\ \ref{fig:3qs}(a).
However, the conditions of constancy of volume and 
continuous function and first derivative at $z_1$ and $z_2$ reduces it to six numbers.
By introducing an auxiliary unit of distance
\begin{equation}
u=\left[\frac{1}{2}(a_1^2+a_2^2)\right]^{1/2},
\end{equation}
one can define three symmetric coordinates $\sigma_i$ and three reflection-asymmetric coordinates $\alpha_i$ as
\begin{equation}
\begin{aligned}
 &\sigma_1 = \frac{l_2-l_1}{u}, \qquad   &&\alpha_1 = \frac{1}{2}\frac{(l_1+l_2)}{u}, \\
 &\sigma_2 = \frac{a_3^2}{c_3^2},  &&\alpha_2 = \frac{(a_1^2-a_2^2)}{u^2}, \\
 &\sigma_3 = \frac{1}{2}\left(\frac{a_1^2}{c_1^2}+\frac{a_2^2}{c_2^2}\right), &&\alpha_3 = \frac{a_1^2}{c_1^2} - \frac{a_2^2}{c_2^2}.
\end{aligned}
\end{equation}
The coordinate $\alpha_1$ is furthermore determined by requiring that the center of mass to be at the origin,
so there are only five independent shape coordinates.

\begin{figure}[htbp!] 
\centering    
\includegraphics[width=0.7\textwidth]{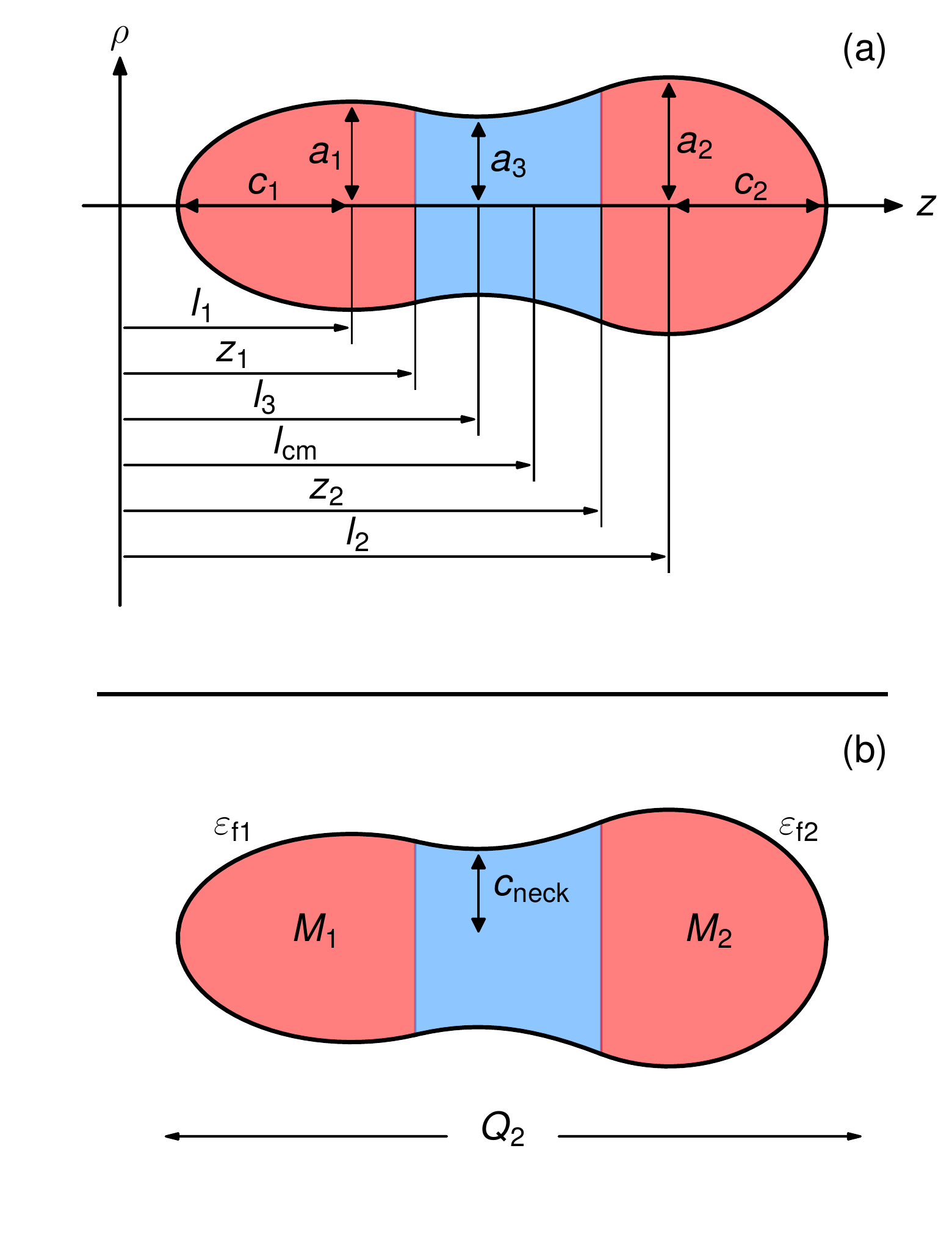}
\caption[Three-quadratic-surface parametrization]{
Shape described by three smoothly joined portions of quadratic surfaces of revolution.
(a) Each surface is specified by the position $l_i$ of its center, its transverse semiaxis $a$, and its semi-symmetry
axis $c$. The middle hyperboloid of revolution joins smoothly with the two end spheroids at $z_1$ and $z_2$. 
The location $l_{\rm cm}$ of the center of mass of the drop is also shown.
(b) Parametrization in terms of five independent shape coordinates: charge quadrupole moment $Q_2$; neck $c_{\rm neck}$; 
nascent-fragment deformations $\varepsilon_{\rm f1}$, $\varepsilon_{\rm f2}$; and 
mass asymmetry $\alpha=(M_1-M_2)/(M_1+M_2)$, where $M_1$ and $M_2$ are the volumes inside the end-body quadratic surfaces, 
were they completed to form closed-surface spheroids.
}
\label{fig:3qs}
\end{figure}

The shapes in 3QS coordinates can be defined in five more familiar shape coordinates (see Fig.\ \ref{fig:3qs}(b)):
(1) elongation coordinate expressed in terms of the quadrupole moment $Q_2$ of the nuclear charge distribution. 
The dimensionless elongation parameter $q_2$ is defined as
\begin{equation}
q_2=\frac{4\pi Q_2}{3ZR^2_A},
\end{equation}
where $R_A$ is the nuclear radius; (2) neck radius $c_{\rm neck}$; 
(3) left nascent-fragment deformation $\varepsilon_{\rm f1}$; (4) right nascent-fragment deformation $\varepsilon_{\rm f2}$; and 
(5) mass asymmetry $\alpha=(M_1-M_2)/(M_1+M_2)$, where $M_1$ and $M_2$ are the volumes inside the end-body quadratic surfaces, 
were they completed to form closed-surface spheroids.
The five shape coordinates are denoted as 
$\boldsymbol{\chi}=(q_2,c_{\rm neck},\varepsilon_{\rm f1},\varepsilon_{\rm f2},\alpha)$.

\section{Macroscopic models}
\label{sec:mac_models}
The macroscopic part of the energy, $E_{\text{mac}}$, can be obtained from the LDM or refinements such as the finite-range droplet model (FRDM) or 
the finite-range liquid-drop model (FRLDM) \cite{moller81:a,moller81:b,moller16:a}. 

\subsection{Liquid-drop model}
\label{sec:ldm_model}
The LDM has its origin in the semi-empirical mass model, usually attributed to von
Weizäcker \cite{weizacker35:a} and Bethe and Bacher \cite{bethe36:a}.
This model relies on the analogy with an incompressible liquid drop and the formula for the nuclear binding energy includes five contributions:
\begin{enumerate}
\item \textit{Volume term}: The nuclear forces show saturation, which means that an individual nucleon interacts only with its nearest neighbours. 
This gives a binding-energy term that is proportional to the number of nucleons, i.e. a volume term (since the nuclear radius is proportional to $A^{1/3}$).
\item \textit{Surface term}: The nucleons on the surface are less tightly bound than those in the interior because they have no neighbours outside the surface. 
This decrease in binding energy gives a negative energy term proportional to the surface area $4\pi R^2=4\pi A^{2/3}$.
\item \textit{Coulomb term}: The $Z$ protons in the nucleus repel each other due to the Coulomb force,
which is proportional to $Z^2$ and inversely proportional to the radius $R\propto A^{1/3}$.
This energy term is negative because the nucleons are less tightly bound than they would be without the repulsion. 
\item \textit{Asymmetry term}: 
Due to the Pauli principle
it is more favourable to have an approximately equal number of protons and neutrons.
A negative energy term corresponding to the difference $|N-Z|$ is therefore needed. 
The best agreement with observed binding energies is obtained if this term is proportional to $(N-Z)^2/A$.
\item \textit{Pairing term}: Nuclei with an even number of nucleons are systematically more bound than those with an odd one.
This gives rise to a pairing term $E_{\rm oe}$, where $E_{\rm oe}=\Delta$ for even-even nuclei, $E_{\rm oe}=0$ for odd-mass nuclei and $E_{\rm oe}=-\Delta$ for odd-odd nuclei.
\end{enumerate}
The total nuclear binding energy $E_{\rm B}$ is then given by
\begin{equation}
\label{eq:semi_empirical_binding}
E_{\rm B}=a_{\rm V}A-a_{\rm S}A^{2/3}-a_{\rm C}\frac{Z^2}{A^{1/3}}-a_{\rm A}\frac{(N-Z)^2}{A} + E_{\rm oe},
\end{equation}
called the semi-empirical binding formula. 
The constants are chosen to make this formula best fit the observed binding energies of nuclei.

Soon after fission was discovered, Bohr and Wheeler \cite{bohr39:a} suggested a generalization of the semi-empirical mass model
to describe the shape changes in the division of the nucleus.
Only the Coulomb and surface energies were assumed to depend on deformation, so that
\begin{equation}
\label{eq:ldm_bohr}
E_{\rm B}=a_{\rm V}A - a_{\rm S}A^{2/3}B_{\rm S}(\alpha) - a_{\rm C}\frac{Z^2}{A^{1/3}}B_{\rm C}(\alpha) - a_{\rm A}\frac{(N-Z)^2}{A} + E_{\rm oe},
\end{equation}
where $B_{\rm S}(\alpha)$ and $B_{\rm C}(\alpha)$ are the ratios of the surface and Coulomb energies at deformation $\alpha$ to that for spherical shape.

It has also been argued that the surface term should be isospin-dependent, 
so that the surface term is written as 
$-a_{\rm S}\left(1-\kappa_{\rm S}I^2\right)A^{2/3}B_{\rm S}(\alpha)$,
where $I=(N-Z)/A$ is the relative neutron excess.
The macroscopic energy is then with Eq.\ (\ref{eq:binding_energy}) given by
\begin{equation}
\begin{aligned}
E_{\text{mac}}&=ZM_{\text{H}}+NM_{\text{n}} \qquad  &&\bigg( \begin{aligned} &\text{masses of $Z$ hydrogen} \\ &\text{atoms and $N$ neutrons} \end{aligned} \bigg) \\ 
&-a_{\rm V}\left(1-\kappa_{\rm V}I^2\right)A \qquad &&(\text{volume and volume-asymmetry energies}) \\ 
&+a_{\rm S}\left(1-\kappa_{\rm S}I^2\right)A^{2/3}B_{\rm S}(\alpha) &&(\text{surface and surface-asymmetry energy}) \\
&+a_{\rm C}\frac{Z^2}{A^{1/3}}B_{\rm C}(\alpha) &&(\text{Coulomb energy}) \\
&- E_{\rm oe} &&(\text{pairing energy}),
\end{aligned}
\end{equation}
where $\kappa_{\rm V}=a_{\rm A}/a_{\rm V}$.

\subsection{Finite-range liquid-drop model}
The surface-energy term in the LDM accounts for the fact that nucleons at the surface have fewer neighbours than the nucleons at the center.
However, the nuclear force has a range of about 1 fm.
Thus, the surface-energy term in the LDM is inadequate for systems with strong variations in the nuclear surface.
The typical example is fission where the nucleus develops a small neck before it splits.
The surface nucleons in the neck region then obtain some binding due to their interaction with the nucleons on the other side of the neck.

\begin{figure}[b] 
\centering    
\includegraphics[width=0.8\textwidth]{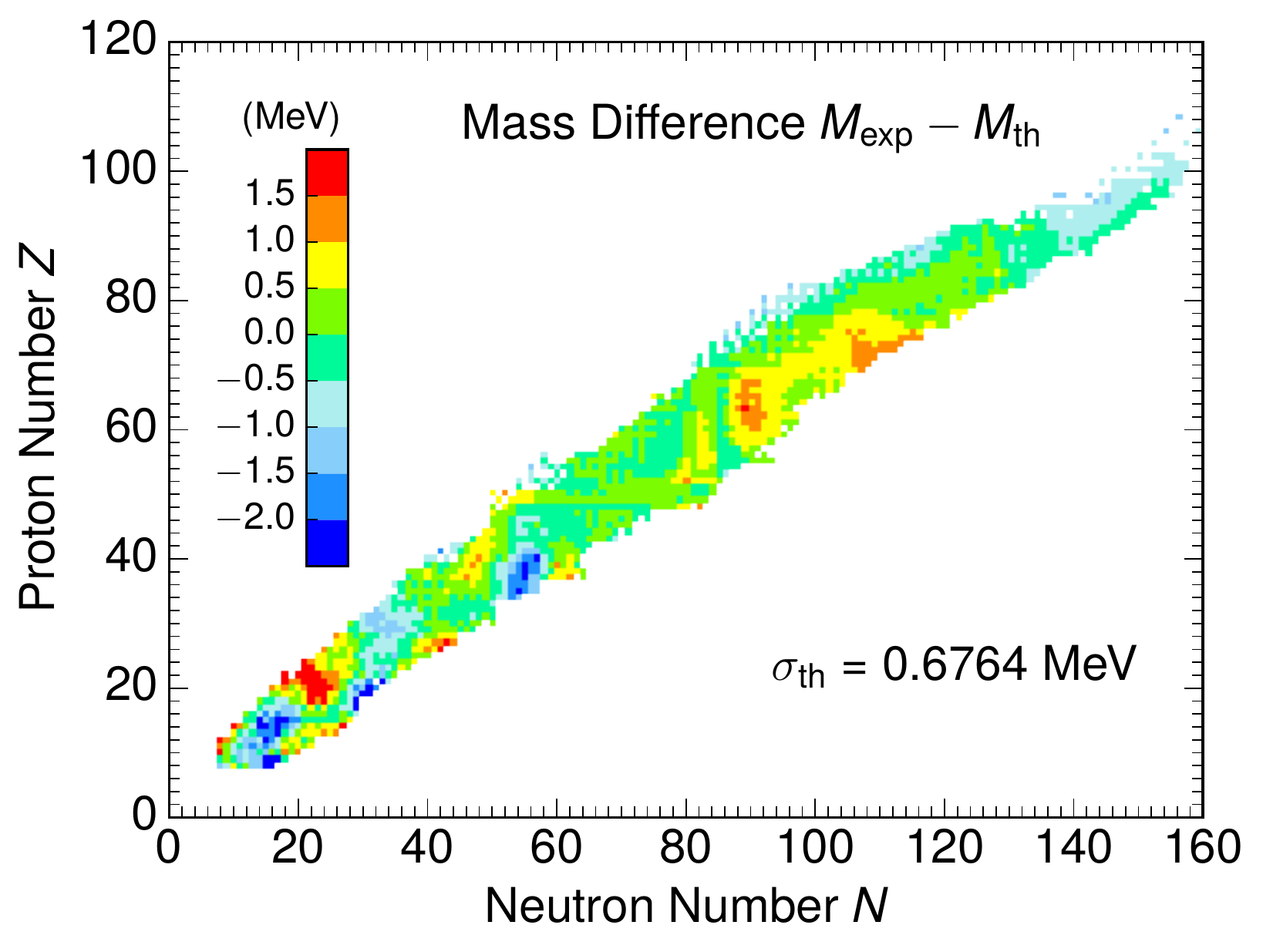}
\caption[Mass difference $M_{\rm exp}-M_{\rm th}$ in FRLDM]{
Difference between experimental masses from the AME2012
evaluation \cite{wang12:a} and masses calculated in the FRLDM \cite{moller16:a}
for ground-state shapes,
with an error $\sigma_{\rm th}=0.6764$ MeV.
}
\label{fig:frdlm2012_exp}
\end{figure}

The effect of the finite range of the nuclear force on the surface energy is one of the additional effects taken into account in the FRLDM \cite{moller16:a},
which has provided reliable predictions of a large number of nuclear-structure properties for all nuclei between the proton and neutron drip lines.
Figure \ref{fig:frdlm2012_exp} shows the difference between experimental masses from the AME2012
evaluation \cite{wang12:a} and masses calculated in the FRLDM
(with microscopic corrections included)
for nuclei in their ground-state shapes. The resulting error is $\sigma_{\rm th}=0.6764$ MeV.
Although the preferred model of ground-state masses is the FRDM \cite{moller16:a},
with an error of $\sigma_{\rm th}=0.5728$ MeV,
the FRLDM is more suitable for the very deformed shapes occurring in fission.

The macroscopic energy in the FRLDM is expressed as
\begin{equation}
\begin{aligned}
E_{\text{mac}}&=ZM_{\text{H}}+NM_{\text{n}}  \quad &&\bigg( \begin{aligned} &\text{masses of $Z$ hydrogen} \\ &\text{atoms and $N$ neutrons} \end{aligned} \bigg) \\ 
&-a_{\rm V}\left(1-\kappa_{\rm V}I^2\right)A &&(\text{volume and volume-asymmetry energies}) \\
&+a_{\rm S}\left(1-\kappa_{\rm S}I^2\right)B_1A^{2/3} &&(\text{surface energy}) \\
&+a_{\rm C}\frac{Z^2}{A^{1/3}}B_3 &&(\text{Coulomb energy}) \\
&-E_{\rm oe} &&(\text{average pairing energy}) \\
&+a_0A^0 &&(\text{$A^0$ energy}) \\
&-c_4\frac{Z^{4/3}}{A^{1/3}} &&(\text{Coulomb exchange correction}) \\
&+f(k_{\rm F}r_{\rm p})\frac{Z^2}{A} &&\bigg( \begin{aligned} &\text{proton form-factor correction} \\ &\text{to the Coulomb energy} \end{aligned} \bigg) \\ 
&-c_a(N-Z) &&(\text{charge-asymmetry energy}) \\
&+E_{\rm W} &&(\text{Wigner energy}) \\  
&-a_{\rm el}Z^{2.39} &&(\text{energy of bound electrons})
\end{aligned}
\end{equation}
The quantity $B_1$ is a generalization of the surface area of the nucleus at the
actual shape to the surface area of the nucleus at the spherical shape, 
which also accounts for the effect of the finite range of the nuclear force.
$B_3$ is the relative Coulomb energy.
The pairing term $E_{\rm oe}$ in this expression contains separate average pairing gaps for protons and neutrons, $\overline{\Delta}_{\rm n}$ and $\overline{\Delta}_{\rm p}$,
and an average neutron-proton interaction energy $\delta_{\rm np}$.
For further details of the additional terms see Ref.\ \cite{moller16:a}.

\section{Microscopic corrections}
\label{sec:mic_models}
Strutinsky observed that the deviation of the binding energy from the 
macroscopic LDM prediction was large for nuclei with a smaller than average single-particle level density above the Fermi surface.
He therefore proposed a method \cite{strutinsky1967:a,strutinsky1968:a} in which these shell effects are
considered as small deviations from a uniform single-particle level spectrum.
This deviation is then added as a correction to the macroscopic energy.
The effect of pairing can also be treated as a correction in a similar way
and the total microscopic correction is then given by the sum of both corrections,
\begin{equation}
E_{\text{mic}}=E_{\text{sh}}+E_{\text{pc}}.
\end{equation}
Both terms are calculated separately for protons and neutrons and then summed, i.e.
\begin{equation}
E_{\text{sh}}=E^{\rm neut}_{\text{sh}}+E^{\rm prot}_{\text{sh}}, \qquad E_{\text{pc}}=E^{\rm neut}_{\text{pc}}+E^{\rm prot}_{\text{pc}}.
\end{equation}

\subsection{Folded-Yukawa single-particle potential}
\label{sec:folded_yukawa}
The Strutinsky method employs the single-particle energies from a phenomenological mean-field potential.
The single-particle potential felt by a nucleon is given by
\begin{equation}
V = V_1+V_{\rm s.o.}+V_{\rm C},
\end{equation}
where $V_1$ is the spin-independent nuclear part of the potential,
$V_{\rm s.o.}$ is the spin-orbit potential,
and $V_{\rm C}$ is the Coulomb potential for protons.
In the FRLDM the $V_1$ term is calculated in terms of the folded-Yukawa potential
\begin{equation}
V_1(\mathbf{r}) = -\frac{V_0}{4\pi {a_{\rm pot}}^3} \int_V \frac{e^{-|\mathbf{r}-\mathbf{r}'|/a_{\rm pot}}}{|\mathbf{r}-\mathbf{r}'|/a_{\rm pot}} d^3r',
\end{equation}
where the integration is over the volume of the
shape,
and where $a_{\rm pot}$ is the range of the Yukawa function.

\subsection{Shell correction}
The shell-correction energy $E_{\text{sh}}$ is defined as the difference between
the sum of the actual single-particle energies in the mean-field potential
and the single-particle energies corresponding to a smearing of the actual single-particle energies.
The expression for neutrons is (analogous expression hold for protons)
\begin{equation}
E^{\rm neut}_{\text{sh}}=\sum_{i=1}^Ne_i - \tilde{E}^{\rm neut},
\end{equation}
where $e_i$ are the calculated single-particle energies.
The quantity $\tilde{E}^{\rm neut}$
is the smooth single-particle energy sum
obtained by smearing the calculated energies $e_i$ over an energy range $\gamma$
using a Gaussian function modified with a polynomial of order $p$.
The values of the two constants $p$ and the range $\gamma$ are chosen
by requiring stability of the results.

\subsection{Pairing correction}
The pairing correction is defined in terms of a pairing correlation energy $E_{\rm pair}$ and an average pairing correlation energy $\tilde{E}_{\rm pair}$.
For neutrons it is given by (analogous expression hold for protons)
\begin{equation}
E^{\rm neut}_{\rm pc}=E^{\rm neut}_{\rm pair}-\tilde{E}^{\rm neut}_{\rm pair}.
\end{equation}
The pairing corrections are the differences between the
pairing energies calculated in the Lipkin-Nogami approximation
\cite{lipkin60:a,nogami64:a,pradhan73:a} and the average pairing energies calculated as discussed
in Ref.\ \cite{moller92:c}.

\chapter{Nuclear level densities}\label{ch:level_densities}

Nuclear excited energy levels display a discrete spectrum for low excitation energies. 
The nuclear level density is defined as the number of levels per unit energy at a certain excitation energy.
In other words it is the number of different ways in which individual nucleons can be placed in the various single particle orbitals such that 
the excitation energy lies in the range $E$ to $E+dE$. 
It increases rapidly with excitation energy.

\section{Fermi-gas level density}\label{sec:fermi}
The simplest type of model for calculations of level densities is the Fermi-gas (FG) model,
which is derived for a uniform single particle spectra.
For an excitation energy $E^\ast$ this results in 
a level density $\rho_{\rm FG}(E^\ast)\sim e^{2\sqrt{aE^\ast}}$ \cite{bethe36:b}.
The level-density parameter $a$ is given by $a=\pi^2/(6g_0)$, 
where $g_0$ is the density of single-particle states \cite{levdens_param:a}.

For a deformed nucleus with shape $\boldsymbol{\chi}$ the formula becomes \cite{uhrenholt13:a}
\begin{equation}
\label{eq:levdens_fg}
\rho_{\rm FG}(E^\ast(\boldsymbol{\chi}),I) = \frac{2I+1}{48}\left(\frac{\hbar^2}{2\mathcal{J}}\right)^{1/2}(E_{\rm intr})^{-3/2}\text{exp}\left(2\sqrt{aE_{\rm intr}}\right),
\end{equation}
where
\begin{equation}
E_{\text{intr}}=E^\ast(\boldsymbol{\chi})-I(I+1)\hbar^2/2\mathcal{J}(\boldsymbol{\chi}),
\end{equation}
is the intrinsic energy of a state with angular momentum $I$ and 
moment of inertia $\mathcal{J}(\boldsymbol{\chi})$.

\subsection{Back-shifted Fermi-gas level density}
Actual single particle spectra display irregular structures due to microscopic effects. 
These effects can approximately be taken into account by introducing a back-shift in the energy.
The relevant FG level density is then given by
\begin{equation}
\label{eq:levdens_bfg}
\rho_{\rm BFG}(E^\ast(\boldsymbol{\chi}),I) = \frac{2I+1}{48}\left(\frac{\hbar^2}{2\mathcal{J}}\right)^{1/2}(\tilde{E}_{\rm intr})^{-3/2}\text{exp}\left(2\sqrt{a\tilde{E}_{\rm intr}}\right),
\end{equation}
where $\tilde{E}_{\rm intr}$ is a back-shifted excitation energy.
Ignatyuk has suggested the following expression for the back-shifted excitation energy \cite{ignatyuk79:a}
\begin{equation}
\label{eq:levdens_backshifted_energy}
\tilde{E}_{\rm intr} = E_{\rm intr} + \left(1-e^{-E_{\rm intr}(\boldsymbol{\chi})/E_{\rm d,sh}}\right)E_{\rm sh}
+ \left(1-e^{-E_{\rm intr}(\boldsymbol{\chi})/E_{\rm d,pc}}\right)E_{\rm pc},
\end{equation}
where $E_{\rm d,sh}$ and $E_{\rm d,pc}$ are the damping interval for the shell energy $E_{\rm sh}$ and the pairing energy $E_{\rm pc}$ , respectively.

\subsection{Effective level density}\label{sec:effective_levdens}
Similarly to the back-shifted energy in Eq.\ (\ref{eq:levdens_bfg}), an ``effective'' level density was introduced in Ref.\ \cite{randrup13:a}
to take account of the gradual decrease in microscopic effects as the nuclear excitation energy is raised.
For a nucleus with shape $\boldsymbol{\chi}$ and excitation energy $E^\ast(\boldsymbol{\chi})$,
an effective excitation energy $E^\ast_{\rm eff}(\boldsymbol{\chi})$ is defined as
\begin{equation}
E^\ast_{\rm eff}(\boldsymbol{\chi}) = E^\ast(\boldsymbol{\chi}) + \left[ 1-\mathcal{S}\left(E^\ast(\boldsymbol{\chi})\right) \right]
E_{\rm mic}(\boldsymbol{\chi}),
\end{equation}
where $E_{\rm mic}(\boldsymbol{\chi})$ is the microscopic part of the potential energy.
The suppression function $\mathcal{S}$ is required to be equal to one at excitation energy zero,
and should converge to zero for large $E^\ast(\boldsymbol{\chi})$.
The function employed reads
\begin{equation}
\mathcal{S}\left(E^\ast(\boldsymbol{\chi})\right) = \frac{1+e^{-E_1/E_0}}{1+e^{(E^\ast(\boldsymbol{\chi})-E_1)/E_0}},
\end{equation}
with parameters $E_0=15$ MeV and $E_1=20$ MeV.
The effective level density is obtained by inserting the effective excitation energy
into the simple FG expression,
\begin{equation}
\rho_{\rm eff}(E^\ast) \sim \text{exp}\left(2\sqrt{aE^\ast_{\rm eff}}\right),
\end{equation}
with level-density parameter $a=A/(8\text{ MeV})$.

\section{Combinatorial level density method}\label{sec:combinatorial_levdens}
A method was developed in Ref.\ \cite{uhrenholt13:a} for microscopic
calculations of level densities in deformed nuclei. 
For a specified shape $\boldsymbol{\chi}$, the single-particle levels for protons and neutrons 
are obtained by solving the Schrödinger equation in the associated
folded-Yukawa potential (see Sec.\ \ref{sec:folded_yukawa}).
For each many-body state, blocked BCS calculations for neutrons and protons separately provide 
the energies of the intrinsic many-body states $E_i^{\rm n}(\boldsymbol{\chi})$ and $E_i^{\rm p}(\boldsymbol{\chi})$.

All shapes considered have axial symmetry. Rotation is treated by considering the diagonal contribution from the collective rotation in the particles+rotor model.
This gives the rotational contribution $E^{\text{rot}}_i$ to the energy,
\begin{equation}
E_i^{\text{rot}}(I,\boldsymbol{\chi}) = \frac{(I(I+1)-K_i(\boldsymbol{\chi})^2)}{2\mathcal{J}_i(\boldsymbol{\chi})},
\end{equation}
where $\mathcal{J}_i(\boldsymbol{\chi})$ is the pairing and shape-dependent moment of inertia \cite{bengtsson86:a}.

The total energy of a state is then given by
\begin{equation}
E_i(I,\boldsymbol{\chi}) = E_i^{\rm n}(\boldsymbol{\chi}) + E_i^{\rm p}(\boldsymbol{\chi}) + E_i^{\rm rot}(I,\boldsymbol{\chi}).
\end{equation}
The level density for a fixed angular momentum $I$ is obtained by counting the states $E_i(I,\boldsymbol{\chi})$
in a bin of width $\Delta E$ centered around $E_b$,
\begin{equation}
\rho(E_b,I)=\frac{1}{\Delta E}\int^{E_b+\Delta E/2}_{E_b-\Delta E/2}\sum_i\delta(E-E_i(I,\boldsymbol{\chi}))dE.
\end{equation}
The bin width is in the present studies taken as $\Delta E=200\text{ keV}$. 

\subsection{Extrapolation to high excitation energies}\label{sec:levden_extrapolation}
Since it is too time consuming to calculate level densities at all required excitation energies,
an approximation is performed. 
The calculated combinatorial level density is employed up to an excitation energy $E^\ast\approx6$ MeV
and then smoothly continued upwards by an analytical expression.
The following formula was used in Ref.\ \cite{ward17:a},
\begin{equation}
\label{eq:extrapol_old}
\rho(E^\ast(\boldsymbol{\chi}),I)=C(\boldsymbol{\chi})\tilde{E}_{\text{intr}}^{-3/2}\text{exp}\left(2\sqrt{a\tilde{E}_{\text{intr}}}\right),
\end{equation}
where $\tilde{E}_{\text{intr}}$ is the back-shifted intrinsic excitation energy in Eq.\ (\ref{eq:levdens_backshifted_energy}),
and where the level-density parameter is expressed as $a=A/e_0$.
The constant $C(\boldsymbol{\chi})$ was determined by continuity with the corresponding microscopic value at the matching energy 
$E^\ast(\boldsymbol{\chi})=\text{5.9 MeV}$.
The three parameters $e_0$, $E_{\text{d,sh}}$, $E_{\text{d,pc}}$ were determined by a nonlinear least-squares fit of 
the logarithm of the extrapolated values to the corresponding microscopically calculated level densities for $^{236}\text{U}$.

For high excitation energies, the level densities should however approach the FG expression in Eq.\ (\ref{eq:levdens_bfg}).
In order to obtain the correct asymptotic behavior of the level densities
the extrapolation was therefore slightly modified to the following expression
\begin{equation}
\label{eq:extrapol_new}
\begin{aligned}
&\rho(E^\ast(\boldsymbol{\chi}),I)= \\
=&\left(1+C(\boldsymbol{\chi})e^{-E^\ast(\boldsymbol{\chi})/E_{\text{d,sh}}}\right) 
\frac{2I+1}{48}\left(\frac{\hbar^2}{2\mathcal{J}}\right)^{1/2} \tilde{E}_{\text{intr}}^{-3/2}\text{exp}\left(2\sqrt{a\tilde{E}_{\text{intr}}}\right),
\end{aligned}
\end{equation}
where the asymptotic value is approached at the same rate as the damping of the shell effects.


\part{\color{red}{Fission and fusion}}\label{part:fission}

\chapter{Potential-energy surfaces}\label{ch:fission_barriers}
As already recognized shortly after its discovery, the fission process can be described as an evolution of the nuclear shape from
a roughly spherical shape into two separate fragments. 
One therefore has to calculate the potential energy for the various shapes the nucleus can undertake on its way to fission.

There are two ways to construct these potential-energy surfaces. 
One method is to solve a self-consistent mean-field problem with appropriate constraints on the nuclear shape (see e.g. Ref.\ \cite{schunck16:a}).
The other method, which is the one followed here, is to first define a class of shapes that are supposed to contain the relevant shapes
and then employ the macroscopic-microscopic method to calculate the potential energy of those shapes.

\section{Liquid-drop model}
The most basic features of fission can be understood by considering only the macroscopic part of the energy
as in Sec.\ \ref{sec:ldm_model}.
Since only the Coulomb term and the surface term 
depend on the deformation,
the total deformation-dependent energy is then simply the sum of these two terms, 
\begin{equation}
E(\alpha_2)=E_{\rm S}(\alpha_2)+E_{\rm C}(\alpha_2),
\end{equation}
where the deformation parameter $\alpha_2$ characterizes the elongation of the fissioning nucleus.
The deformation-dependence of the Coulomb energy and the surface energy is shown in Fig.\ \ref{fig:ldm_barrier}(a) for nucleus $^{236}$U.
The surface area is minimal for a sphere, while deformed shapes have larger surfaces. Therefore, the surface energy increases with deformation.
The Coulomb energy is on the other hand minimal when the nucleons are far apart and maximal for spherical shapes. Therefore, the Coulomb energy decreases with deformation.
The counteracting effects of the two terms give rise to a fission barrier in the potential energy as a function of the elongation shape parameter
(Fig.\ \ref{fig:ldm_barrier}(b)). 
(Note the difference in scale in the two figures.)
The fission barrier which the nucleus has to overcome in order to fission, has a maximum which is referred to as the saddle point.
Beyond the saddle, the repulsive Coulomb force between the nascent fragments overtake the surface tension.

\begin{figure}[b] 
\centering    
\includegraphics[width=0.8\textwidth]{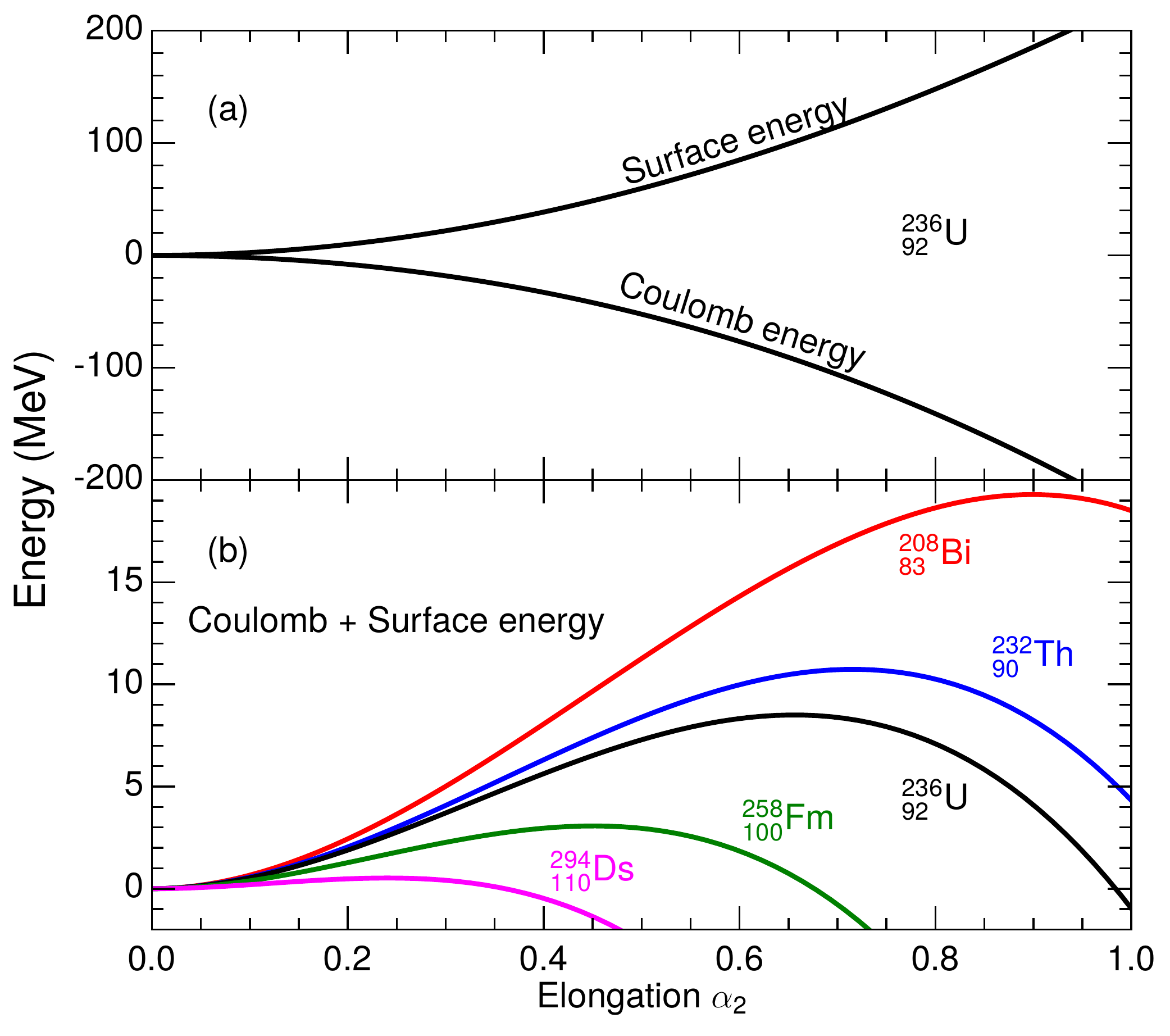}
\caption[Fission barriers with LDM]{
Surface energy and Coulomb energy in the LDM for nucleus $^{236}$U (a) 
and the sum of the both terms for different nuclei (b)
as a function of elongation coordinate $\alpha_2$.
}
\label{fig:ldm_barrier}
\end{figure}

The fissioning nucleus usually start with a small deformation (close to spherical). 
In order to fission, the nucleus then has to pass the barrier. 
This can occur either spontaneously, where the nucleus tunnels through the
barrier, or through induced fission where the nucleus gain energy from an inducing
particle so that it can overcome the barrier.
The point when the nucleus splits into two fragments is 
called the scission point.

The surface energy and the Coulomb energy also have different dependencies on
the number of nucleons, which means that the fission barrier is different for each
nucleus as shown in Fig.\ \ref{fig:ldm_barrier}(b). 
The fission barrier is rather large for light nuclei, which are therefore very stable against fission.
As one moves to heavier nuclei the fission barrier becomes ever smaller in both height and width.
Accordingly, it will require correspondingly less excitation 
for the nucleus to overcome the barrier, as well as tunnel through the barrier,
and undergo fission. 
Ultimately, for sufficiently heavy nuclei, the stability against deformation is lost altogether.

\section{Finite-range liquid-drop model}
\label{sec:potsurf_frldm}

\begin{figure}[b]
\centering    
\includegraphics[width=1.0\textwidth]{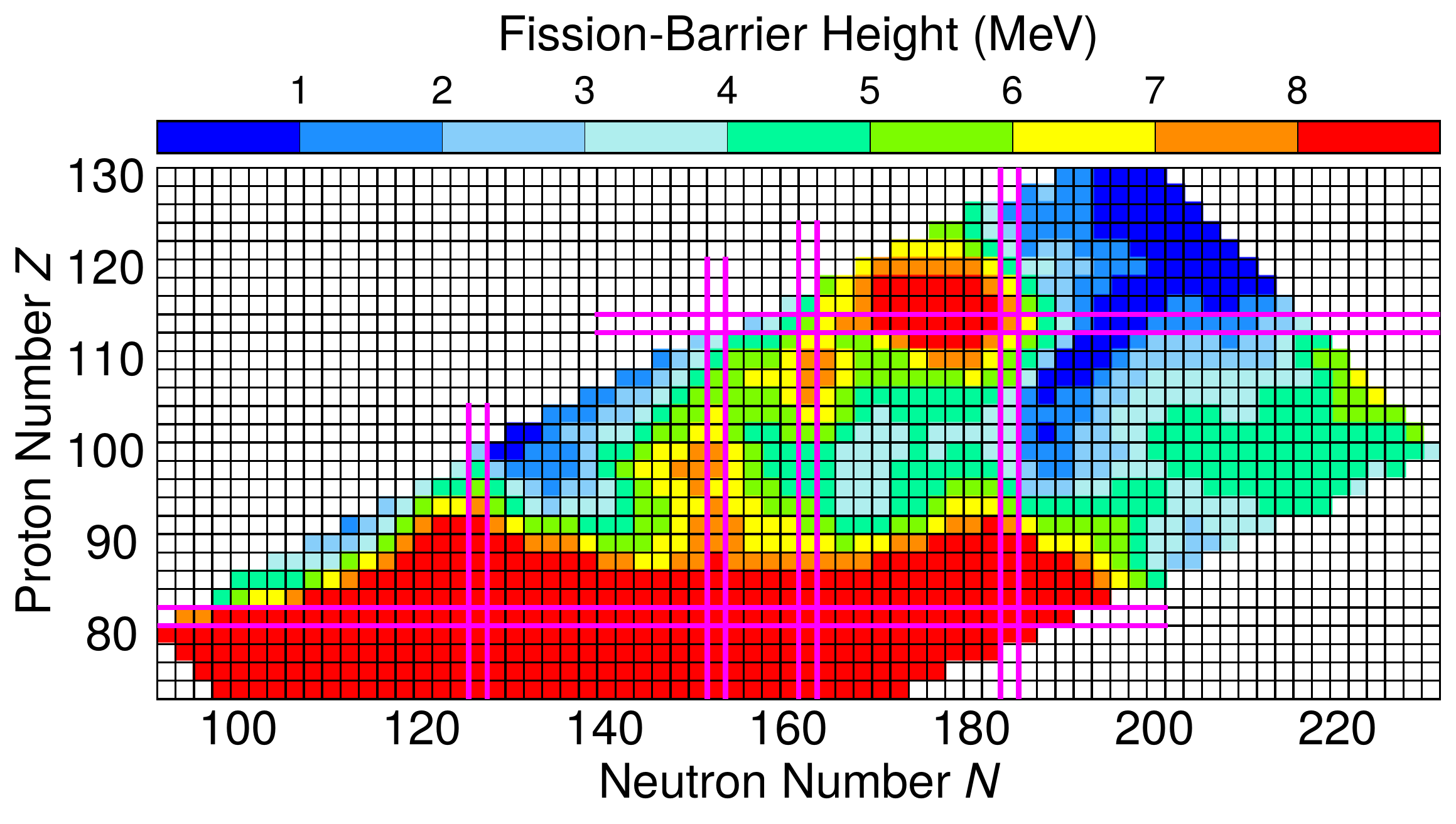}
\caption[Chart of fission-barrier heights in FRLDM]{
Fission-barrier heights versus proton and neutron number calculated within the FRLDM.
Pairs of magenta parallel lines indicate magic neutron and proton numbers in the model ($N=126,184$ and $Z=82,114$).
See Ref.\ \cite{moller15:a} for further details.
}
\label{fig:fission_barriers_chart}
\end{figure}

Despite the success of the LDM, it has clear limitations to explain experimental observations,
in particular the asymmetric mass-division seen in certain nuclei.
The smooth macroscopic deformation energy in the LDM is modified as a result of the shell effects, 
leading to more complicated structures in the barriers.
It can in particular lead to a double-humped barrier with a secondary minimum in which a long-lived shape isomeric state may be hosted.
Microscopic effects, in particular those that lower the ground-state energy, 
can also increase the barrier for superheavy nuclei, thus giving rise
to the possibility for those nuclei be more stable against fission.
It was furthermore proposed that certain experimental data could best be explained if the
barrier exhibited a triple-humped structure \cite{moller74:a}.
In the early studies of fission,
the potential energy was usually calculated only for a few hundred nuclear shapes.
Several million shapes are however needed to fully account for all possible fission paths.

In the present studies, the potential energy $U(\boldsymbol{\chi})$ is calculated within the FRLDM 
in a grid of more than 6 million shapes given in the 3QS parametrization, as described in Ch.\ \ref{ch:mac_mic_models}.
Several potential-energy studies within the FRLDM can be found in Refs.\ \cite{moller01:a,moller04:a,ichikawa09:a,moller09:a,ichikawa12:a,moller15:a,ichikawa19:a}.
The five-dimensional potential-energy landscape of a nucleus can be analyzed with the immersion techniques described in Ref.\ \cite{moller09:a}
in order to identify various saddle points.
Figure \ref{fig:fission_barriers_chart} shows calculated fission-barrier heights for even-even nuclei in the FRLDM \cite{moller15:a}.
Higher barrier heights are seen around $^{252}_{100}\text{Fm}_{152}$, $^{270}_{108}\text{Hs}_{162}$, and $^{298}_{114}\text{Fl}_{184}$
due mostly to strong ground-state shell effects.

\subsection{Fission modes}\label{sec:fission_modes}
In order to explain the observation that fission could occur both
in a symmetric and an asymmetric way,
Turkevich and Niday \cite{turkevich51:a} proposed in 1951
two different channels through which fission occurs.
These are referred to as fission modes and
differ both in mass and kinetic energy of the resulting fragments.

Additional fission modes were further introduced by Brosa \cite{brosa90:a} where values of the total kinetic energy (TKE) of the fragments
are associated with different shapes at scission.
The asymmetric mode is denoted the standard (St) mode since it is the dominating mode in low-energy fission of actinides (see Sec.\ \ref{ch:mass_yields}).
The St mode is further split into two modes called Standard 1 (St1) and Standard 2 (St2). 
They are ascribed to two different shell effects in the heavy fragment;
the St1 mode is attributed to the doubly magic nucleus $^{132}$Sn,
while the St2 mode is attributed to the deformed magic neutron number $N=88$ \cite{wilkins76:a}.
The symmetric mode is called superlong (SL) mode, where the name refers to a more elongated scission shape and thus lower TKE.

\begin{figure}[t]

 \begin{center} 
 \includegraphics[width=0.9\textwidth]{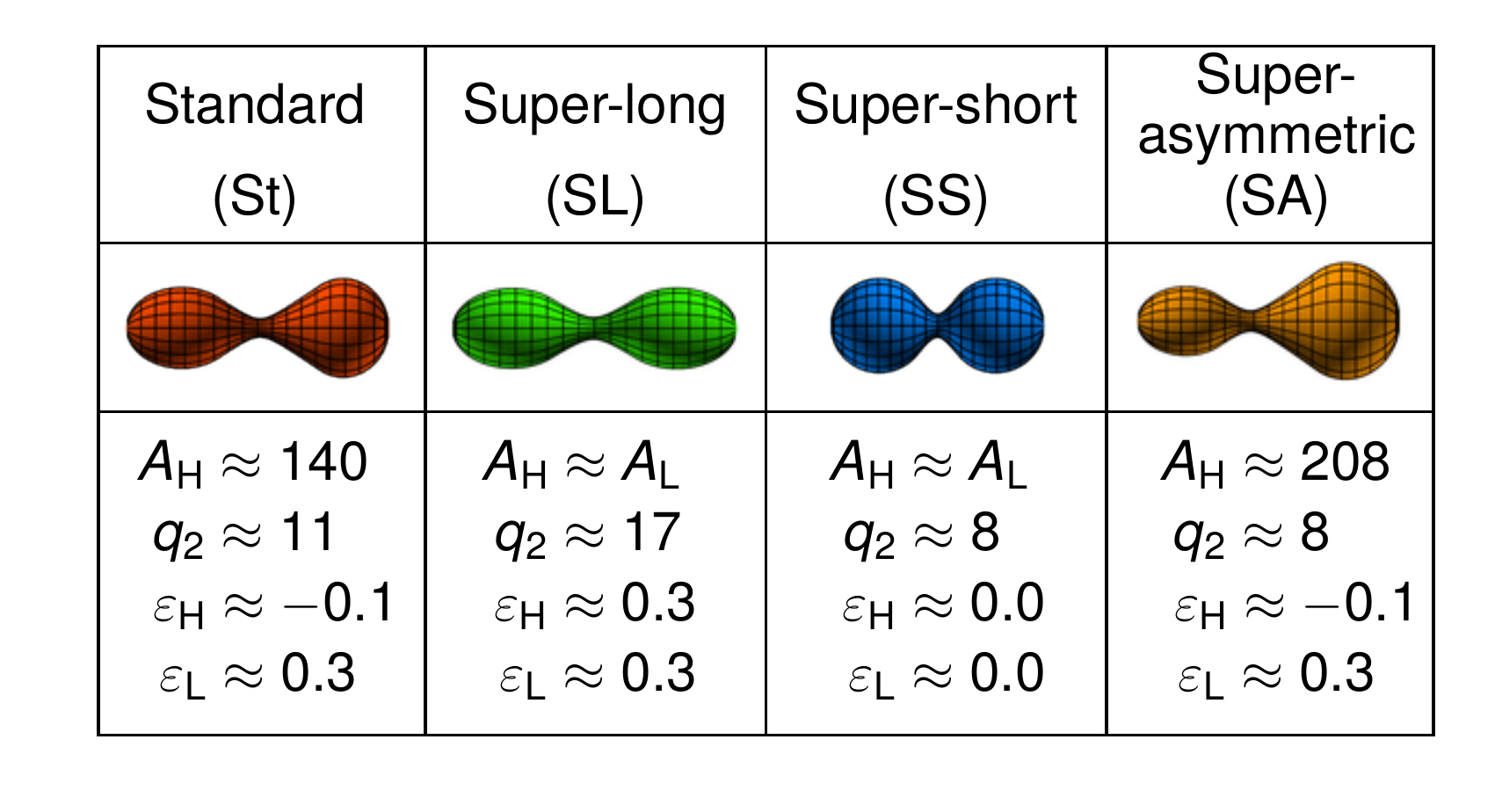} 
 \caption[Fission-mode shapes]{
 Typical shapes at scission of fission modes identified in the calculations.
Also given are typical scission values of the elongation $q_2$ of the fissioning nucleus and
deformations $\varepsilon_{\rm L/H}$ and mass numbers $A_{\rm L/H}$ of the fragments.
 }
\label{fig_shapes_modes}  
 \end{center} 
 \hspace{16mm}
 \vspace{-7mm}
\end{figure}

For fission of nuclei around $^{258}$Fm it was observed that it mainly split symmetrically with a very high TKE.
This was argued to arise due to both fragments being close to doubly magic nucleus $^{132}$Sn,
resulting in a very compact scission shape and thus a high TKE value \cite{hulet89:a}.
This type of fission is therefore said to occur in a super-short (SS) fission mode.

In the calculations presented here it is also obtained that nuclei in the superheavy region, $106\leq Z\leq 114$ and $162\leq N\leq 176$
fission in a very asymmetric way, where the heavy fragment is found to be close to $^{208}$Pb with a corresponding light fragment (see Sec.\ \ref{ch:mass_yields}). 
This type of fission is denoted as a 
super-asymmetric (SA) mode\footnote{The term ``super-asymmetric mode'' is also used by Brosa but in fission of $^{252}$Cf where the heavy-fragment mass number is 161 \cite{brosa90:a}.}.

Figure \ref{fig_shapes_modes} shows typical shapes at scission for the fission modes identified in the calculations (see Ch.\ \ref{ch:scission_quantities}).
Typical values are also given for the elongation $q_2$ of the fissioning nucleus and
deformations $\varepsilon_{\rm L/H}$ and mass numbers $A_{\rm L/H}$ of the fragments.
In the calculations we do not distinguish between the St1 and St2 modes.

The different modes generally follow different paths after the first saddle of the potential-energy surface.
Beyond the fission isomer, different ``fission valleys'' can be determined in
the calculated potential-energy surface \cite{moller09:a}. 
This is done by determining, for each value of elongation coordinate $q_2$, minima
in the corresponding restricted space in the four other shape coordinates.
Valleys are then defined as a sequence of similar minima that persist for successive $q_2$ values
where the shape coordinates change gradually between neighbouring $q_2$ values.

\subsection*{$^{236}$U}\label{sec:fission_valleys_u236}
Figure \ref{fig:epot_vs_q2_236u} shows two fission valleys for the nucleus $^{236}$U as a function of elongation $q_2$.
The asymmetric valley (red curve) and the symmetric valley (green curve) exhibit similar values
for the fragment masses and deformations to that of the St mode and the SL mode, respectively.
The two valleys are separated by a ridge (black curve with triangles).
The ground state is located at $q_2\approx0.7$ (not shown)
with energy $E_{\rm gs}$ (horizontal dot-dashed line).
The asymmetric St path exhibit a steep slope beginning around $q_2\approx9$, where the neck rapidly decreases.
Scission in the St mode then typically occurs slightly further out in elongation at $q_2\approx11$.
The symmetric SL path continues further out in $q_2$ until the neck starts to shrink;
fission along this path therefore occur at a larger elongation around $q_2\approx17$.

\begin{figure}[hbt!]
\centering    
\includegraphics[width=0.7\textwidth]{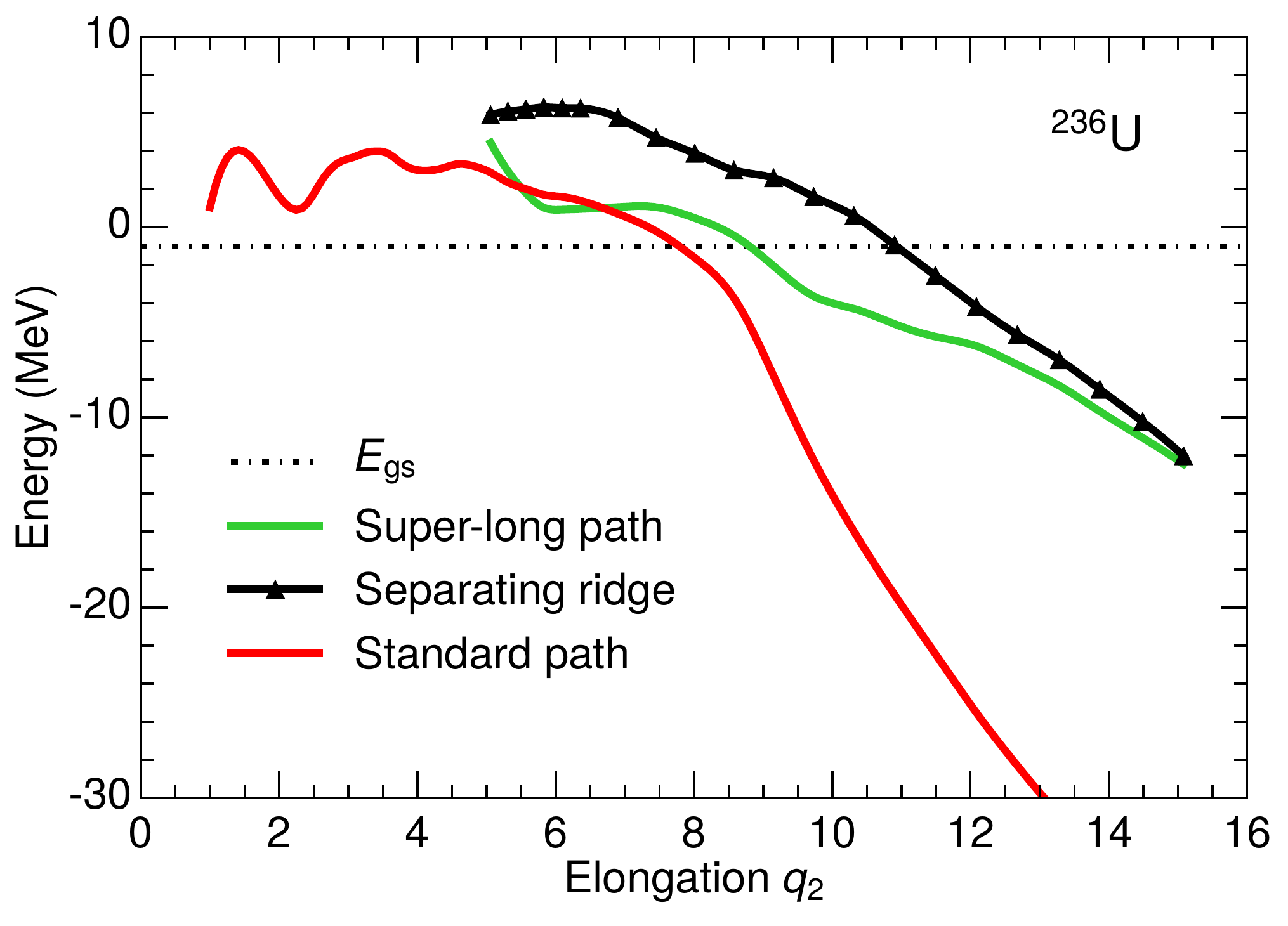}
\caption[Fission valleys in $^{236}$U]{
The potential energy along the fission paths for the St mode (solid red) and the SL mode (solid green),
shown as function of the elongation $q_2$ for $^{236}$U.
The ridge separating the two paths is shown by the black curve with triangles.
The ground state is located at $q_2\approx0.7$ (not shown)
with energy $E_{\rm gs}$ (horizontal dot-dashed line).
}

\label{fig:epot_vs_q2_236u}
\end{figure}

\begin{figure}[t]
\centering    
\includegraphics[width=0.9\textwidth]{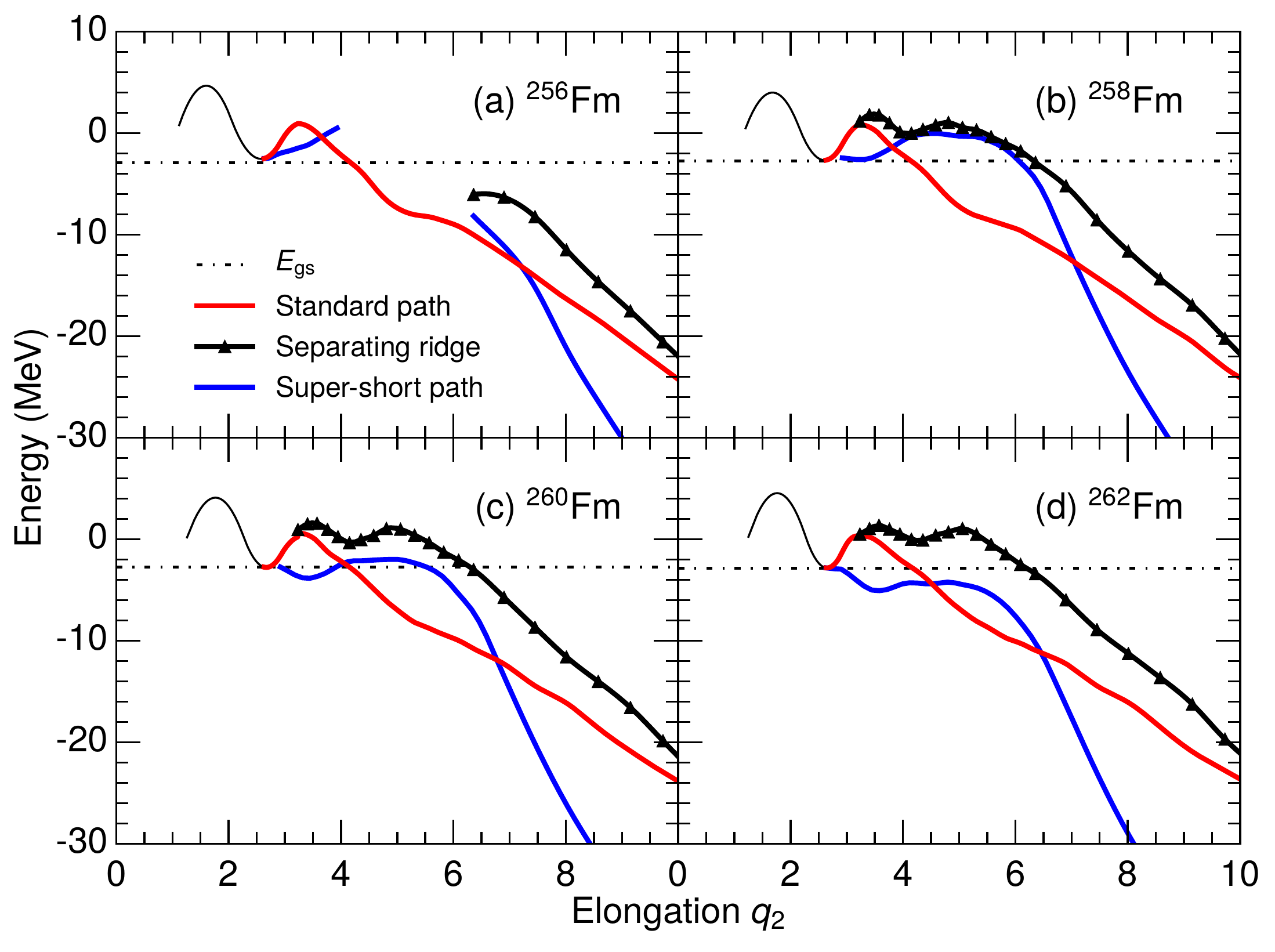}
\caption[Fission valleys in $^{256,258,260,262}$Fm]{
The potential energy along the fission paths for the St mode (solid red) and the SS mode (sold blue),
shown as function of the elongation $q_2$ for 
(a) $^{256}$Fm, (b) $^{258}$Fm, (c) $^{260}$Fm and (d) $^{262}$Fm.
The ridge separating the two paths is shown by the black curve with triangles.
The ground-state energy is shown by the horizontal dot-dashed line. 
The ground-state minimum ($q_2\approx0.75$) is separated by the first
barrier (black thin line) from the isomeric minimum ($q_2\approx2.5$).
}

\label{fig:epot_vs_q2_fm}
\end{figure}

\subsection*{$^{256-262}$Fm}\label{sec:fission_valleys_fm}
Figure \ref{fig:epot_vs_q2_fm} shows fission valleys for even $^{256-262}$Fm as a function of elongation $q_2$.
The thin black line shows the first saddle connecting the ground state (not shown)
and the $2^{\rm nd}$ minimum.
The symmetric SS path (blue curves) and the asymmetric St path (red curves)
exhibit similar fragment masses and deformations as the SS mode and the St mode, respectively.
The St valley is observed in all four isotopes,
while the SS valley is fully developed first in $^{258}$Fm and heavier isotopes.
The potential energy in the SS valley dramatically decreases already at $q_2\approx6$
with an associated shrinking of the neck radius.
This valley therefore leads to a very compact scission shape.
The SS valley in $^{258}$Fm is rather fragile since the ridge to the St valley
is less than 1 MeV at $q_2\approx4$.
The depth of the SS valley gradually increases when approaching $^{264}$Fm,
which corresponds to a split with exactly two doubly-magic $^{132}$Sn.

\subsection*{$^{274}$Hs}\label{sec:fission_valleys_hs274}
Figure \ref{fig:epot_vs_q2_274hs} shows fission valleys for $^{274}$Hs as a function of $q_2$;
the very asymmetric SA valley (orange curve) and a symmetric valley (dashed black curve).
The SA valley corresponds to a spherical heavy fragment near $^{208}$Pb
and a prolate deformed light fragment.
The neck has almost vanished for $q_2\gtrsim6$
and the valley approximately corresponds to a pure Coulomb interaction between the two fragments.
This valley is similar to the ``fusion valleys'' discussed for neighbouring nuclei in Ref.\ \cite{ichikawa05:a}
and in Ch.\ \ref{ch:fusion}.

\begin{figure}[t!] 
\centering    
\includegraphics[width=0.7\textwidth]{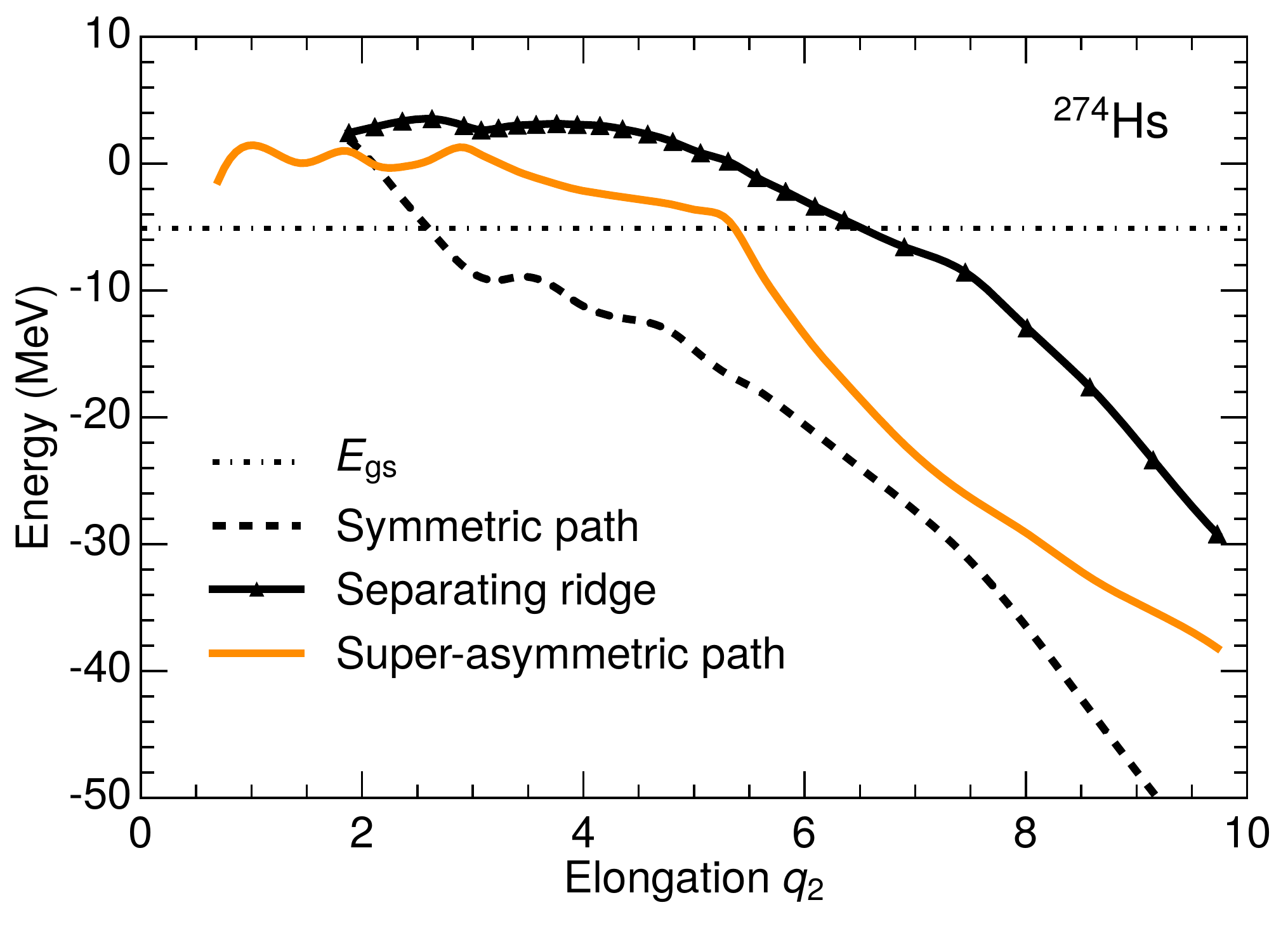}
\caption[Fission valleys in $^{274}$Hs]{
The potential energy along the fission paths for the SA mode (solid orange) and 
the symmetric path (dashed black),
shown as function of the elongation $q_2$ for $^{274}$Hs.
The ridge separating the two paths is shown by the black curve with triangles.
The ground state is located at $q_2\approx0.5$ (not shown)
with energy $E_{\rm gs}$ (horizontal dot-dashed line).
}
\label{fig:epot_vs_q2_274hs}
\end{figure}

\chapter{Fission dynamics}\label{ch:fission_dynamics}
Although many attributes of the fission process can be understood by studying the potential-energy landscape of the fissioning nucleus,
nuclear fission is the result of a complicated dynamical evolution of a small many-body system.
Much progress have been made within microscopic self-consistent approaches,
though such calculations are computationally rather heavy (see e.g. Ref.\ \cite{schunck16:a} for a review).

The Metropolis random-walk approach employed in the present studies offers an efficient method to simulate the fission process, 
avoiding the complications of dealing with the time-dependence explicitly.
It is based on the classical Langevin formalism, which is a stochastic differential equation describing
the time evolution of the macroscopic degrees of freedom of a system.
It is suitable when the time scale of the macroscopic degrees of freedom
is much larger than the microscopic degrees of freedom.

\section{Stochastic dynamics formalism}

\subsection{The Langevin equation}
The Langevin description of fission is based on the assumption that 
the system exhibits
two different time scales;
one being associated with the slow motion of the collective shape degrees of freedom 
and the other with the rapid motion of the intrinsic degrees of freedom.
Let the $N$-dimensional variable $\boldsymbol{\chi}=\{\chi_i\}$ specify a particular nuclear shape ($N=5$ in the present studies).
We then want to determine how $\boldsymbol{\chi}$ evolves from a near-spherical shape of the fissioning nucleus towards 
two separated fragments. 
The local gradient of the potential energy $U(\boldsymbol{\chi})$ provides a driving force, 
$\mathbf{F}(\boldsymbol{\chi})=-\partial U(\boldsymbol{\chi})/\partial\boldsymbol{\chi}$, 
that will seek to change the shape so that the potential energy is minimized.
A shape change is also associated with a rearrangement of the nucleons inside the nucleus. 
The associated kinetic energy is assumed to be of normal form, $K=\frac{1}{2}\sum^N_{ij}M_{ij}\dot{\chi}_i\dot{\chi}_j$, where $M_{ij}(\boldsymbol{\chi})$ 
is the $N\times N$ inertial-mass tensor. 

If there were no coupling between the shape variables $\boldsymbol{\chi}$ and the internal degrees 
of freedom, then the shape dynamics would be conservative. The time evolution of the collective shape then follows from the Lagrangian function,
\begin{equation}
\mathcal{L}(\boldsymbol{\chi},\dot{\boldsymbol{\chi}})=\frac{1}{2}\sum_{ij}M_{ij}(\boldsymbol{\chi})\dot{\chi}_i\dot{\chi}_j-U(\boldsymbol{\chi}),
\end{equation}
and the associated collective momentum is given by
\begin{equation}
p_i(\boldsymbol{\chi},\dot{\boldsymbol{\chi}})=\frac{\partial}{\partial \chi_i}\mathcal{L}(\boldsymbol{\chi},\dot{\boldsymbol{\chi}}).
\end{equation}

However, the interaction between the macroscopic and the microscopic degrees of freedom causes the shape variables to experience a 
dissipative force characterized by the $N\times N$ dissipation tensor $\boldsymbol{\gamma}(\boldsymbol{\chi})$.
It consist of a friction force 
$-\sum_j\gamma_{ij}\dot{\chi}_j(t)$ 
and a stochastic term $\xi_i(\boldsymbol{\chi},\dot{\boldsymbol{\chi}},t)$.

The equation of motion for the time evolution of the nuclear shape is then obtained by equating the rate of momentum change with the forces acting,
 \begin{equation}
 \label{eq:langevin}
 \dot{p}_i = -\frac{\partial}{\partial\chi_i} U(\boldsymbol{\chi}) + \frac{1}{2}\sum_{jk}\dot{\chi}_j\dot{\chi}_k\frac{\partial}{\partial\chi_i}M_{jk}(\boldsymbol{\chi}) 
 - \sum_j\gamma_{ij}\dot{\chi}_j(t) + \xi_i(\boldsymbol{\chi},\dot{\boldsymbol{\chi}},t).
 \end{equation}
The calculation of the time evolution of the nuclear shape parameters $\boldsymbol{\chi}(t)$ then requires knowledge of three distinct quantities:
$U(\boldsymbol{\chi})$, $\mathbf{M}(\boldsymbol{\chi})$ and $\boldsymbol{\gamma}(\boldsymbol{\chi})$.
The Langevin equation in Eq.\ (\ref{eq:langevin}) can be solved directly by starting from the specified initial state 
and generating a large number of shape evolutions (see e.g. Refs.\ \cite{sierk17:a,usang19:a}).

\subsection{Collective inertia}
The mass tensor $M_{ij}$ describes the inertia of a nucleus with respect to changes of its deformation.
A microscopic method for calculating the mass tensor is the Generator coordinate method \cite{schunck16:a}.

For a fission path described by a parameter $s$, the collective inertia is given by
\begin{equation}
\label{eq:inertia_general}
M(s)=\sum_{ij}M_{ij}(s)\frac{d\chi_i}{ds}\frac{d\chi_j}{ds}.
\end{equation}
A common approximation of the mass tensor is to simply assume that the flow is incompressible and irrotational. 
A semi-empirical formula for the inertial mass in the elongation direction 
for calculating SF half-lives was derived in Ref.\ \cite{randrup76:a}
\begin{equation}
\label{eq:inertia}
M(r)=\mu\left(1+k\frac{17}{15}\text{exp}\left[-a\left(r-\frac{3}{4}\right)\right]\right),
\end{equation}
where $r$ is the center-of-mass distance (in terms of the radius for the spherical nucleus $R_0=1.2A^{1/3}$) between the two halves.
This expression decreases with increasing values of $r$ and approaches the reduced mass $\mu=M_{\rm f1}M_{\rm f2}/(M_{\rm f1}+M_{\rm f2})$ 
appropriate to separated fragments at large distances.
For small values of $r$, the inertia is expected to be higher than the hydrodynamical 
irrotational-flow result due to microscopic effects
associated with single-particle level crossings. 
The asymptotic value at small $r$ is then taken into account by relating the inertia $M(r)$ to the inertia $M^{\rm irr}$ corresponding
to irrotational flow.
The parameter $k$ accounts for the increase of the inertia above the
hydrodynamical value, while the constant $a$ determines how fast the inertia approaches the asymptotic value.

Nuclei near $^{258}$Fm that fission in the SS mode reach scission at a much smaller center-of-mass distance $r$ than
the nuclei considered for Eq.\ (\ref{eq:inertia}).
Correspondingly, 
the reduced mass should be reached at a smaller $r$.
A modified expression was therefore used in Ref.\ \cite{moller94:b} to account for this effect
\begin{equation}
\label{eq:inertia_fm}
M(r)=\mu\left(1+f(r,r_{\rm sc})k\frac{17}{15}\text{exp}\left[-a\left(r-\frac{3}{4}\right)\right]\right),
\end{equation}
where  
\begin{equation*}
f(r,r_{\rm sc}) = \begin{cases}
\left(\frac{r_{\rm sc}-r}{r_{\rm sc}-0.75}\right)^2 & r\leq r_{\rm sc}, \\
0 & r\geq r_{\rm sc},
\end{cases}
\end{equation*}
and where $r_{\rm sc}=1.59$ correspond to the center-of-mass distance for two touching spheres at scission.

The expressions for the inertial masses in Eqs.\ (\ref{eq:inertia}) and (\ref{eq:inertia_fm}) are used in 
calculations of tunneling probabilities as described in Sec.\ \ref{sec:spontaneous_fission},
where the values $k=16$ and $a=\frac{128}{51}$ are employed \cite{moller94:b}.

\subsection{Brownian shape dynamics}
Since the dissipation associated with the nuclear shape dynamics is relatively strong, 
the shape changes will be relatively slow.
It might then be reasonable to ignore inertial forces altogether by putting the inertias to zero.
This assumption of overdamped motion is supported by
recent time-dependent density functional calculations of the whole
fission process \cite{bulgac16:a,bulgac19:a,bulgac19:b}, 
while other calculations describe a fading away of the dissipation in the last stages of fission, when a thin
neck develops towards scission \cite{tanimura15:a}.
The equation of motion is then reduced to
\begin{equation}
0=-\partial U(\boldsymbol{\chi})/\partial\boldsymbol{\chi}-\boldsymbol{\gamma}(\boldsymbol{\chi})\cdot\dot{\boldsymbol{\chi}}
+\boldsymbol{\xi}(\dot{\boldsymbol{\chi}},\boldsymbol{\chi},t).
\end{equation}
This is the Smoluchowski limit describing Brownian motion.
It is however more complicated than the usual Brownian motion; it occurs in $N$ dimensions, 
the medium is anisotropic ($\boldsymbol{\gamma}$ is not diagonal) and non-uniform ($\boldsymbol{\gamma}$ depends on $\boldsymbol{\chi}$), 
and the body is situated in an external potential, $U(\boldsymbol{\chi})$.

\section{Metropolis walk method}
Studies \cite{randrup11:b} suggest that the fission-fragment charge yield is rather insensitive to anisotropies in $\boldsymbol{\gamma}$, 
presumably because a large degree of equilibration takes place in the course of the strongly damped evolution. 
If one assumes that the dissipation tensor is isotropic, i.e. proportional to the unit tensor 
for any shape $\boldsymbol{\gamma}(\boldsymbol{\chi})=\gamma(\boldsymbol{\chi})\mathbf{I}$, 
then the shape evolution can be simulated by a Metropolis walk on the potential energy lattice \cite{randrup11:a}.

For a system with total energy $E_{\rm tot}$,
all of the local excitation energy $E^\ast(\boldsymbol{\chi})=E_{\rm tot}-U(\boldsymbol{\chi})$ 
is assumed to go to the intrinsic degrees of freedom described by the level density $\rho(E^\ast)$.
The steps in the random walk are determined by the Metropolis algorithm. From the current shape $i$ a neighbouring candidate shape $j$
is selected at random. The probability $P_{i\rightarrow j}$ to accept a step is given by
\begin{equation}
\label{eq:metropolis_prob}
P_{i\rightarrow j}=\text{min}\left[1,\rho_j(E^\ast(\boldsymbol{\chi}_j))/\rho_i(E^\ast(\boldsymbol{\chi}_i)))\right],
\end{equation}
where $\rho_{i(j)}(E^\ast)$ is the level density for the shape $\boldsymbol{\chi}_{i(j)}$.

The Metropolis walks are performed on the five-dimensional potential-energy surfaces described in Sec.\ \ref{sec:potsurf_frldm}.
The walks are usually started in the ground state or the isomer minimum.
The asymmetry $\alpha$ is assumed to be frozen in when the neck radius has shrunk to $c_{\rm neck}=2.5$ fm \cite{randrup11:b}. 
Subsequently, the system reaches a scission configuration at $c_{\rm neck}=c_{\rm sc}=1.5$ fm where 
the value of the potential energy at this shape $U(\boldsymbol{\chi})$, the mass asymmetry coordinate $\alpha$, 
the elongation $q_2$, and the two fragment deformations $\varepsilon_{\rm f1}$ and $\varepsilon_{\rm f2}$ are registered.
The choice of $c_{\rm sc}=1.5$ fm is discussed in Sec.\ \ref{sec:tke}. 
The proton and neutron numbers, $Z$ and $N$, are determined by requiring the same $Z/N$ ratio as for the fissioning nucleus.
In the present studies, only fragments with even $Z$ and $N$ are considered.
The random walks are then repeated to obtain the fission-fragment distributions.
Typically $10^5$ number of walks are performed to obtain convergence of the results.

The probabilities determining the steps in Eq.\ (\ref{eq:metropolis_prob}) are evaluated
with the microscopic combinatorial level densities in Sec.\ \ref{sec:combinatorial_levdens}
for $^{235}\text{U}(\text{n},\text{f})$,
while other calculations are evaluated with the effective level densities in Sec.\ \ref{sec:effective_levdens}.
(A comparison between the two methods can be found in Ref.\ \cite{ward17:a}.)

The random walk will usually tend to stay inside the fission barrier for many steps before eventually passing it. 
To speed up the calculations in studies of large number of nuclei,
the potential energy is augmented by a bias term, $V_{\rm bias}=V_0Q^2_0/Q_2^2$ \cite{randrup11:a}, 
where $Q_2$ is the quadrupole moment of the fissioning nucleus and $Q_0$ is the average ground-state quadrupole moment of deformed actinide nuclei.
For small $Q_2$, $V_{\rm bias}$ will encourage increases of $Q_2$, 
while it will have less effect for more deformed shapes closer to scission.
The resulting yields are generally not sensitive to variations in the bias strength
(an exception is the special case $^{258}$Fm where there is a subtle competition between two fission modes).
An alternative formulation of the Metropolis random-walk method was recently introduced in Ref.\ \cite{verriere21:a}, 
which does not require a bias potential to speed up calculations.

\subsection{Spontaneous-fission simulation}\label{sec:spontaneous_fission}
For spontaneous fission (SF) one finds forbidden regions of the elongation $q_2$, 
where the total energy is below the potential energy. 
This problem of forbidden regions is approached in the studies by performing the Metropolis
random walks with the addition of a small amount of energy $\Delta E$, 
so that the walks may pass over the forbidden region. 
On the outer side of the forbidden region, this energy is given back, and the dynamics continues to scission. 
In this way, the random-walk algorithm effectively selects different shape configurations as starting
points after the barrier in the SF simulation.

Most nuclei exhibit a single dominating mode for both SF and for low excitation energies.
The fission paths are then similar in both cases.
However, if there are two competing modes, as in $^{258}$Fm,
the fission path is very sensitive to small changes in energy (see Fig.\ \ref{fig:epot_vs_q2_fm}(b)).
A simple estimate of the tunneling ratio comparing the two fission paths is obtained by 
performing one-dimensional WKB tunneling calculations from the isomeric minimum into the different valleys.
The penetration probability $P$ through the fission barrier is calculated within the WKB formalism \cite{froman65:a} as
\begin{equation}
P=\frac{1}{1+e^{S(L)}},
\end{equation}
where $S(L)$ denotes the action integral along the one-dimensional trajectory $L$,
\begin{equation}
S(L)=\frac{2}{\hbar}\int^{b}_{a}\sqrt{2M(q_2)[U(q_2)-E_{\rm gs}]}dq_2,
\end{equation}
with $a$ and $b$ the entrance and exit point for the tunneling region.
The quantities $U(q_2)$ and $E_{\rm gs}$ denote the potential energy and the ground-state energy, respectively.
The inertial mass $M(q_2)$ for the St and SS paths are calculated with Eqs.\ (\ref{eq:inertia}) and (\ref{eq:inertia_fm}), respectively,
where it is expressed in terms of $q_2$ by a linear fit between $r$ and $q_2$.
This yields that about 75\% of the flux in $^{258}$Fm tunnels to the SS valley and 25\% to the St valley. 
The strong favouring of tunneling to the SS valley is due to the low mass parameter along this tunneling path.

\chapter{Scission quantities}\label{ch:scission_quantities}



\section{Mass distributions}\label{ch:mass_yields}
Figure \ref{fig:mass_yield} shows calculated mass yields for neutron-induced fission of $^{235}$U
for different incident-neutron energies $E_{\rm n}$.
For fission of actinides, such as uranium and plutonium, 
the yield distribution has two peaks showing that the split where one of the fragments is heavier is the most common, according the St mode.
When the energy of the incoming neutron is increased, these peaks decrease while the amount of symmetric yield increases.
This is because the influence of the shell structure, associated with the asymmetric St mode, decreases
with increasing excitation energy.
The yield eventually acquires a Gaussian shape corresponding to a LDM behaviour.

\begin{figure}[t] 
\centering    
\includegraphics[width=0.6\textwidth]{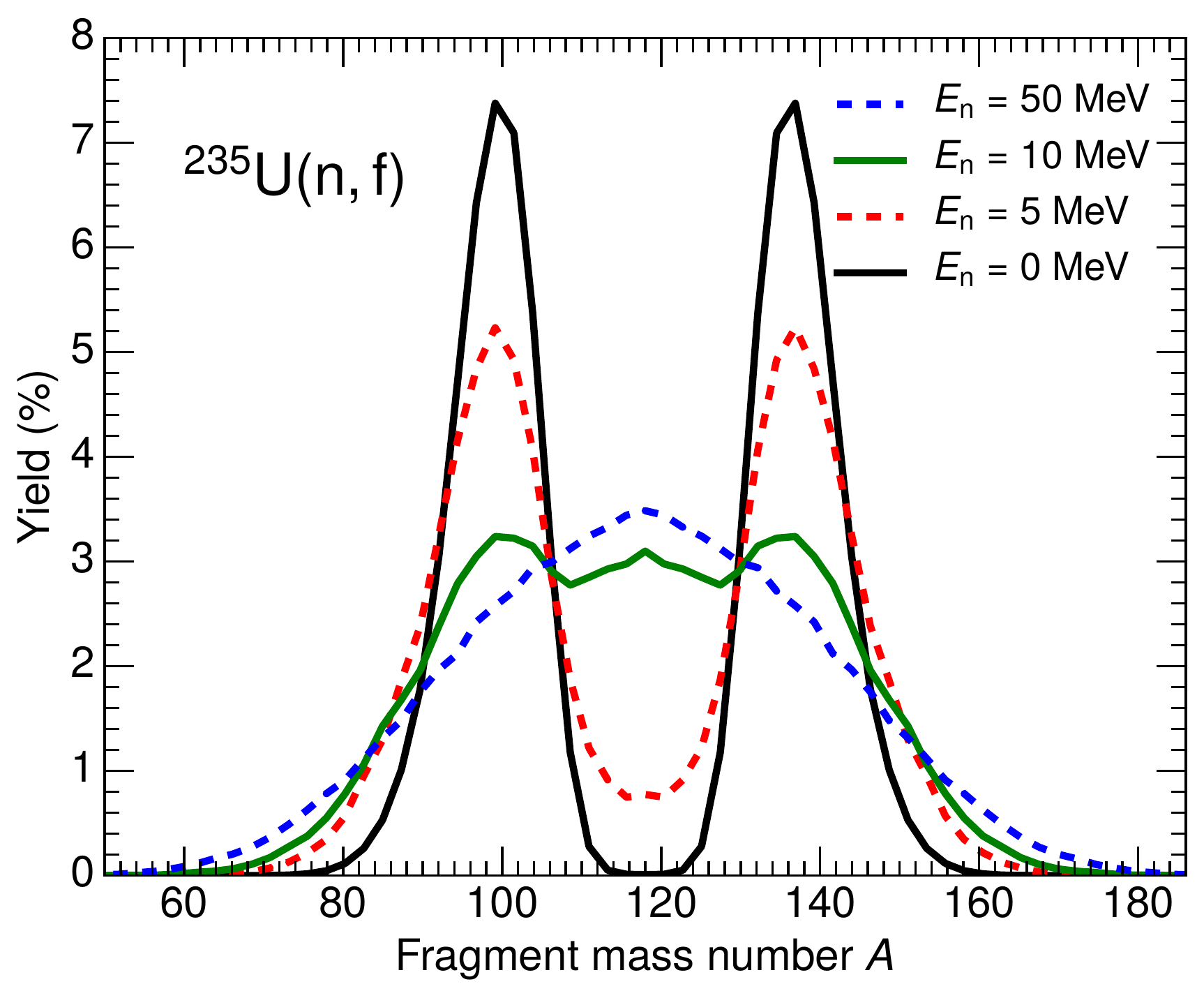}
\caption[Fission-fragment mass yields in $^{235}\text{U}(\text{n},\text{f})$]{
Calculated fission-fragment mass yields in $^{235}\text{U}(\text{n},\text{f})$ for incident-neutron 
energies $E_{\rm n}=0$, 5, 10, and 50 MeV.
}
\label{fig:mass_yield}
\end{figure}

\begin{figure}[t] 
\centering    
\includegraphics[width=1.0\textwidth]{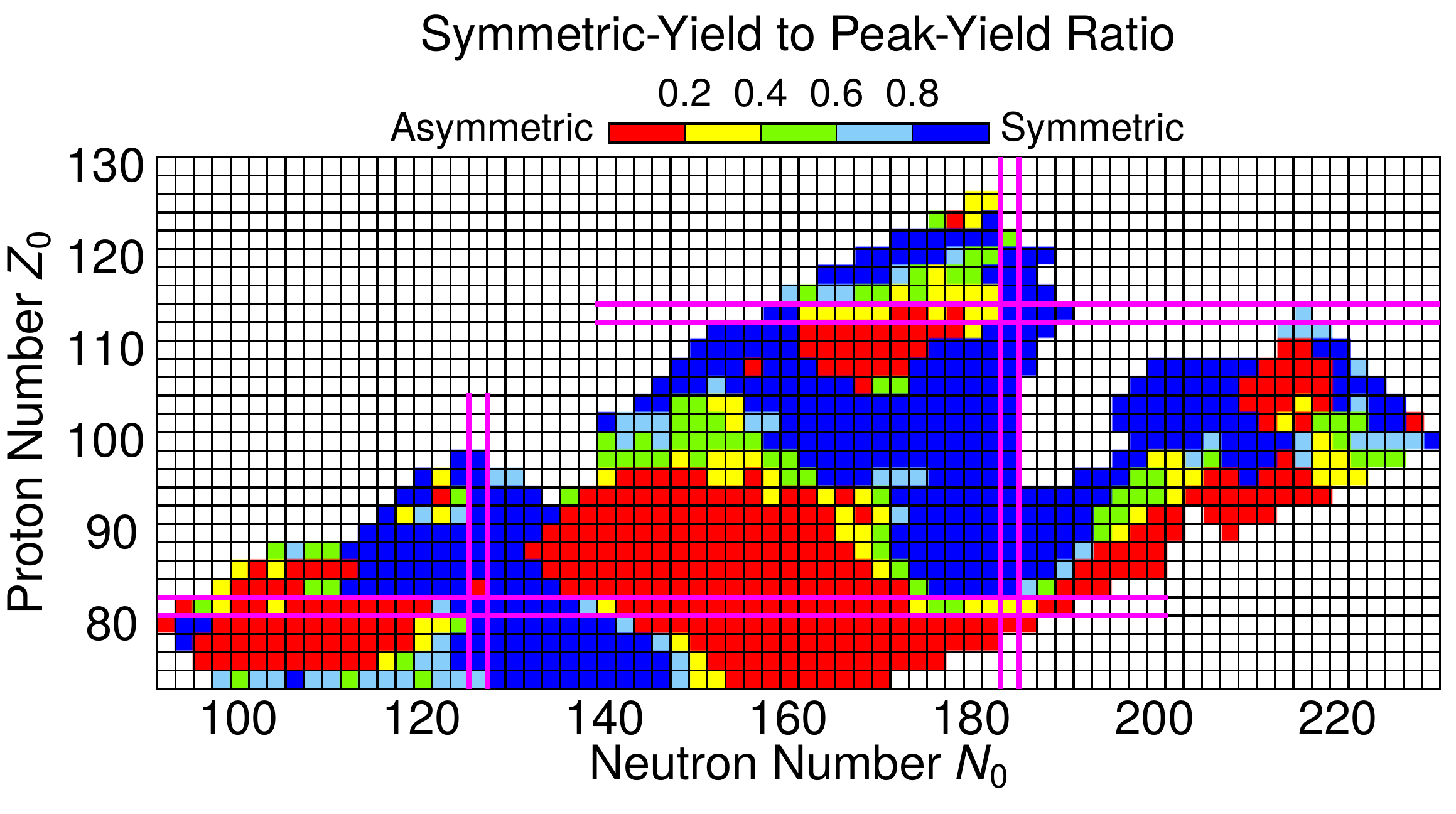}
\caption[Chart of fission-fragment mass yields]{
Calculated symmetric-yield to peak-yield ratios versus $N_0$ and $Z_0$
for fissioning nuclei between the proton and neutron
drip lines and $74 \le Z_0 \le 126$, for even-even nuclei. Nuclei with barriers
calculated to be lower than 3 MeV are not included (cf. Fig.\ \ref{fig:fission_barriers_chart}).
Pairs of magenta parallel lines indicate magic neutron and proton numbers in the model 
($N=126,184$ and $Z=82,114$). The figure is taken from Paper IV.
}
\label{fig:yield_chart}
\end{figure}

\begin{figure}[t] 
\centering    
\includegraphics[width=0.6\textwidth]{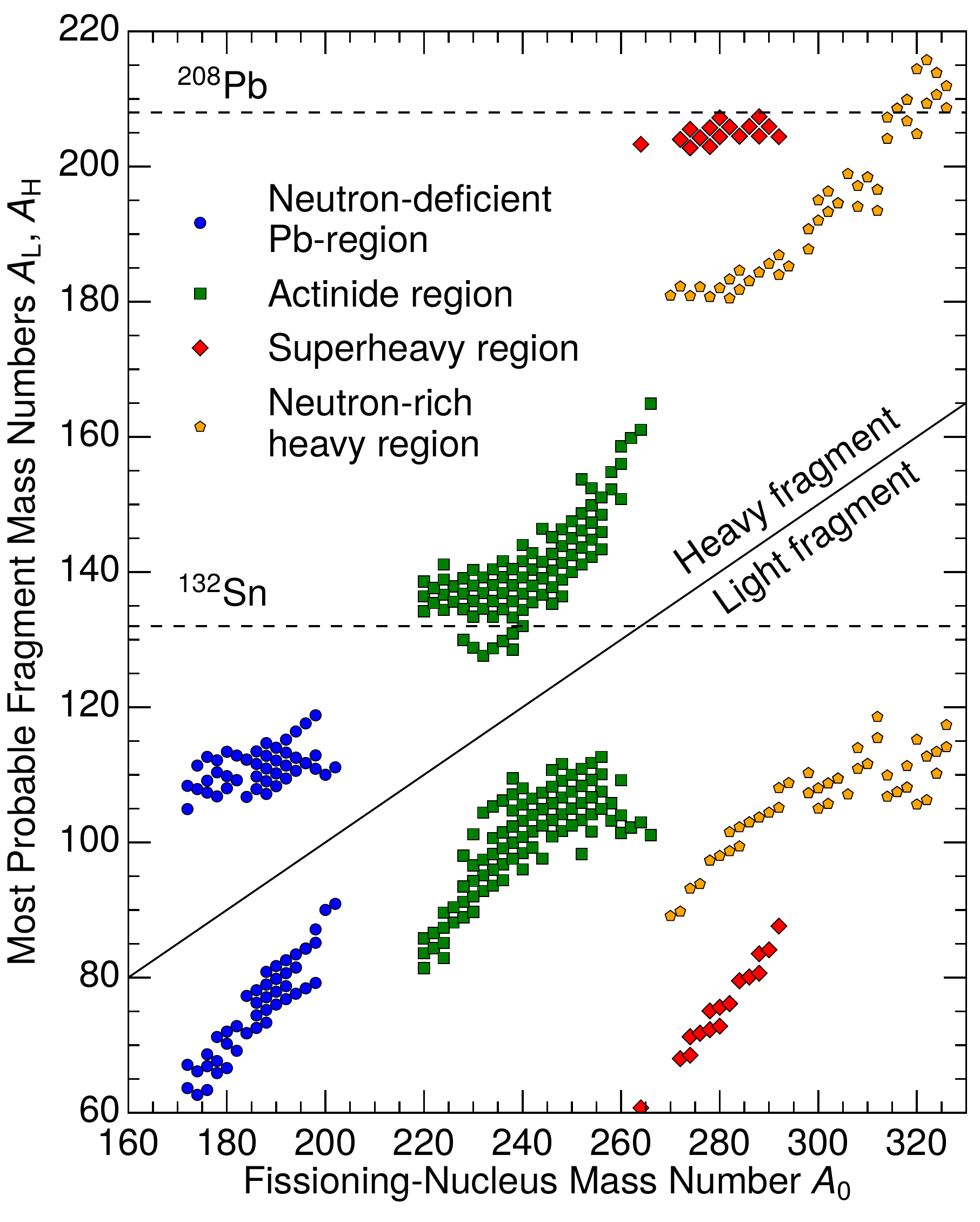}
\caption[Systematics of fission-fragment mass numbers]{
Calculated fission-fragment mass numbers following fission of nuclei in
neutron-deficient Pb-region ($74\leq Z_0\leq 86$, $92\leq N_0\leq 126$), actinide ($74\leq Z_0\leq 96$, $132\leq N_0\leq 186$),
superheavy ($106\leq Z_0\leq 114$, $156\leq N_0\leq 178$) and neutron-rich ($82\leq Z_0\leq 110$, $188\leq N_0\leq 218$). 
Only nuclei with asymmetric fission and with a symmetric-yield to peak-yield ratio less than 0.2
are included (red squares in Fig.\ \ref{fig:yield_chart}).
The figure is taken from Paper IV.
}
\label{fig:fragmasses}
\end{figure}

\begin{figure}[htbp!]
\centering    
\includegraphics[width=0.6\textwidth]{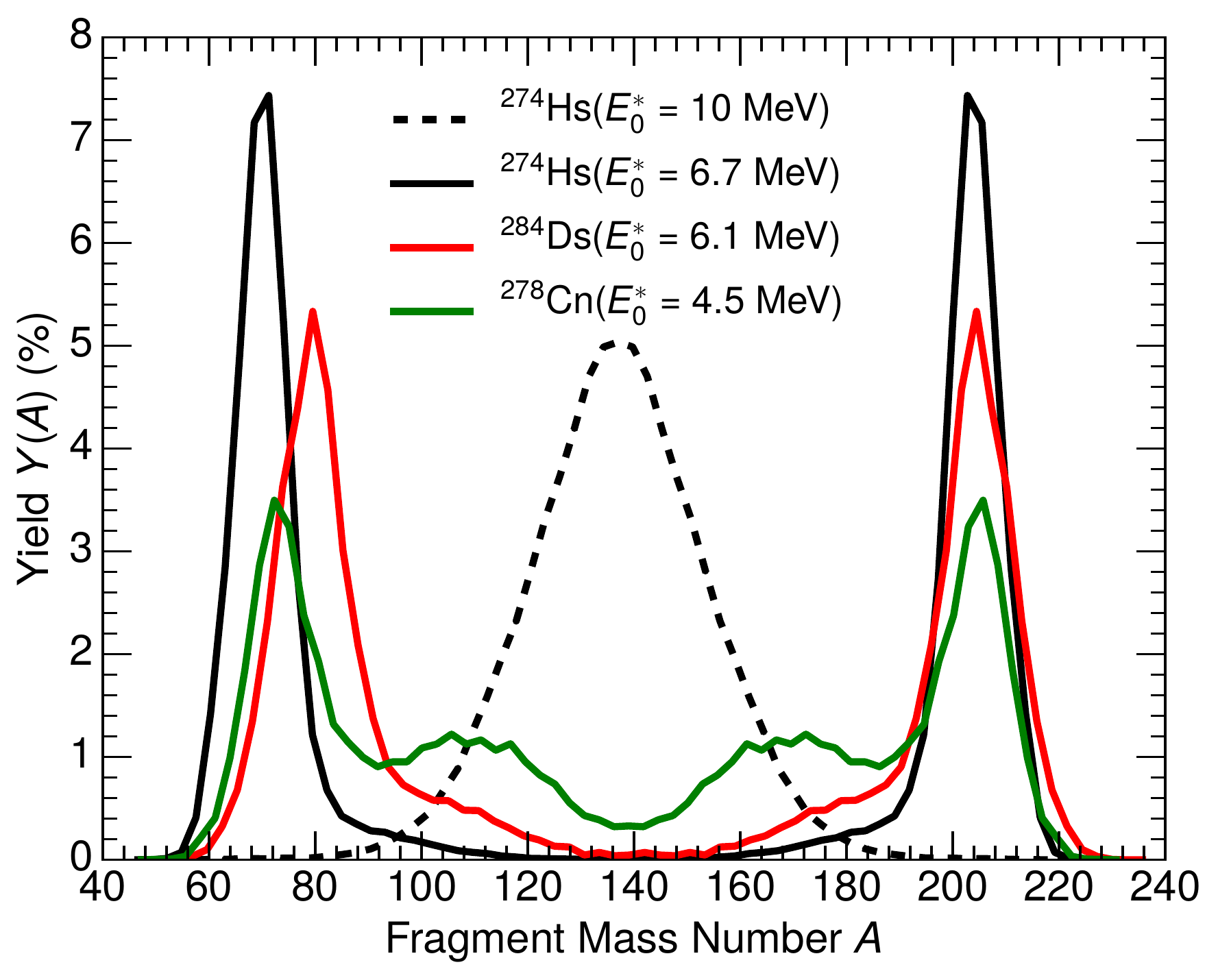}
\caption[Fission-fragment mass yields in $^{274}$Hs, $^{278}$Cn, $^{284}$Ds]{
Calculated fission-fragment mass yields in $^{274}$Hs, $^{278}$Cn, $^{284}$Ds
for excitation energies $E^\ast_0=6.7$ MeV, 4.5 MeV, 6.1 MeV (solid lines), respectively.
Also shown is the calculated yield for $^{274}$Hs with $E^\ast_0=10$ MeV (black dashed line).
}
\label{fig:fig_massyield_hs274}
\end{figure}

\begin{figure}[htbp!]
\centering    
\includegraphics[width=0.6\textwidth]{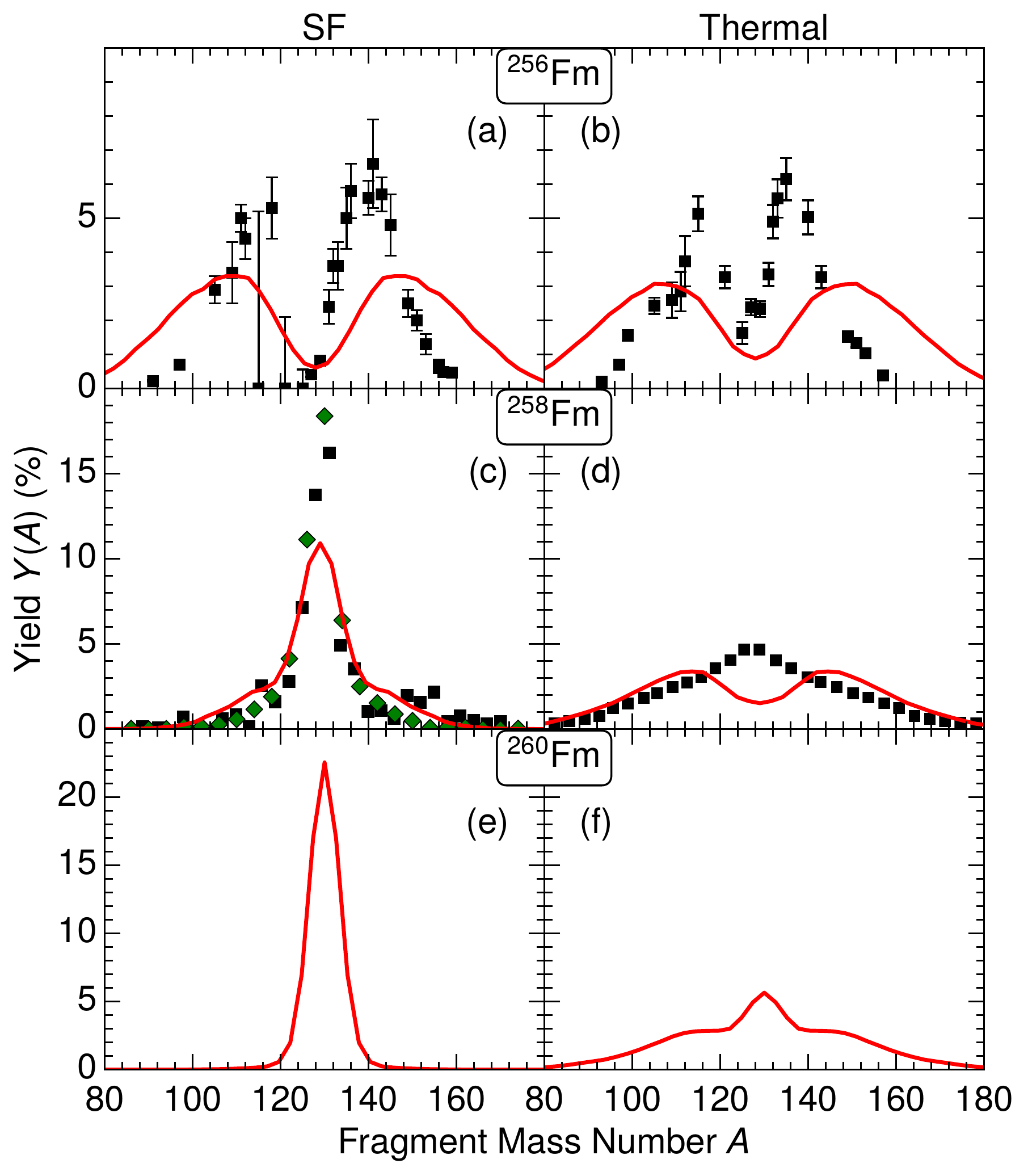}
\caption[Fission-fragment mass yields in $^{256,258,260}$Fm 
]{
Calculated fission-fragment mass yields for SF (left panel) and thermal energies (right panel)
for $^{256}$Fm (a,b), $^{258}$Fm (c,d) and $^{260}$Fm (e,f). 
Results from calculations are shown by solid red lines and 
results from data are shown by black squares and green diamonds taken from 
Refs.\ \cite{gindler77:a,flynn72:a,flynn75:a,hulet89:a,hoffman80:a,john71:a}. 
The experimental data all represent post-neutron evaporation mass yields. 
}
\label{fig:mass_yields_fm}
\end{figure}

The random-walk model was in Paper IV used to perform systematic yield calculations of 896 even-even nuclei between the drip lines from $Z=74$ to $Z=126$.
Initial excitation energies $E^\ast_0$, with respect to the ground state of the fissioning nucleus $(Z_0,N_0)$,
were chosen just sufficiently above the barrier to obtain reasonable computing times. 
Figure \ref{fig:yield_chart} show the ratio between the calculated mass yield at symmetry and the maximum on the yield curve;
red squares correspond to asymmetric yields while blue squares correspond to symmetric yields.
Four regions of nuclei fissioning asymmetrically are identified:
neutron-deficient ($74\leq Z_0\leq 86$, $92\leq N_0\leq 126$),
actinide ($74 \leq Z_0\leq 96$, $132\leq N_0\leq 186$), superheavy ($106\leq Z_0\leq 114$, $156\leq N_0\leq 178$) 
and neutron-rich ($82\leq Z_0\leq 110$, $188\leq N_0\leq 218$).

There are no experimental studies of the whole region shown in Fig.\ \ref{fig:yield_chart} but studies of 70 nuclei 
from $Z_0=85$ to $Z_0=94$ were presented in Ref.\ \cite{schmidt00:a}. 
It was suggested there that the transition between symmetric fission in the
lighter actinide region and asymmetric for heavier actinides is at $A_0\approx 226$. 
It is further stated that this is somewhat surprising since one would expect both protons and neutrons to affect what
regions fission symmetrically or asymmetrically. 
However, in Ref.\ \cite{schmidt00:a} fission mass distributions across the line
$A_0\approx 226$ are obtained for only a few proton numbers, namely $Z_0=89$, 90 and 91. 
Here, and in Ref.\ \cite{moller15:b}, which covers a larger, contiguous region of nuclei than the experimental work, the results
show that both protons and neutrons affect asymmetry. 
Particularly interesting is that above $Z_0\approx 88$ ($N_0\approx 132$) the calculated transition
line is clearly not a constant mass number $A_0$, but approximately a
constant $N_0-Z_0$ for a range of about eight proton numbers.  
This prediction has yet to be tested experimentally. 

The result in Fig.\ \ref{fig:yield_chart} is illustrated in a complementary way in Fig.\ \ref{fig:fragmasses} where heavy and light fragment
mass pairs are plotted as coloured symbols for nuclei that fission asymmetrically.
Few experimental data exist in the neutron-deficient Pb-region,
but asymmetric fission of $^{180}_{80}\text{Hg}_{100}$ was reported in Ref.\ \cite{andreyev10:a}.
Figure \ref{fig:fragmasses} shows that asymmetric fission of actinides with mass number from $A_0\approx220$ to $A_0\approx246$
correspond to divisions in the St mode with a heavy-fragment mass number that stays relatively constant around $A_{\rm H}\approx140$.
Consequently, the light mass increases so the heavy/light fragment mass difference decreases as the fissioning system becomes heavier. 
This is also seen in measurements (see e.g. Fig.\ 4 in Ref.\ \cite{flynn72:a}).

Striking in Fig.\ \ref{fig:fragmasses} is the abrupt transition to a small region (corresponding to $Z_0\approx 110, N_0\approx 166$ in Fig.\ \ref{fig:yield_chart})
of very large differences between the heavy and light fragment masses. 
This corresponds to the SA mode with a division into a $^{208}$Pb-like heavy fragment and the corresponding partner.
Mass yields of nuclei $^{274}_{108}\text{Hs}_{166}$, $^{278}_{112}\text{Cn}_{166}$, $^{284}_{110}\text{Ds}_{174}$ in this region are shown
as solid lines in Fig.\ \ref{fig:fig_massyield_hs274},
where it is seen that the heavy-fragment peak stays constant,
while the light-fragment peak changes depending on the mass of the fissioning nucleus.
Secondary peaks are also observed around $A_{\rm L}$:$A_{\rm H}=108$:$170$ for $^{278}_{112}\text{Cn}_{166}$.
A $^{208}$Pb-like fission fragment in the superheavy region is compatible with results based on density functional theory \cite{warda18:a,matheson19:a,scamps19:a},
whereas calculations using a pre-scission point model predicts divisions into a $^{132}$Sn-like light fragment and the corresponding partner \cite{carjan19:a}.
The competition between these two fission modes was further analysed recently within the Langevin approach in Ref.\ \cite{ishizuka20:a}.
For excitation energy $E^\ast_0=10$ MeV,
strongly asymmetric peaks due to $^{208}$Pb were obtained in the region $Z_0=120-122$,
while the influence of $^{208}$Pb was found to be negligible for lighter nuclei and with symmetric yields dominating.
The SA mode is however fragile as can be seen in the potential-energy paths in Fig.\ \ref{fig:epot_vs_q2_274hs};
increasing the energy to $E^\ast_0\gtrsim8$ MeV
makes it possible to cross the ridge to the symmetric valley
and, correspondingly, to a larger amount of symmetric yield.
This is shown Fig.\ \ref{fig:fig_massyield_hs274},
where fission of $^{274}_{108}\text{Hs}_{166}$ with $E^\ast_0=10$ MeV
results in a symmetric mass yield (black dashed line).

Figure \ref{fig:mass_yields_fm} shows the calculated mass yields (solid red curves) of $^{256,258,260}$Fm, 
for SF ($E^\ast_0=0$ MeV) (left panel), and for excitation energies $E_0^\ast$ corresponding to thermal-neutron induced fission ($E^\ast_0\approx S_{\rm n}$) (right panel).
For these nuclei, a transition from asymmetric to symmetric mass yield is obtained
corresponding to the north-east part of the actinide region in Fig.\ \ref{fig:yield_chart}.
For $^{256}$Fm ($Z_0=100$, $N_0=156$) the St mode is the dominating mode corresponding to asymmetric yield distribution,
where calculations yield a slightly broader distribution than data.
The yield does not change much when the energy is increased,
though the amount of symmetric yield increases slightly.
In SF of $^{258}$Fm there is a mixture of the symmetric SS mode and the asymmetric St mode,
with fractions 55\% and 45\%, respectively.
Although the mass yield is symmetric, it is not as narrow as in $^{260}$Fm
where the SS mode completely dominates for SF. 
When the energy is increased, the strong shell effects associated with the SS mode decreases,
and the amount of asymmetric yield increases in $^{258}$Fm and $^{260}$Fm.
For thermal fission, the SS mode has fully disappeared for $^{258}$Fm in the calculations.
The broad symmetric yield in the data indicates however that there is some fraction of the SS mode left for $^{258}$Fm,
similar to calculations for thermal fission in $^{260}$Fm in which 25\% of the SS mode is still present.

Further results of mass and charge yields calculated within the random-walk model
can be found in Refs.\ \cite{randrup11:a,randrup11:b,moller12:a,randrup13:a,moller14:b,moller15:b,ward17:a,moller17:a,moller15:c,schmitt21:a}.

\section{Shapes of scission fragments}\label{sec:scission_shapes}
In addition to the mass asymmetry coordinate $\alpha$, which determines fragment mass numbers $A_{\rm L}$ and $A_{\rm H}$,
the scission configurations are also characterized by the elongation $q_2$ of the fissioning nucleus and 
the spheroidal deformations $\varepsilon_{\rm L}$ and $\varepsilon_{\rm H}$ of the fragments.
These deformation parameters affects the subsequent kinetic energy and excitation energy of the fragments,
discussed in Ch.\ \ref{ch:energy_release_fission}.

\begin{figure}[t] 
 \begin{center} 
 \includegraphics[width=0.6\textwidth]{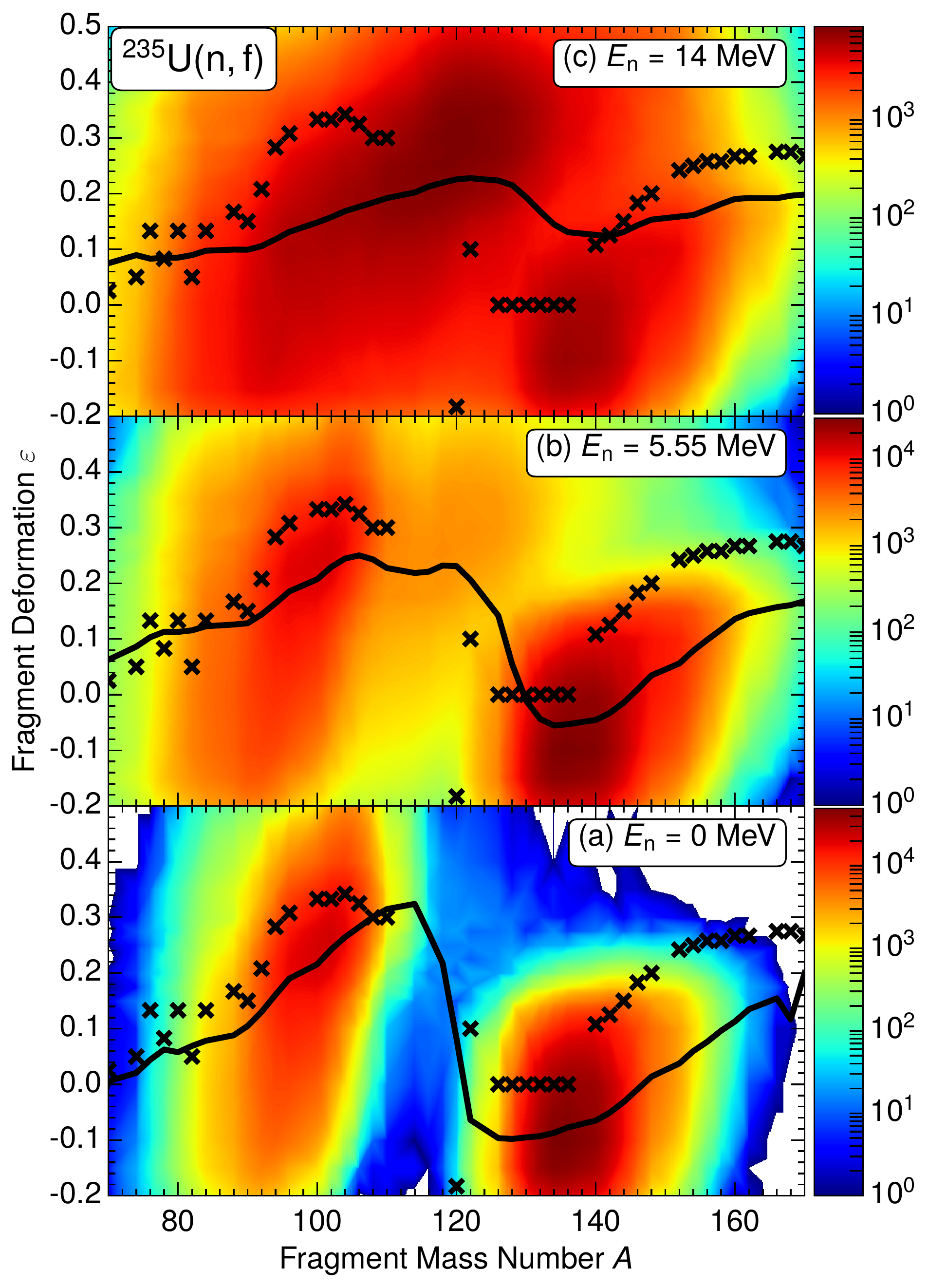} 
 \caption[Fragment deformations in $^{235}{\rm U}(\text{n},\text{f})$ for $E_{\rm n}=0$, 5.55, 14 MeV]{
Contour plots (on a logarithmic scale) in the plane of the fragment mass number $A$ and 
fragment deformation $\varepsilon$ in  $^{235}\text{U}(\text{n},\text{f})$ for incident-neutron energies: (a) $E_{\rm n}=0$ MeV, 
(b) $E_{\rm n}=5.55$ MeV, and (c) $E_{\rm n}=14$ MeV.
Average deformations are shown as solid black curves and ground-state deformations
are shown as black crosses.
 }
\label{fig:eps_vs_a_u236}  
 \end{center} 
\end{figure}

\begin{figure}[t] 
 \begin{center} 
 \includegraphics[width=0.7\textwidth]{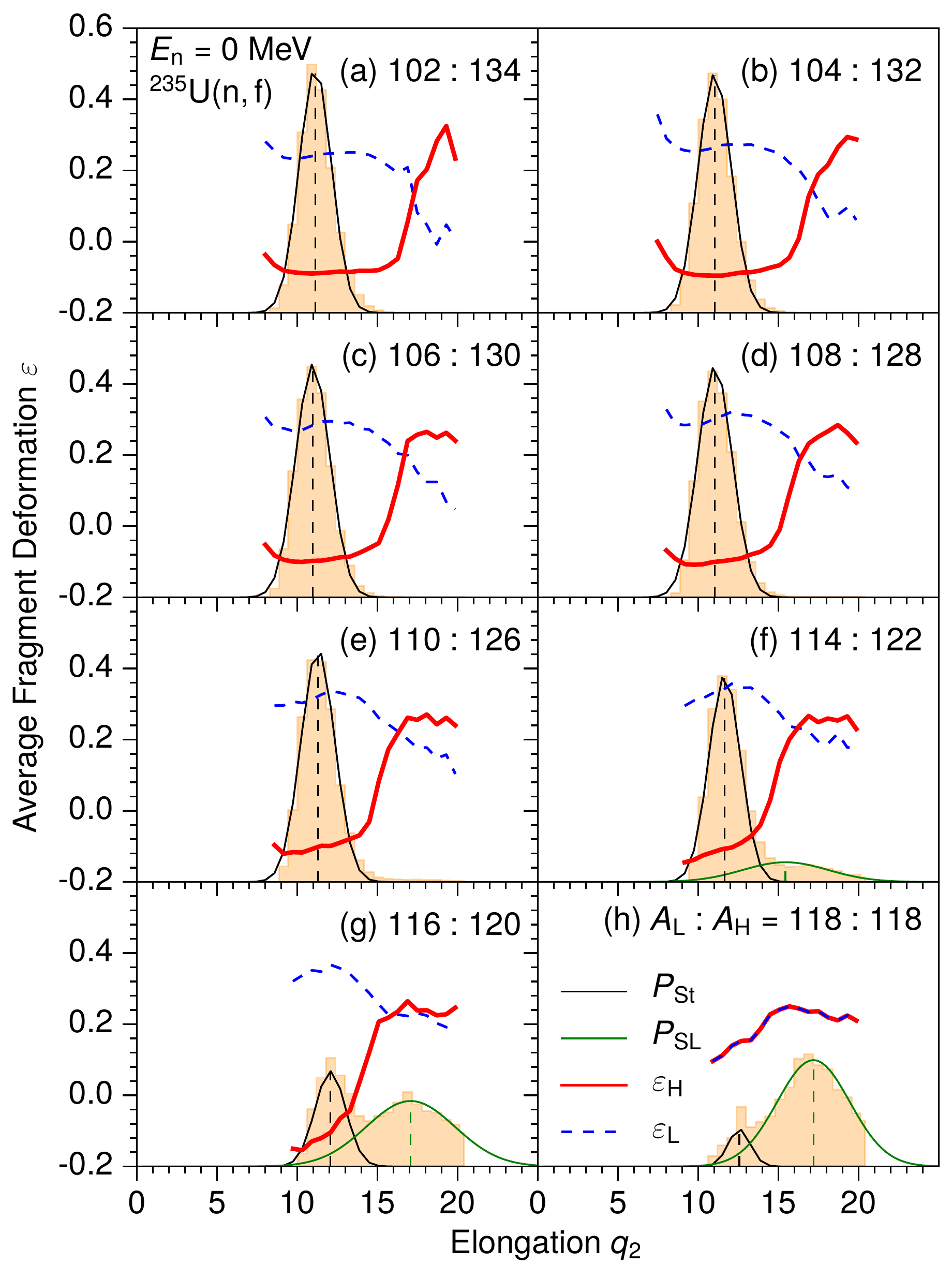} 
 \caption[Fragment deformations vs. $q_2$ in $^{235}\text{U}(\text{n},\text{f})$ for $E_{\rm n}=0$ MeV]{
Average light and heavy-fragment deformations at scission, $\varepsilon_{\rm L}$ (thin blue dashed curves) 
and $\varepsilon_{\rm H}$ (thick red curves), versus elongation $q_2$ for selected mass-splits in $^{235}\text{U}(\text{n}_{\rm th},\text{f})$. 
The orange histograms show the calculated $q_2$ distributions, $P(q_2)$, with fitted Gaussian distributions
corresponding to the St (black curve) and SL (green curve) fission modes.
}
\label{fig:fig_eps_vs_q2_u236_e0}  
 \end{center} 
\end{figure}

\begin{figure}[t] 
 \begin{center} 
 \includegraphics[width=0.7\textwidth]{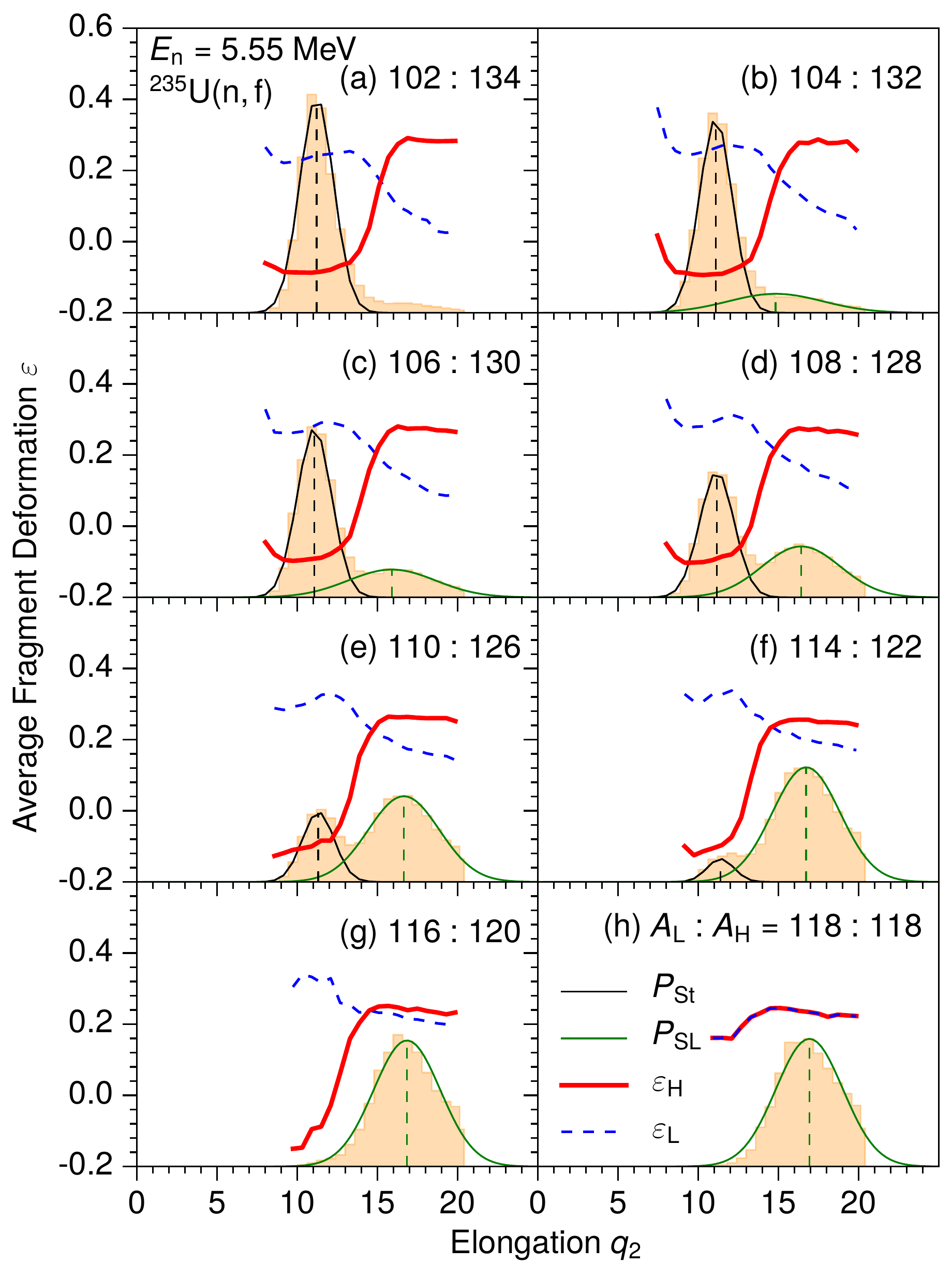} 
 \caption[Fragment deformations vs. $q_2$ in $^{235}\text{U}(\text{n},\text{f})$ for $E_{\rm n}=5.55$ MeV]{
 Same as Fig.\ \ref{fig:fig_eps_vs_q2_u236_e0} but for $E_{\rm n}=5.55$ MeV.
 }
\label{fig:fig_eps_vs_q2_u236_e5}  
 \end{center} 
\end{figure}

Figure \ref{fig:eps_vs_a_u236} shows
a contour plot of the location of the endpoints projected onto fragment deformation $\varepsilon$ and fragment 
mass number $A$ in $^{235}\text{U}(\text{n},\text{f})$ for different
incident-neutron energies $E_{\rm n}$.
Both the calculated average fragment deformations at scission (solid black curves) 
and the ground-state deformations (black crosses) display a saw-tooth
behaviour as a function of fragment mass for $E_{\rm n}=0$. 
However, the scission deformations tend to be below the values of the ground-states deformations, 
towards more oblate shapes. 
The St mode is dominating for this energy with mass peaks $A_{\rm L}$:$A_{\rm H}\approx100$:$136$
and deformations $\varepsilon_{\rm L}$:$\varepsilon_{\rm H}\approx0.25$:-$0.1$.
The SL mode with two prolate fragments are obtained to some degree for the most symmetric mass splits.
At higher energies, the SL mode spreads to more asymmetric mass divisions.
This results in a bimodal distribution in the heavy-fragment deformation $\varepsilon_{\rm H}$ for $A_{\rm H}\approx132$,
with one peak at $\varepsilon_{\rm H}\approx-0.1$ due to the St mode and one peak at $\varepsilon_{\rm H}\approx0.3$
due to the SL mode.

Figure \ref{fig:fig_eps_vs_q2_u236_e0} shows the 
average light and heavy-fragment deformations at scission, $\varepsilon_{\rm L}$ (thin blue dashed curves) 
and $\varepsilon_{\rm H}$ (thick red curves), as a function of elongation $q_2$ for eight mass-splits in 
$^{235}\text{U}(\text{n},\text{f})$ for neutron energy $E_{\rm n}=0$ MeV. 
The orange histograms show the obtained $q_2$ distributions, $P(q_2)$, with fitted Gaussian distributions
identified as the St (black curve) and SL (green curve) modes.
Similar behaviour is obtained for all mass-splits
with deformations
$\varepsilon_{\rm L}$:$\varepsilon_{\rm H}\approx0.3$:-$0.1$ for small $q_2$ values 
and $\varepsilon_{\rm L}$:$\varepsilon_{\rm H}\approx0.3$:$0.3$ for larger $q_2$ values.
The St mode is seen to be dominating except for the most symmetric mass splits $A_{\rm L}=A_{\rm H}=118$.
Figure \ref{fig:fig_eps_vs_q2_u236_e5} is similar to Fig.\ \ref{fig:fig_eps_vs_q2_u236_e0}, 
but for $E_{\rm n}=5.55$ MeV.
The average fragment deformations
are very similar to those for $E_{\rm n}=0$ MeV,
but the fractions of the $q_2$-distributions changes,
where the SL mode is seen to spread to more asymmetric splits.
This corresponds to the bimodal distribution also seen in \ref{fig:eps_vs_a_u236}(b) for $A_{\rm H}\approx132$.

\begin{figure}[t] 
 \begin{center} 
 \includegraphics[width=0.6\textwidth]{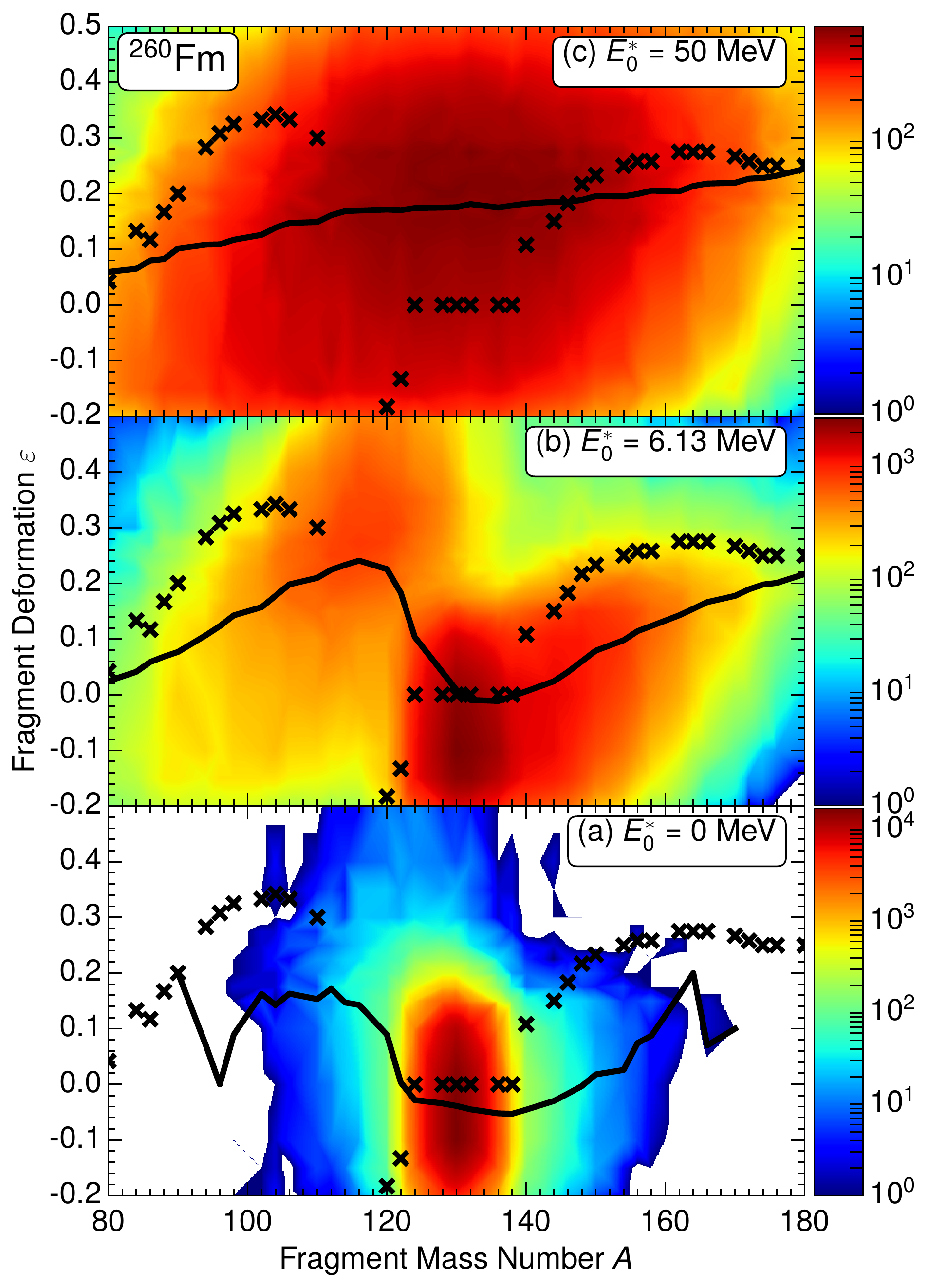} 
 \caption[Fragment deformations in $^{260}{\rm Fm}$ for $E^\ast_0=0$, 6.13, 50 MeV]{
Contour plots (on a logarithmic scale) in the plane of the fragment mass number $A$ and 
fragment deformation $\varepsilon$ in fission of $^{260}\text{Fm}$ for 
initial excitation energies: (a) $E_0^\ast=0$ MeV, (b) $6.13$ MeV, (c) and $50$ MeV.
Average deformations are shown as solid black curves and ground-state deformations
are shown as black crosses.
 }
\label{fig:eps_vs_a_fm260}  
 \end{center} 
\end{figure}

\begin{figure}[t] 
 \begin{center} 
 \includegraphics[width=0.7\textwidth]{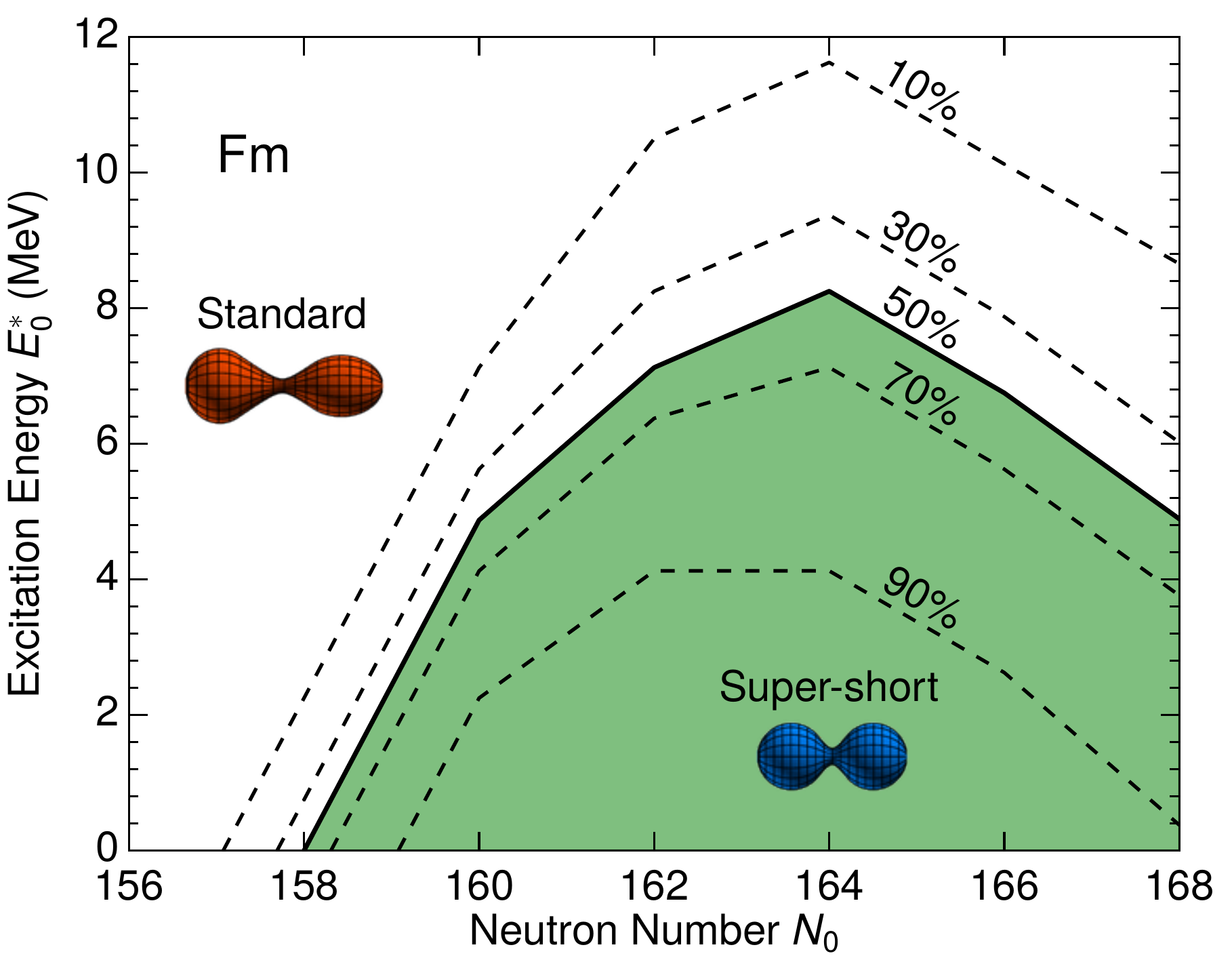} 
 \caption[Phase diagram of fission modes in fermium isotopes]{
The calculated relative contribution by the SS fission mode in fermium isotopes, 
as a function of the neutron number $N_0$ and the initial excitation energy $E_0^\ast$, 
presented in the form of a contour plot, on which the solid line
marks the 50\% contour where the SS mode and the
St mode contribute equally. Typical scission shapes for
the two modes are also shown.
The figure is taken from Paper VI.
 }
\label{fig:phase_diagrams}  
 \end{center} 
\end{figure}

\begin{figure}[t]
 \begin{center} 
 \includegraphics[width=1.0\textwidth]{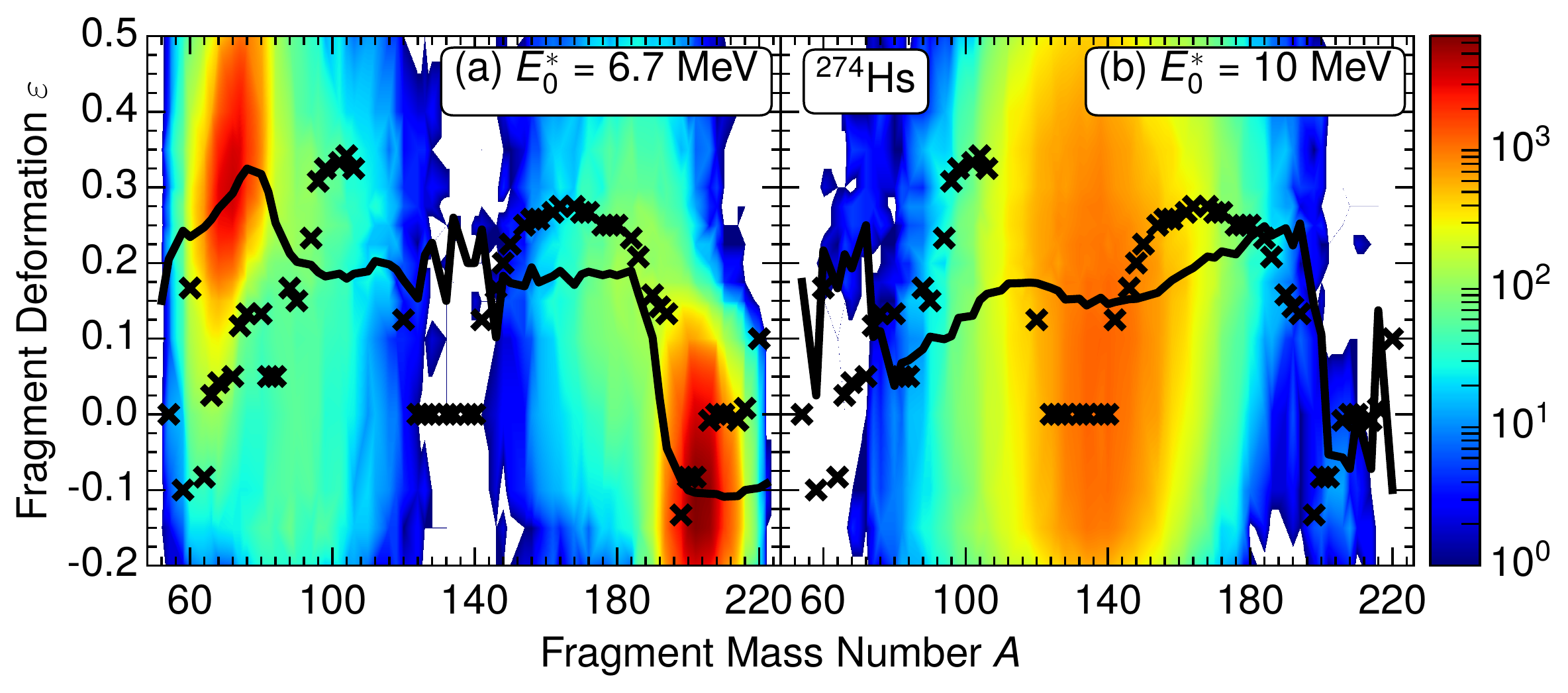} 
 \caption[Fragment deformations in $^{274}{\rm Hs}$ for $E^\ast_0=6.7$, 10 MeV]{
Contour plots (on a logarithmic scale) in the plane of the fragment mass number $A$ and 
fragment deformation $\varepsilon$ in fission of $^{274}{\rm Hs}$ for initial excitation energies
$E_0^\ast=6.7$ MeV (a) and $E_0^\ast=10$ MeV (b).
Average deformations are shown as solid black curves and ground-state deformations
are shown as black crosses.
 }
\label{fig:eps_vs_a_hs274}  
 \end{center} 
\end{figure}

Figure \ref{fig:eps_vs_a_fm260} shows
a contour plot of the location of the endpoints projected onto fragment deformation $\varepsilon$ and fragment 
mass number $A$ in $^{260}\text{Fm}$ for different excitation energies $E_0^\ast$.
For $E_0^\ast=0$ MeV, most of the events correspond to the SS mode with 
symmetric mass-split and two spherical fragments.
At $E_0^\ast=6.13$ MeV (thermal fission),
the amount of asymmetric events increases due to the St mode
and show a similar saw-tooth behaviour as for $^{235}\text{U}(\text{n}_{\rm th},\text{f})$ in \ref{fig:eps_vs_a_u236}(c).
When the excitation is increased further, the shell effects will eventually disappear
and the most probable split correspond two prolate fragments.

The fraction of the SS mode is shown
in Fig.\ \ref{fig:phase_diagrams} for even fermium isotopes $^{254-268}$Fm 
undergoing fission at different excitation energies. 
It is seen how the SS mode dominates at $E_0^\ast=0$ MeV for $^{258-268}$Fm 
(and probably disappears for $N_0>172$). 
But the mode is fragile and quickly disappears in $^{258}$Fm as the excitation energy is increased. 
For $^{260}$Fm the transition from a dominance of the SS
mode to a dominance of the St mode appears at a slightly higher energy, about 5 MeV. 
For $^{264}$Fm the SS mode survives to highest excitation energy, as is
expected since the symmetric fission in this case corresponds to two doubly-magic $^{132}$Sn nuclei.

The calculated variation in the stability of the SS mode with increasing excitation energy 
can be understood from the potential-energy structure in Fig.\ \ref{fig:epot_vs_q2_fm} in Sec.\ \ref{sec:potsurf_frldm}. 
At low energies the walks are restricted to the SS valley, while
the ridge towards the asymmetric St valley may be crossed at higher energies. 
The small slope of the energy along the SS valley between $q_2\approx 4$ and $q_2\approx 6$, seen for $^{258,260,262}$Fm, 
leads to an almost diffusive dynamics, implying several possibilities to cross the ridge. 
Once a transition to the St valley is made, the probability to return
to the SS valley is very small, since the dynamics
along the St valley is strongly driven towards scission by the large energy slope.
The potential energy along the SS valley decreases with 
increasing neutron number and reaches its smallest values for $N_0=164$. 
Because the energy of the ridge is rather independent of $N_0$, 
transitions from the SS valley to the St valley require higher energies with increasing neutron
number, with a maximum for $N_0=164$, thus making the SS mode in those isotopes correspondingly more
resilient.

Figure \ref{fig:eps_vs_a_hs274} shows the fragment deformations in fission of
$^{274}\text{Hs}$ for energies $E_0^\ast=6.7$ MeV (a) and $E_0^\ast=10$ MeV (b).
Both energies yield similar average values,
whereas the majority of events are changed from asymmetric to symmetric mass-splits when the energy is increased.
The heavy fragment near $^{208}$Pb obtains a slightly oblate shape $\varepsilon_{\rm H}\approx-0.1$,
while the light fragment becomes prolate $\varepsilon_{\rm L}\approx0.3$,
identified as the SA mode.
This is similar to the results obtained within the Langevin approach in Ref.\ \cite{ishizuka20:a},
where the fragment generally obtain a prolate shape,
except near $^{208}$Pb where a spherical shape is the most probable.

\chapter{Energy release}\label{ch:energy_release_fission}

Large amount of energy is released in fission due to the mass-energy relation $E=mc^2$.
The majority of the energy is released in terms of kinetic energy of the two fragments,
while some of the energy goes into excitation energy in the fragments.
The excitation energy is subsequently dissipated through emission of particles.

\section{Energies in the fission process}\label{sec:energies_fission}
The energies involved in the fission process is schematically illustrated in Fig.\ \ref{fig:energies}. 
The energy curve (thick solid line) describes a path through the five-dimensional potential-energy surface that is ended at the scission point. 
After scission, the potential energy drops drastically mainly due to the Coulomb repulsion between the two fission fragments. 

\begin{figure}[htbp!]
\centering
\includegraphics[width=0.7\linewidth]{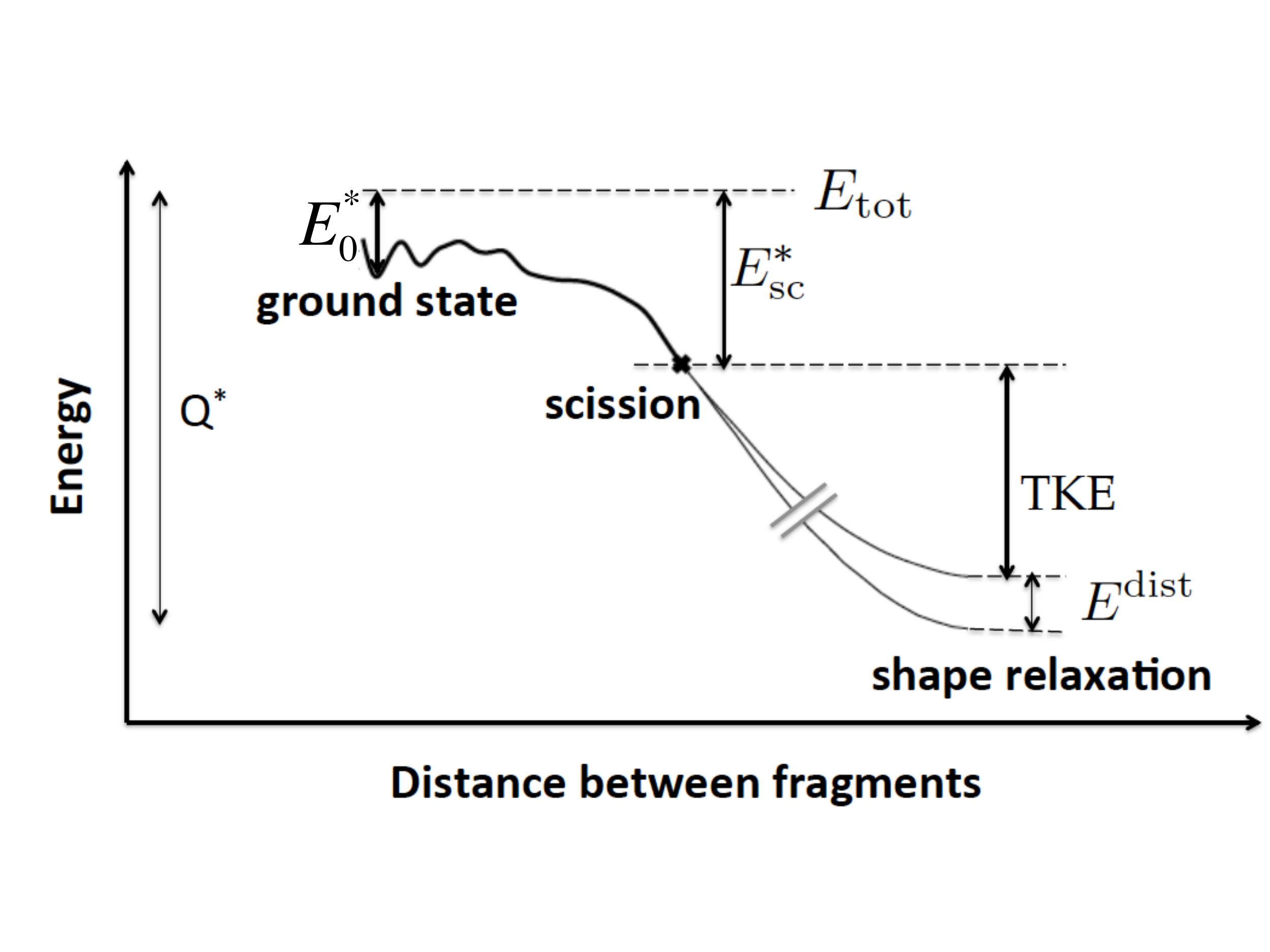}
\vspace{-0.8cm}
\caption[Energies in the fission process]{
Illustration of the energies in the fission process. The thick curve shows the minimal
energy of the fissioning nucleus along a typical diffusive path on the structured potential-energy 
landscape, until scission is reached, marked by the cross on the figure. At scission, the intrinsic
energy, $E^\ast_{\text{sc}}$, is shared between the fragments. Beyond scission, the process is not explicitly
calculated, and only energy conservation is applied (schematically denoted by thin lines). Beyond
scission, the fragments are subject to Coulomb repulsion, which is subsequently converted to kinetic
energy, and the distortion energy in each fragment is relaxed within the fragment. 
The totally available energy, $Q^\ast$, is composed
of the initial excitation energy $E^\ast_0$ and
the $Q$-value for the considered fission process. 
This energy is converted into TKE of the fragments and intrinsic excitation energy in the fragments. The intrinsic excitation energy
has two contributions, (a) the excitation energy at scission $E^\ast_{\text{sc}}$, and (b) the distortion
energy $E^{\text{dist}}$.
}
\label{fig:energies} 
\end{figure}

The nucleus is initially excited by the energy $E^\ast_0$, where in neutron induced fission, $(\text{n},\text{f})$, this corresponds to
\begin{equation}
E_0^\ast = S_{\text{n}}+E_{\text{n}},
\end{equation}
where $S_{\text{n}}$ is the neutron separation energy of the fissioning nucleus and $E_{\text{n}}$ the kinetic energy of the incoming neutron. 
For thermal fission, $(\text{n}_{\rm th},{\rm f})$, in which $E_{\rm n}\approx0$ MeV, 
the initial excitation energy is $E^\ast_0=S_{\rm n}$.
Together with the $Q$-value, this is the totally available energy
in the fission process,
\begin{equation}
\label{eq:qval}
Q^\ast = E_0^\ast + M(Z_0,N_0) - M(Z_{\rm L},N_{\rm L})-M(Z_{\rm H},N_{\rm H}),
\end{equation}
where $M(Z_0,N_0)$ is the ground-state mass of the parent nucleus,
and $M(Z_{\rm L},N_{\rm L})$ and $M(Z_{\rm H},N_{\rm H})$ are the ground-state masses of the
light- and heavy fragments, respectively.

In the random-walk model, the collective kinetic energy associated with the shape
evolution is assumed to be negligible prior to scission.
The excitation energy at the scission point, $E^*_{\text{sc}}$, is then the energy difference between the total energy $E_{\text{tot}}$ 
and the potential energy $U(\boldsymbol{\chi}_{\text{sc}})$ at the scission configuration,
\begin{equation}
\label{eq:Esciss}
E^\ast_{\mathrm{sc}}=E_{\mathrm{tot}}-U(\boldsymbol{\chi}_{\text{sc}}).
\end{equation}
This excitation energy at scission is to be shared between the two fragments,
\begin{equation}
E^\ast_{\mathrm{sc}}=E^{\rm intr}_{\rm L}+E^{\rm intr}_{\rm H},
\end{equation}
and is discussed in Sec.\ \ref{sec:fragment_eexc}.

After time to equilibrate, the accelerated fission fragments relax their respective shapes to ground-state
deformations and thereby gain a distortion energy, 
which is calculated as the energy difference between the
fragment mass at scission and the ground-state mass, 
\begin{equation}
E^{\text{dist}}_i=M_i(\varepsilon^{\text{sc}}_i)- M_i(\varepsilon^{\text{gs}}_i),
\label{eq:distortion}
\end{equation}
where $i=\text{L}$ or H. 
The total excitation energy $E^\ast_i$ of a fragment is then composed by the two parts, intrinsic excitation energy and distortion energy, 
\begin{equation}
\label{eq:tot_frag_eexc}
E^\ast_i=E_i^{\text{intr}} + E_i^{\text{dist}}.
\end{equation}
The sum of the excitation energy of the light- and heavy fragment makes up the 
the total excitation energy (TXE),
\begin{equation}
\label{eq:txe}
\text{TXE}=E^\ast_{\rm L} + E^\ast_{\rm H}.
\end{equation}

In total, the available energy is thus shared between the TKE and the TXE of the fragments, i.e.,
\begin{equation}
\label{eq:qval2}
Q^\ast=\text{TKE}+\text{TXE}.
\end{equation}
The $Q^*$-value is calculated from Eq.\ (\ref{eq:qval}) and the TXE-value from Eq.\ (\ref{eq:txe}). 
The TKE-value is then obtained from Eq.\ (\ref{eq:qval2}),
which corresponds to TKE of the fragments before neutron emission.
After neutron emission, the fragment masses decrease somewhat,
and consequently, the TKE decreases slightly (if the fragment speed is assumed to be unchanged).
Calculated results in the thesis represent TKE before neutron emission.

\section{Fragment kinetic energy}
\label{sec:tke}

The dependence of the average TKE on the heavy-fragment mass number is shown in Fig.\ \ref{fig:tke_vs_af_u236}(a) for $^{235}\text{U}(\text{n}_{\rm th},\text{f})$.
Two different scission scenarios are considered. The first scenario uses $c_{\rm sc}=2.5$ fm, 
the neck radius at which the mass division is assumed to freeze out. 
The corresponding scission configurations are relatively compact and the resulting TKE
values exceed the experimental data significantly for all divisions. 
This suggests that the fragments maintain contact for a while after their masses have been determined.
Therefore we assume that the effective loss of contact occurs later or when the neck radius
has shrunk further to $c_{\rm sc}=1.5$ fm.
The corresponding shapes are then more elongated and, consequently, the resulting TKE values are lowered. 
Furthermore, the internal excitation energies are higher and the emerging fragments have larger
quadrupole moments.

\begin{figure}[b]
 \begin{center} 
 \includegraphics[width=0.8\linewidth]{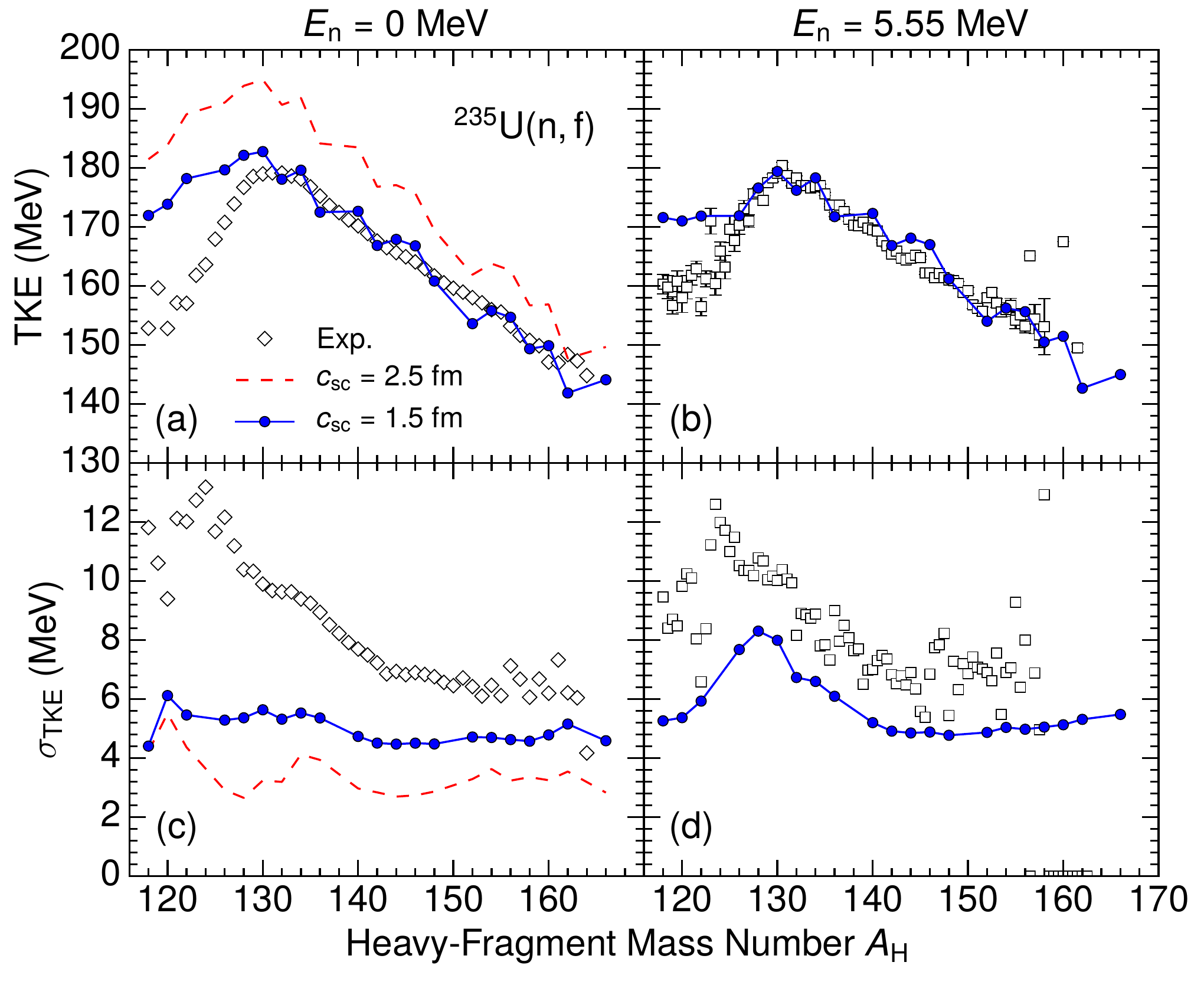} 
 \caption[TKE vs. $A$ in $^{235}{\rm U}(\text{n},\text{f})$ for $E_{\rm n}=0$ and 5.55 MeV]{
Panel (a) shows the average pre-neutron
TKE versus the heavy-fragment mass number $A_{\rm H}$ for $^{235}\text{U}(\text{n}_{\rm th},\text{f})$ for
two values of the scission neck radius, $c_{\rm sc}=1.5$ fm (filled circles
connected by solid blue lines) and $c_{\rm sc}=2.5$ fm (dashed red line). In
(c) the calculated width of the TKE distribution, $\sigma_{\rm TKE}$, is shown for
the same two values of $c_{\rm sc}$. Panels (b) and (d) are similar to (a) and
(c), but are for a higher incident-neutron energy, $E_{\rm n}=5.55$ MeV, and
only results for the adopted scission radius, $c_{\rm sc}=1.5$ fm, are shown.
Measured values of TKE are shown for thermal fission \cite{adili13:a} (open
diamonds) and for $E_{\rm n}=5.55$ MeV \cite{muller84:a} (open squares).
 The figure is taken from Paper V.
 }
\label{fig:tke_vs_af_u236} 
 \end{center} 
\end{figure}

\begin{figure}[b] 
 \begin{center} 
 \includegraphics[width=0.7\linewidth]{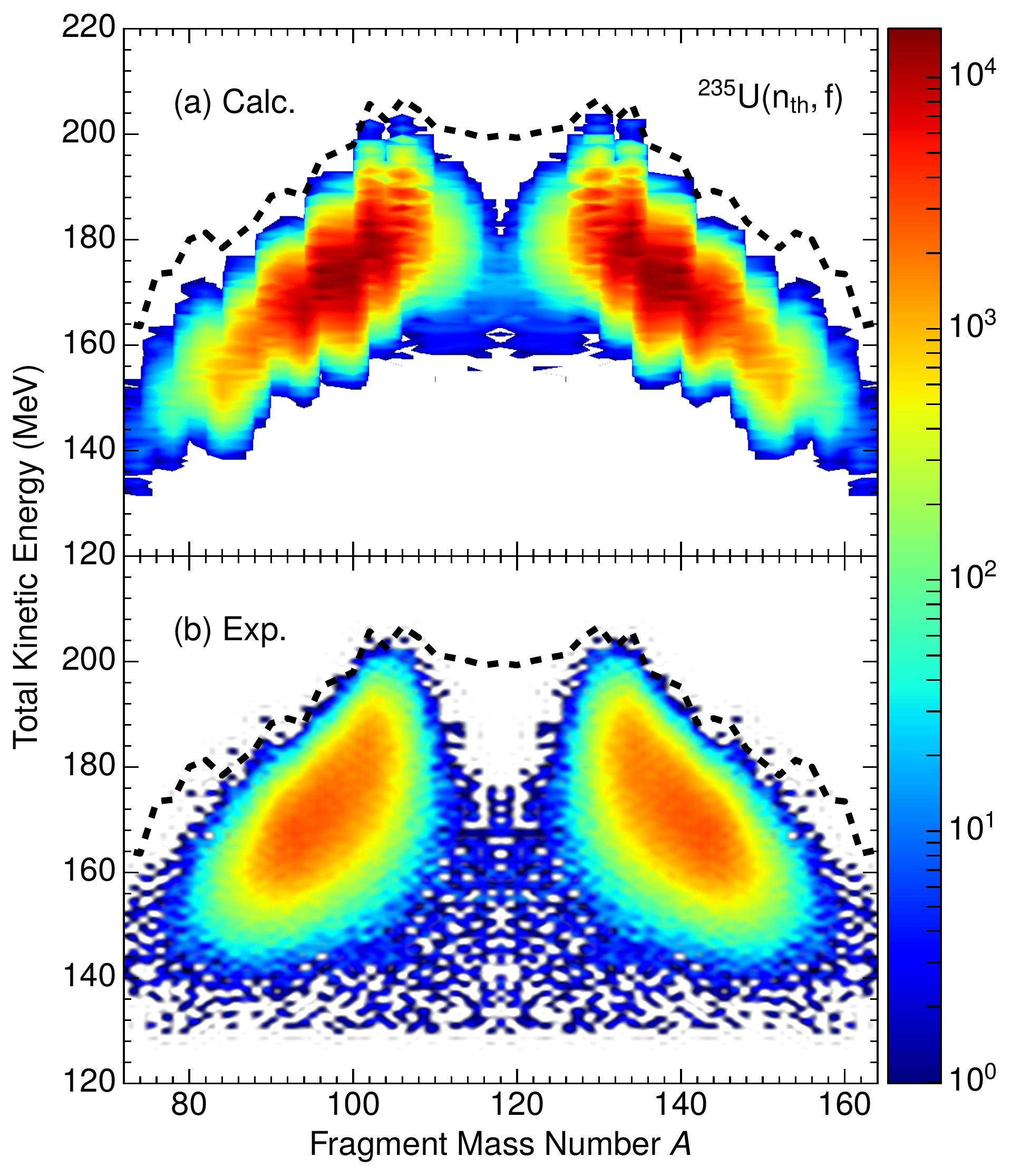} 
 \caption[Contour figure of TKE in $^{235}{\rm U}(\text{n}_{\rm th},\text{f})$]{
Number of scission events in log scale versus fission-fragment mass number $A$ and pre-neutron TKE for $^{235}\text{U}(\text{n}_{\rm th},\text{f})$.
(a) Calculated results with $c_{\rm sc}=1.5$ fm. 
(b) Experimental data from Ref.\ \cite{adili13:a}. The dashed curve shows $Q^\ast$ values for different fragment masses. 
The experimental number of events is scaled to the same number of events as calculated.
The jaggedness of the calculated contour plot is due to the selection of even-even mass numbers, and the finiteness of the calculational grid in TKE. 
The experimental results contain a certain degree of smearing due to uncertainty in measured mass numbers.
The figure is taken from Paper V.
 }
\label{fig:tke_236} 
 \end{center} 
\end{figure}

In the region beyond $A_{\rm H}\geq132$, where the average TKE exhibits a steady decrease with $A_{\rm H}$, 
the experimental data are very well reproduced by the calculations. 
However, in the more symmetric region, the extracted TKE values exceed
the measured values considerably, by up to 20 MeV. 
This discrepancy may be due to the extraordinary elongation of the associated scission shapes which may not be adequately
described within the 3QS shape family in terms of which the potential-energy surface has been calculated. 
An insufficient elongation leads to an over-prediction of TKE and an underestimate of the intrinsic excitation energy.

\begin{figure}[t] 
 \begin{center} 
 \includegraphics[width=\linewidth]{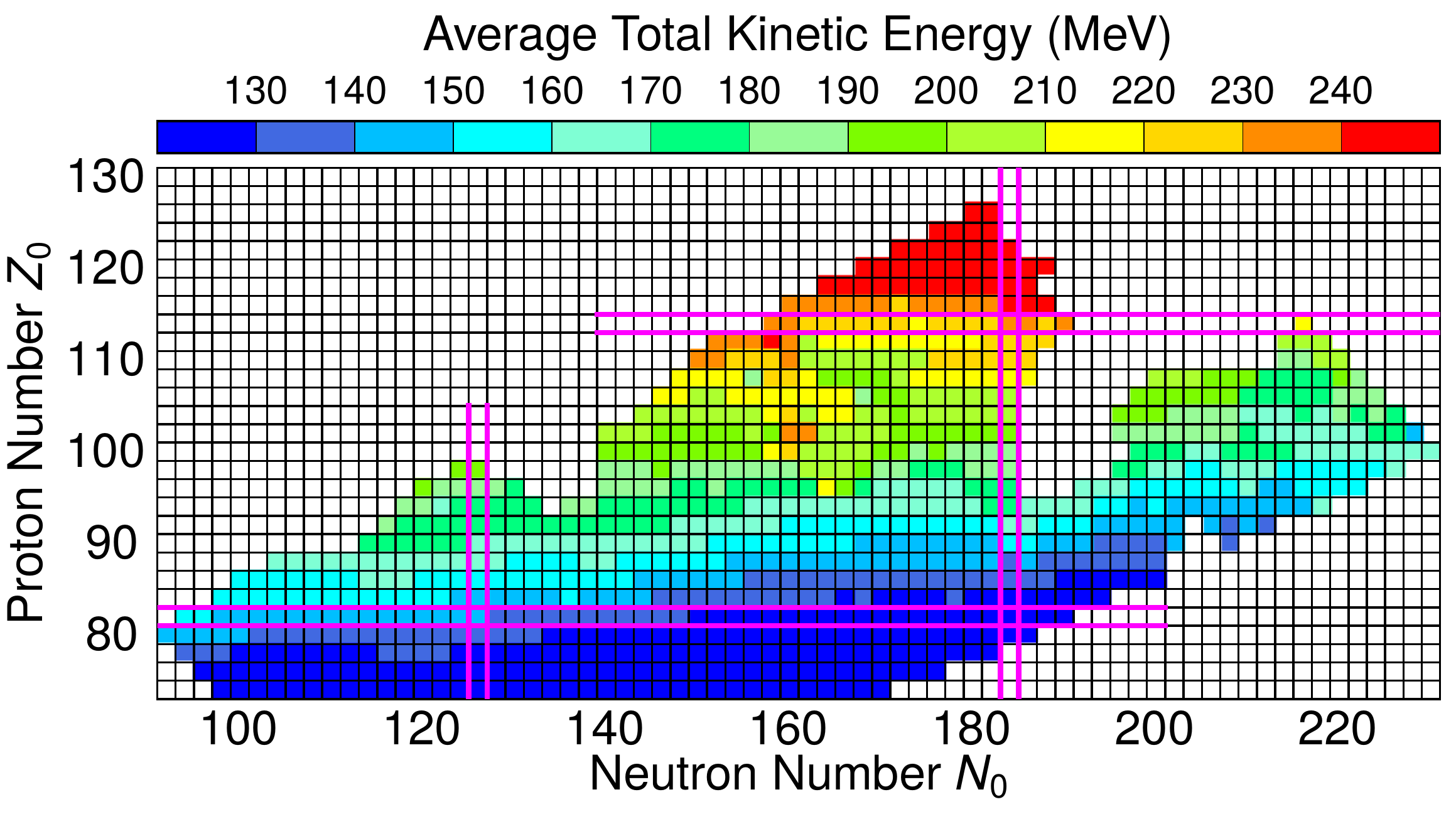} 
 \caption[Chart of fission-fragment total kinetic energy]{
 Calculated average fission-fragment TKE following fission of
 heavy nuclei. The figure is taken from Paper IV.
 }
\label{fig:tkeav} 
 \end{center} 
\end{figure}
\begin{figure}[htbp!]
 \begin{center} 
 \includegraphics[width=\linewidth]{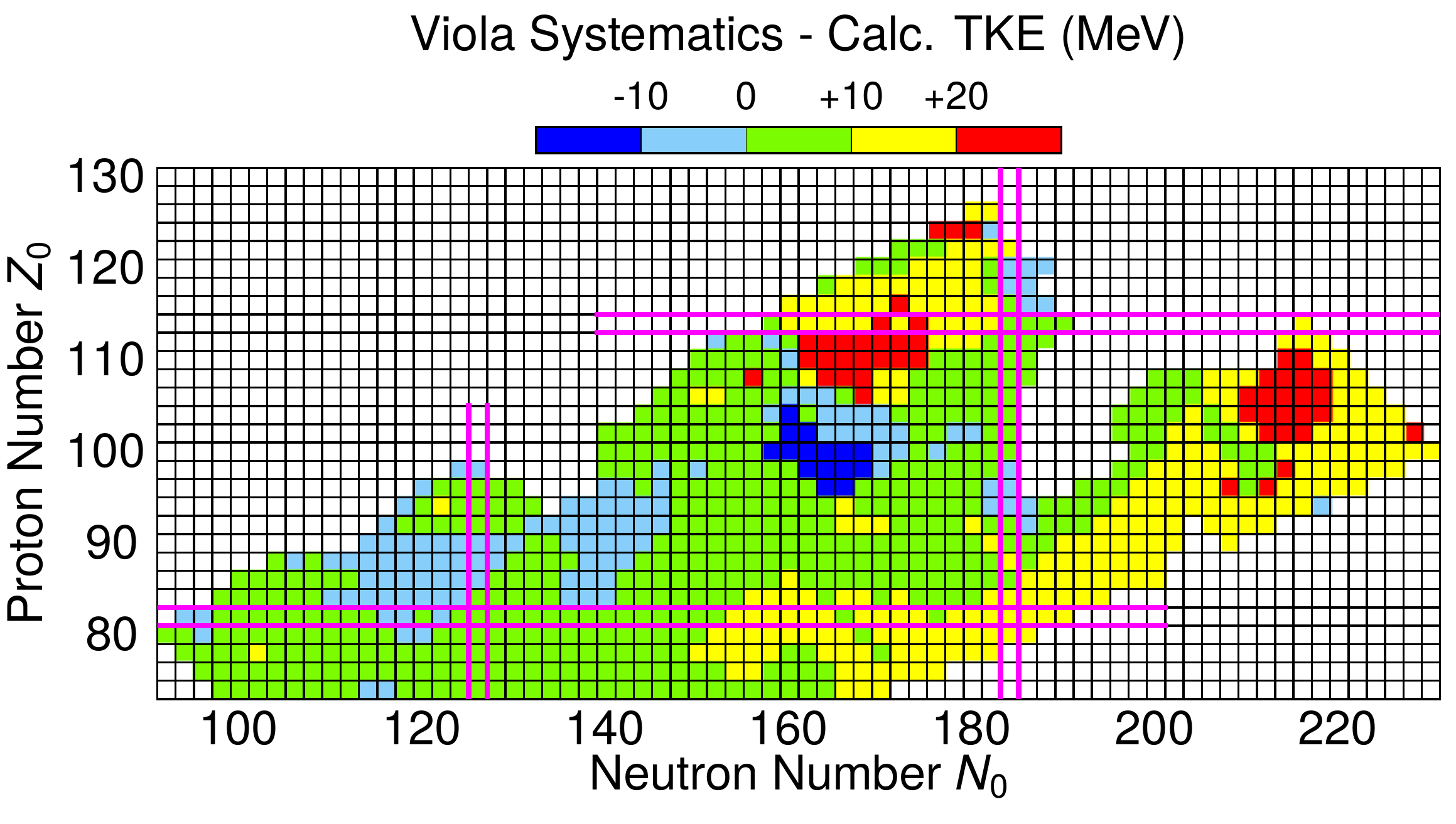} 
 \caption[Viola TKE systematics minus calculated average TKE]{
 Viola TKE systematics minus calculated average TKE. The figure is taken from Paper IV.
 }
\label{fig:vmctke} 
 \end{center} 
\end{figure}
\begin{figure}[b]
 \begin{center} 
 \includegraphics[width=0.7\textwidth]{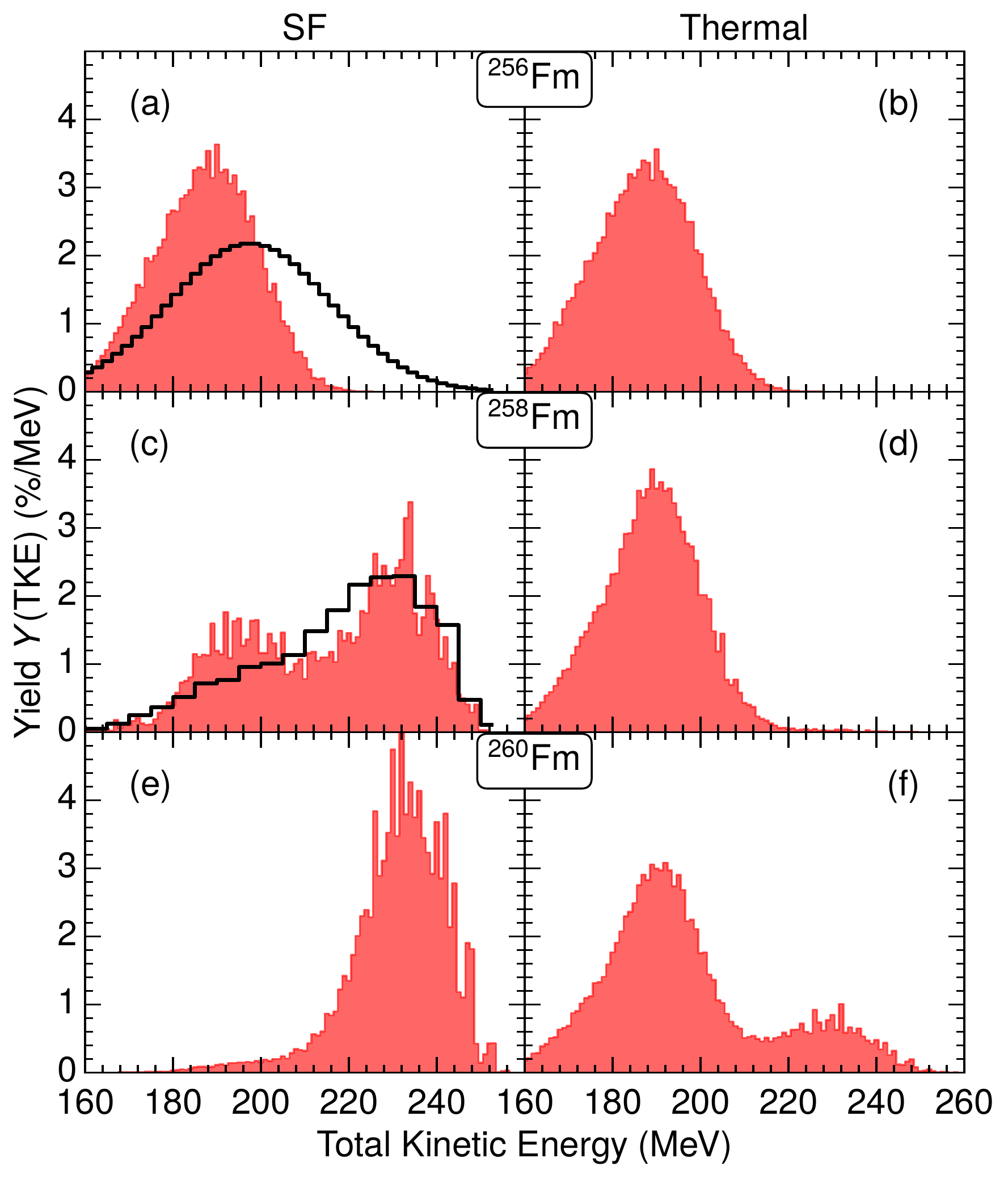} 
\caption[TKE yields in $^{256,258,260}$Fm for SF and thermal energy]{
TKE yields for SF (left panel) and thermal energies (right panel)
for $^{256}$Fm (a,b), $^{258}$Fm (c,d), and $^{260}$Fm (e,f). 
Results from calculations are shown by red histograms and results from data is shown by black histograms 
taken from Ref.\ \cite{hoffman80:a} ($^{256}$Fm, unspecified if before or after neutron emission)
and Ref.\ \cite{hulet89:a} ($^{258}$Fm, post-neutron emission).
   }
\label{fig:tke_yields_fm}  
 \end{center} 
\end{figure}

\begin{figure}[b] 
 \begin{center} 
 \includegraphics[width=1.0\linewidth]{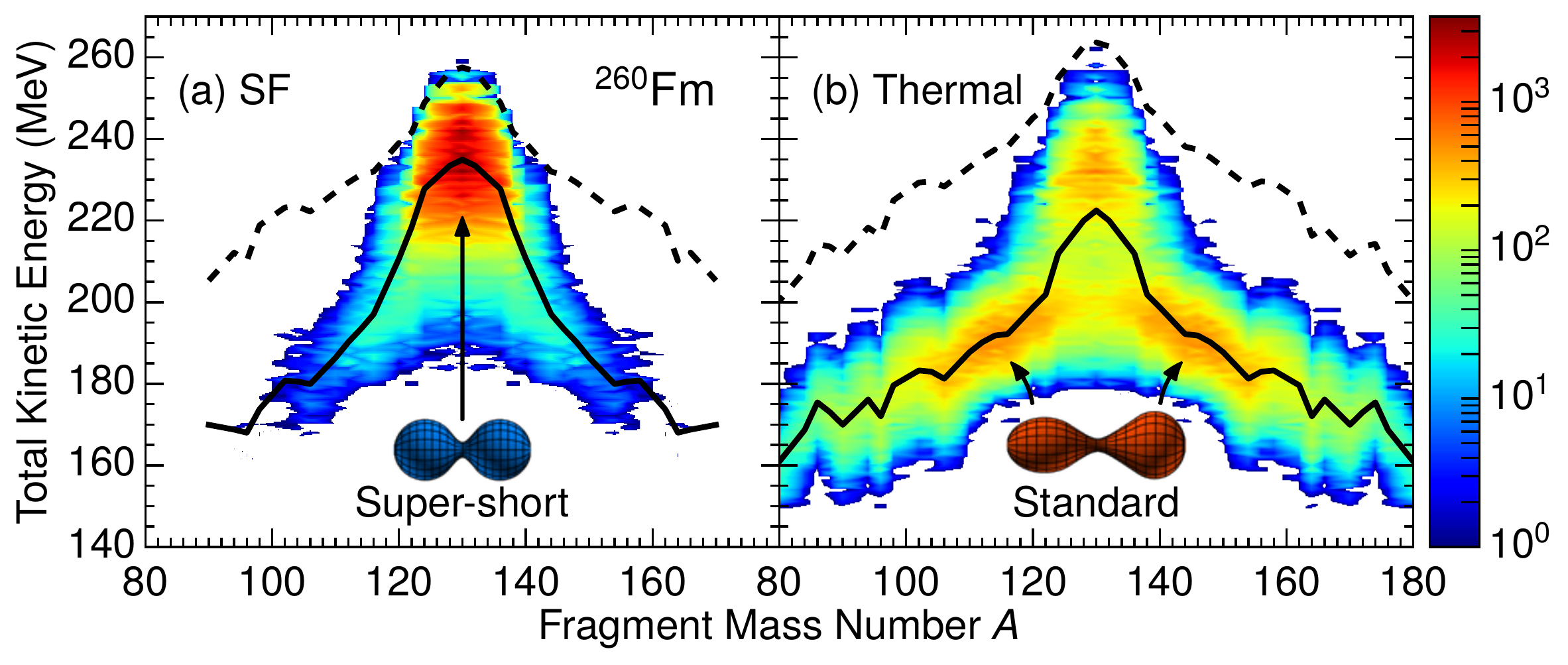} 
 \caption[Contour figure of total kinetic energy in $^{260}$Fm]{
Contour plots (on a logarithmic scale) for $^{260}$Fm
in the plane of the fragment mass number $A$ and TKE based on SF (a)
or thermal (b) fission events. Also shown are $Q^\ast$ (dashed lines) and the average TKE (solid lines) for each $A$.
Typical scission shapes are shown for the compact, symmetric
SS and the elongated, asymmetric St modes.
The figure is taken from Paper VI.
 }
\label{fig:tke_fm260} 
 \end{center} 
\end{figure}

\begin{figure}[t] 
 \begin{center} 
 \includegraphics[width=0.6\linewidth]{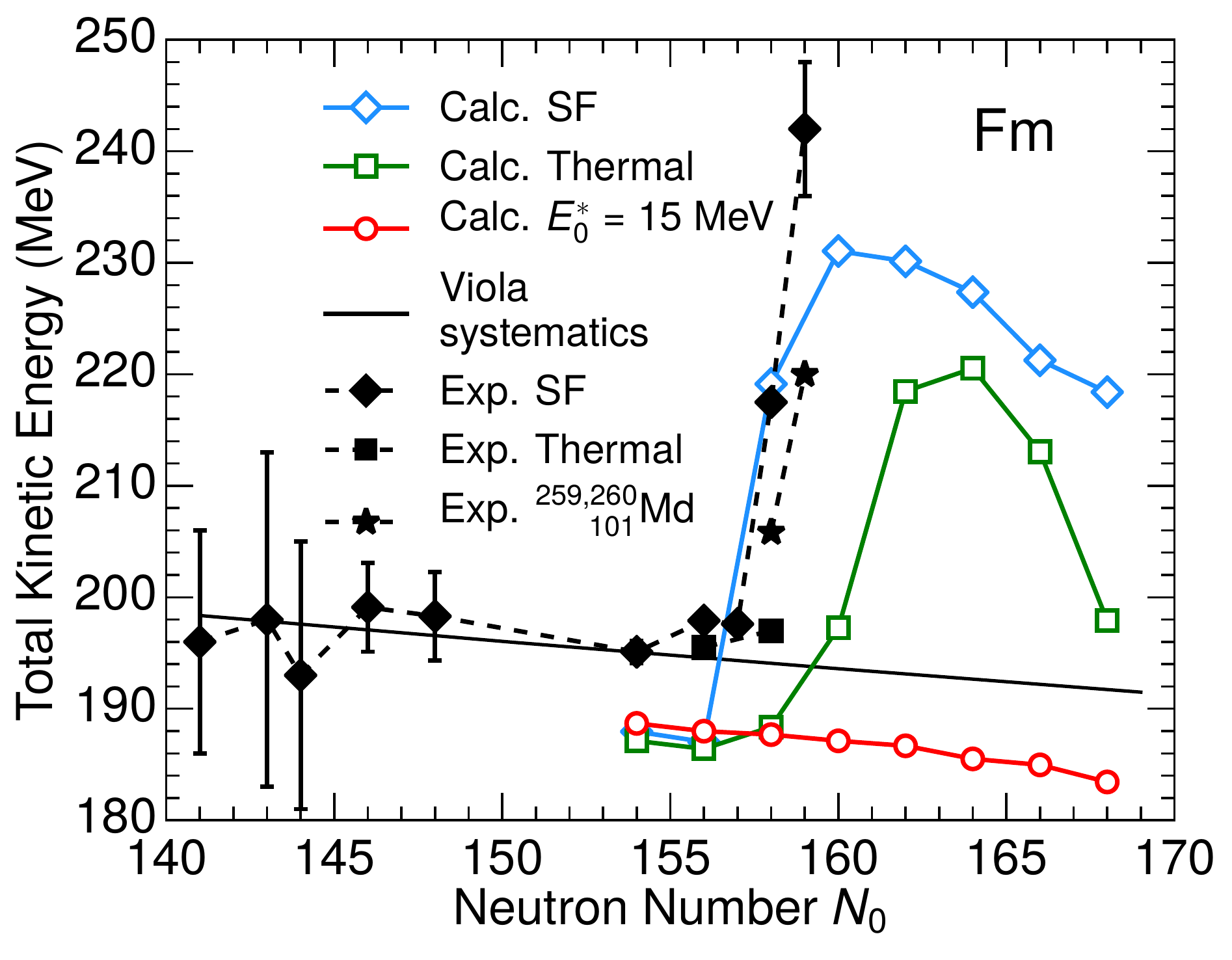} 
 \caption[Energy dependence of TKE in fission of fermium isotopes]{
Average TKE from fission of fermium isotopes versus neutron number $N_0$. 
For even $^{254-268}$Fm, theoretical results (before neutrons emission from fragments)
are shown as open symbols for SF (blue diamonds), thermal fission (green squares)
and $E_0^\ast=15$ MeV (red circles). Experimental data
are shown as filled black diamonds (SF) \cite{hulet89:a,khuyagbaatar08:a,hoffman80:b,gindler77:a,unik73:a,balagna71:a,hulet80:a} and 
filled black squares (thermal fission) \cite{john71:a,ragaini74:a}.
Data for $^{259}$Fm show most probable TKE value.
Some of the data correspond to TKE of fragments before neutrons emission and some to after neutron emission
(for $^{246,248}$Fm the difference is about 3 MeV, which is within the error bars \cite{hoffman80:b}).
The solid line shows Viola systematics \cite{viola85:a} given by Eq.\ (\ref{eq:violatke}).
The black stars show SF data for neighbouring nuclei $^{259,260}_{\;\;\;\;\;\; 101}\text{Md}$ \cite{hulet89:a}.
The figure is taken from Paper VI.
 }
\label{fig:fig_tke_vs_n_fm} 
 \end{center} 
\end{figure}

As seen in Fig.\ \ref{fig:tke_vs_af_u236}(c), the width of the TKE distribution
for a given $A_{\rm H}$ is underestimated in the calculations for both
scission conditions, but the agreement with data is generally
better for the adopted value, $c_{\rm sc}=1.5$ fm, than for $c_{\rm sc}=2.5$ fm. 
For the employed value, the underestimation is about 50\% in the region of symmetric fission and decreases with
increasing fission-fragment asymmetry. 
The underestimation may (at least in part) be due to the fact that the calculations
include only even-even fragment pairs having (approximately) equal $N/Z$ ratios, namely, that of $^{236}$U. 
As a consequence of this restriction, there is only one $(N,Z)$ combination for a
given $A$, whereas the actual fission process populates several
combinations and thus leads to a broader TKE distribution.

Figures \ref{fig:tke_vs_af_u236}(b) and \ref{fig:tke_vs_af_u236}(d) show the average TKE and the width $\sigma_{\rm TKE}$, 
respectively, versus the heavy-fragment mass number
for incoming neutron of energy $E_{\rm n}=5.55$ MeV. 
The measured TKE values are very well reproduced not only for $A_{\rm H}\geq132$, as was the case for thermal fission [see Fig.\ \ref{fig:tke_vs_af_u236}(a)], but
even down to $A_{\rm H}=126$, leaving only a rather narrow region
around symmetry with a significant overestimate, by up to 10 MeV.
Also the calculated widths agree better with data for energetic than for thermal neutrons [Fig.\ \ref{fig:tke_vs_af_u236}(d)], though the
calculated values generally are too small. Furthermore, the
data exhibits a maximum in $\sigma_{\rm TKE}$ at small $A_{\rm H}$ values and
the calculations yield a similar feature. 
This maximum is related to the bimodal structure in the heavy-fragment deformation around $A_{\rm H}\approx130$,
discussed in Sec.\ \ref{sec:scission_shapes}.

Figure \ref{fig:tke_236} shows contour plots of the calculated (a) and 
measured (b) number of fission events with respect to fragment mass number $A$ and TKE for $^{235}\text{U}(\text{n}_{\rm th},\text{f})$.
The restriction of fixed $N/Z$ ratio implies that there is a definite $Q^\ast$ value for each mass division, $Q^\ast(A)$, 
which represents the maximum possible TKE value attainable in the calculations. 
It is shown by the dashed curve in both panels of Fig.\ \ref{fig:tke_236} and it can be seen
that the bulk of the events lie well below this boundary, both
theoretically and experimentally. In reality a particular mass
division can lead to fragments with a variety of $N/Z$ ratios,
leading to a corresponding range of $Q^\ast$ values for each $A$.

The calculated average TKE values for the entire region of study in Fig.\ \ref{fig:yield_chart}
are shown in Fig.\ \ref{fig:tkeav},
where the TKE values are seen to  generally increase for heavier nuclei.
More easily interpretable is the difference between the Viola TKE systematics \cite{viola85:a}
\begin{equation}
\label{eq:violatke}
\text{TKE}^{\rm (Viola)}=0.1189 Z_0^2/A_0^{1/3}+7.3 \text{ MeV},
\end{equation}
and the actually calculated average TKE.
This is illustrated in Fig.\ \ref{fig:vmctke}. 
In their main features, the Viola TKE systematics are well reproduced in the calculations.
In the region of very asymmetric fission below $Z_0=114$ ($N_0\approx$ 162--174) the TKE is lower than the systematics. 
Also in the heavy neutron-rich region ($N_0\approx$ 210--220), where
we obtain asymmetric fission yields, the average TKE is lower than given by the Viola systematics.

\subsection*{$^{254-268}$Fm}\label{sec:tke_fm}
Large values of the TKE are seen near $^{258}$Fm due to the presence of the SS mode
and are also seen experimentally \cite{hoffman80:a,hulet80:a,hulet86:a,hulet89:a,wild90:a}.
Figure \ref{fig:tke_yields_fm} shows calculated TKE distributions (red histograms) compared to data (black curves) for the
isotopes $^{256}$Fm, $^{258}$Fm, and $^{260}$Fm for SF (left panel) and thermal fission (right panel). 
Fission of $^{256}$Fm exhibits a single Gaussian distribution peaked at $\text{TKE}\approx190$ MeV,
corresponding to fission in the St mode for both energies.
The mixture of the St mode and SS mode for SF in $^{258}$Fm results in
a bimodal distribution, which is also seen in the data.
In $^{260}$Fm, the SS mode is dominating for SF
and this yields a single-humped distribution peaked at the higher value $\text{TKE}\approx230$ MeV.
For thermal fission, there is a mixture of the two modes also in $^{260}$Fm
and a bimodal distribution is obtained.

Figure \ref{fig:tke_fm260} shows the number of scission events for $^{260}$Fm
versus fragment mass number and TKE for
(a) SF and (b) $E_0^\ast=6.13$ MeV (thermal fission). 
For SF most events are seen to occur in the SS mode with symmetric fragment masses with large TKE values. 
Increasing the excitation energy to $E^\ast_0=6.13$ MeV gives a very different distribution of TKE-values.
Compared to SF the number of events in the SS mode has decreased, and many events now occur 
in the St mode at low TKE with an asymmetric division of fragment masses.
Projecting the scission events shown in Fig.\ \ref{fig:tke_fm260} on the TKE-axis yields the
two TKE distributions shown in \ref{fig:tke_yields_fm}(e) and (f).
Typical shapes of the two modes are shown in Fig.\ \ref{fig:tke_fm260}(a) and (b).

Figure \ref{fig:fig_tke_vs_n_fm} shows the average TKE for even $^{254-268}$Fm versus the neutron number $N_0$
of the fissioning nucleus for three different excitation energies,
and provides an experimental correspondence to the phase diagram in Fig.\ \ref{fig:phase_diagrams}.
Experimental data for the lighter isotopes $N_0<158$ are reasonably described
by the linear behaviour of the Viola systematics.
For SF the SS mode starts playing a role at $N_0=158$, where TKE is seen to increase suddenly,
both in data and calculations.
A similar increase is also seen in data för neighbouring nuclei $^{259,260}$Md (black stars).
For thermal fission ($E_0^\ast\approx6$ MeV) the SS mode
dominates for $^{260-268}$Fm resulting in large TKE-values.
And finally, for 15 MeV all nuclei are found to fission in
the St mode, and the TKE-values are small for all isotopes
with similar linear behaviour as the Viola systematics.

\subsection*{$^{274}$Hs}\label{sec:tke_hs}
Figure \ref{fig:tke_hs274} shows calculated scission events for $^{274}$Hs versus fragment mass number and TKE for
(a) $E^\ast_0=6.7$ MeV and (b) $E^\ast_0=10$ MeV.
The average TKE values are shown as solid black lines, while the total available energy $Q^\ast$
are shown as dashed black lines.
The compact scission shapes around $A_{\rm L}$:$A_{\rm H}\approx66$:$208$, associated with the SA mode, result in large TKE values that
are close to the maximum possible value $Q^\ast$.
The corresponding average TXE of the fragments, given by the difference between the curves for $Q^\ast$ and TKE,
is only about 20 MeV for the SA mode.
Conversely, the TXE value for the symmetric mass-splits in Fig.\ \ref{fig:tke_hs274}(b) is about 80 MeV.
The change from asymmetric to symmetric fission when the energy is increased then leads to 
an increase in average TXE of about 60 MeV. 
As will be discussed in Ch.\ \ref{ch:particle_emission}, this in turn results in a large difference in the number of neutrons emitted from the fragments.

\begin{figure}[hbt!] 
 \begin{center} 
 \includegraphics[width=1.0\linewidth]{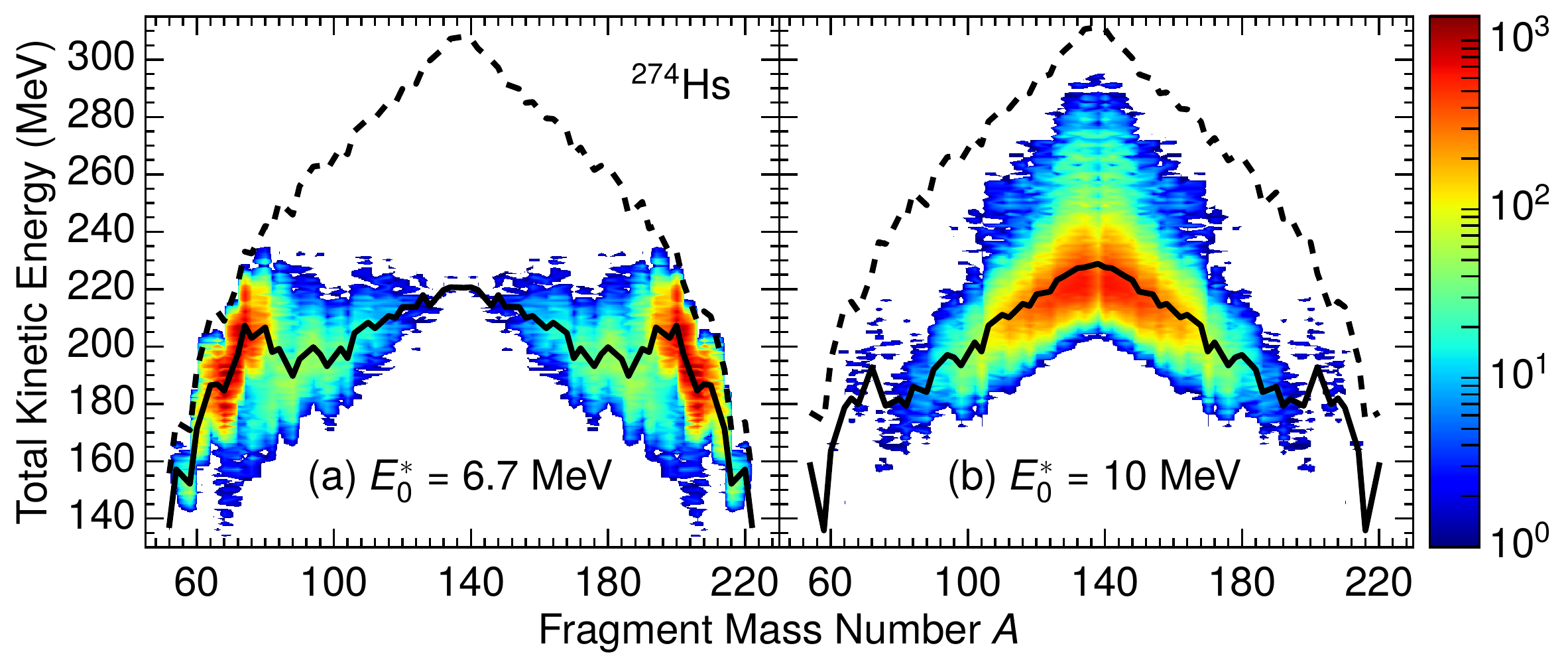} 
 \caption[Contour figure of total kinetic energy in $^{274}$Hs]{
 Contour plots (on a logarithmic scale) for $^{274}$Hs
in the plane of the fragment mass number $A$ and TKE 
for excitation energies $E^\ast_0=6.7$ MeV (a) and 10 MeV (b).
Also shown are $Q^\ast$ (dashed lines) and the average TKE (solid lines) for each $A$.
 }
\label{fig:tke_hs274} 
 \end{center} 
\end{figure}

\section{Fragment excitation energy}\label{sec:fragment_eexc}
After the fissioning nucleus has split, the resulting fission fragments relax back to their ground-state deformations.
The associated distortion energies are then converted into statistical fragment excitations after scission,
and it is only the intrinsic excitation energy $E^\ast_{\text{sc}}$ in the fissioning nucleus that is shared between the nascent fragments.
The distortion energy, obtained with Eq.\ (\ref{eq:distortion}), and the intrinsic excitation energy then makes up the total excitation
in the fragment which subsequently goes to evaporation of neutrons.

\begin{figure}[b] 
 \begin{center} 
 \includegraphics[width=0.7\textwidth]{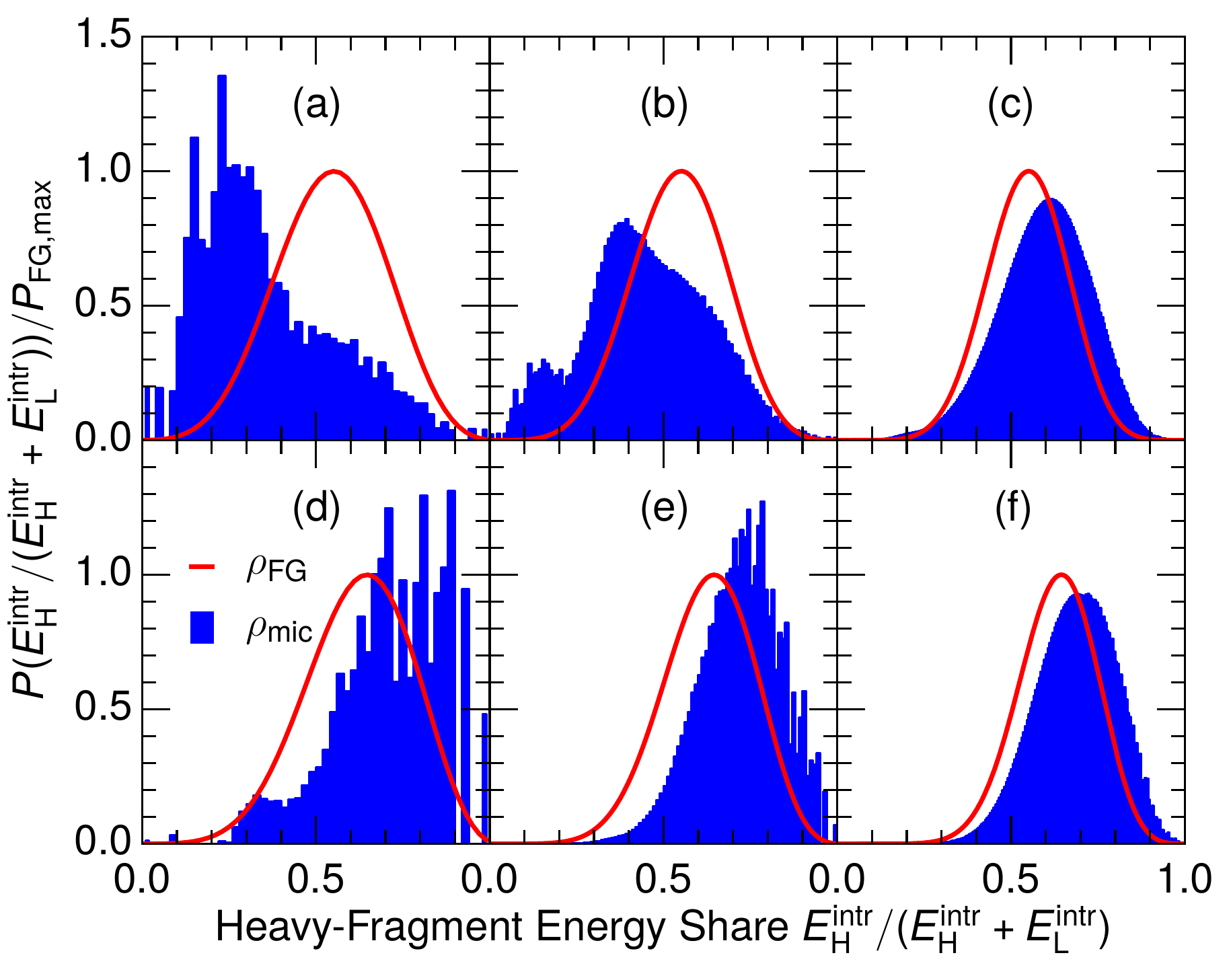} 
 \caption[Energy-partition distributions in $^{235}\text{U}(\text{n},\text{f})$]{
 The distribution function $P(E^{\rm intr}_{\rm H};E^\ast_{\rm sc})$ for the total excitation of
 the heavy fragment in $^{235}\text{U}(\text{n},\text{f})$ for two different divisions,
 either $(N,Z,\varepsilon)_{\rm H}=(80,50,−0.1)$ and $(N,Z,\varepsilon)_{\rm L}=(64,42,0.3)$ (top
 panels) or $(N,Z,\varepsilon)_{\rm H}=(92,60,0.1)$ and $(N,Z,\varepsilon)_{\rm L}=(52,32,0.1)$ 
 (bottom panels), and three different values of the available energy at scission $E^\ast_{\rm sc}$=10 (left column), 
 20 (center column), 40 (right column) MeV.
 The distributions obtained from microscopic (blue histograms)
 and FG (solid red curves) level densities are normalized to the
 maximum value of the FG result. 
 The figure is taken from Paper I.
 }
\label{eexc_partition}  
 \end{center} 
\end{figure}

\begin{figure}[b]
 \begin{center} 
 \includegraphics[width=0.6\textwidth]{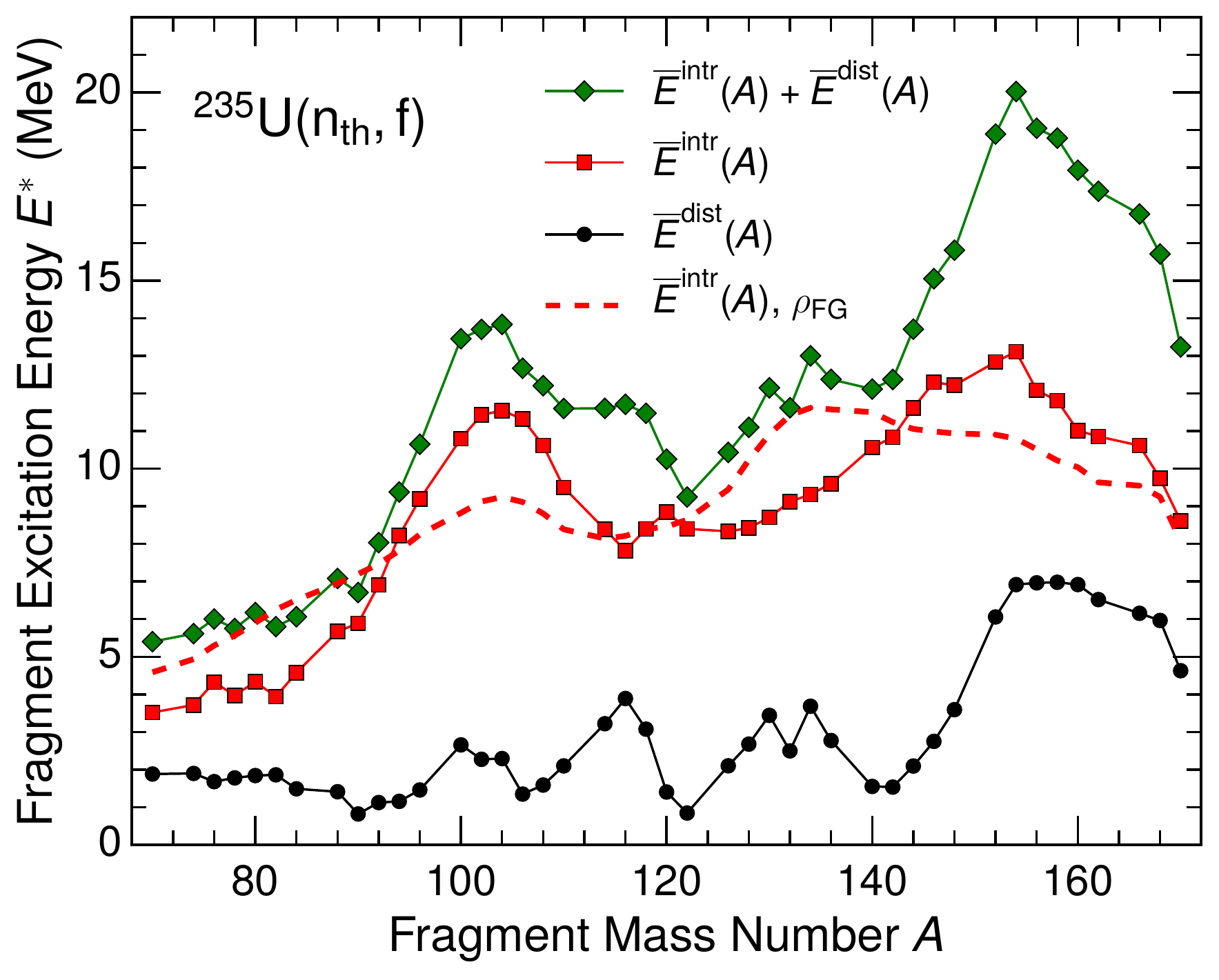} 
 \caption[Fragment excitation energy vs. $A$ in $^{235}{\rm U}(\text{n}_{\rm th},\text{f})$]{ 
 The mean fragment distortion energy $\overline{E}^{\rm dist}(A)$ (black circles),
 the mean intrinsic excitation energy $\overline{E}^{\rm intr}(A)$ (red squares), and the
 sum, $\overline{E}^{\rm dist}(A)+\overline{E}^{\rm intr}(A)$ (green diamonds) are shown as functions of the
 fragment mass number $A$ for $^{235}\text{U}(\text{n}_{\rm th},\text{f})$.
 The red dashed curve show $\overline{E}^{\rm intr}(A)$ with the simple FG level density (red dashed curve). 
 The figure is taken from Paper I.
 }
\label{eexc_vs_af_235u}  
 \end{center} 
\end{figure}
\begin{figure}[b]
 \begin{center} 
 \includegraphics[width=0.7\textwidth]{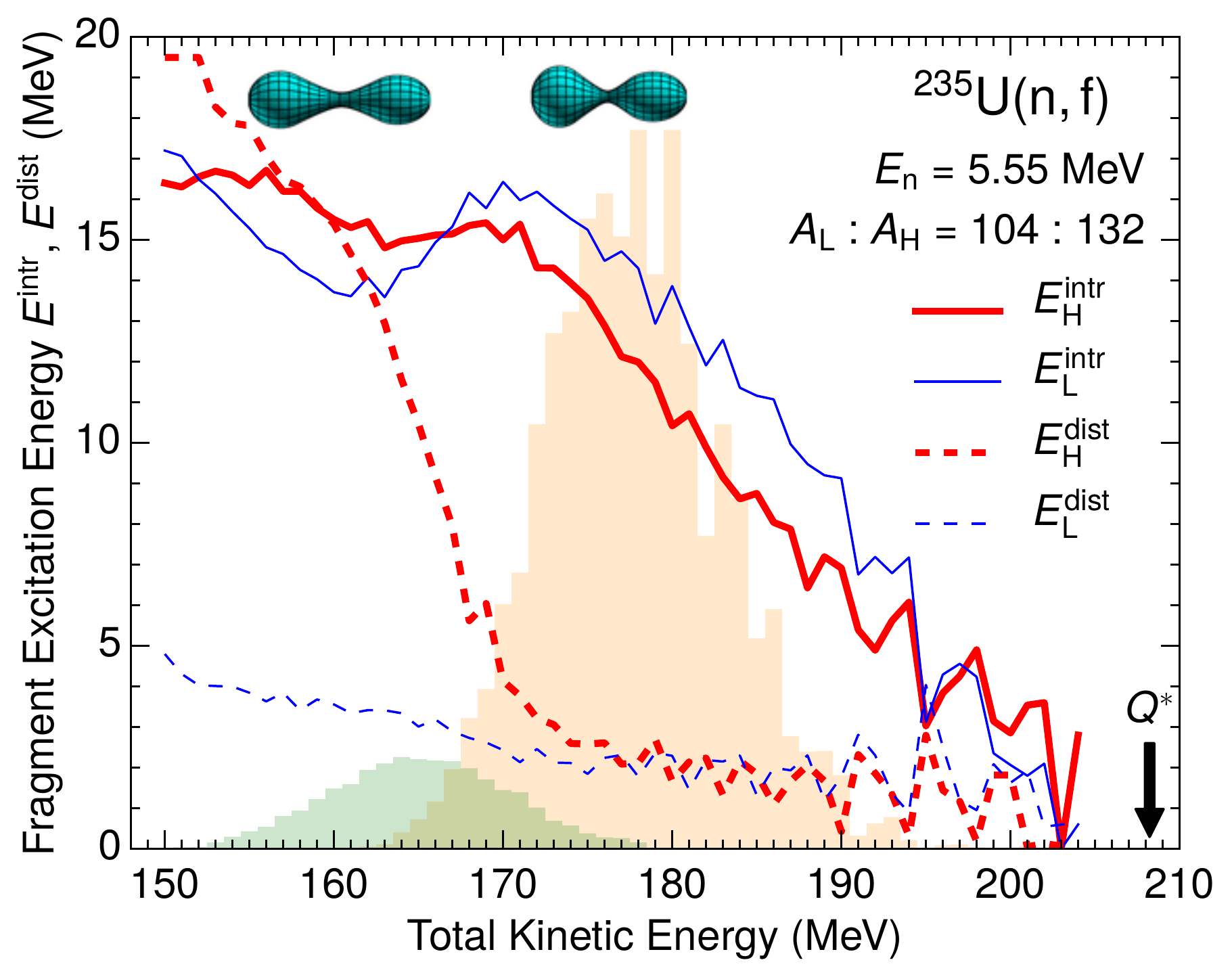} 
 \caption[Bimodal fission in $^{235}{\rm U}(\text{n},\text{f})$]{
Fragment excitation energy as a function of TKE in $^{235}{\rm U}(\text{n},\text{f})$ for incident-neutron energy $E_{\rm n}=5.55$ MeV
and mass split $A_{\rm L}$:$A_{\rm H}=104$:$132$.
 Probability densities are shown for the two modes: SL (green area), and St (orange area).
Typical scission shapes of the SL and St modes are also shown. 
 The figure is taken from Paper V.
 }
\label{bimodal_235u}  
 \end{center} 
\end{figure}

It has long been puzzling that the heavy fragment emits fewer neutrons
than the light fragment in low-energy fission of actinides (see Ch.\ \ref{ch:particle_emission}).
This indicates that the heavy fragment obtains a smaller amount of excitation energy than the light fragment,
which appears to differ from simple statistical expectations.
If the two nascent fragments are assumed to be in thermal equilibrium at scission,
the excitation energy is divided between the two fragments according to their level densities.
Employment of the simplified FG level density $\rho_{\rm FG}(E^\ast)\sim \text{exp}(2\sqrt{aE^\ast})$ as in previous studies \cite{madland82:a,lemaire05:a,kornilov07:a},
results in an energy division in proportion of the mass ratio of the fragments, $E^{\rm intr}_{\rm L}/E^{\rm intr}_{\rm H}=A_{\rm L}/A_{\rm H}$.
However, as was recently pointed out by Schmidt and Jurado \cite{schmidt10:a}, 
the simplified FG level density may be misleading at low energies where structure effects tend to be significant.
They instead used the constant-temperature level density $\rho(E^\ast)\sim \text{exp}(E^\ast/T)$ \cite{gilbert65:a},
where the initial temperature $T$ of a fragment was parametrized with an expression containing a shell-effect term.
This leads to an ``energy-sorting mechanism'',
in which energy flows from the hot to the cold fragment,
and all intrinsic excitation energy is found in the fragment with the lower initial temperature $T$.
When the heavy fragment is close to the $^{132}$Sn, its temperature is increased due to strong shell effects,
and the excitation energy then goes to the light fragment.

It the present studies, we also assume that thermal equilibrium is obtained at scission,
but divide the the total intrinsic excitation energy available at scission $E^\ast_{\text{sc}}$
based on microscopically calculated level densities of the fragments at their scission shapes.
The intrinsic excitation energy of the heavy fragment, $E^{\rm intr}_{\rm H}$,
is then governed by the following micro-canonical distribution,
\begin{equation}
\label{eq:energy_distrib}
P(E^{\rm intr}_{\rm H};E^\ast_{\rm sc})\sim \tilde{\rho}_{\rm H}(E^{\rm intr}_{\rm H};\varepsilon^{\rm sc}_{\rm H})
\cdot\tilde{\rho}_{\rm L}(E^\ast_{\rm sc}-E^{\rm intr}_{\rm H};\varepsilon^{\rm sc}_{\rm L}),
\end{equation}
where $E^{\rm intr}_{\rm L}=E^\ast_{\rm sc}-E^{\rm intr}_{\rm H}$ due to energy conservation, 
and where $\tilde{\rho}_{i}(E^\ast_{i};\varepsilon^{\rm sc}_{i})=\tilde{\rho}(N_i,Z_i,E^\ast_{i};\varepsilon^{\rm sc}_{i})$
is the effective density of states (defined below) of a nucleus with neutron and proton numbers $N_i$ and $Z_i$, 
spheroidal deformation $\varepsilon_i$, and an excitation energy of $E^\ast_i$, with $i=\text{H}$, L.

The fragment level densities $\rho_i(E^\ast_i,I_i;\varepsilon_i^{\rm sc})$ 
at the corresponding fragment shapes at scission
are calculated by employing the combinatorial method
described in Sec.\ \ref{sec:combinatorial_levdens}.
Since we are interested in the energy distribution only, 
we sum over the fragment angular momentum, $I_i$, to obtain the effective density of states entering in Eq.\ (\ref{eq:energy_distrib}),
\begin{equation}
\tilde{\rho}_{i}(E^\ast_{i};\varepsilon^{\rm sc}_{i})=\sum_{I_i}(2I_i+1)\rho_i(E^\ast_i,I_i;\varepsilon_i^{\rm sc}).
\end{equation}

Figure \ref{eexc_partition} shows the energy distribution $P(E^{\rm intr}_{\rm H};E^\ast_{\rm sc})$ at three different values 
of the total available energy $E^\ast_{\rm sc}$ for two different mass
divisions of $^{236}$U having $(A_{\rm H}$:$A_{\rm L})=(130$:$106)$ and (152:84)
The energy distribution is calculated with both the microscopic level density discussed above and the simple 
macroscopic FG level density, $\rho_{\rm FG}(E^\ast)\sim \text{exp}(2\sqrt{aE^\ast})$ with $a=A/(8\text{ MeV})$. 
Both yield rather broad distributions due to the smallness of the nuclear system. 
The macroscopic form yields smooth Gaussian-like distributions peaked at $E^{\rm intr}_{\rm H}/E^{\rm intr}_{\rm L}=A_{\rm H}/A_{\rm L}$, whereas the
microscopic form yields irregular distributions that may have qualitatively different appearances, 
especially at lower values of $E^\ast_{\rm sc}$ where quantal structure effects are most significant. In particular,
it is possible that one fragment receives all the available energy
with the partner fragment being left without excitation. Although
the probability for this decreases quite rapidly with increasing $E^\ast_{\rm sc}$, this feature is in dramatic contrast to the macroscopic result.

For each scission configuration obtained at the end of the
Metropolis walk, the excitation energies of the nascent fragments
are sampled from the appropriate microscopic partition distribution. 
For $^{235}\text{U}(\text{n}_{\rm th},\text{f})$, the resulting mean intrinsic excitation energy $\overline{E}^{\rm intr}(A)$ is shown in Fig.\ \ref{eexc_vs_af_235u}
as a function of the fragment mass number $A$, 
together with the mean fragment distortion energy $\overline{E}^{\rm dist}(A)$, as well as the sum of these two quantities
which represents the total excitation energy of the fragment.
The figure also shows the mean excitation energy obtained with the simple
FG level density to illustrate the effect of the microscopic level densities.

Figure \ref{bimodal_235u} shows the separate contributions to the final
fragment excitation energy from their intrinsic and distortion
energies at scission, displayed versus the resulting TKE for $A_{\rm L}$:$A_{\rm H}$=104:132
in $^{235}{\rm U}(\text{n},\text{f})$ with $E_{\rm n}=5.55$ MeV.
The black arrow indicates the maximum possible kinetic energy allowed by the $Q^\ast$-value.
For such an event to take place the fragments have to be emitted in their ground-state deformations
and with no intrinsic excitation energy;
all available energy would then be transferred into kinetic energy of the fragments.
The majority of the events correspond to fission in the St mode (orange histogram)
centered at $\text{TKE}\approx178$ MeV,
while the SL mode (green histogram) also is present in this case for lower values at $\text{TKE}\approx164$ MeV.

The results in Fig.\ \ref{bimodal_235u} can be understood
by comparing with Fig.\ \ref{fig:fig_eps_vs_q2_u236_e5}(b),
where the fragment deformations vs.\ $q_2$ are shown for the same mass split.
The elongation $q_2$ roughly corresponds to the TKE of the fragments,
while the fragment deformation, relative its ground-state shape,
corresponds to the distortion energy.
Since the heavy fragment $A_{\rm H}=132$ is spherical in the ground state,
the small deformations at small $q_2$ (large TKE) results in
low distortion energy, while the large deformation $\varepsilon_{\rm H}\approx0.3$
at large $q_2$ (low TKE) results in large amounts of distortion energy.

\chapter{Neutron evaporation from fragments}\label{ch:particle_emission}
The majority of the excitation energy in the fragments is dissipated by emission of neutrons.
The relaxation of the fragments from their scission shapes to their ground-state shapes is assumed to occur at a shorter
time scale than the subsequent neutron evaporation \cite{vandenbosch73:a}.
The neutrons are therefore evaporated from fragments in their ground-state deformations.
The fragments may also release the excitation energy through other decay channels, such as photon emission,
but neutron emission is typically the fastest decay.
When all of the excitation energy for neutron emission has been exhausted,
the remaining excitation energy is used to emit photons.
Only neutron emission is considered in the present studies.

\section{Formalism}
The random-walk calculations give rise to a set of fission fragments having different proton and neutron numbers. 
Associated with each fragment is a probability function that describes how common it is to find the fragment with a specific excitation energy.
Considering a single fragment in its ground state with a certain excitation energy we assume that it cools by evaporating neutrons. 
Several neutrons may be evaporated in sequence until the energy of the remaining nucleus falls below the neutron separation energy threshold. 
Concerning the energy available for neutron emission, we assume that the rotational energy $E_{\text{rot}}$
of the initial fission fragment can not be used for neutron emission. 

How many neutrons that evaporates will also depend on how much kinetic energy each neutron flies away with. 
The amount of kinetic energy taken away by a neutron follows a statistical distribution. 
This distribution is obtained by considering the probability per unit time of a mother nucleus emitting a neutron with kinetic energy between 
$\epsilon_{\rm n}$ and $\epsilon_{\rm n}+d\epsilon_{\rm n}$ \cite{weisskopf37:a}:
\begin{equation}
N\sigma\left(E^\ast_{\rm A},\epsilon_{\rm n}\right)\epsilon_{\rm n}\rho_{\rm A-1}\left(E^\ast_{\rm A-1}\right)d\epsilon_{\rm n}.
\end{equation}

For our purpose $N$ can be considered a constant since it is independent of $\epsilon_{\rm n}$.  
$\rho_{A-1}$ is the level density of the daughter nucleus after neutron evaporation evaluated at energy $E^\ast_{A-1}=E^\ast_{A}-S_{\text{n}}-\epsilon_{\rm n}$, 
where $E^\ast_{A}$ is the initial excitation energy and $S_{\text{n}}$ denotes the neutron separation energy of the evaporating nucleus. 
In this formula, the emission probability is expressed in terms of the cross section $\sigma\left(E^\ast_{A},\epsilon_{\rm n}\right)$
for the inverse problem of compound nucleus formation through neutron capture. 
The energy dependence of this process can be taken into account using optical-model potentials \cite{madland82:a}.
However, in the present approach we adopt the original estimate \cite{weisskopf37:a} of an energy independent cross section. 
The resulting formula can then be expressed as
\begin{equation}
\label{eq:neutron_spectra}
\tilde{N}\epsilon_{\rm n}\rho_{A-1}\left(E^\ast_{A}-S_{\text{n}}-\epsilon_{\rm n}\right)d\epsilon_{\rm n}.
\end{equation}
This gives the relative probabilities for neutrons evaporating with different kinetic energies. 
We assume that $E^\ast_{A-1}>E_{\text{rot}}$,
i.e. that the rotational energy of the initial fission fragment is not converted into neutrons. 
This gives the possible range of neutron kinetic energies as $0<\epsilon_{\rm n}<E^\ast_{A}-S_{\text{n}}-E_{\text{rot}}$. 

\begin{figure}[htbp!]
\centering
\includegraphics[width=0.6\linewidth]{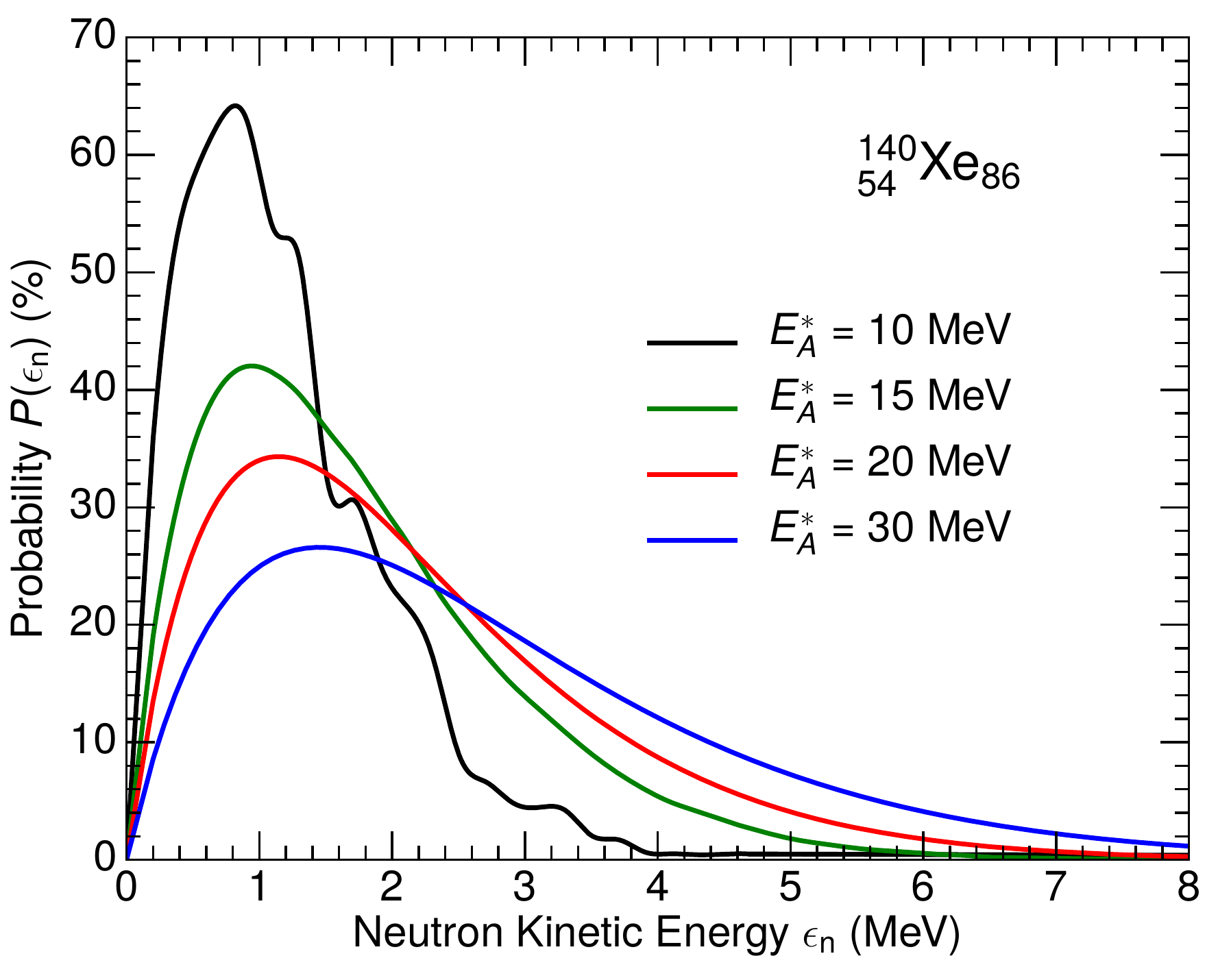}
\caption[Neutron energy spectrum]{
Probability density distribution $P(\epsilon_{\rm n})$ for the kinetic energy of evaporated neutron from nucleus $^{140}_{54}\text{Xe}_{86}$.
Results are shown for different initial excitation energies $E^\ast_{\rm A}$.
Irregularities in the distribution are seen due to structure effects in the level density at low energies.
}
\label{fig:ekin_prob} 
\end{figure}

A neutron kinetic energy can then be drawn from a random ensemble having Eq.\ (\ref{eq:neutron_spectra}) as a probability distribution function \cite{randrup09:a}. 
Fig.\ \ref{fig:ekin_prob} shows the probability distribution function for nucleus $^{140}_{54}\text{Xe}_{86}$,
which is a typical fragment in fission of $^{236}$U.
The peak in the distribution is located at $\epsilon_{\rm n}\approx1$ MeV for the lowest energy.
Irregularities in the distribution are seen due to structure effects in the level density at low energies.
For higher excitation energies, the distributions become wider and the location of the peak increases slightly.

If the resulting energy of the daughter nucleus is sufficient, the evaporation process is repeated until neutron emission is no longer energetically possible. 
New neutron energies are drawn from the corresponding ensemble and the process is restarted and repeated through the chain until a converged value for 
the average number of emitted neutrons is obtained.
Since the initial fission fragment nucleus has a distribution of excitation energy, the event-by-event simulation must be performed for several initial
energies and the resulting number of emitted neutrons are then weighted together accordingly. 
Summing the number of emitted neutrons from all fission fragments and weighting with their fission yields gives the total neutron 
multiplicity of the nucleus.

\section{Neutron multiplicities}
Figure \ref{fig:nu_vs_af} shows the calculated mean neutron multiplicity $\bar{\nu}(A)$ in $^{235}\text{U}(\text{n}_{\rm th},\text{f})$
together with experimental data from a variety of experiments. 
The jaggedness in the calculations arise due to the restriction of even $Z$ and $N$ in the fragments,
which leads to jumps in the neutron separation energy between neighbouring fragments.
The observed sawtooth behavior (minima at $A\approx76$, 126 and maxima at $A\approx 110$, 156) is reasonably well reproduced by the calculation and
arises from a combined effect of the behavior of the 
neutron separation energy $S_{\rm n}(A)$, which displays a jump near $A=132$ due to the 
closed shells at $Z=50$ and $N=82$, and the behavior of the total fragment energy (see Fig.\ \ref{eexc_vs_af_235u}). 
The shortfall of $\bar{\nu}(A)$ in the region around $A=110$ arises from
the fact that the calculations lead to too  large TKE values for mass divisions near symmetry (see Fig.\ \ref{fig:tke_vs_af_u236}(a)),
and thus too low excitation energy.
Figure \ref{fig:nu_prob} shows the total neutron multiplicity distribution for incident neutron with
thermal energy. The calculated distribution capture the overall behaviour,
though the experimental data shows a slightly broader distribution.

\begin{figure}[b]
\centering
\includegraphics[width=0.6\linewidth]{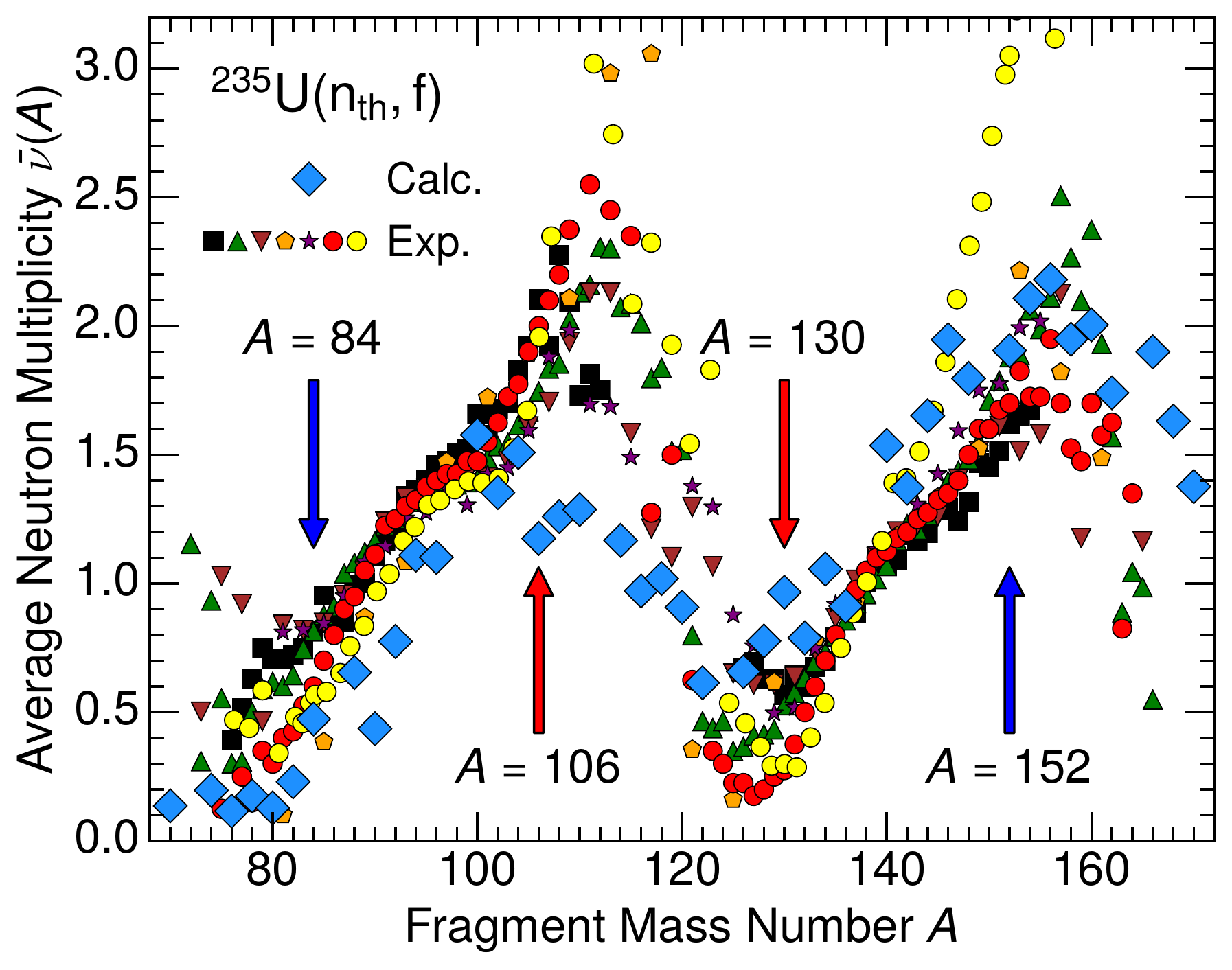}
\caption[$\bar{\nu}$ vs. $A$ in $^{235}\text{U}(\text{n}_{\rm th},\text{f})$]{
Calculated mean neutron multiplicity $\bar{\nu}(A)$ (blue diamonds) for $^{235}\text{U}(\text{n}_{\text{th}},\text{f})$ as a
function of the mass number $A$ of the primary fission fragment,
which is compared to a variety of experimental data \cite{nishio98:a,apalin65:a,vorobyev10:a,batenkov05:a,maslin67:a,gook18:a}. 
The red/blue arrows point to the mass divisions selected in Fig.\ \ref{eexc_partition}. 
The figure is taken from Paper I.
}
\label{fig:nu_vs_af} 
\end{figure}

\begin{figure}[htbp!]
\centering
\includegraphics[width=0.6\linewidth]{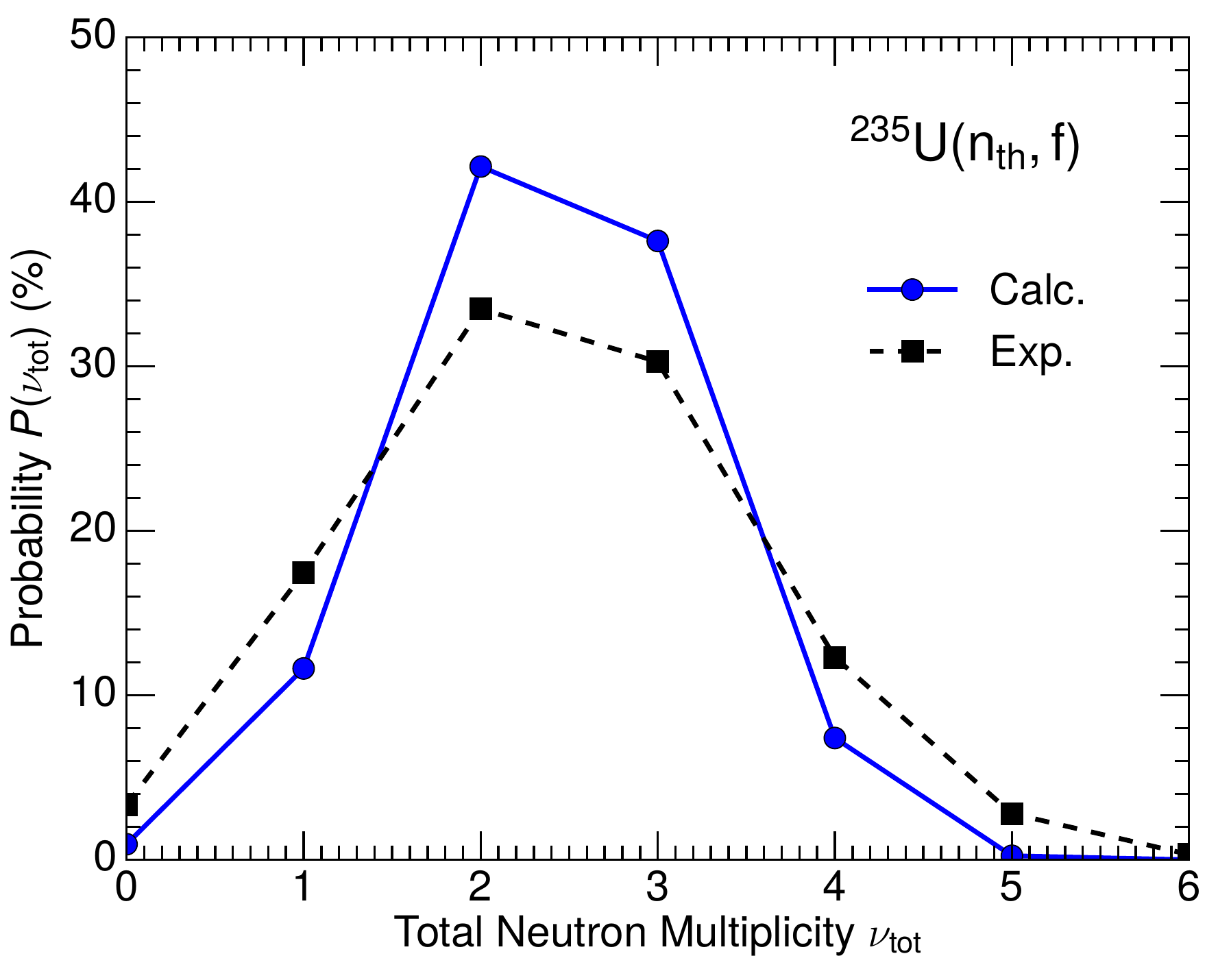}
\caption[$P(\nu_{\rm tot})$ in $^{235}{\text{U}(\text{n}_{\rm th},\text{f})}$]{
Calculated multiplicity distribution for total emitted neutrons $\nu_{\rm tot}$ (solid line,
blue circles) compared to experimental data \cite{boldeman85:a} (dashed line, black squares) in $^{235}\text{U}(\text{n}_{\rm th},\text{f})$.
The figure is taken from Paper II.
}
\label{fig:nu_prob} 
\end{figure}

\begin{figure}[t]
\centering
\includegraphics[width=0.6\linewidth]{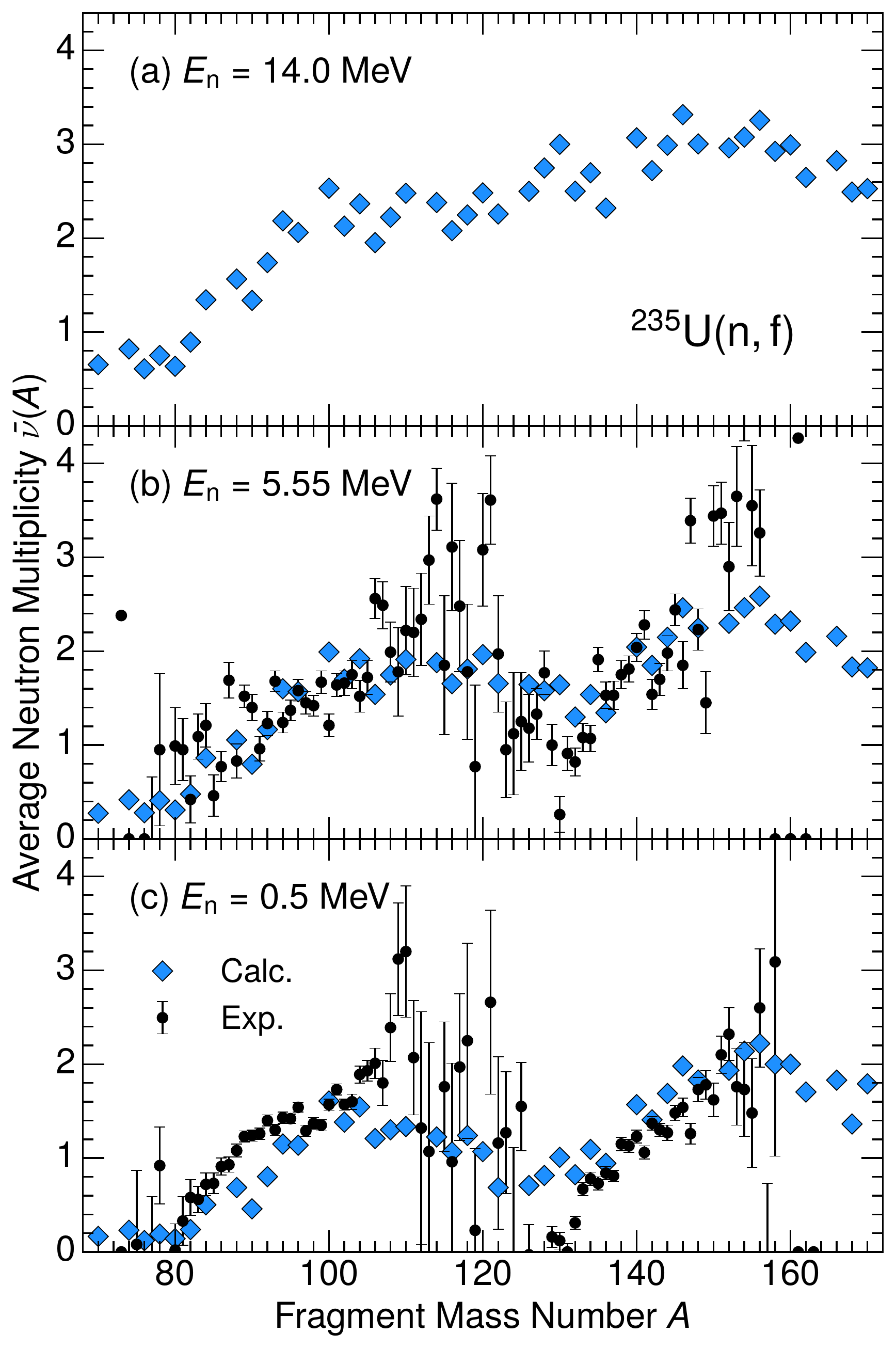}
\caption[$\bar{\nu}$ vs. $A$ in $^{235}\text{U}(\text{n},\text{f})$ for $E_{\rm n}=0.5$, 5.55, 14 MeV]{
For $^{235}\text{U}(\text{n},\text{f})$ is shown the calculated mean neutron multiplicity as a 
function of the mass number of the primary fission fragment, $\bar{\nu}(A)$, 
for three different incident neutron energies $E_{\rm n}$: 
0.5 MeV (c), 5.55 MeV (b), 14 MeV (a). The experimental data from Ref.\ \cite{muller84:a}
are also shown.
The figure is taken from Paper I.
}
\label{fig:nu_vs_af_fast} 
\end{figure}

\begin{figure}[t] 
\centering
\includegraphics[width=1.0\linewidth]{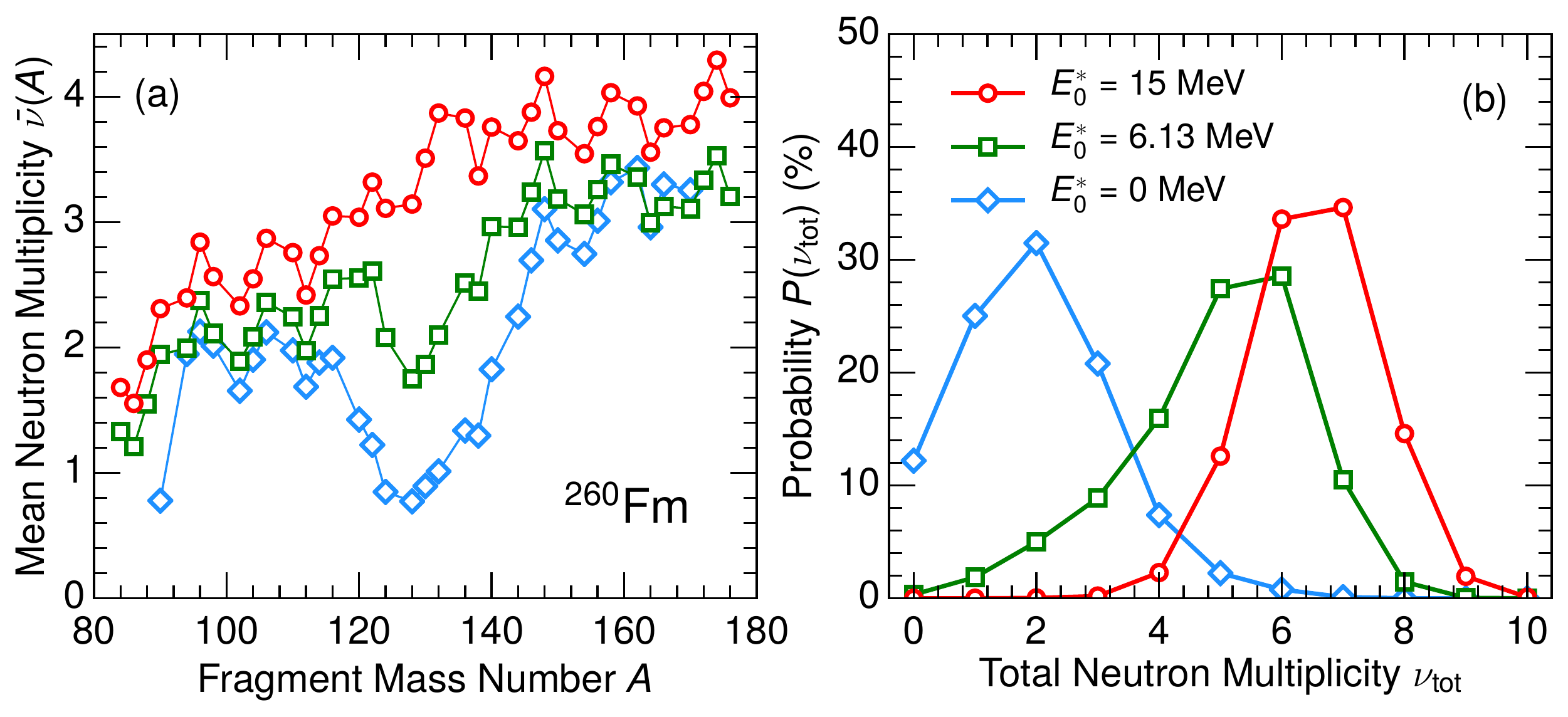}
\caption[$\bar{\nu}(A)$ and $P(\nu_{\rm tot})$ in $^{260}$Fm for $E_0^\ast=0$, 6.13, 15 MeV]{
Calculated mean neutron multiplicity $\bar{\nu}(A)$ (a) and total neutron multiplicity distribution $P(\nu_{\rm tot})$ (b) 
in fission of $^{260}$Fm for initial excitation energies $E_0^\ast=0$ (blue diamonds), 6.13 (green squares), 15 MeV (red circles).
}
\label{fig:fig_nu_both_fm260} 
\end{figure}

\begin{figure}[hbt!] 
 \begin{center} 
 \includegraphics[width=0.6\linewidth]{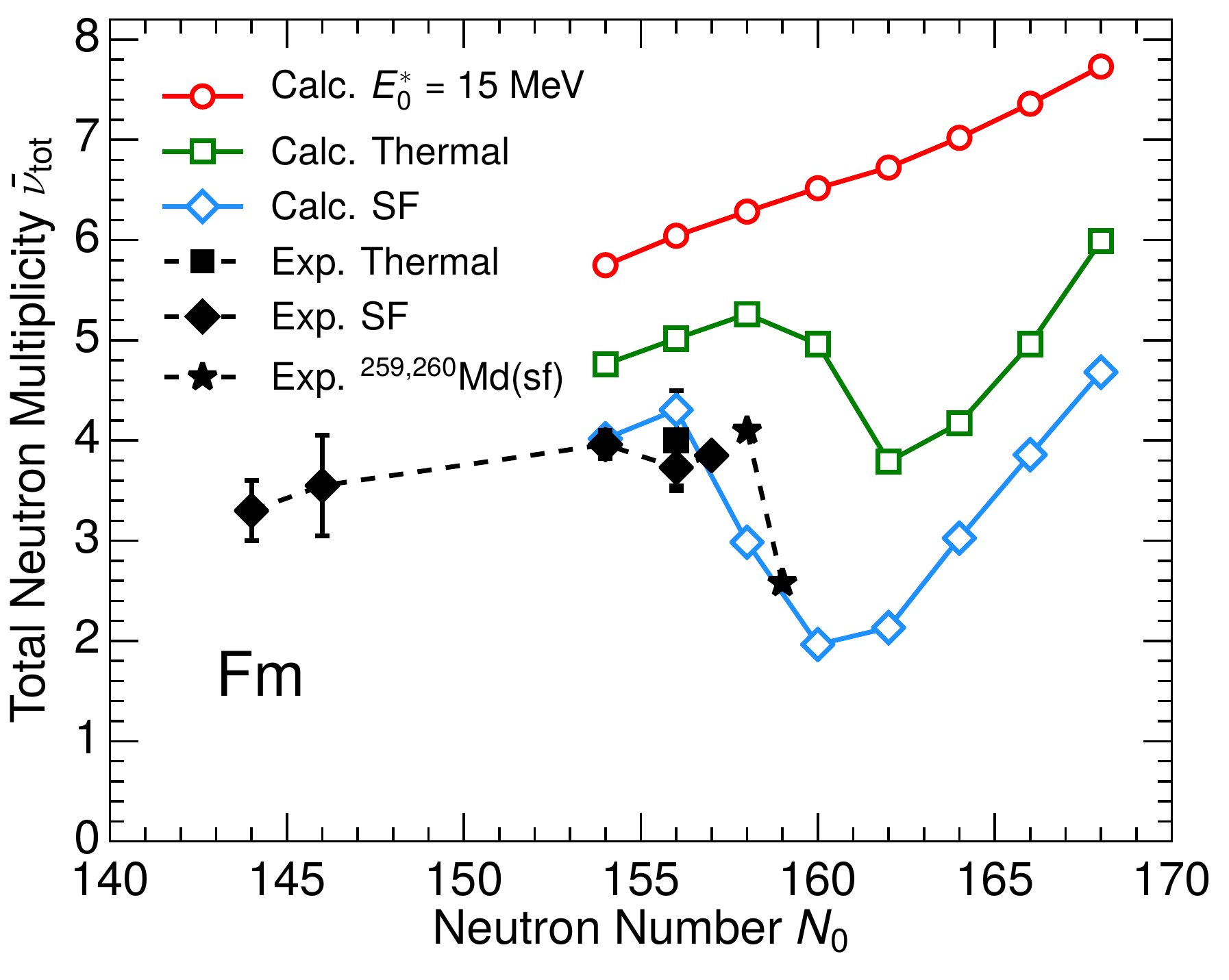} 
 \caption[Energy dependence of $\bar{\nu}_{\rm tot}$ in fission of fermium isotopes]{
Average total neutron multiplicity $\bar{\nu}_{\rm tot}$ from fission of fermium isotopes versus neutron number $N_0$. 
For even $^{254-268}$Fm theoretical results
are shown as open symbols for SF (blue diamonds), thermal fission (green squares)
and $E_0^\ast=15$ MeV (red circles). Experimental data are shown as
black diamonds (SF) \cite{svirikhin10:a,svirikhin12:a,lazarev77:a,hoffman80:c} and black squares (thermal fission) \cite{flynn72:b}.
The black stars show SF data for neighbouring nuclei $^{259,260}\text{Md}$ ($Z_0=101$, $N_0=158,159$) taken from Refs.\ \cite{wild90:a,svirikhin15:a}.
The figure is taken from Paper VI.
 }
\label{fig:fig_nu_vs_n_fm} 
 \end{center} 
\end{figure}

\begin{figure}[b]
\centering
\includegraphics[width=1.0\linewidth]{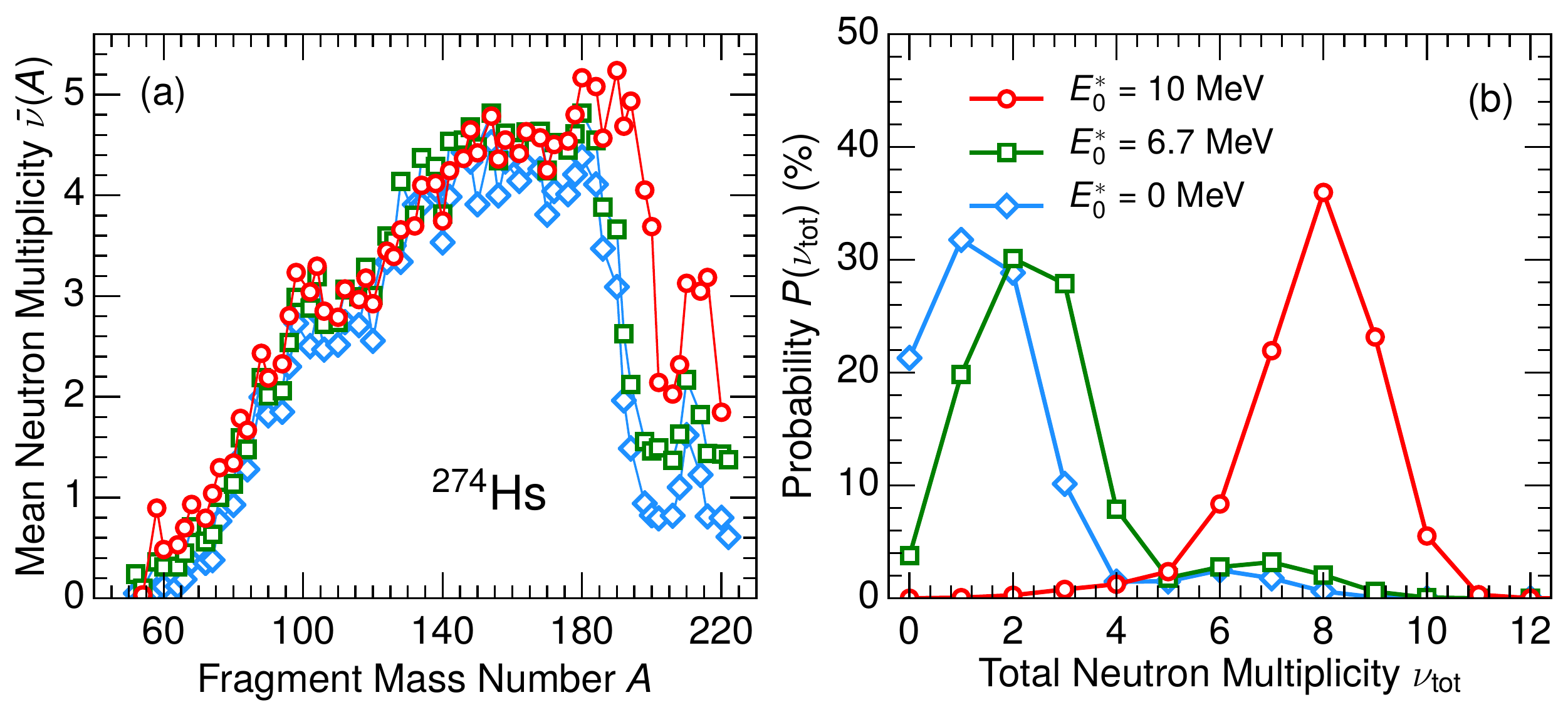}
\caption[$\bar{\nu}(A)$ and $P(\nu_{\rm tot})$ in $^{274}$Hs for $E_0^\ast=0$, 6.7, 10 MeV]{
Calculated mean neutron multiplicity $\bar{\nu}(A)$ (a) and total neutron multiplicity distribution $P(\nu_{\rm tot})$ (b) 
in fission of $^{274}$Hs for initial excitation energies $E_0^\ast=0$ (blue diamonds), 6.7 (green squares), 15 MeV (red circles).
}
\label{fig:fig_nu_both_hs274} 
\end{figure}

The energy dependence of the neutron multiplicity is illustrated in Fig.\ \ref{fig:nu_vs_af_fast} which shows $\bar{\nu}(A)$ 
resulting from fission at three different incident neutron energies.
In the region around $A\approx130$, the low neutron
multiplicity occurring for thermal fission grows rather rapidly with
increasing neutron energy, causing the sawtooth feature of $\bar{\nu}(A)$ to weaken. 
One reason for this behavior is due to the decrease of the strong negative shell correction at higher excitation energy for fragments in
this mass region, increasing the level density and thus the share of the excitation energy taken up by the heavy fragment at scission.
Another reason is the SL mode entering the mass-asymmetric division at higher energies,
resulting in large distortion energies for fragments $A\approx130$,
as seen in Fig.\ \ref{bimodal_235u}.
On the contrary, the neutron multiplicity from the light fragments is affected less by the increase in energy. 

Figure \ref{fig:fig_nu_both_fm260}(a) shows average neutron multiplicity $\bar{\nu}(A)$
as a function of fragment mass number $A$ in fission of $^{260}$Fm for 
excitation energies $E_0^\ast=0$ MeV, 6.13 MeV, and 15 MeV.
For SF, low neutron multiplicity is obtained in the symmetric region because of the SS mode.
When the energy is increased, the largest change is seen in the symmetric region where the decrease of the SS mode
result in an increase in the excitation energy, and thus the number of neutrons emitted.
There are no experimental data available for $^{260}$Fm,
but measurements have been performed for SF of $^{260}_{101}\text{Md}_{159}$ in Ref.\ \cite{wild90:a}.
This nucleus fission primarily in the SS mode
and differ only in one proton and neutron compared to $^{260}_{100}\text{Fm}_{160}$.
The measured values of $\bar{\nu}(A)$ show large similarities with calculated results for $^{260}$Fm(SF) (see Fig.\ 10 in Ref.\ \cite{wild90:a}).

Figure \ref{fig:fig_nu_both_fm260}(b) shows the total neutron multiplicity distribution for $^{260}$Fm for 
excitation energies $E_0^\ast=0$ MeV, 6.13 MeV, and 15 MeV.
The calculated results for $E_0^\ast=0$ MeV also agree well with the SF results in $^{260}_{101}\text{Md}$ (see Fig.\ 5 in Ref.\ \cite{wild90:a}),
where the average neutron multiplicity is measured to 2.58.
It is also obtained in the calculations that there is a substantial probability for both fragments in $^{260}$Fm(SF) to emit no neutrons. 
This is referred to as cold fission since it correspond to fragments with very low excitation energy \cite{gonnenwein94:a}.
The calculated probability of zero neutrons for $^{260}$Fm is 12\%,
while the corresponding measured probability in $^{260}_{101}\text{Md}_{159}$(SF) is about 9\%.

The average total neutron multiplicity $\bar{\nu}_{\rm tot}$ are shown in Fig.\ \ref{fig:fig_nu_vs_n_fm}
for even $^{254-268}$Fm 
versus the neutron number $N_0$ of the fissioning nucleus for three different excitation energies.
Similarly to the calculated TKE values in Fig.\ \ref{fig:fig_tke_vs_n_fm},
it also provides an experimental correspondence to the phase diagram in Fig.\ \ref{fig:phase_diagrams}.
Since the available energy either goes to TKE or excitation energy,
the behaviour of the TKE as a function of $N_0$ in Fig.\ \ref{fig:fig_tke_vs_n_fm} is mirrored in
the neutron multiplicities in Fig.\ \ref{fig:fig_nu_vs_n_fm}.
The SS mode with the compact scission shape of two spherical $^{132}$Sn
yields both low distortion energy as well as low intrinsic excitation energy,
thus also low neutron multiplicity.
Increasing the energy leads to a an increase of the St mode with a higher neutron multiplicity.
The monotonic increase in neutron multiplicity versus $N_0$
comes mainly from the fact that higher neutron number of the fissioning nucleus
results in more neutron-rich fragments with lower neutron-separation energies;
each neutron therefore costs less to emit.

Figure \ref{fig:fig_nu_both_hs274}(a) shows average neutron multiplicity $\bar{\nu}(A)$
as a function of fragment mass number $A$ in $^{274}$Hs for 
excitation energies $E_0^\ast=0$, 6.7, and 10 MeV.
All three curves show similar behaviour with low neutron multiplicities around $A_{\rm H}\approx208$.
Results for the two lower energies, $E_0^\ast=0$ MeV and $E_0^\ast=6.7$,
both exhibit the very asymmetric yield of the SA mode.
The increase in energy from $E_0^\ast=0$ MeV to $E_0^\ast=6.7$
leads to an increase of number of neutrons from fragments $A_{\rm L}\approx66$ and $A_{\rm H}\approx208$ 
of about 0.3 and 0.7 neutrons, respectively.
The total neutron multiplicity thus increases with roughly one neutron,
as is also seen in the probability distribution in \ref{fig:fig_nu_both_hs274}(b).
On the other hand, for $E_0^\ast=10$ MeV the most probable mass-split is symmetric, $A_{\rm L}$:$A_{\rm H}\approx137$:137,
with neutron multiplicities $\nu_{\rm L}\approx\nu_{\rm H}\approx4$.
The increase in energy of only 3.3 MeV thus results in a drastic change in the total neutron multiplicity from about 2.5 to 8.
This is similar to the change of mode in the fermium isotopes.
In this case however, increase in energy leads to a change from asymmetric to symmetric fission.

The general increase in $\bar{\nu}_{\rm tot}$ with mass number $A_0$
comes mainly from the increase of the $Q$ value as the mass of the fissioning nucleus increases.
A simple estimate of the $Q$ value can obtained with the LDM formula in Eq.\ (\ref{eq:semi_empirical_binding}).
The difference in binding energy only involves the surface and Coulomb terms,
and if one assumes a symmetric mass-split, $A_{\rm L}=A_{\rm H}=A_0/2$, the following expression is obtained
\begin{equation}
Q=0.37E_{\rm C}-0.26E_{\rm S}\text{ MeV}, 
\end{equation}
where $E_{\rm C}=0.7103 Z_{\rm 0}^2/A_{\rm 0}^{1/3}$ and $E_{\rm S}=17.80 A_{\rm 0}^{2/3}$ 
are used for the Coulomb energy and surface energy of a spherical shape (in units of MeV), respectively \cite{green54:a}.
If the nucleus is initially excited by the energy $E^\ast_0$, 
the total available energy is $Q^\ast=E^\ast_0+Q$.
Subtracting $Q^\ast$ by the TKE Viola systematics in Eq.\ (\ref{eq:violatke}) gives the following estimate of the TXE in the fragments
\begin{equation}
\text{TXE} = E^\ast_0 + 0.1439\cdot Z_{\rm 0}^2/A_{\rm 0}^{1/3} - 4.628\cdot A_{\rm 0}^{2/3} - 7.3 \text{ MeV}.
\end{equation}
This expression decreases for more neutron-rich nuclei, corresponding to less energy available for neutron emission.
However, a more neutron-rich fissioning nuclues will have more neutron-rich fragments;
each neutron will cost less to emit.
Assuming simply a constant proton-to-mass ratio of $Z_{\rm 0}/A_{\rm 0}=0.39$ (corresponding to $^{236}$U)
and that each neutron costs 8 MeV to emit, 
one obtains the following simple expression for the average total neutron multiplicity
\begin{equation}
\label{eq:nu_systematics}
\bar{\nu}_{\rm tot}(A_{\rm 0},E^\ast_0) = E^\ast_0/8 + 0.00274\cdot A_{\rm 0}^{5/3} - 0.5785\cdot A_{\rm 0}^{2/3} - 0.9125.
\end{equation}

\begin{figure}[t]
\centering
\includegraphics[width=1.0\linewidth]{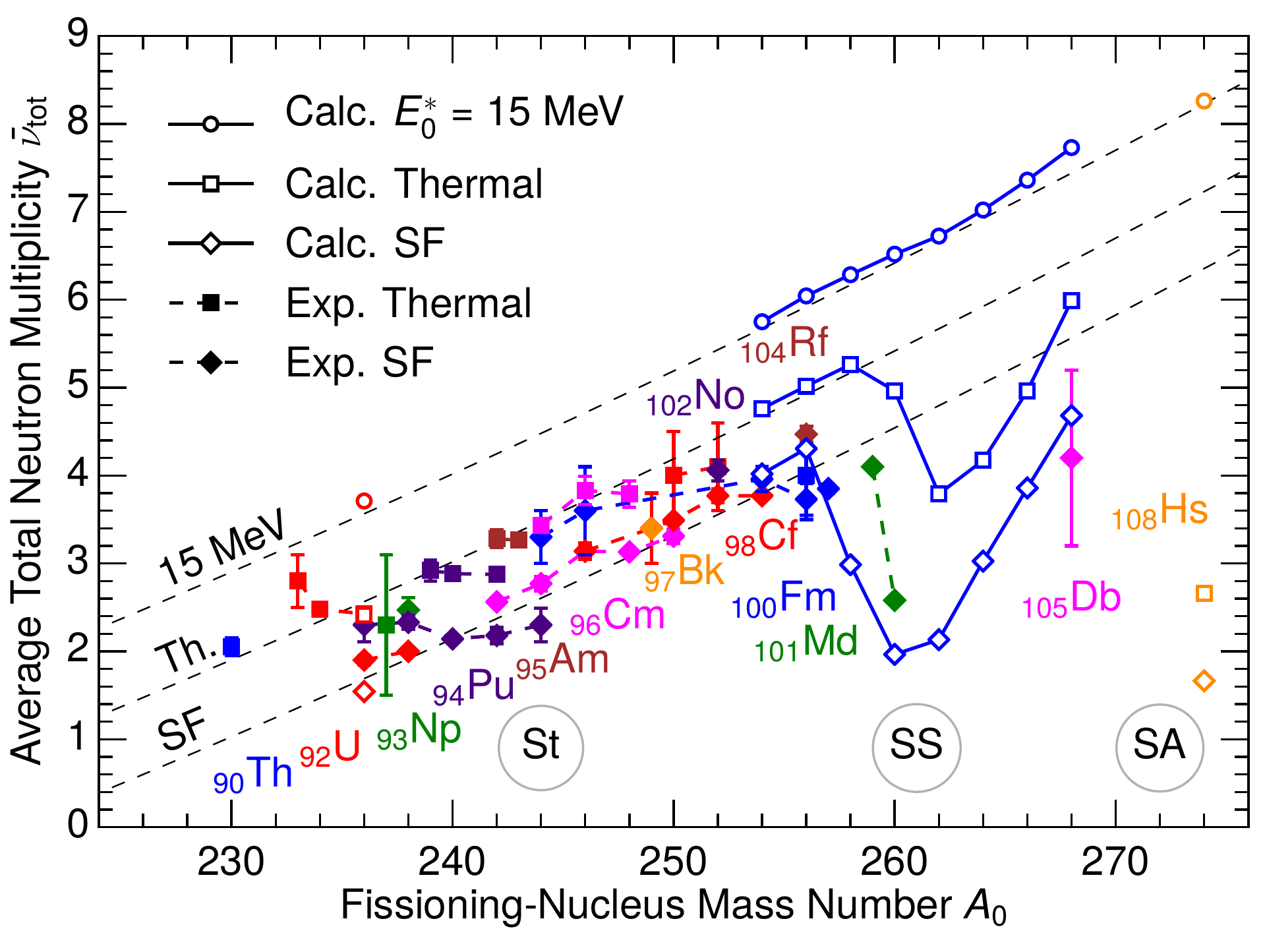}
\caption[Neutron multiplicity systematics]{
Total neutron multiplicity $\bar{\nu}_{\rm tot}$ as function of the mass number $A_0$ of the fissioning nucleus.
Filled symbols correspond to experimental data for SF (diamonds) and thermal fission (squares)
from Refs.\ \cite{condo71:a,thierens80:a,hicks56:a,huanqiao84:a,orth71:a,kroshkin70:a,lazarev77:a,
vorobyev05:a,kosyakov72:a,hoffman80:c,svirikhin10:a,svirikhin12:a,svirikhin15:a,
wild90:a,svirikhin16:a,svirikhin09:a,haddad89:a,jaffey70:a,lindner90:a,flynn75:b,zhuravlev74:a}.
Calculations are shown as open symbols for SF (diamonds),
thermal fission (squares), and $E^\ast_0=15$ MeV (circles)
for fissioning nuclei $^{236}$U, even $^{254-268}$Fm, and $^{274}$Hs.
Black dashed lines show the systematics with Eq.\ (\ref{eq:nu_systematics})
for energies $E^\ast_0=0$, 7 and 15 MeV.
Regions where the St, SS and SA modes are present in calculations of low-energy fission are also indicated.
}
\label{fig:fig_nutot_vs_acn} 
\end{figure}

Fig.\ \ref{fig:fig_nutot_vs_acn} shows average total neutron multiplicity $\bar{\nu}_{\rm tot}$
versus the mass number $A_{\rm 0}$ of the fissioning nucleus,
where filled symbols correspond to experimental data for SF (diamonds)
and thermal fission (squares).
Calculations are shown as open symbols for SF (diamonds),
thermal fission (squares), and $E^\ast_0=15$ MeV (circles)
for fissioning nuclei $^{236}$U, even $^{254-268}$Fm, and $^{274}$Hs.
The three black dashed lines show the systematics with Eq.\ (\ref{eq:nu_systematics})
for energies $E^\ast_0=0$, 7 and 15 MeV.
For SF, both data and calculations are described reasonably well by the $\bar{\nu}_{\rm tot}$-systematics curve
up to $A_0\approx258$, where the appearance of the SS mode result in lower neutron multiplicities. 
Calculations for $^{274}$Hs with low excitation energies also result
in large deviations due to the compact SA mode.
For a fissioning nucleus with an initial excitation energy $E^\ast_0$, 
the extra energy compared to SF generally goes to excitation energy of the fragments.
The neutron multiplicity for thermal fission is then increased by about one neutron,
corresponding to the black dashed curve in the middle.
For $E^\ast_0=15$ MeV, when the influences of both the SS mode and the SA mode have disappeared,
calculations agree well with the systematic behaviour.
This is reasonable since the simple expression was based on the LDM,
which is assumed to be valid at higher energies when shell effects are negligible.

\section{Correlations between neutron multiplicities and TKE}
Since the Metropolis random-walk method is an event-by-event model,
it can describe correlations between various quantities.
One example is the correlation between number of neutrons emitted and the TKE of the fragments.
This was recently measured by G{\"o}{\"o}k et al.\ \cite{gook18:b}
and was studied in Paper V.
A phenomenological deterministic
model of prompt neutron emission was also recently applied to the
same problem, yielding very good agreement with data \cite{tudora19:a}.

\begin{figure}[b] 
 \begin{center} 
 \includegraphics[width=0.6\textwidth]{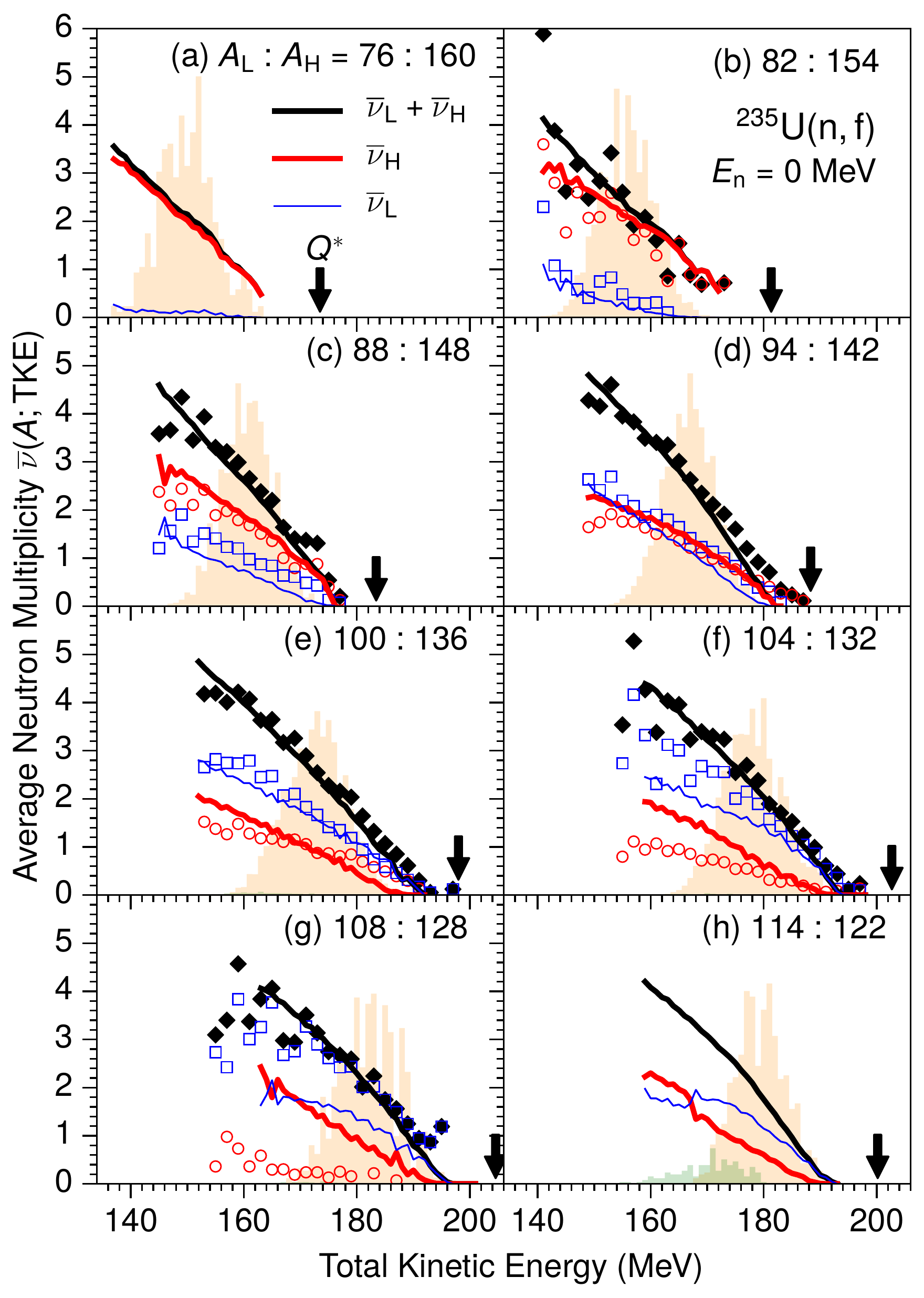} 
 \caption[$\bar{\nu}(A,\text{TKE})$ in $^{235}\text{U}(\text{n},\text{f})$ for $E_{\rm n}=0$ MeV]{
Average multiplicity of neutrons evaporated from the light or heavy fragment in $^{235}\text{U}(\text{n}_{\rm th},\text{f})$ 
as a function of TKE, $\bar{\nu}_{\rm L}(A,\text{TKE})$, and $\bar{\nu}_{\rm H}(A,\text{TKE})$, 
for eight mass divisions. 
Calculated: $\bar{\nu}_{\rm L}$ (thin blue lines), $\bar{\nu}_{\rm H}$ (thick red lines), $\bar{\nu}_{\rm L}+\bar{\nu}_{\rm H}$ (black lines). 
Measured \cite{gook18:b}: $\bar{\nu}_{\rm L}$ (open blue squares), $\bar{\nu}_{\rm H}$ (open red circles), $\bar{\nu}_{\rm L}+\bar{\nu}_{\rm H}$ (black squares). 
The histograms show the calculated TKE distributions, $P(\text{TKE})$, for the St mode (orange) and the SL mode (green). 
The arrows point to the $Q^\ast$ values. 
The figure is taken from Paper V.
   }
\label{fig:nu_vs_tke_nth}  
 \end{center} 
\end{figure}

\begin{figure}[t] 
 \begin{center} 
 \includegraphics[width=0.6\textwidth]{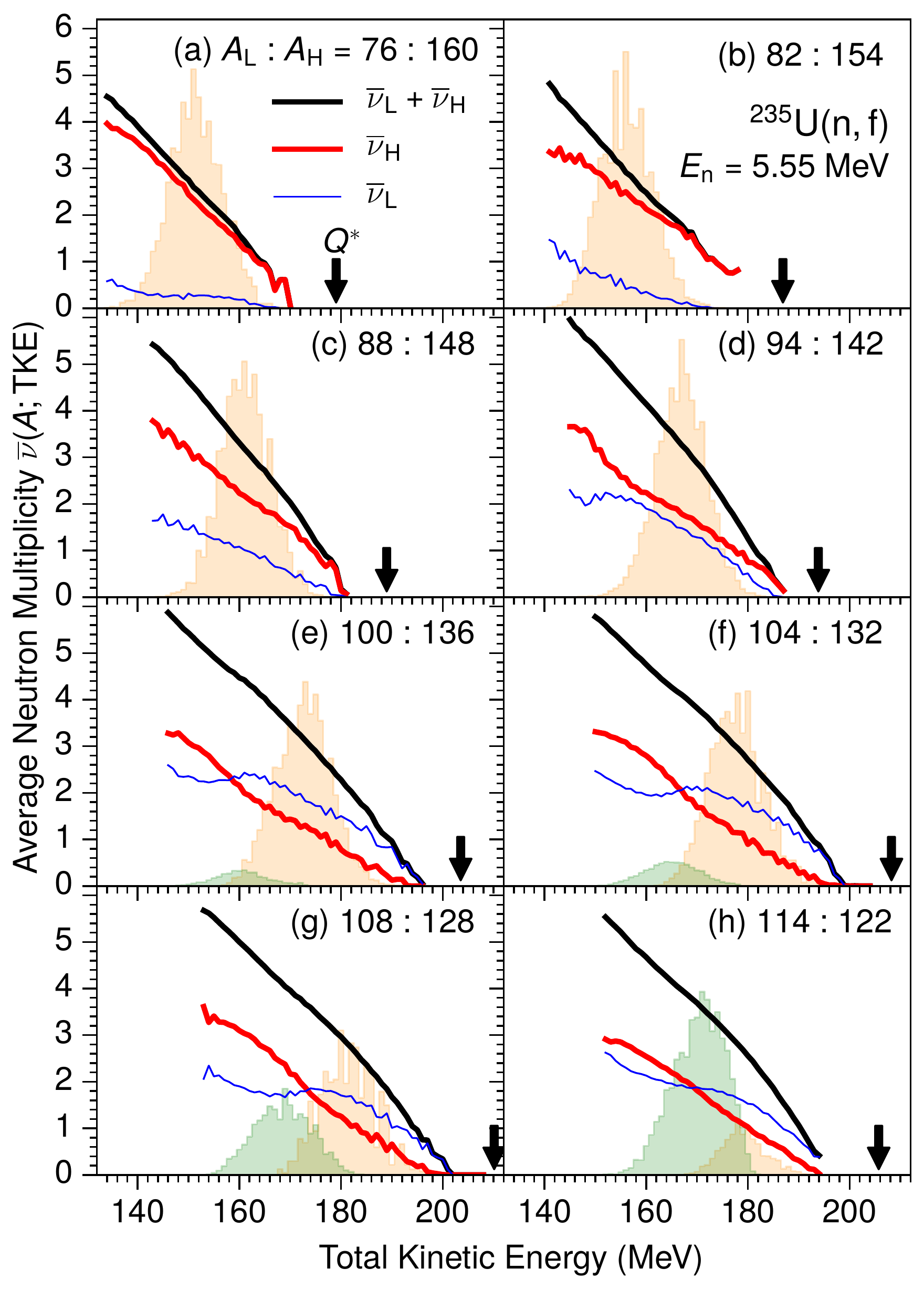} 
 \caption[$\bar{\nu}(A,\text{TKE})$ in $^{235}\text{U}(\text{n},\text{f})$ for $E_{\rm n}=5.55$ MeV]{
Same as Fig.\ \ref{fig:nu_vs_tke_nth} but for $E_{\rm n}=5.55$ MeV.
   The figure is taken from Paper V.
   }
\label{fig:nu_vs_tke_e5}  
 \end{center} 
\end{figure}

Figure \ref{fig:nu_vs_tke_nth} shows the calculated average neutron multiplicity from the light and the heavy fragments for specified TKE,
$\nu(A_{\rm L};\text{TKE})$ and $\nu(A_{\rm H};\text{TKE})$, as well as their sum, 
for eight mass divisions in $^{235}\text{U}(\text{n}_{\rm th},\text{f})$. 
Also shown
are the experimental results \cite{gook18:b}.
The calculated TKE distributions, $P(\text{TKE})$, are also shown for the St mode (orange) and the SL mode (green). 
The calculated multiplicities, $\bar{\nu}_{\rm L}(\text{TKE})$ and $\bar{\nu}_{\rm H}(\text{TKE})$, 
agree well with the measured values for asymmetric divisions, $A_{\rm H}\geq136$. 
For the more symmetric mass splits the calculations substantially overestimate the number of neutrons emitted from the heavy fragment, 
most severely for the mass split 108:128. 
This interval in fragment mass number coincides
with that of the discrepancy found for the average TKE (see discussion in Sec.\ \ref{sec:tke}).
For very asymmetric divisions the heavy fragment receives most of the excitation energy and, 
as a result, it contributes almost all of the neutrons. 
Closer to symmetry however, more neutrons are instead emitted from the light fragment.

Figure \ref{fig:nu_vs_tke_e5} shows the TKE-gated mean neutron multiplicity $\bar{\nu}(A;\text{TKE})$ 
for the same eight mass divisions as in Fig.\ \ref{fig:nu_vs_tke_nth}, but for incident neutron energy $E_{\rm n}=5.55$ MeV. 
When the energy of the incoming neutron increases, the $Q^\ast$-value increases correspondingly by the same amount.
We find that the average TKE changes very little and most of the additional energy goes to TXE. 
The four most asymmetric division, Figs.\ \ref{fig:nu_vs_tke_e5}(a)-(c),
is similar to the thermal result for
both $\bar{\nu}_{\rm L}$ and $\bar{\nu}_{\rm H}$, except for an overall
increase due to the increased excitation of the primary fission fragment.
This smooth evolution with $E_{\rm n}$ may be contrasted with the
behavior for the four least asymmetric divisions in Figs.\ \ref{fig:nu_vs_tke_e5}(d)-(h),
where qualitative changes are apparent. 
It is especially noticeable that $\bar{\nu}(A_{\rm L};\text{TKE})$ 
and $\bar{\nu}(A_{\rm H};\text{TKE})$ cross so the heavy fragment becomes dominant at low TKE, and the light fragment emits most neutrons
at higher TKE. 
This is due to the appearance of the SL mode (green histograms) for more asymmetric mass splits at low TKE,
as seen in Fig.\ \ref{bimodal_235u}.
This prediction has yet to be tested experimentally.

\chapter{Fusion-quasifission dynamics}\label{ch:fusion}

\section{Introduction}\label{sec:intro_fusion}
The heaviest element occurring in Nature is plutonium with 94 protons.
Heavier elements have however been synthesized in various reactions (see e.g. \cite{hofmann02:b,chemey19:a}).
The primary method for producing superheavy elements (SHE) with $Z\geq104$ has been through heavy-ion fusion reactions.
For relatively light colliding nuclei, reaching the Coulomb barrier
means the shape configuration is inside the fission saddle point and
leads automatically to the formation of the compound nucleus. 
However, for heavier nuclei at the Coulomb barrier the
nucleus will find itself outside the fission saddle point, so for colliding energies close to the Coulomb barrier
a compound nucleus is rarely formed \cite{moller76:a,moller03:a}.
The composite nucleus can then diffuse over the inner saddle to the ground state,
but also undergo a re-separation process, 
referred to as quasifission (QF).
This results in a decrease in the production rate of SHE.

The production of a SHE in a fusion reaction therefore proceeds in three steps (see Fig.\ \ref{fig:fusion_schematic}):
(1) contact of the surfaces of the colliding nuclei, 
(2) formation of
the fused nucleus by evolving to a compact shape, in competition with QF, 
(3) survival of the fused nucleus after cooling by neutron evaporation.
The total cross section for the resulting evaporation residue is then given by the product of the three steps.

\begin{figure}[t] 
 \begin{center} 
 \includegraphics[width=0.8\linewidth]{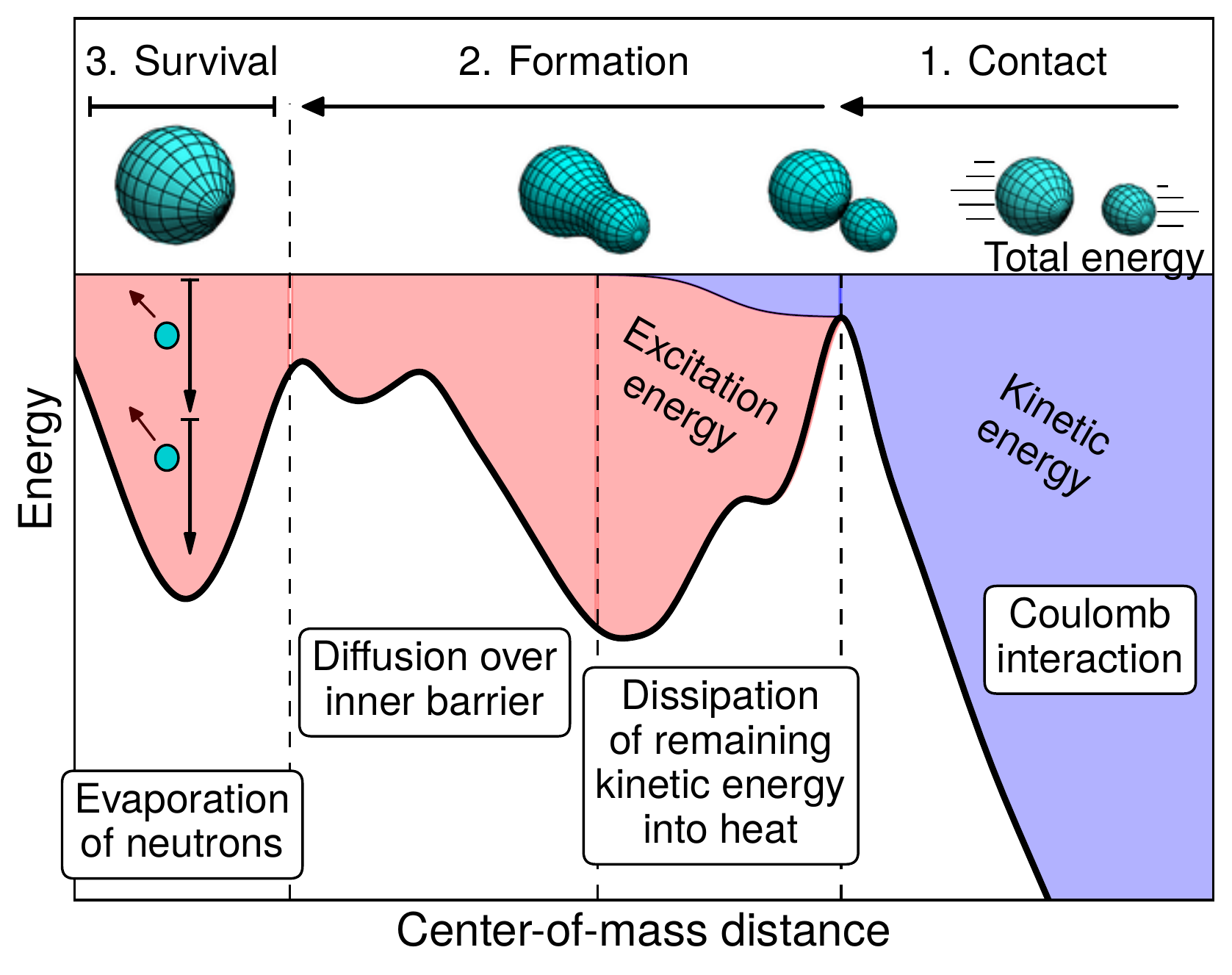}
 \caption[Schematic depiction of the fusion process]{
The three steps in the fusion process: 
(1) contact of the surfaces of the colliding nuclei, 
(2) formation of a compact compound nucleus,
(3) survival of the compound nucleus against fission by neutron evaporation.
 }
\label{fig:fusion_schematic} 
 \end{center} 
\end{figure}

At present, both cold and hot fusion reactions have been used to produce SHE.
In cold fusion with a $^{208}$Pb target, the compound nucleus is formed with relatively low excitation 
energy so that the survival probability of the compound nucleus against fission in step (3) is maximized. 
On the other hand, in hot fusion with a $^{48}$Ca target, the formation probability of the
compound nucleus in step (2) is maximized instead.

In the present studies we only consider the compound-nucleus formation in step (2).
The considered systems correspond to different projectiles on a $^{208}$Pb-target
typical in cold-fusion reactions,
with main focus on the reaction $^{50}\text{Ti}+^{208}$Pb
resulting in compound nucleus $^{258}$Rf.

\section{Method}\label{sec:formation_method}
In the fusion-by-diffusion (FBD) model by Swiatecki et al.\ \cite{swiatecki03:a,swiatecki05:a},
it is assumed that the composite system slips into a fission valley after reaching a contact configuration in step (1).
It can then diffuse over the saddle to form a compound nucleus,
which is described by solving the one-dimensional Smoluchowski equation.
It accounts for experimental cross sections reasonably well,
but it does not give any information regarding the shape evolution or the events leading to QF.
As discussed in Ch.\ \ref{ch:fission_dynamics}, the Smoluchowski equation emerges from the general Langevin description in the highly dissipative limit 
and can be approximately simulated by means of a random walk on the associated
multidimensional potential-energy surface $U(\boldsymbol{\chi})$.

\begin{figure}[t] 
 \begin{center} 
 \includegraphics[width=0.9\linewidth]{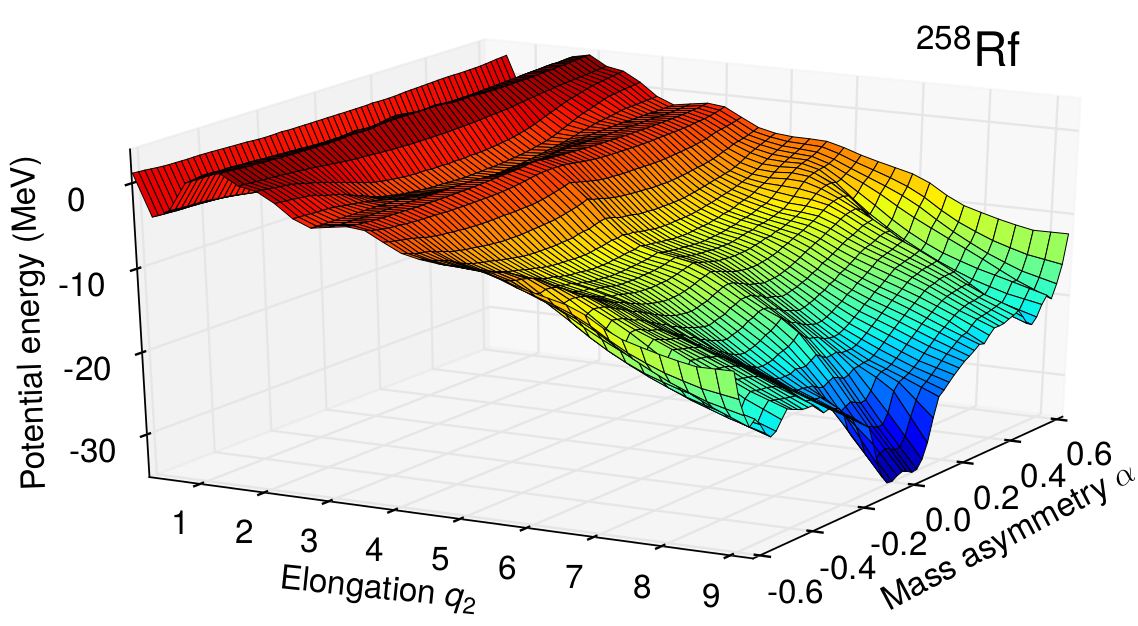} 
 \caption[Potential energy vs.\ $(\alpha,q_2)$ for $^{258}$Rf]{
 Potential energy for compound nucleus $^{258}$Rf projected on the mass asymmetry $\alpha$
 and elongation $q_2$ obtained by minimization with respect to the other shape coordinates.
 A valley in $q_2$-direction is seen around the mass asymmetry $\alpha=0.6$, 
corresponding to fusion with projectile $^{50}$Ti and target $^{208}$Pb.
 }
\label{fig:fig38_rf258} 
 \end{center} 
\end{figure}

\begin{figure}[bt]
\centering
\includegraphics[width=0.8\linewidth]{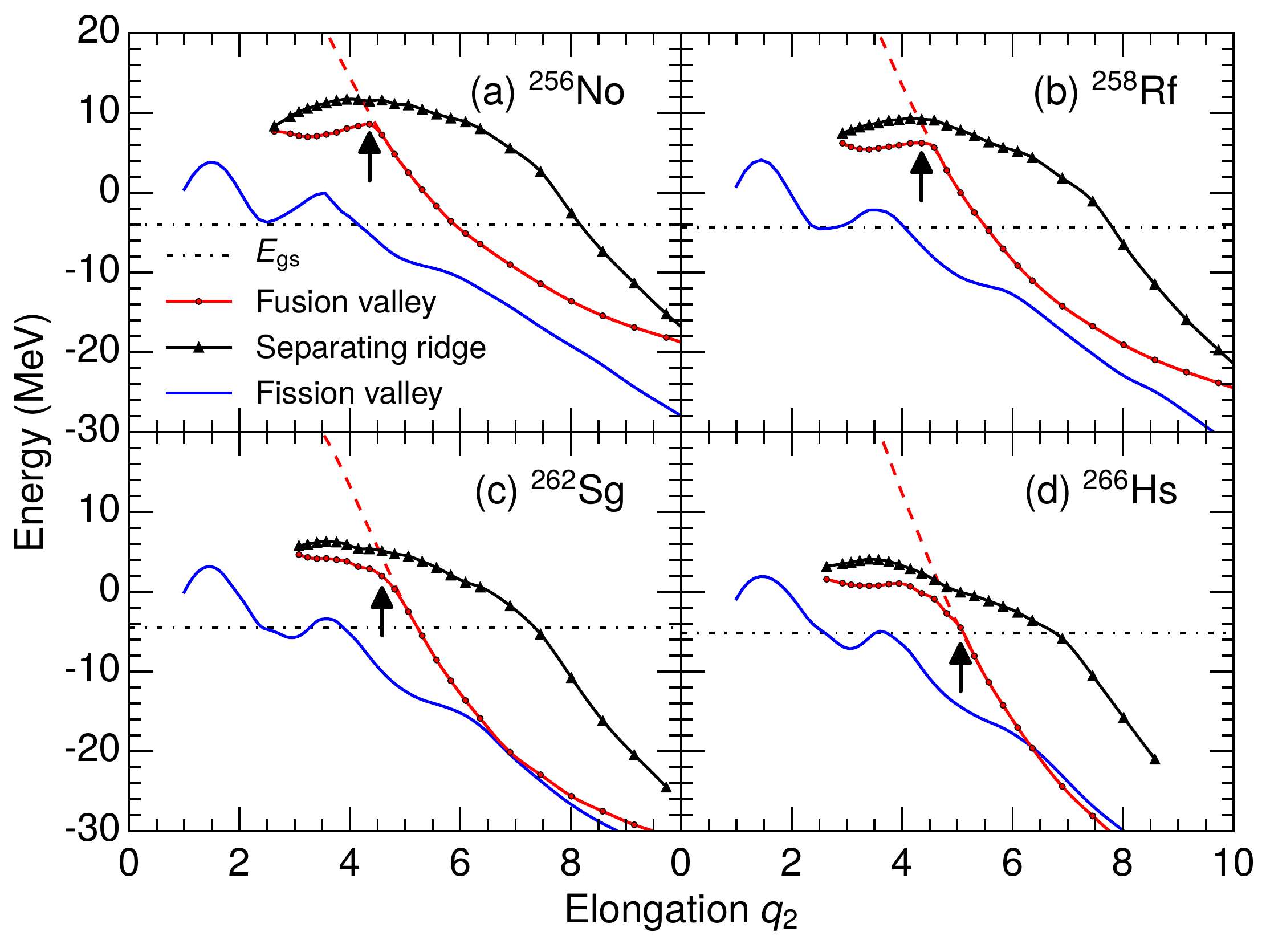}
 \caption[Potential energy vs.\ $q_2$ for $^{256}$No, $^{258}$Rf, $^{262}$Sg, $^{266}$Hs]{
 Potential energy as a function of elongation $q_2$ for compound nuclei (a) $^{256}$No, (b) $^{258}$Rf, (c) $^{262}$Sg, (d) $^{266}$Hs. 
 The blue curves correspond to the minimum potential energy for the mainly symmetric fission valleys,
 while the red curves correspond to the asymmetric valleys.
 The Coulomb energies are shown as red dashed lines.
 The ridge between the two valleys is shown
 as a black curve with triangles.
 The arrows indicate the contact points $q_2^{(\text{cont})}$.
  The figure is taken from Paper VII.
} 
\label{fig:epot1d_fusion} 
\end{figure}

\begin{figure}[t] 
 \begin{center} 
 \includegraphics[width=0.6\linewidth]{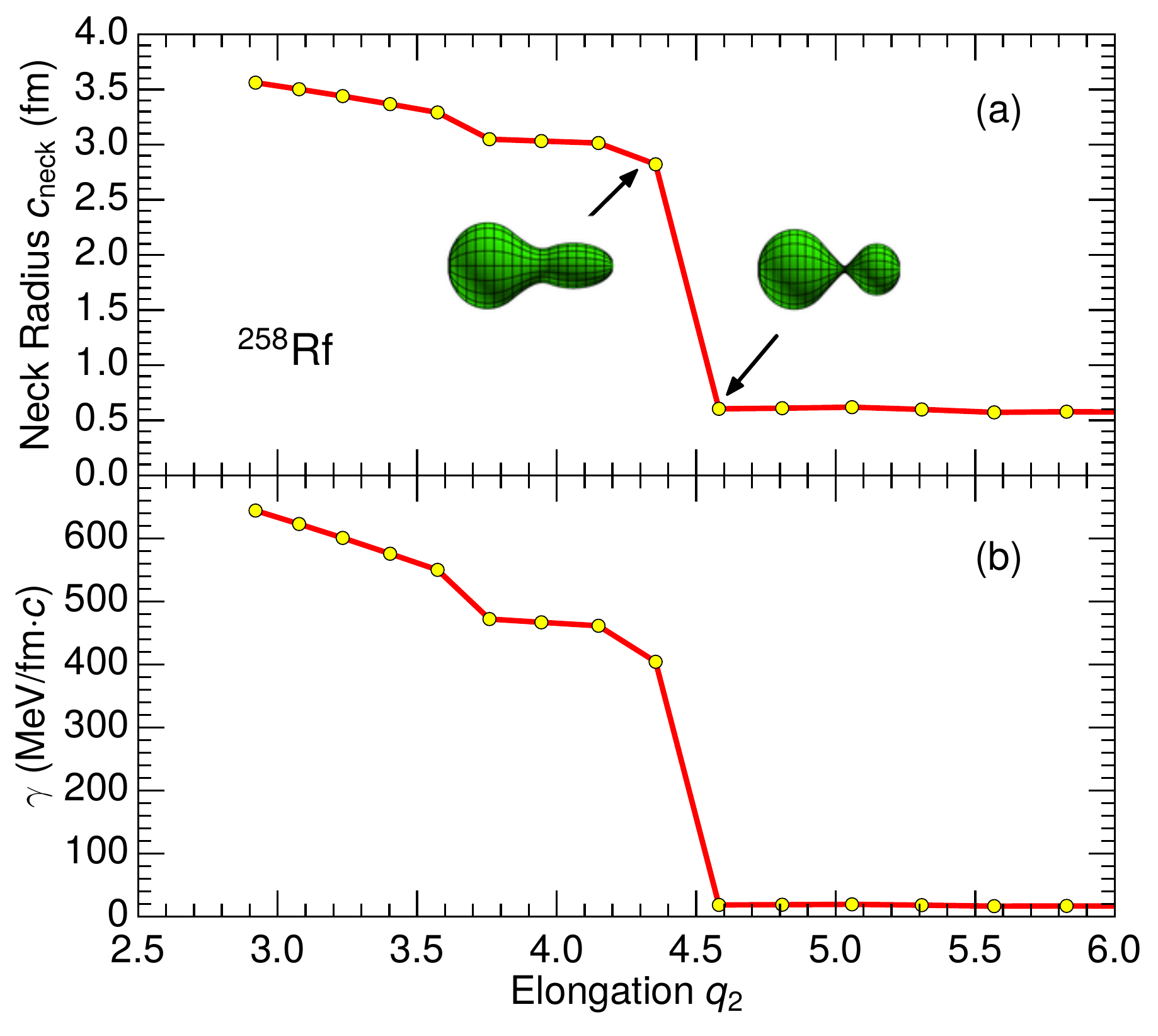} 
  \caption[Neck radius and friction coefficient along fusion valley in $^{258}$Rf]{
 Neck radius $c_{\rm neck}$ (a) and the friction coefficient $\gamma$ (b)
 as a function of elongation $q_2$ along the asymmetric valley for $^{258}$Rf.
Shapes of the system before and after contact are also shown.
 The figure is taken from Paper VII.
 }
\label{fig:friction} 
 \end{center} 
\end{figure}

\begin{figure}[bt]
\centering
\includegraphics[width=0.6\linewidth]{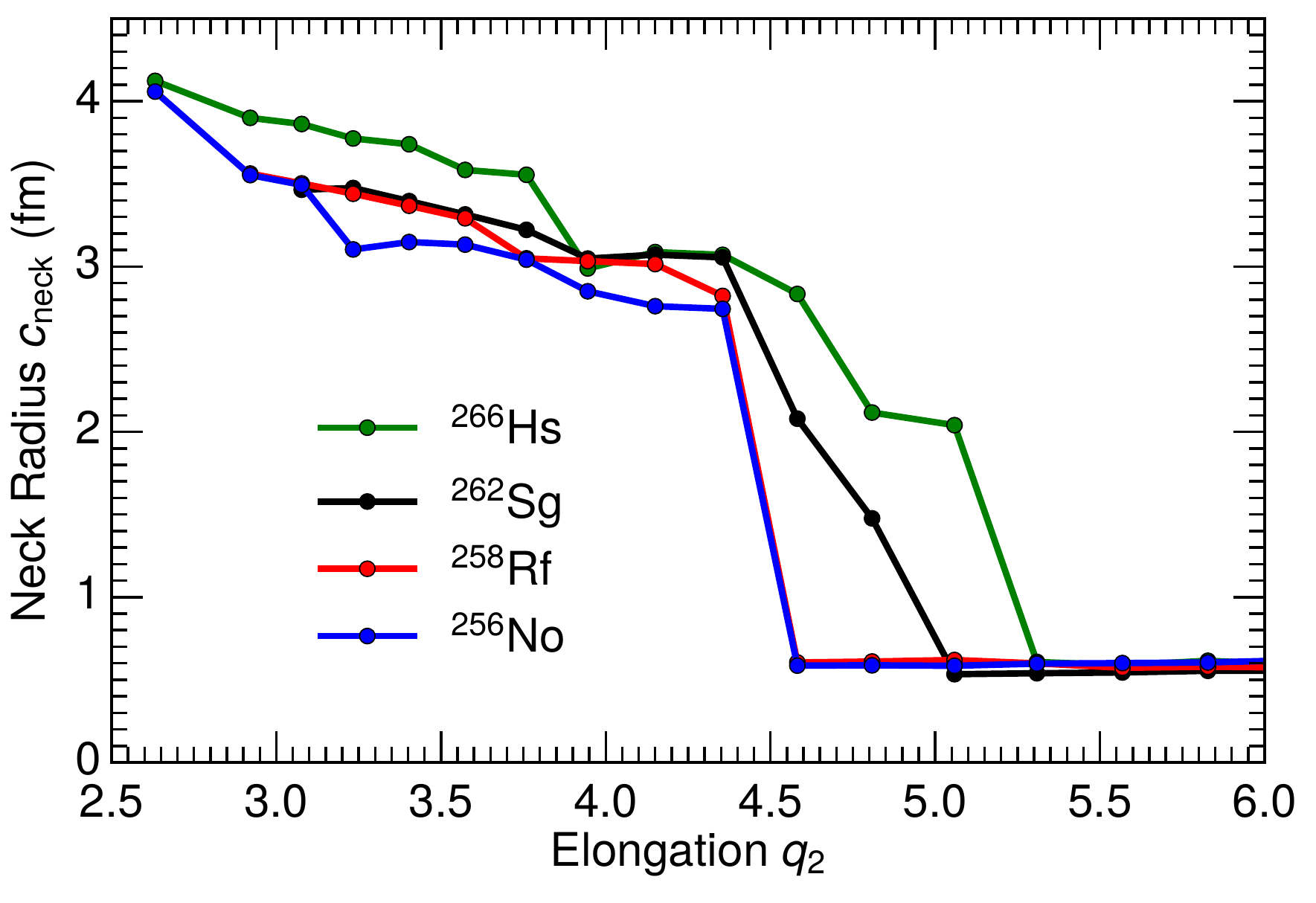}
   \caption[Neck radii along fusion valley for $^{256}$No, $^{258}$Rf, $^{262}$Sg, $^{266}$Hs]{
 Neck radius $c_{\rm neck}$ as a function of elongation $q_2$
 along the asymmetric fusion valley for compound nuclei $^{256}$No, $^{258}$Rf, $^{262}$Sg, and $^{266}$Hs.
 The figure is taken from Paper VII.
} 
\label{fig:neck}
\end{figure}

Figure \ref{fig:fig38_rf258} shows the potential energy projected on the mass asymmetry $\alpha$ and elongation $q_2$ obtained 
by minimization with respect to the other shape coordinates for $^{258}$Rf.
A valley in the potential energy along $q_2$ is seen around the mass asymmetry $\alpha=0.6$, 
corresponding to fusion with projectile $^{50}$Ti and target $^{208}$Pb.

It can however be misleading to simply minimize the potential energy,
since the shape coordinates must also change gradually,
and it is more suitable to analyze the potential energy
using the immersion method discussed in Sec.\ \ref{sec:potsurf_frldm}.
This method is used to obtain the curves in Fig.\ \ref{fig:epot1d_fusion}, which shows potential energy as a function of the quadrupole-moment parameter $q_2$
for compound nuclei $^{256}$No (a), $^{258}$Rf (b), $^{262}$Sg (c), and $^{266}$Hs (d)
corresponding to reactions with a $^{208}$Pb-target 
and projectiles $^{48}$Ca, $^{50}$Ti, $^{54}$Cr, and $^{58}$Fe, respectively.
In addition to the mainly symmetric fission valley, there is an asymmetric valley
which bears a strong resemblance to the target and projectile shapes, thus denoted as fusion valley.
The well separated nuclei only interact via the Coulomb potential,
giving rise to the smooth behaviour of the asymmetric valley for large $q_2$
(though in the 3QS parametrization the target and projectile are always connected with a non-zero neck).
This smooth behaviour is extrapolated to smaller $q_2$ values by fixing the four other coordinates (red dashed line).
However, when the nuclei come close to each other the nuclear interaction will start to have effect,
and it will be more energetically favourable to develop a neck.
This is called the contact point $q_2^{(\text{cont})}$ and is shown as black arrows in Fig.\ \ref{fig:epot1d_fusion}.
 While the $^{208}$Pb-target is near spherical for all $q_2$ values in the asymmetric valley,
 the projectile changes from small deformation to about $\varepsilon_{\rm P}\approx0.4$ inside the contact point.

We assume that the colliding nuclei will slip into the asymmetric valley after overcoming the Coulomb barrier.
The kinetic energy remaining at contact will quickly be dissipated into intrinsic excitation energy
due to strong friction when the neck develops.
The friction force is proportional to the velocity, $\vec{F}_{\rm fric}=-\gamma \vec{v}$,
where the friction coefficient can be estimated with the window friction formula \cite{blocki78:a}
\begin{equation}
\gamma \approx16\pi c_{\rm neck}^2\text{ MeV}/\text{fm}\cdot c,
\end{equation}
where $c_{\rm neck}$ is the neck radius of the compound nucleus.
Figure \ref{fig:friction}(a) shows the value of the neck radius along the asymmetric valley for $^{258}$Rf,
where it is seen how the neck changes from a practically zero neck
to a value around $c_{\rm neck}\approx3$ fm,
corresponding to the contact point $q_2^{(\text{cont})}$.
This rapid increase in neck radius is also in agreement with the ``neck zip'' discussed in the FBD model \cite{swiatecki03:a,swiatecki05:a}.
The nuclear interaction implies a substantial lowering of the energy.
The shapes before and after the contact are also shown in Fig.\ \ref{fig:friction}(a),
where it is seen how the neck zip is associated with a dramatic shape change of the projectile from near spherical
to a very large prolate deformation due to the onset of the short-range nuclear attraction.

The corresponding friction coefficient $\gamma$ along the asymmetric valley is shown in Fig.\ \ref{fig:friction}(b).
It is negligible before contact, whereas it increases to about
$\gamma\approx500\text{ MeV/fm}\cdot c$ after the contact point, due to the neck zip.

Figure \ref{fig:neck} shows the neck radius along the asymmetric fusion valley for all the four nuclei considered.
The systems correspond to different projectiles, whereas the target is $^{208}$Pb in all cases.
The center-of-mass distance for the contact configuration of the target and projectile
therefore increases for reactions with heavier projectiles.
Correspondingly, the contact point where the neck develops corresponds to a larger $q_2$ for heavier systems.

The starting configuration for the diffusion over the inner saddle is expected to begin at a more compact shape when the 
kinetic energy of the colliding nuclei is increased.
In a recent study \cite{boilley19:a} it was estimated that the starting elongation should be proportional to the velocity $\sqrt{E_{\rm cm}-B_{\rm cont}}$,
where $E_{\rm cm}$ the kinetic energy of the colliding nuclei in the center-of-mass frame,
and $B_{\rm cont}$ is the Coulomb barrier for reaching a contact configuration.
This is in accordance with the assumption of a strong, constant friction.
Here, we employ a simple expression for the starting value of $q^{\rm (st)}_2$ based on a similar energy dependence, 
\begin{equation}
\label{eq:start_q2}
q_2^{(\text{st})}(E^\ast_{\rm CN}) = q_2^{\rm (cont)} - d\sqrt{E^\ast_{\rm CN}-E_0},
\end{equation}
where $E_0=U(\boldsymbol{\chi}^{(\text{cont})})-E_{\rm gs}$, and where we adopt the value $d=0.27\text{ MeV}^{-1/2}$. 
The other four shape coordinates are then determined by the asymmetric valley
to form the starting point $\boldsymbol{\chi}^{(\text{st})}$.

From the starting point $\boldsymbol{\chi}^{(\text{st})}$, the system can diffuse across the inner saddle to form a compound nucleus.
The shape changes are selected by the Metropolis method using the effective level density described in Sec.\ \ref{sec:effective_levdens}.
The excitation energy $E_{\rm CN}^\ast$ of the compound nucleus is given by 
\begin{equation}
E_{\rm CN}^\ast = E_{\rm cm} + Q,
\end{equation}
where 
\begin{equation}
Q=(M_{\rm P}+M_{\rm T}-M_{\rm CN})c^2,
\end{equation}
and where $M_{\rm CN}$, $M_{\rm P}$, and $M_{\rm T}$ are the masses of the compound nucleus, the projectile nucleus,
and the target nucleus, respectively.
The masses are calculated within the same macroscopic-microscopic model that is used
to obtain the potential-energy surfaces.

The walks are stopped and binned as a fusion event if it evolves inside the inner saddle point
(located at $q_2\approx1.5$, see Fig.\ \ref{fig:epot1d_fusion}).
This defines a formation of the compound nucleus in our model.
The walks can also lead to QF, which in the model is defined when the neck radius becomes smaller than $c_{\rm sc}=1.5$ fm,
as in the fission calculations.
Then the walk is stopped and binned as a QF event.
The fusion probability $P_{\rm form}$ is then obtained as the number of fusion events divided by the total number of events, i.e.
\begin{equation}
P_{\rm form}=\frac{\text{Number of fusion events}}{\text{Total number of events}},
\end{equation}
where the total number of events correspond to the $10^5$ walks performed.

The QF mass yield $Y_{\rm QF}(A)$ is defined as the percentage of QF events resulting in
fragment-mass number $A$. The yield is normalized to 200\% because each QF event results in two fragments.
The proton and neutron numbers, $Z$ and $N$, are determined by requiring the same $Z/N$ ratio as for the compound nucleus.
In the present study only fragments with even $Z$ and $N$ are considered.
Calculation of the total kinetic energy ($\text{TKE}_{\rm QF}$) of QF fragments is performed as specified for fission in Sec.\ \ref{sec:energies_fission}.
In this case however, the total available energy is given by the initial kinetic energy $E_{\rm cm}$ of the projectile and the target nuclei,
which in QF is shared between the $\text{TKE}_{\rm QF}$ and the total excitation energy ($\text{TXE}_{\rm QF}$) of the QF fragments
\begin{equation}
 E_{\rm cm} = \text{TKE}_{\rm QF} + \text{TXE}_{\rm QF}.
\end{equation}

\begin{figure}[t]
\centering
\includegraphics[width=0.9\linewidth]{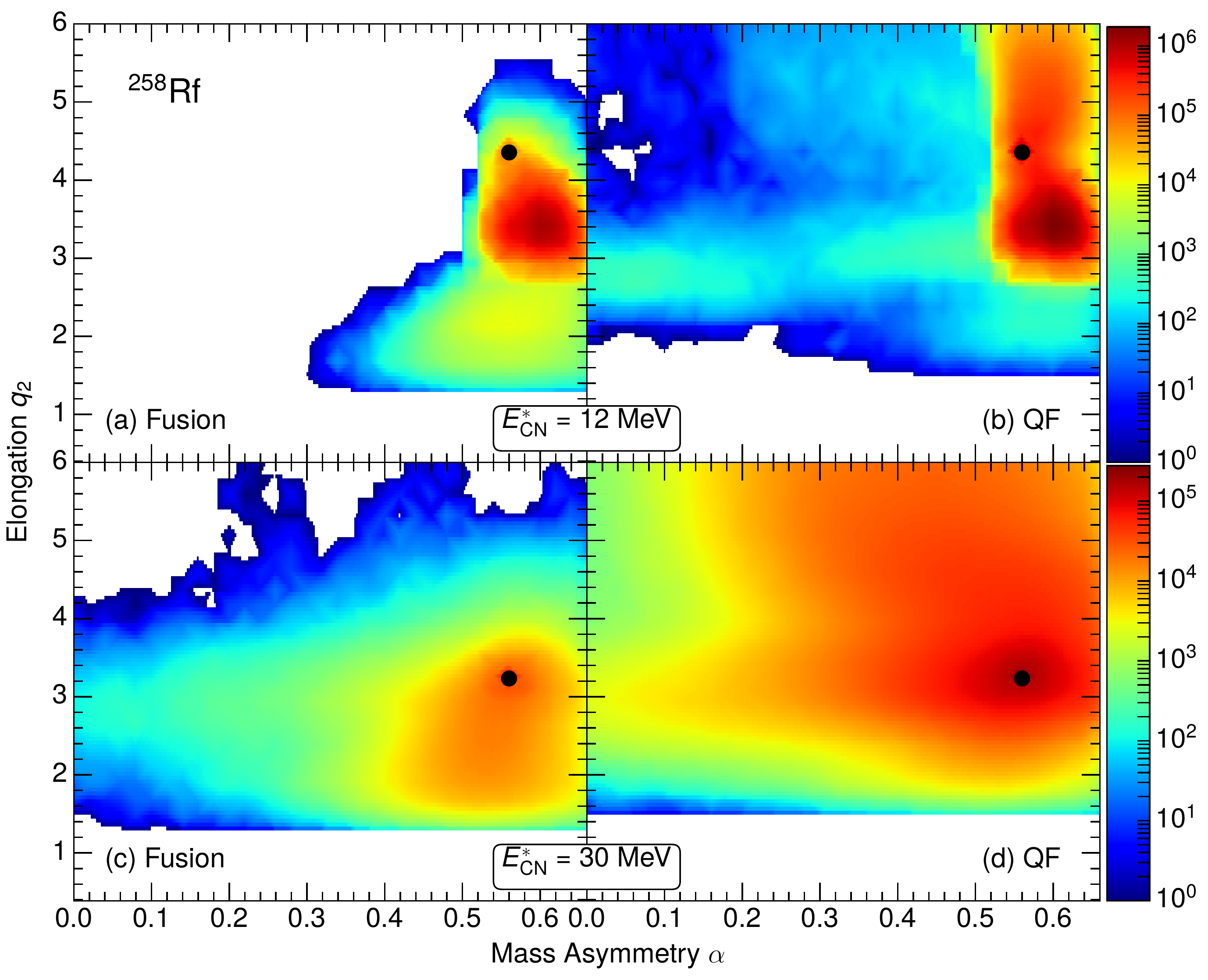}
\caption[Fusion and QF trajectories in $^{258}$Rf for $E^\ast_{\rm CN}=12$ and 30 MeV]{
Number of visits to grid points with a given combination of asymmetry $\alpha$ and elongation $q_2$ for $^{258}$Rf.
For $E^\ast_{\rm CN}=12$ MeV, tracks are shown resulting in: (a) fusion (8 \%) and (b) QF (92 \%).
For $E^\ast_{\rm CN}=30$ MeV, tracks are shown resulting in: (c) fusion (25 \%). (d) QF (75 \%).
Circles mark the starting point.
The inner saddle point is located at $q_2\approx1.5$.
The figure is taken from Paper VII.
}
\label{fig:trajs_both}
\end{figure}

\begin{figure}[b]
\centering
\includegraphics[width=0.6\linewidth]{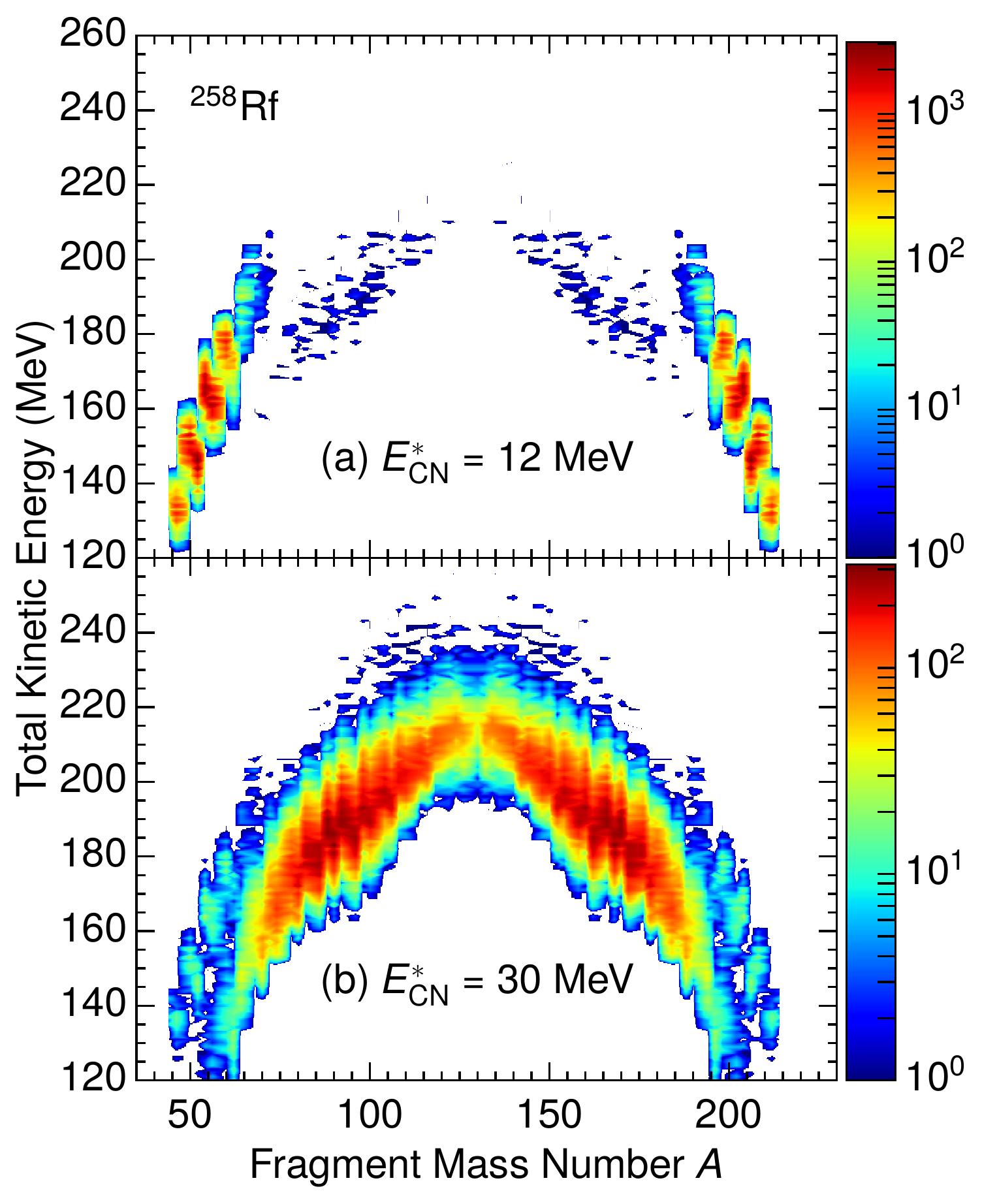}
   \caption[QF events vs.\ $(A,\text{TKE}_{\rm QF})$ in $^{258}$Rf for $E^\ast_{\rm CN}=12$ and 30 MeV]{
Calculated number of QF events vs. $(A,\text{TKE}_{\text{QF}})$ (log-scale) 
in the reaction $^{50}{\rm Ti}+^{208}$Pb$\rightarrow^{258}$Rf
for compound-nucleus excitation energy $E^\ast_{\rm CN}=12$ MeV (a) and $E^\ast_{\rm CN}=30$ MeV (b).
 The figure is taken from Paper VII.
}
\label{fig:tke_a_258rf} 
\end{figure}

\section{Simulations}\label{sec:formation_results}
Figure \ref{fig:trajs_both} shows for $^{258}$Rf with energy $E^\ast_{\rm CN}=12$ MeV,
the total number of visits to sites with a given combination of
asymmetry $\alpha$ and elongation $q_2$ for walks leading to fusion (a) and to QF (b).
The calculated probabilities of fusion and QF for this energy are 8\% and 92\%, respectively.
The starting point for the walks is determined from Eq.\ (\ref{eq:start_q2}) and is marked as a black circle.
For this energy, it is not possible to pass the ridge to the fission valley until $q_2\approx3$.
The majority of the walks leading to QF therefore keep the mass asymmetry as the initial projectile-target system.
A large number of visits are spent around $q_2\approx3.5$
due to a minor local minimum in the asymmetric valley as seen in Fig.\ \ref{fig:epot1d_fusion}(b).
Once the walk passes this minimum, the ridge to the fission valley disappears.
The fusion walks then approach smaller values of the mass asymmetry towards the fission valley as seen in Fig.\ \ref{fig:trajs_both}(a),
until the the inner saddle is crossed ($q_2\approx1.5$).
However, a few walks that pass this point go out again 
corresponding to more symmetric QF mass splits.

Figures \ref{fig:trajs_both}(c) and (d) is similar to Fig.\ \ref{fig:trajs_both}(a) and (b),
but for excitation energy $E^\ast_{\rm CN}=30$ MeV.
The calculated probabilities of fusion and QF in this case are 25\% and 75\%, respectively.
The increase in fusion probability comes primarily from the more compact starting point,
and thus fewer steps are needed to pass the inner saddle.
For this energy it is possible to cross the ridge to the fission valley already at the starting point.
Trajectories leading to fusion, as well as QF, can therefore explore a wider range of shapes.

The formation probabilities of 8\% and 25\% for the energies $E^\ast_{\rm CN}=12$ MeV and 30 MeV
compare reasonably well with the data  for the reaction $^{50}\text{Ti}+^{208}$Pb
presented in Ref.\ \cite{naik07:a}, where the values 2\% and 19\% were reported 
for energies $E^\ast_{\rm CN}=14.2$ MeV and $E^\ast_{\rm CN}=32.7$ MeV, respectively.

Figure \ref{fig:tke_a_258rf} shows contour plots of the number of QF 
events in $^{50}\text{Ti}+^{208}$Pb with respect to fragment mass number $A$ and $\text{TKE}_{\rm QF}$ for
the two energies $E^\ast_{\rm CN}=12$ MeV (a) and $E^\ast_{\rm CN}=30$ MeV (b).
For the lower energy it is seen that the majority of the events 
correspond to mass numbers in a rather small region around projectile and target mass numbers,
whereas the values of $\text{TKE}_{\rm QF}$ varies quite a lot from around 120 MeV to 200 MeV.
The mass peak obtained in the calculations is located at $A_{\rm H}\approx204$ with 
an average value of $\text{TKE}_{\rm QF}\approx166$ MeV.
Measurements for this reaction with energy near the Coulomb barrier
show that the majority of the QF correspond to asymmetric mass-splits $A_{\rm L}$:$A_{\rm H}\approx50$:$208$,
but with an average $\text{TKE}_{\rm QF}$-value of roughly 200 MeV
(Fig.\ 23(a) in Ref.\ \cite{itkis15:b}).
This difference in $\text{TKE}_{\rm QF}$ indicates that the QF fragments should be somewhat more compact than what is obtained in the calculations.
It might be that the strong shell effects of $^{208}$Pb in the SA fission mode
is not as present as for other SHE, e.g. in fission of $^{274}$Hs where higher TKE is obtained in the SA mode (see Fig.\ \ref{fig:tke_hs274}).
Compared to $^{258}$Rf, the $N/Z$ ratio of $^{274}$Hs is somewhat closer to the ratio of $^{208}$Pb.
The $N/Z$ degree of freedom of fragments may then play a role.

As discussed earlier, the amount of symmetric QF events increases when the energy is increased, which is also seen in Fig.\ \ref{fig:tke_a_258rf}(b).
It is in Ref.\ \cite{kozulin14:a} discussed that the average $\text{TKE}_{\rm QF}$ of excited nuclei ($E^\ast_{\rm CN}>40$ MeV)
has a parabolic dependence on the fragment mass and independent of the excitation energy. 
This correspond to when the shell effects have disappeared and is well described by the LDM 
with symmetric Gaussian-like shape mass and energy distributions.
This parabolic behaviour is also similar to the calculated results in Fig.\ \ref{fig:tke_a_258rf}(b).

The obtained results in this first study  compare reasonably well with the data for the reaction with
projectile $^{50}$Ti and target $^{208}$Pb.
It would be interesting to apply the method to other $^{208}$Pb-reactions, as well as to hot-fusion reactions with a $^{40}$Ca-target.
These investigations might give new insights into the dynamics of formation process in fusion,
as well as a way to further test the model for the description of fusion.

\chapter{Outlook}\label{ch:outlook_fission_fusion}
The fission process seems to be reasonably well described 
in the Metropolis random-walk model, which is based on the assumption of overdamped shape evolution.
This assumption is also supported by
recent time-dependent density functional calculations \cite{bulgac16:a,bulgac19:a,bulgac19:b}.

The event-by-event nature of the random-walk model allows for correlation studies,
where correlations between TKE and neutron multiplicity were calculated in the present studies.
Another example of a correlation study would be to investigate the energy spectrum of evaporated neutrons
from fragments with specified mass number and TKE.
For cases with TKE close to the maximum possible values, corresponding to very low excitation energy,
there are few energy levels to decay to in the resulting daughter nucleus,
and microscopic structure effects are more prominent.
While this is possible to study theoretically, this can however be more challenging to perform experimentally.

The results in the symmetric fission region for $^{235}\text{U}(\text{n},\text{f})$ are less accurate in our present calculations because the scission configurations
encountered for near-symmetric divisions are insufficiently elongated and hence lead to too large TKE values. 
This may be due to limitations in the employed 3QS shape parametrization.
Other models of the fission process, such as those employed in Refs.\ \cite{bulgac16:a,bulgac19:a,scamps18:a,dubray08:a} suggest larger distortions of the
fragments than what is obtained in the present treatment. 
In particular octupole deformations of the fragments have been argued to be important \cite{scamps18:a}.

For an excited nucleus it is energetically possible to emit one or more neutrons before fissioning or gamma-deexcite
to the ground state.
After emitting one or more neutrons, it might still be energetically possible for a resulting daughter nucleus to undergo fission.
This is called $2^{\rm nd}$-chance fission and becomes increasingly more probable with excitation energy.
This can be included in the present model.
The competition between fission and neutron emission is in the transition-state theory \cite{swiatecki08:a}
determined by the level density at the fission saddle point and
the level density in the daughter nucleus after neutron emission.
The $2^{\rm nd}$-chance fission probabilities, as well as higher $n$-chance probabilities, can be calculated with the microscopic combinatorial level density method.
Fission-fragment distributions for an arbitrary energy would then be obtained by
performing random-walks for the initial nucleus ($Z_0$, $N_0)$ ($1^{\rm st}$-chance fission) and the
daughter nuclei; ($Z_0$, $N_0-1)$ ($2^{\rm nd}$-chance fission), ($Z_0$, $N_0-2)$ ($3^{\rm rd}$-chance fission) etc.
The total fragment distributions are then obtained by averaging the $n$-chance distributions with the associated probabilities.
Multi-chance fission calculations with the random-walk model was performed in Ref.\ \cite{moller17:a},
using another method for obtaining the probabilities.
The method was shown to yield good agreement with data for charge yields.

Since the random-walk model is based on the classical Langevin formalism,
it cannot describe tunneling phenomena.
The introduced $\Delta E$-method for simulating SF seems to give reasonable results,
except for $^{258}$Fm where a WKB tunneling calculation is needed.
A more proper description would include dynamical aspects of the simulation, which is challenging.

In the present studies, only even-even fragments with the same proton-to-neutron ratio $Z/N$ as the fissioning nucleus is considered.
An extension of the random-walk model including a proton-neutron asymmetry degree of freedom was developed in Ref.\ \cite{moller15:c}.
This allows for descriptions of odd-even staggering seen in fragment charge yields.
It also allows for calculations of isotopic fragment distributions, which was studied in Ref.\ \cite{schmitt21:a}.
Another method for calculating charge yields was introduced in Ref.\ \cite{verriere21:a},
using particle number projection.
Inclusion of the $Z/N$ degree of freedom of fragments in the calculations of TKE and neutron emission might give new insights
into the properties of fission fragments.

About half of the elements heavier than iron is believed to have been produced in a reaction called the $r$-process \cite{cowan21:a},
which is a succession of rapid neutron-captures, eventually ending in fission.
Thus, fragment distributions and the number of emitted neutrons from these fragments
influence calculations of the final abundances.
Since most nuclei relevant in the $r$-process are very exotic and neutron-rich, 
where experimental data is scarce, the estimated abundance distribution remains rather uncertain.
The random-walk model appears to be suitable for such calculations since it has shown predictive power regarding mass yields,
and give reasonable agreement with data for TKE and neutron multiplicities.

The obtained fusion results in this first study compare reasonably well the data for the reaction with 
projectile $^{50}$Ti and target $^{208}$Pb.
However, the high TKE value in QF for the very asymmetric split seen in measurement 
is not obtained in the calculations.
This indicates that the scission configurations are too elongated.
The $N/Z$ degree of freedom of fragments may play a role for this.
It would also be interesting to apply the method to other
$^{208}$Pb-reactions, as well as to hot-fusion reactions with a $^{40}$Ca-target.
These investigations might give new insights into the dynamics of formation process in fusion,
as well as a way to further test the model for the description of fusion.
The survival probability in the last step in the fusion process corresponds to evaporation of neutrons in competition with fission.
These types of calculations are identical with those regarding multi-chance fission,
and could similarly be calculated with the microscopic combinatorial level densities.

\backmatter
\renewcommand{\bibsection}{\chapter{References}}
\bibliographystyle{unsrt}
\bibliography{refer-crank}  

\end{document}